\pdfoutput=1
\documentclass[a5paper,twoside,openright]{memoir_yann}
\usepackage[a5paper,textwidth=341pt,textheight=510pt,layoutvoffset=15pt]{geometry}

\usepackage{these}
\usepackage{xparse}
\usepackage[utf8x]{inputenc}
\usepackage[T1]{fontenc}
\usepackage[english]{babel}
\usepackage{graphicx}
\usepackage{amsmath}
\usepackage{amsfonts}
\usepackage{stmaryrd} 
\usepackage[usenames,dvipsnames,svgnames,table]{xcolor}
\usepackage{tikz}
\usepackage{tikz-3dplot}
\usetikzlibrary{calc,matrix,arrows,positioning,shapes.geometric}
\usepackage{braket}
\usepackage{nth}
\usepackage{blkarray}
\usepackage{rotating}
\usepackage{amssymb}
\usepackage{graphics}
\usepackage{placeins} 
\usepackage[bookmarks=true,colorlinks=true,linkcolor=blue,backref=page]{hyperref} 
\renewcommand*{\backref}[1]{}
\renewcommand*{\backrefalt}[4]{%
    \ifcase #1 (\textbf{ Not cited}.)%
    \or        (\textbf{Cited on page}~#2.)%
    \else      (\textbf{Cited on pages}~#2.)%
    \fi}
\usepackage[square,numbers]{natbib}
\usepackage{pdfpages}

\usepackage{verbatim}
\usepackage{boxedalign}
\usepackage{empheq}      
\usepackage{etaremune}
\usepackage{gensymb}  
\usepackage{listings}
\usepackage{epstopdf}
\usepackage{enumitem}
\newsubfloat{figure}
\usepackage{subfig}
\usepackage{lettrine}
\usepackage{calligra}

\usepackage{green}
\usepackage{operator}
\usepackage{fancybox}
\usepackage{braket}

 \usepackage[framemethod=tikz]{mdframed}
\newcounter{demo_counter}
\newcounter{remark_counter}

\newcommand\thetheorem{Demonstration ~\arabic{demo_counter}}
\newcommand\theremark{Remark ~\arabic{remark_counter}}

\makeatletter
\mdf@dolist{\mdf@do@stringoption}{%
  {theoremtitle=={}}%
}

\mdf@dolist{\mdf@do@stringoption}{%
  {remarktitle=={}}%
}        

\newrobustcmd\mdfcreatetheoremextratikz{%
  \node[anchor=west,rounded corners,draw=orange,thick,fill=blue!20,xshift=1.4cm,minimum height=.7cm,minimum width=1.4cm] at (P-|O) 
      {~\mdf@frametitlefont{\thetheorem}%
      \ifdefempty{\mdf@theoremtitle}%
      {~}%
      {:~\mdf@theoremtitle~}%
      };
}

\newrobustcmd\mdfcreateremarkextratikz{%
  \node[anchor=west,rounded corners,draw=green,thick,fill=blue!20,xshift=1.4cm,minimum height=.7cm,minimum width=1.4cm] at (P-|O) 
      {~\mdf@frametitlefont{\theremark}%
      \ifdefempty{\mdf@remarktitle}%
      {~}%
      {:~\mdf@remarktitle~}%
      };
}

\makeatother

\newcommand{\doextratikz}[1]{%
  \letcs\mdfcreateextratikz{mdfcreate#1extratikz}%
}

\mdfdefinestyle{theoremstyle}{%
  outerlinewidth=1pt,
  innerlinewidth=3pt,
  roundcorner=5pt,
  linecolor=orange,
  splittopskip=1cm,         
  splitbottomskip=0.5cm,      
  tikzsetting={fill=blue!5},
  innertopmargin=1.2\baselineskip,
  skipabove={\dimexpr0.5\baselineskip+\topskip\relax},
  needspace=3\baselineskip,
  frametitlefont=\sffamily\bfseries,
  settings={\stepcounter{demo_counter}\doextratikz{theorem}},
  }

\mdfdefinestyle{remarkstyle}{%
  outerlinewidth=1pt,
  innerlinewidth=3pt,
  roundcorner=5pt,
  linecolor=green,
  splittopskip=1cm,         
  splitbottomskip=0.5cm,     
  tikzsetting={fill=blue!5},
  innertopmargin=1.2\baselineskip,
  skipabove={\dimexpr0.35\baselineskip+\topskip\relax},
  needspace=2\baselineskip,
  frametitlefont=\sffamily\bfseries,
  settings={\stepcounter{remark_counter}\doextratikz{remark}},
  }

\newenvironment{demo}[1][]
  {\begin{mdframed}[style=theoremstyle,theoremtitle={#1}]
  \relax}{
  \end{mdframed}
  }

\newenvironment{remark}[1][]
  {\begin{mdframed}[style=remarkstyle,remarktitle={#1}]
  \relax}{
  \end{mdframed}
  }

\usepackage{color,calc,graphicx,soul,fourier}
\definecolor{nicered}{rgb}{.647,.129,.149}
\definecolor{ChapBlue}{rgb}{0.00,0.65,0.65}

\makeatletter
\newlength\dlf@normtxtw
\newsavebox{\feline@chapter}
\newcommand\feline@chapter@marker[1][2cm]{%
\sbox\feline@chapter{%
\resizebox{!}{#1}{\fboxsep=1pt%
\colorbox{ChapBlue}{\color{white}\bfseries\sffamily\thechapter}%
}}%
\rotatebox{90}{%
\resizebox{%
\heightof{\usebox{\feline@chapter}}+\depthof{\usebox{\feline@chapter}}}%
{!}{\scshape\so\@chapapp}}\quad%
\raisebox{\depthof{\usebox{\feline@chapter}}}{\usebox{\feline@chapter}}%
}
\newcommand\feline@chm[1][2cm]{%
\sbox\feline@chapter{\feline@chapter@marker[#1]}%
\makebox[0pt][l]{
\makebox[1.7cm][r]{\usebox\feline@chapter}%
}}
\makechapterstyle{daleif1_yann}{
\renewcommand\chapnamefont{\normalfont\Large\scshape\raggedleft\so}
\renewcommand\chaptitlefont{\normalfont\Huge\itshape\color{ChapBlue}}
\renewcommand\chapternamenum{}
\renewcommand\printchaptername{}
\renewcommand\printchapternum{\null\hfill\feline@chm[2.5cm]\par}
\renewcommand\afterchapternum{\par\vskip\midchapskip}
\renewcommand\printchaptertitle[1]{\chaptitlefont\raggedleft ##1\par}
\renewcommand\thechapter {\Roman{chapter}}}
\makeatother
\chapterstyle{daleif1_yann}
\renewcommand{\thesubsubsection}{\thesubsection.\roman{subsubsection}}

\nouppercaseheads
\makepagestyle{mystyle} 
\makerunningwidth{mystyle}{\textwidth}
\makeevenhead{mystyle}{\itshape\thepage}{}{\itshape\leftmark}
\makeoddhead{mystyle}{\itshape\rightmark}{}{\itshape\thepage} 
\makeevenfoot{mystyle}{}{}{} 
\makeoddfoot{mystyle}{}{}{} 
\makeatletter
\makepsmarks{mystyle}{%
  \createmark{chapter}{left}{shownumber}{\@chapapp\ }{.\ }}
\makeatother
\makeheadrule{mystyle}{\textwidth}{0.3pt}

\def \identite{\mbox{l\hspace{-0.50em}1}}
\newcommand{\D}{\mathrm{d}}
\newcommand{\e}{\mathrm{e}}
\newcommand{\mathsym}[1]{{}}
\newcommand{\unicode}[1]{{}}

\newcommand{\pushright}[1]{\ifmeasuring@#1\else\omit\hfill$\displaystyle#1$\fi\ignorespaces} 
\newcommand{\Tr}{\,\mathrm{Tr}}
\newcommand*{\jhat}[1]{#1\kern-0.35em\hat{\phantom{#1}}}
\renewcommand{\Im}{\,\mathrm{Im}}
\renewcommand{\Re}{\,\mathrm{Re}}
\newcommand{\insertfigure}[3]{\begin{figure}[!hbtp]
				\begin{center}{
				  \includegraphics[width=0.97\linewidth]{#1}}
				  \caption{#2}
				  {#3}
				\end{center}
			      \end{figure}
			      }
\newcommand{\cercle}[1]{\tikz\node[circle,draw,thick,inner sep=1pt,outer sep=1pt]{#1};}

\addto\captionsenglish{\renewcommand*\abstractname{Summary}}

\NewDocumentEnvironment{important}{O{black}O{}m}{%
  \setkeys{EmphEqEnv}{#3}%
  \setkeys{EmphEqOpt}{box=\color{red}\fbox,#2}%
  \color{#1}%
  \EmphEqMainEnv}%
  {\endEmphEqMainEnv}

\def \kpara{\vec{k}_{\sslash}}

\renewcommand{\i}{\mathrm{i}}

\makeatletter
\newcounter {subsubsubsection}[subsubsection]
\renewcommand\thesubsubsubsection{\thesubsubsection .\@alph\c@subsubsubsection}
\newcommand\subsubsubsection{\@startsection{subsubsubsection}{4}{\z@}%
                                     {-3.25ex\@plus -1ex \@minus -.2ex}%
                                     {1.5ex \@plus .2ex}%
                                     {\normalfont\normalsize\bfseries}}
\renewcommand\paragraph{\@startsection{paragraph}{5}{\z@}%
                                    {3.25ex \@plus1ex \@minus.2ex}%
                                    {-1em}%
                                    {\normalfont\normalsize\bfseries}}
\renewcommand\subparagraph{\@startsection{subparagraph}{6}{\parindent}%
                                       {3.25ex \@plus1ex \@minus .2ex}%
                                       {-1em}%
                                      {\normalfont\normalsize\bfseries}}
\newcommand*\l@subsubsubsection{\@dottedtocline{4}{10.0em}{4.1em}}
\renewcommand*\l@paragraph{\@dottedtocline{5}{10em}{5em}}
\renewcommand*\l@subparagraph{\@dottedtocline{6}{12em}{6em}}
\newcommand*{\subsubsubsectionmark}[1]{}
\makeatother

\usepackage{hyperref}

\makeatletter
\def\toclevel@subsubsubsection{4}
\def\toclevel@paragraph{5}
\def\toclevel@subparagraph{6}
\makeatother

\begin{document}
\titre{Modeling of Ballistic Electron Emission Microscopy on metal thin films}


\datesout{30 octobre 2014}
\Auteur{Yann}{Claveau}

\Labo{6251 - IPR}
\LaboEtendu{Institut de Physique de Rennes - Département Matériaux et Nanosciences}
\ComposanteUniversitaire{UFR Sciences et Propriétés de la Matière} 

\Rapporteur{Chris}{Ewels}{CR1 CNRS à l'Institut des Matériaux de Nantes}
\Rapporteur{Gian Marco}{Rignanese}{Professeur à l'Université catholique de Louvain}
\Examinateur{Pedro L.}{De Andres}{Directeur de recherche au CSIC-ICMM de Madrid}
\Examinateur[M.]{Fernando}{Flores}{Professeur à Universidad Autonoma de Madrid}         
\Examinateur{Xavier }{Rocquefelte}{Professeur à l'université de Rennes 1}
\Advisor{Sergio}{Di Matteo}{Professeur à l'université de Rennes 1}


\pagenumbering{roman}
%
\includepdf{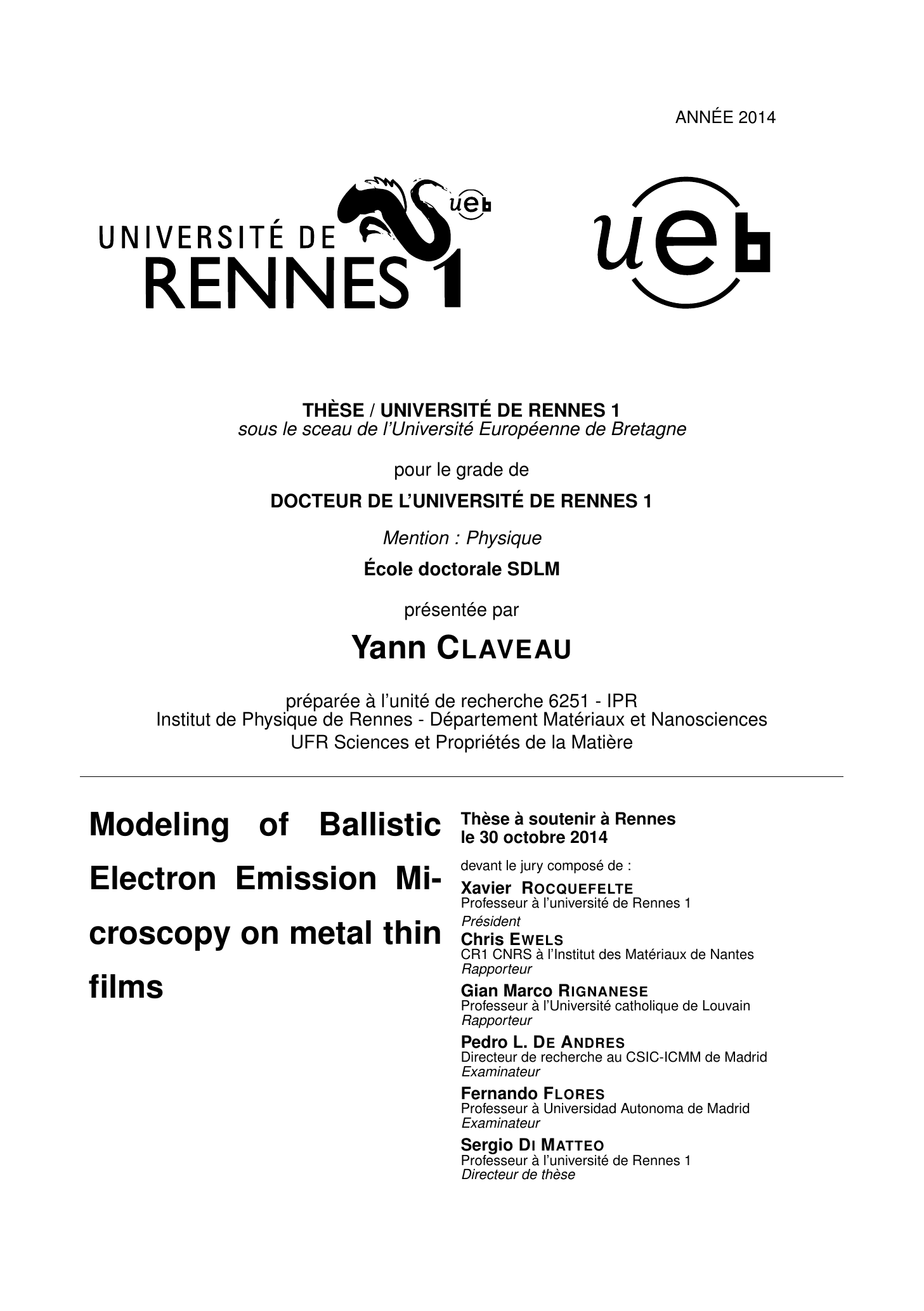}

\newpage \thispagestyle{empty}
\strut
\newpage\thispagestyle{empty}
\strut
\newpage\thispagestyle{empty}

\chapter*{Acknowledgements}
\lettrine[lines=3, lhang=0.35, loversize=0.77, findent=4em,nindent=-1.2em,slope=-1.8em]{H}{ere is the place where I should thank anyone} that has contri-buted in a way or another to this work. As most of them speak french, I shall do it in french. 

Tout d'abord, je voudrais commencer par remercier deux personnes qui ont égayé d'une manière fort agréable un certain vendredi matin 3 octobre : Chris Ewels et Gian-Marco Rignanese. Merci d'avoir accepté de rapporter ce travail et merci pour les rapports du manuscrit. C'est toujours très agréable de lire que l'on a fait du bon travail avant de commencer sa journée. Merci également pour vos critiques parfaitement justifiées dont j'ai essayé de tenir compte au mieux pour cette version finale de ma thèse. Plus particulièrement, je voudrais également remercier Gian-Marco Rignanese pour avoir montré de l'intérêt pour mes travaux lors de notre première rencontre lors d'une summer school au Québec, ainsi que pour son invitation à Louvain.

Merci également à Xavier Rocquefelte d'avoir accepté de juger mon travail à la dernière minute. Et merci pour ses encouragements et son enthousiasme. 

Il me faut également tout particulièrement remercier Fernando Flores et Pedro de Andres. Fernando, gracias por el entusiasmo que has mostrado en tu venida en Rennes para mi proyecto.  Pedro, también gracias por dar me expliqué el funcionamiento de BEEM v2.1. Y finalmente gracias a ambos por vuestra hospitalidad durante mi visita, paciencia y solicitud. También me gustaría dar las gracias a José Ortega y José Ignacio Martínez para el cálculo de las matrices de salto a través Fireball. Ahora voy a dejar de mauling ese idioma.

Tant que nous en sommes aux membres de jury, je remercie également Jean-marc Jancu, Karine Costuas et Jean-Pierre Landesman, directeur de l'IPR, d'avoir accepté d'assister à ma soutenance à mi-parcours. Karine, quand tu veux pour une prochaine aventure des ``théoriciens qui manipent au synchrotron'' ou pour un GdR pour se ``moquer'' de certaines personnes se payant des carreaux de carrelage à 500€ pièce et qui prennent bien soin de le dire à tout le monde. Jean-Pierre, merci pour l'accueil à l'IPR ainsi que pour l'intérêt que tu portes aux non-permanents.

Vient maintenant le tour de mon directeur de thèse\dots Il y a certains moment durant lesquels je suis fier de moi. Le jour où je t'ai demandé si tu pouvais me proposer une thèse en est un. J'ai réellement eu une bonne intuition ce jour là et eu la chance d'avoir un encadrant hors du commun et avec une connaissance de la physique (quantique entre autre) rare et ``non-orthodoxe''.  Sergio, merci pour tout. Je n'aurai pas assez de place ici pour te rendre tous les honneurs que tu mérites mais en résumé : merci pour tout ce que tu m'as appris scientifiquement, sur mon sujet et sur toutes les autres domaines de la physique, merci pour ton soutien en toutes circonstances, tes conseils sur l'enseignement, mais aussi sur la gastronomie italienne, merci d'avoir réussi à dégager du temps sur ton emploi du temps surchargé lorsque j'en avais besoin. Avec Mamaself ce n'étais pas toujours simple. Au passage, merci Christiane et Andrea pour la bonne humeur que vous apportiez à chacun de vos nombreux passages. Attention toutefois à ne pas éffaroucher l'étudiant japonais de Didier qui a dû prendre ma place. Ou peut être est-ce Andrea ? Dans ce cas, Christiane, n'hésite pas. En résumé, Sergio, merci pour tout ça qui conjugué à tes qualités humaines qui font de toi le directeur de thèse idéal. Mais\dots car oui, il y a un ``mais'', je crois qu'il y a quelqu'un d'autre qui mérité bien des remerciements. En réalité, deux personnes, à qui j'ai emprunté un mari pour l'une et un père pour l'autre. Éléonora, Daniel, merci de m'avoir prêté Sergio certains soirs ou week end, notamment lors de l'écriture (trop rapide) de cette thèse. Je n'ai pas pu vous remercier convenablement avant cela, et m'en excuse ! Éléonora, bon courage pour la fin de tes études, et Daniel, futur Jamy ? 

Merci également aux autres théoriciens du groupe : Brice Arnaud, pour m'avoir accueilli deux fois en stage (courageux !) et formé à la physique du solide, merci pour cela. Je suis fier de penser que j'ai su gagner ton respect. Maintenant que je ne suis plus là, j'espère que ton ordinateur ne plantera plus. Dans le pire des cas si cela devait arriver, on verrait ce que l'on peut faire autour d'une ``petite'' bière lors d'un GdR, par exemle. Alain Gellé, merci d'avoir perdu quelques heures pour ne pas visiter une maison pour moi sur Toulouse et merci à Murielle pour la même raison. Didier Sébilleau, merci pour ta gentillesse et les infos insolites que tu nous relaies régulièrement. Merci également pour toute l'aide que tu as pu m'apporter. Sache également, qu'un beau jour, j'espère avoir la même bibliothèque que toi ! Et un grand merci à vous trois du temps que vous m'avez offert pour les diverses répétitions ou problèmes que j'ai pu rencontrer durant ma thèse.

Il me faut également remercier toute l'équipe ``surfaces et interfaces'' et notamment : Pascal Turban, Marie Hervé, Phillipe Shieffer et Sylvain Tricot pour leur encadrement durant mon stage de M2 ainsi que pour les discussions scientifiques tout au long de mon doctorat. Merci pour toutes les choses que j'ai pu apprendre de vous, et merci Phillipe pour toutes les discussions enflammées ``on va changer le monde'', cela me manquera. Je remercie également tout le reste de l'équipe ainsi que les autres personnes du 11E, pour la bonne ambiance et l'accueil chaleureux dont eux seuls ont le secret : Cristelle Mériadec, Bruno Lépine (merci d'avoir parlé en bien de moi à Anne Ponchet), Francine Solal, Sophie Guézo, Soraya Ababou, Alexandra Junay, Gabriel Delhaye, Arnaud Le pottier, Jean-Christophe Le Breton, Denis Morineau, Gilles paboeuf, Sylvie Beaufils et Véronique Vié. Il y a quantités de raisons de vous dire merci, mais la plus importante est l'ambiance générale que vous instiguez à la vie de labo. C'est un plaisir de venir travailler dans de telles conditions.

Je finirai en remerciant toutes les personnes de l'IPR qui m'ont apporté un soutien lors de ma formation à Rennes. Merci donc à tous les enseignants-chercheurs qui nous enseignent la physique pour la plupart avec passion. Mention spéciale pour Franck Thibault que j'ai eu au minimum un semestre par an. Ton humour caustique m'a toujours fait beaucoup rire, et je te remercie pour ton aide lors des ``amphis des lycéens'' ou de ``la fête de la science'' à Dinan. Autre mention particulière pour Phillipe Rabiller qui a toujours offert de son temps pour nous autres étudiants, et notamment pour son aide à la bonne organisation de mon année de césure en 2010 et merci Phillipe de m'avoir emmené maniper à SOLEIL.
Merci également aux ``administratives'' de l'IPR, notamment Valérie Ferri et Nathalie Gicquiaux qui m'ont toujours parfaitement aidé dans mes démarches administratives. Nathalie, excuse-moi encore une fois pour tous les ``états de frais de mission'' que j'ai oubliés de te ramener au retour de chaque mission.

Voilà pour les remerciements de ceux qui m'ont supporté relativement long-temps et qui ont, soit lu ma thèse, soit écouté lors des répétitions plus ou moins mauvaises de mes présentation orales. Excusez -moi pour tout ça ! 

Maintenant parait-il qu'il serait de bon usage de remercier ses proches. Il est vrai que certains n'ont pas le choix de me fréquenter et que pire, d'autres l'ont, mais continue à le faire. Encore mieux, parfois ils posent des questions sur mon sujet de thèse ! Certes ils le regrettent par la suite\dots Mais tout de même, cela mérite d'être salué ici.

Je remercie donc ma famille parce que c'est ma famille et qu'en tant que telle elle répond toujours présente pour rendre service. Merci donc à toutes et à tous : pour avoir gardé les filles quand nous en avions besoin, pour nous avoir aidé à déménager, nourri, logé, blanchi, promené, de vous être intéressé à ce que je fais, de nous avoir aidé avec la maison à Camlez\dots
Et j'en passe bien sûr. 

Merci également à Baptiste, Marina, Christophe, Virginie (je suis très fier d'être le parrain de Robin!), Léo et Marie-Laure. Mis à part les deux derniers qui sont physiciens, vous avez eu le mérite, en plus de m'avoir posé un jour la question ``qu'est-ce que tu fais exactement comme boulo ?'', d'assister à ma soutenance. Enfin mis à part deux d'entre vous, un peu moins courageux il faut l'avouer, qui ne sont venus qu'aux questions\dots Je tairai leur nom. Et puis, je comprends, Christophe et Virginie, que vous ayez eu peur de vous ennuyer ! Je vous ferai une séance de rattrapage au nouvel an. Préparez le rhum et l'armagnac. Léo et Marie-Laure, merci de m'avoir livré cette année une lettre ``postée'' en 2010. Ne changez rien ! Merci également de prolonger sur Toulouse uniquement parce que j'y viens. Ce n'est pas pour ça ? Tant pis, cela me fait plaisir malgré tout.

Et voilà, cet exercice, qui me coûte, il faut bien le dire, s'achève. J'espère n'avoir oublié personne\dots À moins que\dots Je vais peut être remercier Émilie également. Il faut avouer qu'elle me supporte depuis maintenant 13 ans, qu'elle m'a donné deux adorables filles, Lina et Rose, qu'elle a fait en sorte que Lina naisse le même jour qu'elle, de telle sorte que je n'ai que deux dates pour trois à retenir, et qu'elle a pris un congé parental d'éducation pour me permettre de faire un post-doc sur Toulouse. Je dois également la remercier pour avoir assurré avec les filles pendant ma phase de rédaction, durant laquelle j'ai été particulièrement absent. Pour les mêmes raisons, je dois remercier mes filles qui ont été adorables pendant cette période pas si simple. Certains jours elles ne m'ont pas vu du tout. Mais quand j'étais là, c'était plutôt agréable de se faire dorloter. Vous avez pris, toutes les trois, de très bonnes habitudes, ne changez rien !

Ceux qui me connaissent savent que je dis rarement ce genre de chose et que je ne suis pas doué pour ça. J'espère donc que vous apprécierez ces quelques confidences et que vous me pardonnerez si vous espériez-mieux (si vous le méritiez !). Merci à tous.

\newpage\thispagestyle{empty}
\strut
\newpage\thispagestyle{empty}
\strut
\newpage\thispagestyle{empty}

\tableofcontents
\newpage\thispagestyle{empty}
\strut
\newpage
\pagenumbering{arabic}

\chapter*{General introduction}\addcontentsline{toc}{chapter}{General introduction}
\markboth{General introduction}{General introduction}
  \lettrine[lines=3, lhang=0.35, loversize=0.77, findent=1.6em,nindent=-0.7em,slope=-1.4em]{S}{pintronics} (contraction of spin and electronics) is a recent branch in the field of electronics where the spin of the electrons is exploited. 
Its official birth is 1988, after the discovery of the Giant Magneto-Resis-tance (GMR) by Albert Fert and Peter Gründberg~\cite{fert-GMR_PhysRevLett.61.2472,grunberg-GMR_PhysRevB.39.4828}. Since 1997\footnote{The first use of spin-valve sensors in hard disk drive read heads was in the IBM Deskstar 16GP Titan, which was released in 1997 with 16.8 GB of storage}, it is used in our everyday life within the read-heads of the hard disk drive of our computers. 
The GMR effect consists in a significant diminution of the resistance in a thin film made of ferromagnetic and non-magnetic conductive layers, when an external magnetic field is applied. 
For instance, consider that at zero field, both magnetic layers have an anti-parallel magnetization. 
If we apply an external magnetic field in such a way that a reversal of the magnetization is induced and both magnetizations align, then, we observe that the resistance of the heterostructure drops drastically. It is due to the fact that the electrons, whose spin is not aligned with the magnetization of the metal where they propagate, experience more collisions than the ones whose spin is parallel to the magnetization of the metal. Such a system can be conceived as a spin polarizer/analyser: the first ferromagnetic slab polarizes the current and the second ferromagnetic slab analyzes the polarized current. 

It is interesting to notice that the advent of the spintronics could have taken place earlier with the discovery of the Tunneling Magneto-Resistance (TMR) by Jullière in 1975~\cite{Julliere1975225}. 
The TMR is an effect similar to GMR which occurs in a magnetic tunnel junction, whose components consist of two ferromagnets separated by a thin insulator which replaces the non-magnetic spacer of GMR. However, at that time the discovery did not attract a lot of attention. The TMR was rediscovered in the middle of the nineties~\cite{Miyazaki-TMR1995L231,moorea-TMR_PhysRevLett.74.3273}. Another ten years were needed to improve the technique and to observe a TMR reaching several hundred percent at room temperature~\cite{ikeda-TMR_600percent}.

By coupling these GMR/TMR-structures with a semi-conductor, one can control the spin-polarized current which is injected in the semi-conductor~\cite{monsma-spinvalve_transistor_PhysRevLett.74.5260}. A current of electrons whose energy is winthin few eV above the Fermi level is injected from a transmitter within a metallic base which is in contact with a semi-conductor. If their energy is sufficient (these electrons are often called ``hot electrons'', because their energy to overcome the Schottky barrier is much bigger than $k_{B}T$), they can cross the Schottky Barrier at the metal/semi-conductor interface and be collected at the back of the semi-conductor. Such devices are the so-called ``spin-valve'' (see Fig.~\ref{fig_GMR_spinvalve}, Chap.~\ref{chapt_BEES}). 

In this context, the Surfaces and Interfaces team of the Materials and Nanosciences department of the Physical Institute of Rennes (IPR), in particular dr. Pascal Turban, has developed a Ballistic Electron Emission Microscope (BEEM). The principle of this microscope is presented in chapter \ref{chapt_BEES} and in figure \ref{fig_BEEM_principe}. It allows to image metal semi-conductor interfaces and to study structures that holds interesting features for the spintronics.
In the last few years, several researchers from this team investigated the physical effects that govern the magneto-current by studying a model structure Fe/Au/Fe/GaAs with BEEM experiments~\cite{Marie-jap113_quantitative_magnetic_imaging,Marie-apl_spin_filtering}.

These experiments are quite long to carry out. Not only they require a long time to prepare the samples but each spectroscopy experiment takes several hours, and a few days are needed to obtain and analyze a spectrum like in Fig.~\ref{fig_BEES_AuGaAs}.
For these reasons, a reliable theoretical model and a numerical code to quantify it can be very useful: they would help to target a system by making predictions and preselecting the sample to analyze.

The first model to describe a BEEM current, based on the free electron model, was proposed by Kaiser and Bell \cite{kaiser_bell1-PhysRevLett.60.1406,kaiser_bell2-PhysRevLett.61.2368}. It was quite successful to predict the height of the Schottky barrier but it failed to explain the constant behavior of the current in some system, such as Au(111)/Si(001) and Au(111)/Si(111)~\cite{Milliken-AuSi_PhysRevB.46.12826}, as described in chapter \ref{chapt_BEES}. In 1996, F. J. Garcia-Vidal, P.L. de Andres and F. Flores \cite{Garcia-Vidal-PhysRevLett.76.807} proposed a model, based on non-equilibrium approach by means of Keldysh formalism, where electrons propagate within the metal, by taking into account the band structure of the material in which they propagate. Their model was successful to explain the BEEM current behavior for Au(111)/Si(001) and Au(111)/Si(111), for both the intensity and the lateral resolution. 
However, in spite of its success about two decades ago, as it is based on the calculation of semi-infinite slabs, it cannot predict the behavior of electrons in extremely thin metallic films (few layer), or in heterostructures like  spin-valves, that are studied nowadays.

In this context, we have chosen to work again on the model of Garcia-Vidal, F. Flores and P. De Andres~\cite{Garcia-Vidal-PhysRevLett.76.807} and to extend it in order to describe finite structures.  Our idea is:
\begin{enumerate}
 \item to compare the non-equilibrium approach with a simpler equilibrium approach. We ask ourselves if it is possible to interpret experimental results or to make predictions only by considering the band structure.
 \item to provide a direct theoretical support to the experimentalists of our group by means of a user-friendly numerical code that can tell, for instance, what would happen if gold  is replaced by silver in the Fe/Au/Fe/GaAs spinvalve, what would happen if we change the number of layers of iron etc\dots
\end{enumerate}
In order to complete this program, we have decided to work with a tight-binding approach, as in the original work of F. Flores \emph{et al.}. Of course, it would have been possible to use Green functions also within extended non-equilibrium Density Functional Theory calculations. However, for reasons detailed in section \ref{sec_non_eq_calc}, we believe that tight-binding is the most appropriate method to fulfill our objectives.

This manuscript is organized in two parts: the first part is focused on the theoretical and experimental background, and the second on the modeling of Ballistic Electron Emission Spectroscopy (BEES) on metallic films. Unlike the microscopy mode of the BEEM that allows to image buried structures, the spectroscopy mode records the evolution of the BEEM current with respect to the applied bias, as described in chapter \ref{chapt_BEES}.
The first part of this thesis, the general background, corresponds to chapters \ref{chapt_background} and \ref{chapt_BEES}, whereas chapters \ref{chapt_NEPT}, \ref{chapt_BEEM_prog} and \ref{chapt_results}, together with the conclusion, compose the second part where I derive and describe my results. 
In more details, in the first chapter of this thesis, some theoretical background, about equilibrium and non-equilibrium perturbation-theory within second-quantization Green-functions is recalled. Although the reader who is already familiar with this formalism can skip this chapter, it might be useful, in order to get acquainted with the notation that is used in the next chapters. 
Chapter \ref{chapt_BEES} introduces the Ballistic Electron Emission Microscopy and Spectroscopy. We shall see that the existing free-electron models failed to describe some experiments, like Au/Si, and that it is necessary to introduce a new model where the band structure of the material has to be taken into account.
In chapter \ref{chapt_NEPT}, we extend the previous model of F. Flores and P. De Andres' group to thin films by avoiding their decimation technique through a different layer-by-layer perturbation expansion. In this scheme we extend the older approach by considering second and third-nearest neighbor interactions. 
Chapter \ref{chapt_BEEM_prog} is devoted to the presentation of our new BEEM program. After presenting the flow chart, we explain how to format an input file and what is the effect of the key parameters. Whereafter, we present some key subroutines that are required to calculate the Green functions and hence the BEEM current.
Chapter \ref{chapt_results} presents our results obtained with this code and with the simpler equilibrium approach described in chapter \ref{chapt_BEES}. Finally, we draw our conclusions and some perspectives in chapter \ref{chapt_perspectives}.

\chapter{Theoretical background}\label{chapt_background}
  \lettrine[lines=3, lhang=0.35, loversize=0.77, findent=3em,nindent=-0.5em,slope=-1em]{A}{s it is important} to define a common language, and as there are a lot of different notations in non-equilibrium Green function formalism, a general background is presented in this chapter. We start with a quick overview of the second quantization and the derivation of the second-quantization Hamiltonian ; we then introduce Schrödinger, Heisenberg and interaction pictures, and finally move to the fundamental principles of equilibrium and non-equilibrium perturbation theory using Green functions.
  \section{Introduction to second quantization}
  In the usual Schrödinger formalism of non-relativistic quantum-mechanic (it might improperly be called ``first quantization'') the position $x$ of the particle and its impulsion $p$ are replaced by operators $\hat{X}$ and $\hat{P}$ acting on a Hilbert space (see for instance \cite{Feynman-vol3_quantum_mechanics}). Commutation rules of these operators are established by analogy with Hamiltonian-mechanic formulation. Elements or vectors of the Hilbert space describe possible configurations or states
  of a system with a fixed number $N$ of particles. The representation in the coordinate space of such a state is called a wave-function. This wave-function is a probability amplitude,
  that is to say a complex function of the positions $(x_1,\dots,x_N)$ and time $t$, $\psi (x_1,\dots,x_N, t)$, whose square of
  modulus is the probability density of finding the particles at points $(x_1,\dots,x_N)$ and time $ t $. As well known, the wave-function is a solution of the Schrödinger equation, a partial differential equation in space and time.
  
  This approach works well when we deal with a well-definite number of particles. 
  If, however, the interactions are such that the number of particles changes, a better procedure, called ``second quantization'' (the name might be misleading: there is no real quantization, it is just a formal tool), should be introduced. 
  This second quantization formalism is fundamental in relativistic theories, where it was first formulated, because of particle creation and annihilation \cite{Dirac-2nd_quantiz}. 
  Yet, it turns out to be extremely important also in non-relativistic quantum-field theories \cite{fetter,Lifshitz-vol9_cond_mat} in several cases where the number of particles varies, like for Cooper pairs in superconductivity. 
  In our work, it is found to be extremely useful to describe the electron propagation from one metallic layer of the thin film to another, what can be seen intuitively as an electron annihilation and creation from the first layer to the second.

  In order to describe such a process or, more generally, the transitions between states with different numbers of particles, the so-called creation and annihilation operators (or ladder operators) are introduced. Their role is fundamental in the formalism of second quantization. Though in the following we consider fermions, the simplest analogy to understand the physical meaning of ladder operators is in the boson case with the harmonic oscillator \cite[chap. 5]{cohen}. In quantum mechanics, the Hamiltonian operator for a one dimensional harmonic-oscillator is
  \begin{equation}
    H = \frac{\hat{P}_x}{2m} + \frac{1}{2}m\omega^2\hat{X}^2
  \end{equation} 
  where $\hat{X}$ is the position operator and $\hat{P}_x$ is the $x$-component of the impulsion operator of the particle. Since $H$ is time independent, the quantum mechanical study of the harmonic oscillator reduces to the solution of the Schrödinger equation
  \begin{equation}
    H\ket{\psi} = E\ket{\psi}
  \end{equation} 
  where $E$ is the energy associated to an eigenstate $\ket{\psi}$ of the system.
  This is equivalent, in the $x$ representation, to:
  \begin{equation}
    \left[ -\frac{\hbar^2}{2m} \frac{\D^2}{\D x^2} + \frac{1}{2}m\omega^2x^2\right] \psi(x) = E\psi(x)
  \end{equation} 
  The research of eigenvalues and eigenvectors of $\hat{H}$ can be simplified by introducing the ladder operators (for bosons)
  \begin{align}
    \hat{a}           &= \frac{1}{\sqrt{2}} (\hat{X}+\i\hat{P}) \\
    \hat{a}^{\dagger} &= \frac{1}{\sqrt{2}} (\hat{X}-\i\hat{P})
  \end{align}
  with $\hat{X} = \sqrt{\frac{m\omega}{\hbar}}X$ and $\hat{P} = \sqrt{\frac{1}{m\omega\hbar}}P$. Because $\hat{X}$ and $\hat{P}$ obey the canonical commutation relation $[\hat{X},\hat{P}]=\i$, the new operators obey
  \begin{equation}
    [\hat{a},\hat{a}^{\dagger}] = 1
  \end{equation} 
  Another useful formula is
  \begin{align}
    \hat{a}^{\dagger} \hat{a} &= \frac{1}{2} (\hat{X}-\i\hat{P})(\hat{X}+\i\hat{P})\nonumber\\
                 &= \frac{1}{2}(\hat{X}^2+\hat{P}^2-1)
  \end{align}
  Comparing this equation with $\hat{H}=\frac{\hbar\omega}{2}\left(\hat{X}^2+\hat{P}^2 \right)$ we see that
  \begin{align}
    \hat{H} &= \hat{a}^{\dagger} \hat{a} + \frac{1}{2} \nonumber
    \\      &= \hat{a}\hat{a}^{\dagger} - \frac{1}{2}
  \end{align} 
  So that the eigenvectors of the particle-number operator $\hat{n}=\hat{a}^{\dagger} \hat{a}$ are eigenvectors of $\hat{H}$. It is then possible to replace $\hat{H}$ by $\hat{n}$ in the Schrödinger equation
  \begin{equation}
    \hat{n}\ket{v} = v \ket{v}
  \end{equation} 
  The eigenvalues of the quantum harmonic oscillator are thus 
  \begin{equation}
    E_v=\hbar\omega\left(v+\frac{1}{2}\right)
  \end{equation} 
  It is possible to find the eigenvalues of $\hat{n}$ by using commutation relations:
  \begin{align}
    [\hat{n}, \hat{a}]          &= - \hat{a} \\
    [\hat{n},\hat{a}^{\dagger}] &= \hat{a}^{\dagger}
  \end{align}
  which gives
  \begin{align}
    [\hat{n}, \hat{a}]\ket{v} &= \hat{n}( \hat{a}\ket{v})-v( \hat{a}\ket{v}) = - \hat{a}\ket{v} \\
     \hat{n} \hat{a}\ket{v} &= (v-1)  \hat{a}\ket{v}
  \end{align}
  in other words, $(\hat{a}\ket{v})$ is eigenvector of $\hat{n}$ with eigenvalue $(v-1)$. This means that $\hat{a}$ acts on $\ket{v}$ to produce, up to a multiplicative constant, the state $\ket{v-1}$. A similar equations holds for $\hat{a}^{\dagger}$
  \begin{equation}
     \hat{n}\hat{a}^{\dagger}\ket{v} = (v+1) \hat{a}^{\dagger}\ket{v}
  \end{equation} 
  This times, $\hat{a}^{\dagger}$ acts on $\ket{v}$ to produce, up to a multiplicative constant, $\ket{v+1}$. For this reason, $ \hat{a}$ is called a lowering operator and $\hat{a}^{\dagger}$ a raising operator. They lower or raise the energy of a quantity $\hbar\omega$. In quantum field theory, these operators are respectively called "annihilation" and "creation" operators because they destroy and create particles, which correspond to our quantum of energy. There is however a complete analogy between the two cases.

  The fermion case, though conceptually identical, brings in more cumbersome algebra, because of the antisymmetrization requirements of the N-particle wave-function. For this reason, we placed the general treatment in the appendix~\ref{appendix_second_quantization}. We just remind that the Hilbert space on which
  these operators act is what is known as a Fock space, that is to say, a stack of
  infinite Hilbert-spaces communicating through fields and operators and
  comprising the vacuum, a-zero particle space, a one-particle space, a two-particles space,
  etc \dots
  In what follows we describe the second quantization Hamiltonian as we shall use in our work.

  \subsection{Second quantization Hamiltonian}
    The Hamiltonian of a system of $N$ interacting-electrons evolving within a periodic potential $U(\vec{r}_i)$ can be written as
    \begin{align}
    \hat{H}&= \underbrace{\sum_{i=1}^N \left[\frac{-\hbar^2}{2m}\Delta_i + U(\vec{r}_i) \right]}_{\scriptstyle \textrm{monoelectronic Hamiltonian sum } h(\vec{r}_i)} + \underbrace{\frac{1}{2} \sum_{i,j (i\neq j)} V(\vec{r}_i-\vec{r}_j)}_{\scriptstyle\textrm{two-body term}}
    \end{align}
    where $U(\vec{r}_i)=-e\sum_k u(\vec{r}_i-\vec{R}_k)$ is the interaction with the nuclei and $V(\vec{r}_i-\vec{r}_j)= e^2/(4\pi\epsilon_0 |\vec{r}_i-\vec{r}_j|)$ is the Coulomb repulsion. As shown in the appendix~\ref{appendix_second_quantization}, it can be rewritten in second quantization as
    \begin{align}\label{eq_hamiltonien_total_2}
    \hat{H}&=\sum_{k,l} \Braket{k|h|l} \hat{c}^{\dag}_k \hat{c}_l + \frac{1}{2}\sum_{k,l,m,n}\Braket{kl| V | mn}\hat{c}^{\dag}_k\hat{c}^{\dag}_l \hat{c}_n \hat{c}_m \nonumber\\
    &= \sum_{k,l} {T}_{kl} \hat{c}^{\dag}_k \hat{c}_l + \frac{1}{2}\sum_{k,l,m,n}{V}_{klnm}\hat{c}^{\dag}_k\hat{c}^{\dag}_l \hat{c}_n \hat{c}_m
    \end{align} 
    where $\{ \Braket{\vec{r}| l}=\varphi_l(\vec{r})\}$  is a complete basis for the wave-function ($l$ including spin) and where 
    \begin{align}\label{eq_hopping_def}
    \hat{T}_{kl}&=  \Braket{k | h(\vec{r}) | l} \nonumber\\
	        &= \sum_k \underbrace{\Braket{k | -\frac{\hbar^2}{2m} \varDelta_k + U(\vec{r}_k) | l}}_{\varepsilon_0 \delta_{k,l}} + \underbrace{\sum_{i\neq k} \Braket{k | U(\vec{r}_i) | l}}_{-t_{kl}} \nonumber\\
                &= \int \D^3\vec{r} \ \varphi_{k}^*(\vec{r})\ h(\vec{r})\ \varphi_{l}(\vec{r}) 	        
    \end{align}
    and
    \begin{align}
      {V}_{klnm} &= \braket{\varphi_k \otimes \varphi_l | V |\varphi_m \otimes \varphi_n} \nonumber \\
                     &= \int \D^3\vec{r}\int \D^3\vec{r}'\ \varphi_{k}^*(\vec{r}) \varphi_{l}^*(\vec{r}')\ V(\vec{r}-\vec{r}')\ \varphi_{m}(\vec{r})\varphi_{n}(\vec{r}') 
    \end{align}
    The final Hamiltonian can be therefore written as:
    \begin{equation}\label{eq_hamiltonien_total}
    \hat{H}= \sum_{k,l} [\varepsilon_0\delta_{k,l}-t_{kl}]\hat{c}^{\dag}_k \hat{c}_l + \frac{1}{2} \sum_{k,l,m,n} {V}_{klnm} \hat{c}^{\dag}_k \hat{c}^{\dag}_l \hat{c}_m \hat{c}_n
    \end{equation}

    \subsubsection{Tight-binding model}\label{sec_tight_binding_model}
      The tight-binding model consists in making the assumption that the Coulomb interaction of the electrons is negligible compared to their kinetic and lattice energies. In other words, ${V}_{klnm}=0$ in Eq.~\eqref{eq_hamiltonien_total}. The Hamiltonian is then reduced to
      \begin{align}
	\hat{H}_{tb}= \sum_{k,l} [\varepsilon_0\delta_{k,l}-t_{kl}] \hat{c}^{\dag}_{k}  \hat{c}_{l}
      \end{align}
      The physical interpretation of the terms is the following: 
      $\varepsilon_0$ corresponds to the atomic energy or to the orbital energies in the case of multi-orbital atoms (as in the following of this work). It is the on-site energy. The so-called hopping term  $t_{k,l}\hat{c}^{\dagger}_{k} \hat{c}_{l}$ destroys a state characterized by the quantum number $l$ and creates another one with quantum number $k$ with an amplitude $t_{k,l}$. The tight-binding model has been very used in the literature because it allows reproducing electron structure with localized orbitals (i.e., only neighbor interactions are taken into account) of many materials or to model electronic transport, as we shall see below, and has the advantage that it can be  extended in a straightforward manner to describe problems where the electron correlation is not negligible, as done, for instance in the Hubbard model~\cite{Hubbard_I,Hubbard_II,Hubbard_III}.

    \subsubsection{Hubbard model}\label{sec_hubbard_model}
      In his approach, Hubbard supposed that the most important part of the electron-electron interaction is due to the on-site Coulomb repulsion~\cite{Hubbard_I,Hubbard_II,Hubbard_III}. In other words, $V_{klnm}\neq 0$ only when $k$, $l$, $n$ and $m$ all refer to the same site (say, site $\vec{R}_i$). In that case by writing the spin explicitly, the Coulomb repulsion is
      \begin{equation}
         V_{i\sigma i\sigma',i\sigma i\sigma'} = \frac{e^2}{4\pi\varepsilon_0}\int \D^3\vec{r} \D^3\vec{r}' \ \big|\varphi_{i\sigma}(\vec{r})\big|^2 \frac{1}{\big|\vec{r}-\vec{r}\,'\big|} \big|\varphi_{i\sigma'}(\vec{r}\,')\big|^2
      \end{equation}
      and the Hamiltonian writes
      \begin{align}\label{eq_HU}
        \hat{H} &= \hat{H}_{tb} + \hat{H}_U \nonumber\\
          &= \sum_{k,l} [\varepsilon_0\delta_{k,l}-t_{kl}] \hat{c}^{\dag}_{k} \hat{c}_{l} +  U \sum_{i,\sigma} \hat{n}_{i\sigma} \, \hat{n}_{i\bar{\sigma}}
      \end{align}
      where $U=V_{i\sigma i\sigma',i\sigma i\sigma'} /2$ and where $\hat{n}_{i\sigma}=\hat{c}^{\dag}_{i\sigma} \hat{c}_{i\sigma}$ is the number of particle operator of spin $\sigma$ at site $\ket{\vec{R}}_i$. In general, $\hat{c}^{\dag}_i \hat{c}_i$ is the number of particle operator in the state at site $\vec{R}_i$. In the case of, \emph{e.g.}, transition metals, characterized by more than one orbital per site, Eq.~\eqref{eq_HU} should be generalized in order to take into account of the extra degree of freedom~\cite{Hubbard_II}.

\section{Pictures}      
  In the next chapters,  we want to describe Ballistic Electron Emission Microscopy that is a technique based on the Scanning Tunneling Microscope. In this microscopy, the electric field induced by the STM tip can be seen as a weak external perturbation. Hence it seems natural to use a perturbation approach. In this subsection the various representations, or pictures, of quantum mechanics are recalled, namely, Schrödinger, Heisenberg and interaction pictures. As the name suggests, the interaction picture is the natural framework to formulated the perturbation theory, Schrödinger and Heisenberg pictures are a necessary complement to understand it.

  \subsection{Schrödinger and Heisenberg pictures}\label{sec_schrodinger_heisenberg_picture}
    In the Schrödinger picture, only the wave-functions are time dependent, 
    while in the Heisenberg picture it is the operators that hold the time dependence. 
    While the wave-functions in the Schrödinger picture obey the usual Schrödinger equation~\eqref{eq_schro_eq} below, the operators in the Heisenberg representation obey the Heisenberg equation of motion, through the commutator with $\hat{H}$:
    \begin{equation}\label{eq_heisenberg_eq_motion}
     \i\hbar\partial_t\hat{O}_H(t) = [\hat{O}_H(t),\hat{H}_t]
    \end{equation} 
    By definition, for all values of $t$, the expectation value of an operator is the same in both representations:
    \begin{equation}\label{eq_mat_element_equal}
      \Braket{\psi_S(t) | \hat{O}_S  | \psi_S(t) } = \Braket{\psi_H | \hat{O}_H(t)  | \psi_H }
    \end{equation} 
    Where label $S$ refers to the Schrödinger representation and label $H$ to the Heisenberg representation.
    In order to simplify the calculation, let's take $t=0$ the time where both representations coincide:
    \begin{align}
      \hat{O}_H(t=0) &= \hat{O}_S \\
      \Ket{\psi_S(t=0)} &= \Ket{\psi_H}
    \end{align}
    It is useful at this point to introduce the evolution operator $\hat{U}(t,t_0)$ as the operator that leads from the state $\ket{\psi_S(t_0)}$ to the state  $\ket{\psi_S(t)}$:
    \begin{equation}
      \Ket{\psi_S(t)} = \hat{U}(t,t_0) \Ket{\psi_S(t_0)}
    \end{equation}
    Of course, this operator must be related to the Hamiltonian, because for  time dependent phenomena, the Hamiltonian can be seen as the infinitesimal generator of time translations, \emph{i.e.}, it leads from the state $\ket{\psi(t)}$ to the state $\ket{\psi(t+\D t)}$. This is a consequence of the Schrödinger equation for a time dependent Hamiltonian:
    \begin{equation}\label{eq_schro_eq}
      \i\hbar \frac{\partial}{\partial t} \Ket{\psi_S(t)} = \hat{H}(t) \Ket{\psi_S(t)}
    \end{equation}
    Because of the hermiticity of $\hat{H}$ the time derivative $\partial_t \Braket{\psi_S(t)|\psi_S(t)}=0$, i.e., the probability is conserved.

    All this, implies that: $ \hat{U}(t,t_0)$ is a unitary operator: which obeys the following identities:
    \begin{equation}
      \hat{U}(t_0,t_0) = \identite
    \end{equation} 
    and because of the conservation of probability
    \begin{equation}
      \hat{U}^{\dagger}(t,t_0)\hat{U}(t,t_0) = \identite
    \end{equation} 
    So that:
    \begin{equation}
      \hat{U}^{-1}(t,t_0) =  \hat{U}^{\dagger}(t,t_0)
    \end{equation} 
    Furthermore, if the time-reversal invariance can be used, we also have
    \begin{equation}
      \hat{U}^{-1}(t,t_0) =  \hat{U}^{\dagger}(t,t_0) = \hat{U}(t_0,t)
    \end{equation} 
    as
    \begin{equation}
      \hat{U}(t_0,t)\hat{U}(t,t_0) = \identite
    \end{equation} 

    Using the expression for the time dependent wave-function and the equality \eqref{eq_mat_element_equal}, we can move from Schrödinger to Heisenberg representations using the evolution operator as follows:
    \begin{equation}
      \hat{O}_H(t) = \hat{U}^{\dagger}(t,0) \hat{O}_S \hat{U}(t,0)
    \end{equation} 
    This allows to find out the explicit expression of the evolution operator in terms of the Hamiltonian. In fact, one recovers the usual results for time-independent Hamiltonians by noting that in this case, the solution of Schrödinger equation for the evolution operator is
    \begin{equation}
      \hat{U}(t,t_0) = \e^{-\i\hat{H}(t-t_0)/\hbar}
    \end{equation}
    whose general integral form is
    \begin{equation}\label{eq_U_int_form}
      \hat{U}(t,t_0) = \hat{T}\left\{ \exp\left[ -\i \int_{t_0}^{t} \D t' \hat{H}(t') \right] \right\}
    \end{equation}
    where $\hat{T}$ is the time-ordering operator. It orders time dependent operators from right to left in ascending time and adds a factor $(-1)^{p}$ where $p$ is the number of permutation of fermion operators. As we shall see in sections, it is a key operator for the definition of Green functions.

  \subsection{Interaction picture}
    As said above, the interaction picture is the best representation for perturbation theory, \emph{i.e.} when the Hamiltonian is written as $\hat{H}=\hat{H}_0+\hat{V}$ and it is supposed that we can solve the Schrödinger equation for a time-independent $\hat{H}_0$ (but not for $\hat{H}$) and that $\hat{V}$ is a ``small'' perturbation of $\hat{H}_0$. It is an intermediate representation, between Schrödinger and Heisenberg ones, introduced by Dirac (sometimes it is called Dirac representation). In this representation, both operators and wave-functions evolve in time. The wave-functions however develop under the influence of the ``difficult'' interaction part of the Hamiltonian
    \begin{equation}
      \hat{H} = \hat{H_0} + \hat{V}
    \end{equation}
    where $\hat{H_0}$ is time independent as stated above. In this framework, the time-evolution operator $\hat{U}_I(t,0)$ is given by
    \begin{align}\label{eq_Ut0_UI}
      \hat{U}(t,0) &= \e^{-\i\hat{H}_0t/\hbar} \hat{U}_I(t,0) \\
      \hat{U}(0,t) &= \hat{U}_I(0,t) \e^{\i\hat{H}_0t/\hbar} 
    \end{align}
    This operator has the same unitary property that an ordinary time evolution operator. So it is possible to write:
    \begin{equation}
      \hat{U}(t,t_0)= \e^{-\i\hat{H}_0t/\hbar} \hat{U}_I(t,t_0)\e^{\i\hat{H}_0t_0/\hbar} 
    \end{equation} 
    Again, at $t=0$ all the representations coincide. The reason to define the time evolution operator in this way is that, for a small perturbation $\hat{V}$, $\hat{U}_I(t,t_0)$ is close to unity, that is $\hat{U}_I$ encloses the ``smallness'' of the perturbation $\hat{V}$. 
    
    Using again the equality $\ket{\psi_S(0)} = \ket{\psi_H} =\ket{\psi_I(0)}$ and Eq.~\eqref{eq_Ut0_UI}, the matrix elements are:
    \begin{align}\label{eq_equivalence_mat_elements}
      \Braket{\psi_H | \hat{O}_H(t) |\psi_H} &= \Braket{\psi_S(t=0) | \hat{O}_S |\psi_S(t=0)} \nonumber
      \\                               &= \Braket{\psi_S(t=0)| \hat{U}^{\dagger}(t,0) \hat{O}_S \hat{U}(0,t) |\psi_S(t=0)} \nonumber
      \\                               &= \Braket{\psi_H| \hat{U}^{\dagger}_I(t,0) \e^{\i\hat{H}_0t/\hbar} \hat{O}_S  \e^{-\i\hat{H}_0t/\hbar} \hat{U}_I(t,0) |\psi_H} \nonumber
      \\                               &= \Braket{\psi_I(0)| \hat{U}^{\dagger}_I(t,0) \hat{O}_I(t) \hat{U}_I(t,0) |\psi_I(0)}
    \end{align}
    This important result can be interpreted as the fact that the operators in the interaction picture evolve with the $H_0$ part, that we are supposed to know:
    \begin{equation}
      \hat{O}_I(t) = \e^{\i\hat{H}_0t_0/\hbar} \hat{O}_S  \e^{-\i\hat{H}_0t/\hbar}
    \end{equation} 
    while the wave-functions obey
    \begin{equation}\label{eq_evolution_inter_pict}
      \Ket{\psi_I(t)} = \hat{U}_I(t,0) \Ket{\psi_S} 
    \end{equation} 
    That is, the unknown part (but supposed small). We shall see in the next section how to get a closed solution for this problem, at least for a tight-binding Hamiltonian, by means of a Green function approach.

  \section{Green functions}\label{sec_GF}

  This section introduces the concept of Green functions within the second quantization formalism of quantum mechanics. They are also called propagators, as they describe the propagation of an excitation from $(x,t)$ to $(x',t')$. We remind that the Green function method is very useful and widely employed independently of quantum mechanics in the theory of ordinary and partial-differential equations like Poisson equation or Maxwell equations in electrodynamics (see, eg, \cite{jackson-classical_electrodynamics}). In this case, Green functions are used to re-express differential equations as integral equations, to be solved, eventually, by perturbation methods. 
  \emph{Mutatis mutandis}, this method has been used in quantum mechanics to solve the Schrödinger equation in its second-quantization form, as detailed below. In this case, the single-particle Green function allows to find the expectation value of any single-particle operator in the ground state, the ground-state energy and the excitation spectrum of the system \cite[Chap. 3]{fetter}. Green functions are also particularly useful for problems solved by means of perturbation theory as they can be represented diagrammatically through Feynman diagrams~\cite{Lifshitz-vol10_kinetics}.

  The reason why we introduce the Green-function formalism in our work is that the electric current can be expressed in a straightforward way through this formalism, as we shall see in section \ref{section_beem_current}. 

  \subsection{Definition}
  The single-particle Green function is defined in Heisenberg representation by
  \begin{equation}\label{eq_def_GF}
      \i\greennoup{ij\sigma}(t,t') = \frac{\Braket{\Psi_0^H | \hat{T}\left[\hat{c}_{i\sigma}(t) \hat{c}^{\dagger}_{j\sigma}(t') \right] | \Psi_0^H}
					    }{\Braket{\Psi_0^H |\Psi_0^H} }
  \end{equation}
  where $\ket{\Psi_0^H}$ is the Heisenberg ground state of the interacting system satisfying
  \begin{equation}\label{heisenberg_gs}
    \hat{H}\ket{\Psi_0^H}=E \ket{\Psi_0^H}
  \end{equation}
  with the second quantification Hamiltonian of Eq.~\eqref{eq_hamiltonien_total}. We suppose from now on that it is normalized ($\braket{\Psi_0^H|\Psi_0^H}=1$) and remove the denominator in Eq.~\eqref{eq_def_GF}.
  Here, the annihilation $\hat{c}_{i\sigma}(t)$ is a Heisenberg operator with the time dependence
  \begin{equation}\label{heisenberg_op}
    \hat{c}_{i\sigma}(t) = \e^{\i \hat{H}t/\hbar}  \hat{c}_{i} \e^{-\i \hat{H}t/\hbar}
  \end{equation}
  $i$ and $j$ label the components of the field operator. The product $\hat{T}$ represents a generalization of
  \begin{align}
    \hat{T}\left[\hat{c}_{i\sigma}(t) \hat{c}^{\dagger}_{j\sigma}(t') \right] = \left\{ \begin{array}{rl}
										    \hat{c}_{i\sigma}(t) \hat{c}^{\dagger}_{j\sigma}(t')    & t>t'\\
										\pm \hat{c}^{\dagger}_{j\sigma}(t') \hat{c}_{i\sigma}(t)    & t<t'
									      \end{array}
									\right.
  \end{align}
  where the $\pm$ sign refers to bosons/fermions. That's why this product is called time ordering: operators are ordered from right to left in ascending time order and a factor $(-1)^p$ is added for $p$ interchanges of fermion operators, from the original order. From (\ref{heisenberg_gs}) and (\ref{heisenberg_op}) we can write
  \begin{align}
      \i\greennoup{ij\sigma}(t,t') = \left\{ \begin{array}{rl}
						\e^{\i E_0^N(t-t')/\hbar}    \Braket{\Psi_0^H |\hat{c}_{i}\e^{-\i \hat{H}(t-t')/\hbar} \hat{c}^{\dagger}_{j} | \Psi_0^H} & t>t' 
  \\                                            \pm\e^{-\i E_0^N(t-t')/\hbar}\Braket{\Psi_0^H |\hat{c}^{\dagger}_{j}\e^{\i \hat{H}(t-t')/\hbar} \hat{c}_{i} | \Psi_0^H} & t'>t 
					      \end{array}
									\right.
  \end{align}
  The factor $\e^{\i E_0^N(t-t')/\hbar}$ is merely a complex number which can be taken out of matrix element. In contrast, the operator $\hat{H}$ must remain between the field operators.

 \subsection{Retarded and advanced Green functions}\label{sec_ret_adv_gf}
    A very interesting representation of the propagator is the Lehmann representation where the Green function is expressed in frequency space because information about the excitation spectrum can be extracted in a natural way. For this purpose, we re-write Eq.~\eqref{eq_def_GF} in the form (still in Heisenberg picture):
    \begin{multline}\label{eq_GF_pour_lehman}
      \i\greennoup{ij\sigma}(t,t') =\\ \left\{ \begin{array}{rl}
						-\e^{\i E_0^N(t-t')/\hbar}\sum_n    \Braket{\psi_0^{(N)} |\hat{c}_{i}\e^{-\i \hat{H}(t-t')/\hbar}|\psi_n^{(N+1)}} 
						 \Braket{\psi_n^{(N+1)}|\hat{c}^{\dagger}_{j} | \psi_0^{(N)}} &\quad t>0
  \\                                            \ \e^{-\i E_0^N(t-t')/\hbar}\sum_n    
                                                 \Braket{\psi_0^{(N)} | \hat{c}^{\dagger}_{j} \e^{\i \hat{H}(t-t')/\hbar}|\psi_m^{(N-1)}} 
						 \Braket{\psi_m^{(N-1)}| \hat{c}_{i} | \psi_0^{(N)}} &\quad t<0
					      \end{array}
									\right.
    \end{multline} 
    The $\Ket{\psi_m^{(N-1)}}$ and $\Ket{\psi_n^{(N+1)}}$ denote a complete set of eigenstates of the $(N-1)$ and $(N+1)$ electron systems, respectively, characterized by the quantum numbers $m$ and $n$. Their corresponding energies are $E^{N-1}_m$ and $E^{N+1}_n$, while $E_0^N$ is the ground-state energy of the $N$-electron system. 
    Since the volume of the system is kept constant, the change in energy 
    \begin{equation}\label{eq_An}
     A_n= E^{N+1}_n - E^{N}_0
    \end{equation} 
    is the electron affinity. The other difference
    \begin{align}
     I_m = E^{N-1}_m - E^{N}_0
    \end{align}
    is the ionization potential. Introducing these quantities in Eq.~\eqref{eq_GF_pour_lehman}, it gives
    \begin{equation}\label{eq_GF_pour_lehman2}
      \i\greennoup{ij\sigma}(t,t') = \left\{ \begin{array}{rl}
						-\ \e^{\i A_n(t-t')/\hbar}\sum_n    
						 \Braket{\psi_0^{(N)}  |\hat{c}_{i}|\psi_n^{(N+1)}} 
						 \Braket{\psi_n^{(N+1)}|\hat{c}^{\dagger}_{j} | \psi_0^{(N)}} &\quad t>0
  \\                                            \ \e^{-\i I_m(t-t')/\hbar}\sum_n    
                                                 \Braket{\psi_0^{(N)} | \hat{c}^{\dagger}_{j} |\psi_m^{(N-1)}} 
						 \Braket{\psi_m^{(N-1)}| \hat{c}_{i} | \psi_0^{(N)}} & \quad t<0
					      \end{array}
									\right.
    \end{equation} 
    Using the Fourier transform of the Green function 
    \begin{equation}
     \hat{G}_{ij}(t) = \frac{1}{{2\pi}} \int_{-\infty}^{+\infty} \hat{G}_{ij}(\omega) \e^{-\i\omega (t-t')} \D \omega
    \end{equation}
    Eq.~\eqref{eq_GF_pour_lehman2} becomes (see, for example \cite{fetter}):
    \begin{align}\label{eq_G_lehman3}
     \hat{G}_{ij}(\omega) = \sum_n \frac{\alpha_i(n)\alpha_j^{\dagger}(n)}{\hbar\omega-A_n+\i\eta} +
                       \sum_m \frac{\beta_i(m)\beta_j^{\dagger}(m)}{\hbar\omega-I_m-\i\eta} 
    \end{align}
    where $\alpha_i(n)=\braket{\psi_0^{(N)}  |\hat{c}_{i}|\psi_n^{(N+1)}}$ and $\beta_j^{\dagger}(n)= \braket{\psi_0^{(N)} | \hat{c}^{\dagger}_{j} |\psi_m^{(N-1)}} $. $\eta$ is a positive infinitesimal quantity which ensures the correct analytic properties of $\hat{G}_{i,j}(\omega)$. 
    We can introduce the chemical potential by rewriting Eq.~\ref{eq_An}
    \begin{align}
      E_n^{N+1} - E_{0}^{N} &=  E_n^{N+1} -  E_0^{N+1} + E_0^{N+1} - E_0^{N} \nonumber\\
                            &= \varepsilon_{n}^{N+1} + \mu
    \end{align} 
    and in the case of a macroscopic body, as there is a large number of electrons we can write
    \begin{equation}
     \mu = E^{N+1}_n - E^{N}_0 \simeq E^{N-1}_m - E^{N}_0
    \end{equation}
    Thus, Eq.~\eqref{eq_G_lehman3} becomes
    \begin{equation}
	\hat{G}_{ij}(\omega) = \sum_n \frac{\alpha_i(n)\alpha_j^{\dagger}(n)}{\hbar\omega-\mu -\varepsilon_{n}^{N+1}+\i\eta} +
                       \sum_m \frac{\beta_i(m)\beta_j^{\dagger}(m)}{\hbar\omega-\mu +\varepsilon_{n}^{N+1}-\i\eta} 
    \end{equation}    
    From this last equation we can introduce two new Green functions, the so-called retarded and advanced Green functions which are defined, in the Lehman representation, by (see \cite[Sec. 7]{fetter}):
    \begin{align}
      \hat{G}_{ij}^{R}(\omega) &= \sum_n \frac{\alpha_i(n)\alpha_j^{\dagger}(n)}{\hbar\omega-\mu -\varepsilon_{n}^{N+1}+\i\eta} +
                       \sum_m \frac{\beta_i(m)\beta_j^{\dagger}(m)}{\hbar\omega-\mu +\varepsilon_{n}^{N+1}+\i\eta} \\
      \hat{G}_{ij}^{A}(\omega) &= \sum_n \frac{\alpha_i(n)\alpha_j^{\dagger}(n)}{\hbar\omega-\mu -\varepsilon_{n}^{N+1}-\i\eta} +
                       \sum_m \frac{\beta_i(m)\beta_j^{\dagger}(m)}{\hbar\omega-\mu +\varepsilon_{n}^{N+1}-\i\eta}                        
    \end{align} 
    
    The corresponding time-dependent Green functions are (see \cite{fetter}):
    \begin{align}
     \retarded{i,j}{t-t'} & = -\i \theta(t-t')\Braket{\psi_0 | \left\{ \annihilation{i}(t), \creation{j}(t') \right\} |\psi_0} \\
     \advanced{i,j}{t-t'} & =  \i \theta(t'-t)\Braket{\psi_0 | \left\{ \annihilation{i}(t), \creation{j}(t') \right\}|\psi_0}
    \end{align}
    where the curly bracket denotes an anti-commutator and $\theta(t-t')$ the Heaviside function. The retarded Green function is also called a propagator since it gives the wave-function at any time as long as the initial condition is given.

     As said above, those Green functions gives access to spectral quantities, such as the density of states
     \begin{align}
       \rho(\varepsilon) &= \sum_n \delta(\varepsilon-E_n) \nonumber
       \\                &= -\frac{1}{\pi} \Im\Tr \hat{G}^R(\varepsilon)
     \end{align}
     The quantity
     \begin{equation}\label{eq_dos_from_Gr}
       \rho_i(\varepsilon) = -\frac{1}{\pi} \Im \hat{G}_{i,i}^R(\varepsilon)
     \end{equation}
     is the local density of states. It is a relevant quantity in particular when there is no translational invariance. That is what is measured by scanning tunneling microscopes.

  \subsection{Perturbation expansion}\label{sec_pert_expansion}
\begin{figure}
     \centering
     \includegraphics[width=0.9\linewidth]{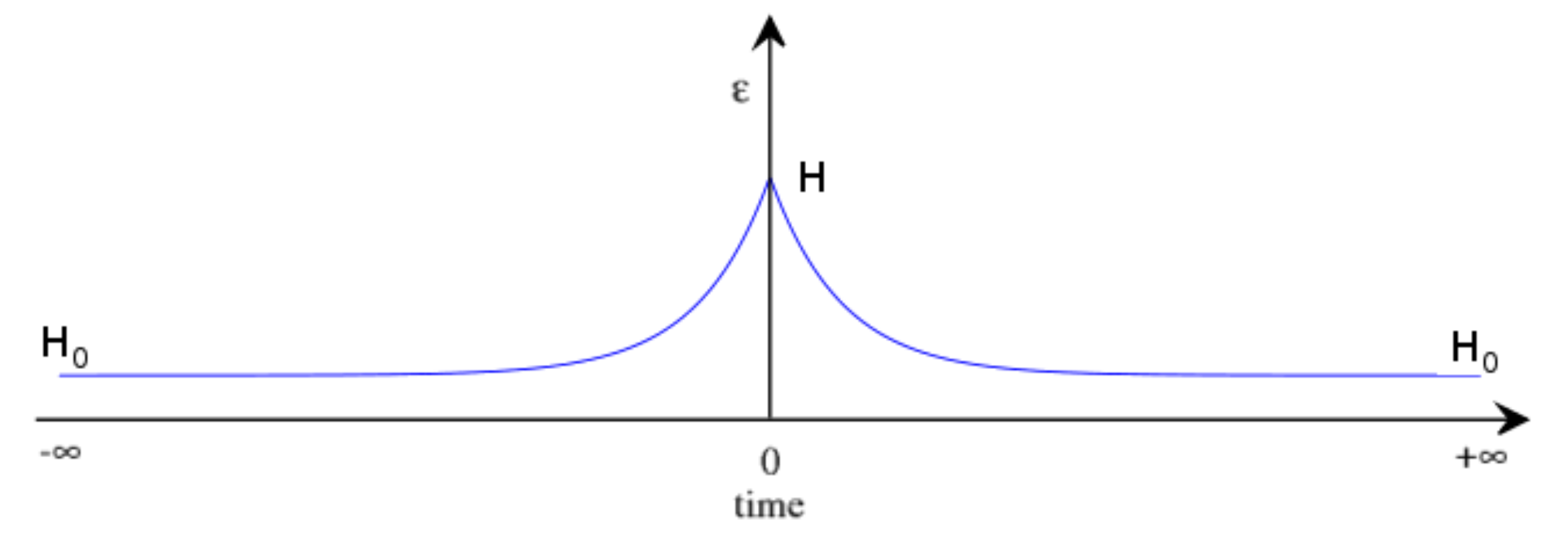}
     \caption{Starting from $t=-\infty$, the perturbation is adiabatically  switch-on until $t=0$, time at which the system is described by the full Hamiltonian $H$. Then the perturbation is adiabatically  switch-off until $t=+\infty$, and the system go back in its original state $\ket{\psi_0}$}\label{fig_adiabatic}
    \end{figure}

    The aim of the perturbation expansion is to generate exact eigenstates of the interacting system, described by $\hat{H}$, from the eigenstates of the non-interacting system, described by $\hat{H}_0$,
    as we suppose to know all about the latter, and from the perturbation $\hat{V}$. In other words, we want to express the Green function of the interacting system 
    \begin{equation*}
      \i\greennoup{ij\sigma}(t-t') = \Braket{\psi_0^H | \hat{T} \left[\hat{c}_{i\sigma}(t) \hat{c}_{j\sigma}^{\dagger}(t')\right] | \psi_0^H}
    \end{equation*} 
    in terms of Green functions of the non-interacting system and $\hat{V}$. In order to do that we rewrite the full Hamiltonian $\hat{H} = \hat{H}_0 +\hat{V}$ and introduce a new time dependent Hamiltonian:
    \begin{equation}
      \hat{H}(t) = \hat{H}_0 + \e^{-\epsilon |t|} \hat{V}
    \end{equation} 
    where $\epsilon$ is a small quantity which allows to switch-on and switch-off the perturbation adiabatically, that is very slowly.\footnote{It should be reminded that originally an adiabatic transformation refers to a thermodynamic transformation with no heat exchange. Roughly speaking, the slow temporal evolution is supposed to keep the state evolution from $\hat{H}_0$ to $\hat{H}$ in a one-to-one correspondence that should mimic an adiabatic transformation.} At very large times, both in the past and in the future, the Hamiltonian reduces to $\hat{H}_0$. At time $t=0$, $\hat{H}$ describes the full interacting-system. This is described in Fig.~\ref{fig_adiabatic}. If the process is slow enough, then any result is independent of $\epsilon$ (adiabatic theorem \cite[Chap. 17, Sec. II.8]{Messiah}).

    As we are interested in a time dependent problem that depends on the small quantity $\epsilon$, we shall use the interaction picture (Eq.~\ref{eq_evolution_inter_pict}):
    \begin{equation}
      \ket{\psi_I(t)} = \hat{U}_{\epsilon}(t,t_0) \ket{\psi_I(t_0)}
    \end{equation}
    
    In the limit $t_0 \rightarrow -\infty$, the Schrödinger-picture state reduces to:
    \begin{equation}
      \Ket{\psi_S(t_0)} = \e^{-\i E_0 t_0/\hbar} \Ket{\phi_0}
    \end{equation}
    where $\Ket{\phi_0}$ is a time-independent eigenstate of the unperturbed Hamiltonian $\hat{H}_0$ with eigenvalue $E_0$. Moreover, as said above, at time $t=0$ all three pictures coincide. Then
    \begin{equation}
       \Ket{\psi_H} =  \Ket{\psi_I(t=0)} = \hat{U}_{\epsilon}(0,-\infty)\Ket{\phi_0}
    \end{equation}
    This last equation is a very important result as it expresses an exact eigenstate of the interacting system in terms of an eigenstate of the non-interacting one. 
    
    Coming back to the definition~\eqref{eq_def_GF} and using the integral form~\eqref{eq_U_int_form}, we have:
    \begin{multline}\label{eq_expansion_of_T}
      \Braket{\psi_0^H | \hat{T} \left[\hat{c}_{i\sigma}(t) \hat{c}_{j\sigma}^{\dagger}(t')\right] | \psi_0^H} = \\
      \sum_{n=0}^{\infty} \frac{(-\i)^n}{n!}\int\D t_1 \dots \int\D t_n \Braket{\phi_0 | \hat{T}\left[ \hat{V}(t_1)\dots \hat{V}(t_n)\ \hat{c}_{i\sigma}(t) \hat{c}_{j\sigma}^{\dagger}(t')\right] | \phi_0}
    \end{multline}
    for $\epsilon\rightarrow 0$. This equation could seem quite complicated due to the time-ordering operator.\footnote{We have even oversimplified it (see \cite{fetter}), as we have assumed that the normalization of $\braket{\psi_{0}^{H}|\psi_{0}^{H}}=1$ implies that of $\ket{\phi_0}$ and this is not generally true. Actually the denominator of Eq.~\ref{eq_def_GF} should also be expanded in the interaction representation, leading to the elimination of the so-called non-connected diagrams of Eq.~\ref{eq_expansion_of_T}. In what follows, we suppose it to be done already.} However, G. C. Wick has built a theorem~\cite{Wick-PhysRev.80.268} which allows to write down such time ordering product by pair, if there are the same number of creation and annihilation operator. Here, as our Hamiltonian is quadratic, we can use this powerful theorem. The proof of it is quite tedious and can be found, for example, in Ref.~\cite[Sec.~8]{fetter}. Here, we will just see how we can use it:    
    \begin{description}
      \item[$n=0$:] $$\hat{T}\left[ \hat{V}(t_1)\dots \hat{V}(t_n)\ \hat{c}_{i\sigma}(t) \hat{c}_{j\sigma}^{\dagger}(t')\right] = \hat{T}\left[\hat{c}_{i\sigma}(t) \hat{c}_{j\sigma}^{\dagger}(t')\right] = \hat{g}_{ij\sigma}^{(0)}$$
      \item[$n=1$:] 
	    \begin{align*}
	      \hat{T}\left[ \hat{V}(t_1)\dots \hat{V}(t_n)\ \hat{c}_{i\sigma}(t) \hat{c}_{j\sigma}^{\dagger}(t')\right] &= 
	             \hat{T}\left[ \hat{V}(t_1)\ \hat{c}_{i\sigma}(t) \hat{c}_{j\sigma}^{\dagger}(t')\right] \\
	       &= \hat{T}\left[\hat{c}_{i\sigma}(t) \hat{c}_{j\sigma}^{\dagger}(t')\right]\hat{V}(t_1) \hat{T}\left[\hat{c}_{i\sigma}(t) \hat{c}_{j\sigma}^{\dagger}(t')\right] \\
	       &= \hat{g}_{ij\sigma}^{(0)} \hat{V} \hat{g}_{ij\sigma}^{(0)}
	    \end{align*}
      \item[$n=2$:] 
	    \begin{multline*}
	      \hat{T}\left[ \hat{V}(t_1)\hat{V}(t_2)\ \hat{c}_{i\sigma}(t) \hat{c}_{j\sigma}^{\dagger}(t')\right] \\
	       = \hat{T}\left[\hat{c}_{i\sigma}(t) \hat{c}_{j\sigma}^{\dagger}(t')\right]\hat{V}(t_1) \hat{T}\left[\hat{c}_{i\sigma}(t) \hat{c}_{j\sigma}^{\dagger}(t')\right] \hat{V}(t_1) \hat{T}\left[\hat{c}_{i\sigma}(t) \hat{c}_{j\sigma}^{\dagger}(t')\right]\\
	       = \hat{g}_{ij\sigma}^{(0)} \hat{V} \hat{g}_{ij\sigma}^{(0)} \hat{V} \hat{g}_{ij\sigma}^{(0)}
	    \end{multline*}
    \end{description}
    Iterating up to infinity we obtain Dyson's equation \ref{eq_dyson_169} that links the full Green function (perturbed) with the unperturbed one and with the perturbation $\hat{V}$ in the following form
    \begin{align}\label{eq_dyson_169}
      \greennoup{ij\sigma}&= \greenzero{ij\sigma} + \greenzero{ij\sigma} \hat{V} \greenzero{ij\sigma} + \greenzero{ij\sigma} \hat{V} \greenzero{ij\sigma} \hat{V} \greenzero{ij\sigma} +\dots \nonumber\\
                          &= \greenzero{ij\sigma} + \greenzero{ij\sigma} \hat{V} \greennoup{ij\sigma}  \nonumber
      \\                  &= \left[{1-\greenzero{ij\sigma} V}\right]^{-1}\greenzero{ij\sigma}
    \end{align}
    whose integral form is
     \begin{equation}\label{eq_dyson_eq}
      \hat{G}_{ij\sigma}(\mathit{t_1,t_1'}) = \greenzero{ij\sigma}({t_1,t_1'}) + \int \D^4 {t_2}\, \D^4 {t_3} \  \greenzero{ij\sigma}({t_1,t_2}) \hat{V}({t_2,t_3}) \greennoup{ij\sigma}({t_3,t_1'}) 
    \end{equation} 
    
    There is a formally simpler approach to derive Dyson equation whose simplicity however hides some important features that we shall use in section \ref{sec_nept_gf}. This approach is shown in Appendix~\ref{appendix_alternate_gf}.

    The Dyson equation is particularly useful because  even if we cannot invert the large matrix $\hat{H}$ to compute $\hat{G}^R$, we have re-expressed it in term of the unperturbed Green function $\greenzero{ij\sigma}$ and the perturbation $\hat{V}$ that are usually easier to evaluate. 
    
    In practice, if we want to know the propagator at a given order, we just stop the above expansion at this order. However, this could lead to misleading results. Moreover, we should be sure that the series \ref{eq_dyson_169} converges, which is not always the case, depending on the perturbation. 
    
    We have seen that this formalism is based on the fact that the perturbation is time independent. However, if it is not the case, then, one has to use another theory called non-equilibrium perturbation theory based on non-equilibrium Green-function (NEGF), or Keldysh Green function, as explained in Sec.~\ref{sec_nept_gf}.

    \subsection{Resolution through equation of motion}\label{sec_resol_through_eom_chapt1}
	
    The Green functions can be also obtained by solving an equation of motion without using the time evolution operator, thereby, avoiding to pass through the interaction picture. 
    This can present some advantages, as we shall see in Chap.~\ref{chapt_NEPT}. We start with the time derivative of Heisenberg operators of the Green function:
    \begin{align}
	\i\hbar\, \partial_t \hat{G}_{ij\sigma}(t-t') &= \left\langle\partial_t\hat{T}\left[\hat{c}_{i\sigma}(t)\hat{c}^{\dagger}_{j\sigma}(t')\right]\right\rangle  + \left\langle\hat{T}\left[\big(\partial_t \hat{c}_{i\sigma}(t)\big)\hat{c}^{\dagger}_{j\sigma}(t')\right]\right\rangle  \label{derivee_green}
    \end{align}
    As the time-ordering operator $\hat{T}$ can be represented by the step function 
    \begin{align}
      \theta(t-t')  \hat{c}_{i\sigma}(t)\hat{c}^{\dagger}_{j\sigma}(t') \quad \mbox{for } t>t' \\
      \theta(t'-t)  \hat{c}^{\dagger}_{j\sigma}(t')\hat{c}_{i\sigma}(t) \quad \mbox{for } t'>t
    \end{align}
    its time derivative is given by a dirac $\delta$-function and one obtains:
    \begin{align}
      \i\hbar\, \partial_t \hat{G}_{ij\sigma}(t-t')=	& \left[\frac{\partial}{\partial t}\theta(t-t')\right] \hat{c}_{i\sigma}(t)\hat{c}^{\dagger}_{j\sigma}(t') - \left[\frac{\partial}{\partial t}\theta(t'-t)\right] \hat{c}^{\dagger}_{j\sigma}(t')\hat{c}_{i\sigma}(t) \nonumber \\
      &+ \left\langle\hat{T}\left[\big(\partial_t \hat{c}_{i\sigma}(t)\big)\hat{c}^{\dagger}_{j\sigma}(t')\right]\right\rangle   \nonumber \\
      =& \delta(t-t') 
      \underbrace{\left[\hat{c}_{i\sigma}(t)\hat{c}^{\dagger}_{j\sigma}(t')+\hat{c}^{\dagger}_{j\sigma}(t')\hat{c}_{i\sigma}(t)\right]
      }_{\left\{\hat{c}_{i\sigma},\hat{c}^{\dagger}_{j\sigma'}\right\}
      =\delta_{ij}\delta_{\sigma\sigma'}
      }  
      +\left\langle\hat{T}\Big[
      \underbrace{\partial_t \hat{c}_{i\sigma}(t)
      }_{ \scriptsize{\lefteqn{=\frac{1}{\i\hbar}\left[\hat{c}_{i\sigma},H\right] \mbox{cf. Eq.~\eqref{eq_heisenberg_eq_motion}}}}}
      \hat{c}^{\dagger}_{j\sigma}(t')
      \Big]\right\rangle
    \end{align}
    The time derivative of the Heisenberg operator $\hat{c}_{i\sigma}(t)$ is obtained through Eq.~\eqref{eq_heisenberg_eq_motion}.
    
    Consider a simple tight binding Hamiltonian 
    \begin{equation}
      \hat{H}= \sum_{ij} t_{ij} \hat{c}^\dagger_{i}\hat{c}_{j} \qquad \mbox{with } t_{ii}=\varepsilon^{(i)}
    \end{equation}
    Using Eq.~\eqref{anti-commutator_3op}, the commutator $\left[\hat{c}_{i\sigma},\hat{H}\right]$ writes
    \begin{align}
      [\hat{c}_{i\sigma},\hat{H}_0]			&=\left[\hat{c}_{i\sigma},\hat{c}^\dagger_{l\sigma}\hat{c}_{m\sigma}\right] = \sum_{lm} \delta_{il} \hat{c}_{m\sigma} t_{lm} \nonumber
      \\								&= \boxed{\sum_m t_{im} \hat{c}_{m\sigma}},
    \end{align}
    Finally, the equation of motion is
    \begin{equation}\label{eq_general_eq_of_motion}\color{red}\fbox{$
      \i\hbar\, \partial_t \hat{G}_{ij\sigma}(t-t') =\delta(t-t') \delta_{ij} + \sum_m t_{im} \hat{G}_{mj\sigma}(t-t')$}
    \end{equation}
    As we are interested by the energy spectrum, we have to use the Fourier transform in time domain of the Green function:
    \begin{align}
	  \hat{G}_{ij\sigma}(\omega) &=  \int_{-\infty}^{+\infty} \hat{G}_{ij\sigma}(t) \e^{\i\omega (t-t')} \D t \nonumber
       \\ \hat{G}_{ij\sigma}(t)      &= \frac{1}{{2\pi}} \int_{-\infty}^{+\infty} \hat{G}_{ij\sigma}(\omega) \e^{-\i\omega (t-t')} \D \omega
    \end{align}
    whose time derivative is
    \begin{align}
      \i\hbar\partial_t \hat{G}_{ij\sigma}(t) &= \frac{\i\hbar}{{2\pi}} \int_{-\infty}^{+\infty} -i\omega \hat{G}_{ij\sigma}(\omega)  \e^{-\i\omega (t-t')} \D \omega \nonumber
      \\                     &= \frac{\hbar\omega}{{2\pi}} \int_{-\infty}^{+\infty} \hat{G}_{ij\sigma}(\omega)  \e^{-\i\omega (t-t')} \D \omega
    \end{align}
    And, the Fourier transform of Dirac delta function is
    \begin{equation}
      \delta(t-t') = \int_{-\infty}^{+\infty} \frac{\D \omega}{{2\pi}} \e^{-\i\omega (t-t')}
    \end{equation}
    Hence, the Fourier transform of Eq.~\eqref{eq_general_eq_of_motion} is
    \begin{multline}
     \int_{-\infty}^{+\infty}\frac{\D \omega}{{2\pi}} \i\hbar\partial_t \left(\hat{G}_{ij\sigma}(\omega) \e^{-\i\omega (t-t')}\right) = \\ \delta_{ij} \int_{-\infty}^{+\infty} \frac{\D \omega}{{2\pi}} \e^{-\i\omega (t-t')} + \sum_m t_{im}  \int_{-\infty}^{+\infty} \frac{\D \omega}{{2\pi}} \hat{G}_{mj\sigma}(\omega) \e^{-\i\omega (t-t')}
    \end{multline}
    Factorizing and regrouping:
    \begin{equation}
          \frac{1}{{2\pi}}\int_{-\infty}^{+\infty} \D \omega\ \e^{-\i\omega (t-t')} \left[ \hbar\omega \hat{G}_{ij\sigma}(\omega) -\delta_{ij}-\sum_m t_{im} \hat{G}_{mj\sigma}(\omega) \right] =0
    \end{equation}
    That implies, because of the completeness of the complex-exponential basis:
    \begin{equation}\label{eq_eom_fourier_chapt1}
       \hbar\omega \hat{G}_{ij\sigma}(\omega) = \delta_{ij} + \sum_m t_{im} \hat{G}_{mj\sigma}(\omega)
    \end{equation}
    If the system is infinite, we can also use the Fourier-transform of the Green functions and the hopping matrices in space domain in order to diagonalize Eq.~\eqref{eq_eom_fourier_chapt1}. In this case, we would obtain:
    \begin{align}
    \hbar\omega \hat{G}_{\vec{k}\sigma}(\omega) &= 1 + t_{\vec{k}} \hat{G}_{\vec{k}\sigma}(\omega) \nonumber\\
      \left(\hbar\omega - t_{\vec{k}}\right) \hat{G}_{\vec{k}\sigma}(\omega) &= 1 
    \end{align}
    where $t_{\vec{k}}\equiv \frac{1}{N}\sum_{{i,j}} t_{i,j} \e^{\i \vec{k}\cdot(\vec{R}_{i}-\vec{R}_{j})}$.
    In the case of metal thin-films, we lose the full 3D periodicity, so we shall not perform the Fourier transform in one of the three directions. Therefore we have to solve directly the equation system \ref{eq_eom_fourier_chapt1}, as shown in the Sec.~\ref{sec_finite_through_eom}.

  \section{Non-equilibrium perturbation-theory and Green functions}\label{sec_nept_gf}


  In non-equilibrium problem, there is no guarantee that the system returns to its initial state at asymptotically large times: this is a fundamental condition to develop perturbation theory as we have seen in sec.~\ref{sec_pert_expansion}. Therefore, perturbation theory cannot be applied along the same lines: any references to asymptotically large times should be avoided in the non-equilibrium theory. This implies, as we shall see below, that a different approach has to be looked for in the adiabatic introduction of the perturbation. Such an approach leads to a new contour for time integration, but, in spite of some conceptual complications, several formal aspects of non-equilibrium perturbation theory (like Dyson's equation) keep an equivalent structure as in equilibrium theory.
  
  The non-equilibrium is formulated as follows. We consider a system evolving under the Hamiltonian
  \begin{equation}
    \hat{H}(t)= \hat{h} + \hat{H}'(t)
  \end{equation} 
  where $h$ is the time independent part of the Hamiltonian, and it can be split in two parts: $\hat{h} = \hat{H}_0 + \hat{H}_i$, where $\hat{H}_0$ is ``simple'' (it can be diagonalized, and hence, Wick's theorem applies) and $\hat{H}_i$ may contain the many body aspects of the problem, and hence requires a special treatment. $ \hat{H}'(t)$ is the external time-dependent perturbation.

  \subsection{Contour-ordered operator}

    \begin{figure}
    \begin{center}{
    \includegraphics[width=0.97\linewidth]{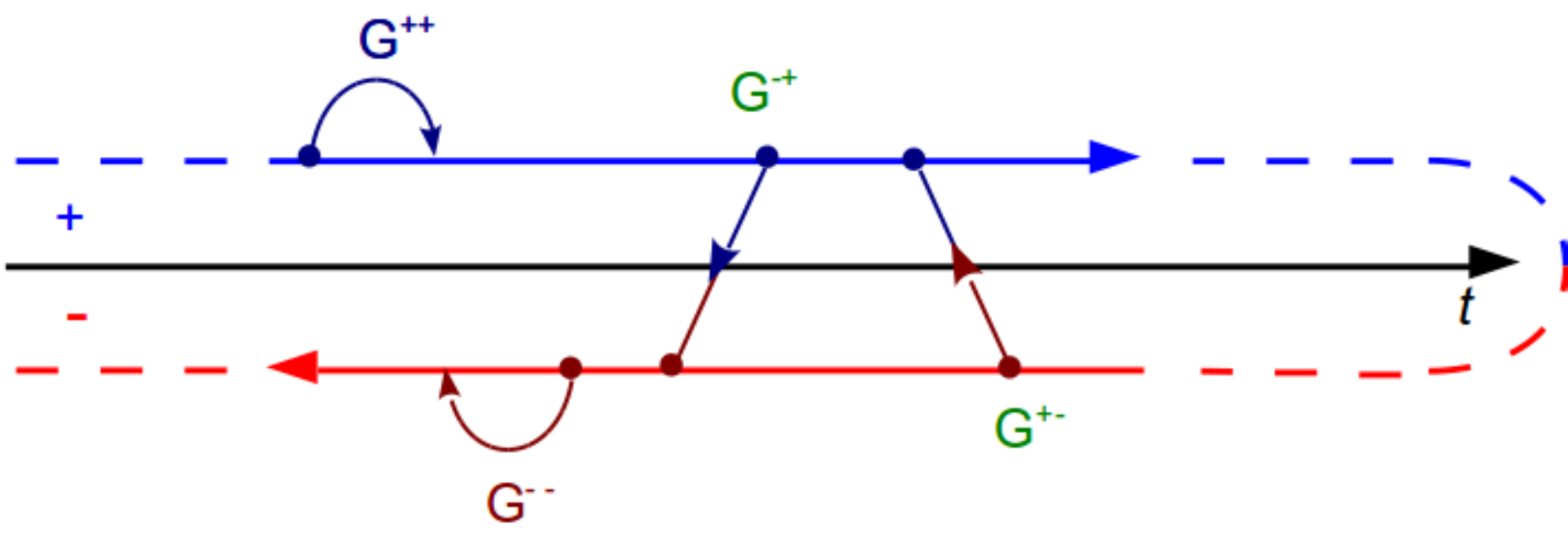}}
    \caption{Contour C. The upper branch is called ``positive branch'' and the lower ``negative branch''. The $\pm$ notation follows~\cite{Pedro-ProgSurfS_66_2001} which is the same as Lifshitz notations~\cite{Lifshitz-vol10_kinetics}, except that the positive and negative branches are exchanged. See Table~\ref{tab_negf_notation} for a summary of the different notations.}
    \label{fig_keldysh_contour}
    \end{center}
    \end{figure}


    As shown in section~\ref{sec_schrodinger_heisenberg_picture} a general operator in Heisenberg picture can be written in terms of interaction-picture operators:
    \begin{equation}\label{eq_Oh_heisenberg_1}
      \hat{O}_H(t) = \hat{U}_h^{\dagger}(t,t_0) \hat{O}_{h}(t)  \hat{U}_h(t,t_0) 
    \end{equation}
    with the unitary operator $\hat{U}_h(t,t_0)$ that determines the state vector at time $t$ in terms of the state vector at time $t_0$:
    \begin{equation}
      \hat{U}_h(t,t_0) = \hat{T}\left\{ \exp\left[ -\i \int_{t_0}^{t} \D t' \hat{H}'_h(t') \right] \right\}
    \end{equation} 
    $\hat{T}$ is the time ordering operator which arranges the latest times to left. $\hat{H}'_h(t)$ is the interaction picture of $\hat{H}'(t)$         
    \begin{equation}\label{eq_Hprime_h}
      \hat{H}'_h(t) = \e^{\i h (t-t_0)} \hat{H}'(t) \e^{-\i h (t-t_0)}
    \end{equation} 

    The following property $\hat{U}_h(t_0,t)=\hat{U}_h^{\dagger}(t,t_0)$ of the evolution operator, allows to rewrite equation \eqref{eq_Oh_heisenberg_1} with a contour-ordered operator as follows:
    \begin{equation}\label{eq_Oh_heisenberg_2}
      \hat{O}_{\hat{H}}(t) = \hat{T}_{C}\left\{ \exp\left[ -\i \int_{C} \D \tau \hat{H}'_h(\tau) \right]\hat{O}_h(t) \right\}
    \end{equation} 
    Where the contour $C$ is represented in Fig.~\ref{fig_keldysh_contour}.

    \begin{demo}[Eq.~\eqref{eq_Oh_heisenberg_1}=Eq.~\eqref{eq_Oh_heisenberg_2}]
    The proof of equivalence Eq.~\eqref{eq_Oh_heisenberg_1} and Eq.~\eqref{eq_Oh_heisenberg_2} is done by:
	\begin{multline}\label{equivalence_deb}
	  \hat{T}_{C}\left\{ \exp\left[ -\i \int_{C} \D \tau \hat{H}'_h(\tau) \right]\hat{O}_h(t) \right\} = \\
		    \sum_{n=0}^{\infty} \frac{(-\i)^n}{n!}\int_C\D\tau_1 \dots \int_C\D\tau_n \hat{T}_{C}\left[ \hat{H}'_h(\tau_1)\dots \hat{H}'_h(\tau_n) \hat{O}_h(t)\right]
	\end{multline} 
	Divide the contour in two branches 
	\begin{equation}
	  \int_C=\int_\rightarrow+\int_\leftarrow
	\end{equation} 
	where $\int_\rightarrow$ goes from $-\infty$ to $+\infty$ and $\int_\leftarrow$ from $+\infty$ to $-\infty$. Replacing contour integral by this sum generates $2^n$ terms. For the demonstration, let's consider one of them:
	\begin{multline}
	  \int_\rightarrow \D\tau_1 \int_\rightarrow\D\tau_2 \int_\leftarrow\D\tau_3 \dots \int_\leftarrow\D\tau_n \hat{T}_{C}\left[ \hat{H}'_h(\tau_1)\dots \hat{H}'_h(\tau_n) \hat{O}_h(t)\right] \\
	    = \int_\leftarrow\D\tau_3 \dots \int_\leftarrow\D\tau_n \hat{T}_\leftarrow \left[ \hat{H}'_h(\tau_3)\dots \hat{H}'_h(\tau_n) \hat{O}_h(t)\right] 
	      \\\times
	     \int_\rightarrow\D\tau_1 \int_\rightarrow\D\tau_2 \hat{T}_\leftarrow \left[ \hat{H}'_h(\tau_1) \hat{H}'_h(\tau_2)\right] 
	\end{multline}
	There are $\binom{n}{m}=n!/[m!(n-m)!]$ combinations of $m$ integrals from $-\infty$ to $+\infty$ (in the above example $m=2$) amongst the $2^n$ terms generated terms. All these terms give the same contribution. Thus we can write
	\begin{multline}
	  \int_C\D\tau_1 \dots \int_C\D\tau_n \hat{T}_{C}\left[ \hat{H}'_h(\tau_1)\dots \hat{H}'_h(\tau_n) \hat{O}_h(t)\right] =\\
	  \sum_{n=0}^{\infty} \frac{n!}{m!(n-m)!} 
	  \times \int_\leftarrow\D\tau_{m+1} \dots \int_\leftarrow\D\tau_n\ \hat{T}_\leftarrow \left[ \hat{H}'_h(\tau_{m+1})\dots \hat{H}'_h(\tau_n) \hat{O}_h(t)\right] 
	  \\ \times \int_\rightarrow\D\tau_1 \dots \int_\rightarrow\D\tau_m\ \hat{T}_\leftarrow \left[ \hat{H}'_h(\tau_1) \dots \hat{H}'_h(\tau_m)\right] 
	\end{multline}
	Replacing $n-m=k$,  both $k$ and $m$ can be summed from 0 to $\infty$ as long as their sum equals $n$. This is achieved by inserting a Kronecker delta:
	\begin{multline}
	  \sum_{m,k=0}^{\infty} \frac{n!}{m!k!}\delta_{n,k+m}\left\{ \int_\leftarrow\D\tau_{1} \dots \int_\leftarrow\D\tau_k \hat{T}_\leftarrow \left[ \hat{H}'_h(\tau_{1})\dots \hat{H}'_h(\tau_k)\right] \right\}\hat{O}_h(t) \\
	  \times\left\{ \int_\rightarrow\D\tau_{1} \dots \int_\rightarrow\D\tau_m \hat{T}_\rightarrow \left[ \hat{H}'_h(\tau_{1})\dots \hat{H}'_h(\tau_m) \right]\right\}
	\end{multline}
	Going back to eq. (\eqref{equivalence_deb}), the $n$-sum is (due to the factor $\delta_{n,k+m}$ and simplifying the $n!$ terms):
	\begin{multline}
	  \hat{T}_{C}\left\{ \exp\left[ -\i \int_{C} \D \tau \hat{H}'_h(\tau) \right]\hat{O}_h(t) \right\}
	  \\= \sum_{k=0}^{\infty}\frac{(-\i)^k}{k!} \int_\leftarrow\D\tau_{1} \dots \int_\leftarrow\D\tau_k \hat{T}_\leftarrow \left[ \hat{H}'_h(\tau_{1})\dots \hat{H}'_h(\tau_k)\right] \hat{O}_h(t) 
	  \\ \times \sum_{m=0}^{\infty}\frac{(-\i)^m}{m!} \int_\rightarrow\D\tau_{1} \dots \int_\rightarrow\D\tau_m \hat{T}_\rightarrow \left[ \hat{H}'_h(\tau_{1})\dots \hat{H}'_h(\tau_m)\right] \hat{O}_h(t)
	\end{multline}
	The factors multiplying $\hat{O}_h(t)$ are $\hat{U}^{\dagger}(t,t_0)$ and  $\hat{U}(t,t_0)$ of Eq.~\eqref{eq_Oh_heisenberg_1}. This demonstrates the equivalence between Eq.~\eqref{eq_Oh_heisenberg_1} and Eq.~\eqref{eq_Oh_heisenberg_2}.

	\end{demo}

    This equivalence shows that the contour-ordering operator is a strong formal tool which allows to develop the non-equilibrium theory along lines parallel to the equilibrium theory. 
    The main difference is that instead of the evolution operator going from $t=-\infty$ to $t=+\infty$, one is forced to consider the evolution operator along the time path depicted in Fig.~\ref{fig_keldysh_contour}. The formal complication introduced by the contour $C$ is that instead of one Green function as in the equilibrium theory, we are forced to introduce four Green functions, as detailed below.

    Similarly to equilibrium theory, a contour-ordered can be defined as:
    \begin{equation}
      \hat{G}(x_1t_1,x_1't_1') \equiv \hat{G}(\mathit{1,1'})\equiv -\i \Braket{\hat{T}_C \left[\annihilation{H}(\mathit{1})\creation{H}(\mathit{1'})\right]}
    \end{equation} 
    the subscript $H$, as before, means that field operators are in Heisenberg picture, and $(\mathit{1})$ is a shorthand notation commonly used for $(x_1,t_1)$.  The contour runs on the real axis from $+\infty$ to $-\infty$.
    This operator works as usual: operators with time labels that occur later on the contour are arranged to the left.

    Contour-ordered Green function is the time ordered Green function of non-equilibrium theory, and possesses as well a perturbation expansion based on Wick's theorem \cite{fetter}. However, as the time labels lie on the contour with two branches, one has also to keep trace of the branch. As sketched above, there are four possibilities which are depicted in Fig.~\ref{fig_keldysh_contour}

    \begin{equation}\label{eq_negf}
      \hat{G}(\mathit{1,1'}) = \left\{\begin{array}{ll}
			\hat{G}^{++}(\mathit{1,1'}) & t_1,t_1'\in C^+ \\
			\hat{G}^{--}(\mathit{1,1'}) & t_1,t_1'\in C^- \\
			\hat{G}^{+-}(\mathit{1,1'}) & t_1,\in C^+,t_1'\in C^- \\
			\hat{G}^{-+}(\mathit{1,1'}) & t_1,\in C^-,t_1'\in C^+ \\
		      \end{array}
    \right.
    \end{equation}
    $\hat{G}^{++}$ and $\hat{G}^{--}$ are respectively the causal (or time ordered) and anti-causal Green functions of non-equilibrium problem:
    \begin{align}
      \hat{G}^{++}(\mathit{1,1'}) &= -\i \Braket{T[\annihilation{H}(\mathit{1})\creation{H}(\mathit{1'})]} \nonumber
      \\           &= -\i \theta(t_{1}-t_{1'}) \Braket{\annihilation{H}(\mathit{1})\creation{H}(\mathit{1'})} + \i \theta(t_{1'}-t_{1}) \Braket{\creation{H}(\mathit{1'})\annihilation{H}(\mathit{1})}
    \\\hat{G}^{--}(\mathit{1,1'}) &= -\i \theta(t_{1'}-t_{1}) \Braket{\annihilation{H}(\mathit{1})\creation{H}(\mathit{1'})} + \i \theta(t_{1}-t_{1'}) \Braket{\creation{H}(\mathit{1'})\annihilation{H}(\mathit{1})}
    \end{align}
    $\hat{G}^{+-}$ and $\hat{G}^{-+}$ are respectively the lesser and greater Green functions of the non-equilibrium problem:
    \begin{align}
      \hat{G}^{+-}(\mathit{1,1'}) &= +\i \Braket{\creation{H}(\mathit{1'})\annihilation{H}(\mathit{1})}
    \\\hat{G}^{-+}(\mathit{1,1'}) &= -\i \Braket{\annihilation{H}(\mathit{1})\creation{H}(\mathit{1'})}
    \end{align}
      Other notations that can be found in the literature are summarized in table~\ref{tab_negf_notation}.
      \begin{table}
	\begin{center}
	  \begin{tabular}[c]{||l|llll||}
	  \hline
	    our notations (as \cite{Pedro-ProgSurfS_66_2001})      & $\hat{G}^{++}(1,1')$    & $\hat{G}^{--}(1,1')$            & $\hat{G}^{-+}(1,1')$    & $\hat{G}^{+-}(1,1')$   \\\hline
	    de Andres' notations \cite{Pedro-ProgSurfS_66_2001}& $\hat{G}^{++}(1,1')$    & $\hat{G}^{--}(1,1')$            & $\hat{G}^{-+}(1,1')$    & $\hat{G}^{+-}(1,1')$   \\\hline
	    Lifshitz' notation \cite{Lifshitz-vol10_kinetics} & $\hat{G}^{--}(1,1')$    & $\hat{G}^{++}(1,1')$            & $\hat{G}^{+-}(1,1')$    & $\hat{G}^{-+}(1,1')$   \\\hline
	    Jauho's notations  \cite{jauho_book}  & $\hat{G}_{c}(1,1')$     & $\hat{G}_{\tilde{c}}(1,1')$     & $\hat{G}^{>}(1,1')$     & $\hat{G}^{<}(1,1')$    \\\hline
	    Caroli's notations \cite{Caroli-JPhysC4_1971}   & $\hat{G}^{c}(1_+,1_+')$ & $\tilde{G}^{c}(1_-,1_-')$ & $\hat{G}^{-}(1_-,1_+')$ & $\hat{G}^{+}(1_+,1_-')$\\\hline
	    Keldysh' notations \cite{keldysh} & $\hat{G}^{c}(1_+,1_+')$ & $\tilde{G}^{c}(1_-,1_-')$ & $\hat{G}^{-}(1_-,1_+')$ & $\hat{G}^{+}(1_+,1_-')$\\\hline
	  \end{tabular}
	  \caption{Summary of notations that can be found in the literature. We choose to follow Pedro de Andres and Fernando Flores' definitions, based on Lifshitz ones. We believe that our notations are the most intuitive: you can easily find them by looking at the picture of fig.~\ref{fig_keldysh_contour}. $\hat{G}^{++}$ connects the positive branch (from $-\infty$ to $+\infty$) with itself, $\hat{G}^{+-}$ connects the negative branch (from $+\infty$ to $-\infty$) with the positive one. Though Lifshitz used the opposite definition of the positive and the negative branches, we have chosen to set the positive branch as the one where time goes in the causal direction.}\label{tab_negf_notation}
	\end{center}
      \end{table}

    By analogy with equilibrium theory, these Green functions are not all independent, and for example: $\hat{G}^{++}+\hat{G}^{--}=\hat{G}^{+-}+\hat{G}^{-+}$. In our case, we shall focus on the ``lesser'' Green function $\hat{G}^{+-}$ because, as we shall see in chapter \ref{chapt_NEPT}, it is directly related to the BEEM current. Moreover, it is possible to link these Green functions with the retarded and advanced Green functions, defined formally in the same way as in the equilibrium case in sec.~\ref{sec_ret_adv_gf}:
    \begin{align}
      \advanced{}{\mathit{1,1'}} &= \i \theta(t_{1'}-t_{1}) \Braket{\left\{\annihilation{H}(\mathit{1})\creation{H}(\mathit{1'})\right\}} \nonumber
      \\                &= \i \theta(t_{1'}-t_{1}) \left[  \hat{G}^{+-}(\mathit{1,1'})-\hat{G}^{-+}(\mathit{1,1'}) \right]
    \\\retarded{}{\mathit{1,1'}} &= -\i\theta(t_{1}-t_{1'}) \Braket{\left\{\annihilation{H}(\mathit{1})\creation{H}(\mathit{1'})\right\}} \nonumber
    \\                  &= -\i\theta(t_{1}-t_{1'}) \left[  \hat{G}^{-+}(\mathit{1,1'})-\hat{G}^{+-}(\mathit{1,1'}) \right] \label{eq_Gr_neq}
    \end{align}
    where curly brackets denote the anticommutator. This leads to $\hat{G}^{R}-\hat{G}^{A}=\hat{G}^{-+}-\hat{G}^{+-}$. The reason why we introduce $\hat{G}^{R}$ and $\hat{G}^{A}$ is that, as we shall see in Eq.~\eqref{eq_keldysh}, it is possible to express $G^{+-}$ (and therefore the current in the metal layer) in terms of retarded and advanced Green functions.

    These new Green functions have to be transformed in such a way that Wick's theorem can be applied. The first step is to repeat the transformation from $H$-dependence to $h$-dependence leading to Eq.~\eqref{eq_Oh_heisenberg_2}:
    \begin{equation}\label{eq_G_after_1st_transformation}
      \hat{G}(\mathit{1,1'})=-\i \Braket{\hat{T}_C\left[S^{C}_{H}\annihilation{h}({1})\creation{h}({1'})\right]}
    \end{equation} 
    with
    \begin{equation}
      S^{C}_{H}= \exp\left[ -\i \int_{C} \D \tau \hat{H}'_h(\tau) \right]
    \end{equation}

    The important features of this results are: it is exact, all time dependence is ruled by the solvable $\hat{H}_0$, and our quadratic Hamiltonian allows to use Wick's theorem~\cite{Wick-PhysRev.80.268}. 
    The general proof of that is quite cumbersome to obtain, but it is nicely described in Ref.~\cite{rammer-RevModPhys.58.323}. 
 
    To summarize, equilibrium and non equilibrium theory are, formally, structurally equivalent. The only (fundamental) difference lays in the replacement of real axis integrals by contour integrals. As these kind of integrals are rather impractical, they have to be replaced by real time ones. This process introduced by Kadanoff and Baym \cite{kadanoff_baym} has been generalized by Langreth and is now known as Langreth' theorem \cite{langreth}. It is presented in the next section.

  \subsection{Langreth theorem}

\begin{figure}
    \begin{center}{
    \includegraphics[width=0.50\linewidth]{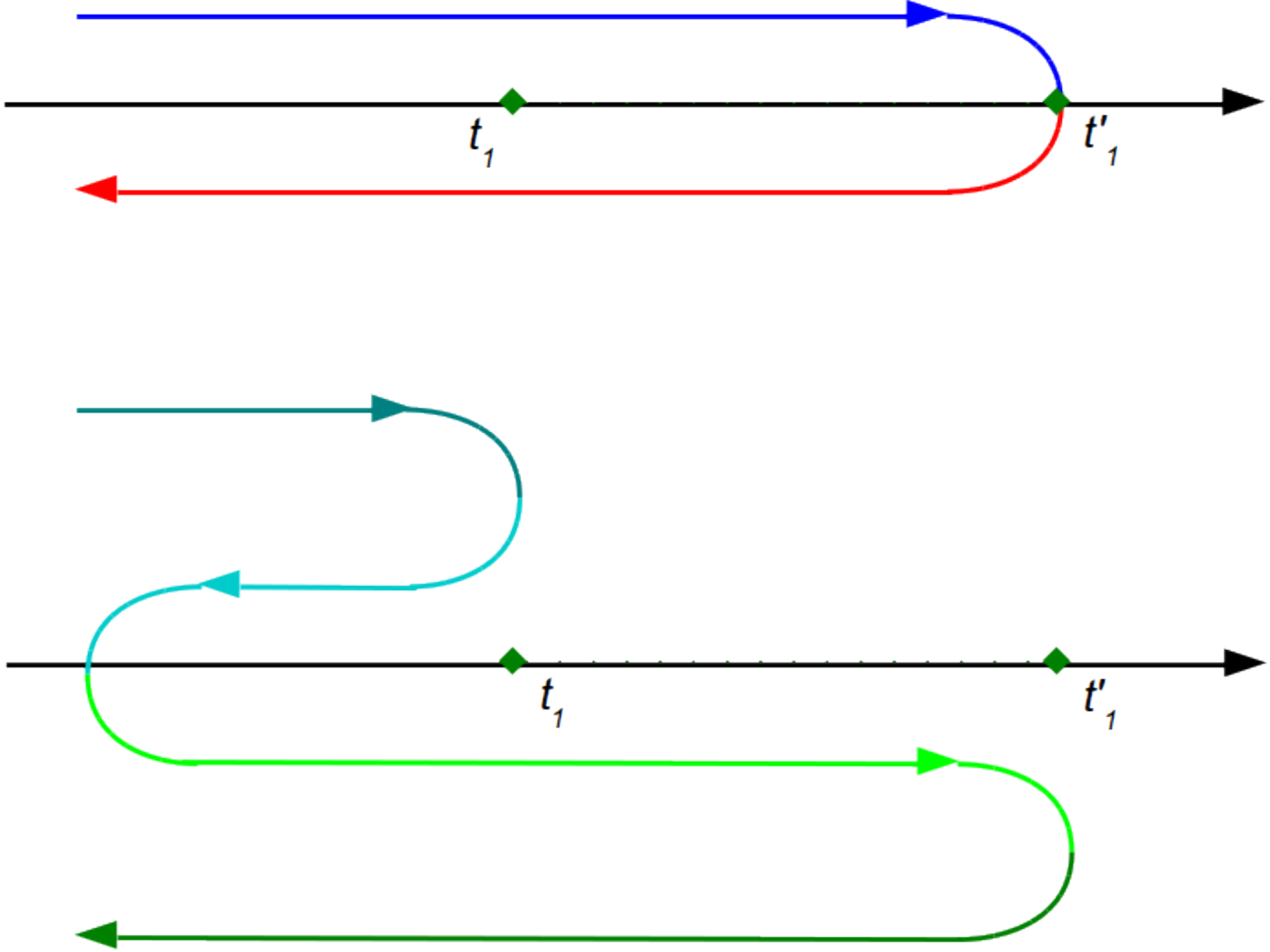}}
    \caption{The above contour $C$ has to be deformed in order to perform the integration. The new contour is is formed of two contours, $C_1$ and $C_{1'}$. The first runs from $-\infty$ to $t_1$ and go back to  $-\infty$. The second goes from $-\infty$ to $t_{1'}$ and go back to  $-\infty$.}
    \label{deformed_contour}
    \end{center}
    \end{figure}

    Dyson equation of a contour-ordered Green function has the same form as the equilibrium function Eq.~\eqref{eq_dyson_eq}:
    \begin{equation}\label{eq_dyson_g+-}
      \hat{G}^{+-}(\mathit{1,1'}) = \hat{g}_0^{+-}(\mathit{1,1'}) + \int \D^4 \mathit{2}\, \D^4 \mathit{3} \  \left[\hat{g}_{0}(\mathit{1,2}) \Sigma(\mathit{2,3}) \hat{G}(\mathit{3,1'}) \right]^{+-}
    \end{equation} 
    In this Dyson equation, we encounter terms with the time structure:
    \begin{equation}
      \hat{C}(t_1,t_{1'})= \int_C \D\tau \hat{A}(t_1,\tau)\hat{B}(\tau,t_{1'})
    \end{equation} 
    and their generalization involving products of three (or more) terms. In order to evaluate this integral assume in a first step that $t_1$ is on the first half of the contour (positive branch) and that $t_{1'}$ is on the other half (negative branch). This corresponds to study $G^{+-}$ in our notation (it corresponds to the lesser ``<'' in the older Kadanoff \& Baym notation). The second step consists in deforming the contour (fig.~\ref{deformed_contour})
    \begin{equation}
      \hat{C}^{+-}(t_1,t_{1'})= \int_{C_1} \D\tau \hat{A}(t_1,\tau)\hat{B}^{+-}(\tau,t_{1'}) + \int_{C_{1'}} \D\tau \hat{A}^{+-}(t_1,\tau)\hat{B}(\tau,t_{1'})
    \end{equation} 
    The $+-$ exponent means that as long as the integration variable $\tau$ is confined on the contour $C_1$ it is less than $t_{1'}$ (in the contour sense). Now, by splitting the integration into two parts, the first term becomes
    \begin{align}
      \int_{C_1} \D\tau \hat{A}(t_1,\tau)\hat{B}^{+-}(\tau,t_{1'}) &= \int_{-\infty}^{t_1}\D t \hat{A}^{-+}(t_1,t)\hat{B}^{+-}(t,t_{1'}) + \int^{-\infty}_{t_1}\D t \hat{A}^{+-}(t_1,t)\hat{B}^{+-}(t,t_{1'})\nonumber
      \\                                              &= \int_{-\infty}^{+\infty}\D t \hat{A}^{R}(t_1,t)\hat{B}^{+-}(t,t_{1'})
    \end{align} 
    using the definition of the retarded Green function Eq.~\eqref{eq_Gr_neq}.
    By doing the same on the second term, an analogous equation arises
    \begin{equation}
      \int_{C_{1'}} \D\tau \hat{A}^{+-}(t_1,\tau)\hat{B}(\tau,t_{1'}) = \int_{-\infty}^{+\infty}\D t \hat{A}^{+-}(t_1,t)\hat{B}^{A}(t,t_{1'})
    \end{equation}
    Finally the first of the Langreth' result is:
    \begin{equation}
      \hat{C}^{+-}(t_1,t_{1'})= \int_{-\infty}^{+\infty}\D t \hat{A}^{R}(t_1,t)\hat{B}^{+-}(t,t_{1'}) + \hat{A}^{+-}(t_1,t)\hat{B}^{A}(t,t_{1'})
    \end{equation}
    This demonstration allows the derivation of a simple recipe for the more general case: any factors that lie on the left of the $+-$ function are retarded. All those on the right are advanced. For instance, the contour time-integral of a product of three functions $\hat{D}=\int_C \hat{A}\hat{B}\hat{C}$ gives:
    \begin{multline}\label{eq_Dpm_time}
      \hat{D}^{+-}(\mathit{1,1'}) =	\int \D^4\mathit{2}\, \D^4\mathit{3} \ \left[\hat{A}^{R}(\mathit{1,2})\hat{B}^{R}(\mathit{2,3})\hat{C}^{+-}(\mathit{3,1'}) + \hat{A}^{R}(\mathit{1,2})\hat{B}^{+-}(\mathit{2,3})\hat{C}^{A}(\mathit{3,1'})\right. \\
                                                                            +\left. \hat{A}^{+-}(\mathit{1,2})\hat{B}^{A}(\mathit{2,3})\hat{C}^{A}(\mathit{3,1'}) \right]
    \end{multline} 
    where the time integrals of the right-hand side are evaluated on the real-time axis. The Langreth theorem therefore provides us with a powerful result to move from contour integrals to real-time integrals and it will be widely applied in the following.
    
    It should be noticed that the Fourier transform of the convolution  in the time-domain~\eqref{eq_Dpm_time} leads to a simple multiplication in the frequency domain
    \begin{equation}
      \hat{D}^{+-}(\omega)= \hat{A}^{R}(\omega)\hat{B}^{R}(\omega)\hat{C}^{+-}(\omega) + \hat{A}^{R}(\omega)\hat{B}^{+-}(\omega)\hat{C}^{A}(\omega) + \hat{A}^{+-}(\omega)\hat{B}^{A}(\omega)\hat{C}^{A}(\omega)  
    \end{equation}
    The reason why we can make this simplification is because the Fourier transform of the Green function is given by:
    \begin{equation}
      \hat{G}_{\omega}^{+-} = \lim_{t_1\rightarrow t_1'} \int \D t_2 \D t_3 \int \frac{\D\omega}{2\pi}  \frac{\D\omega'}{2\pi}  \frac{\D\omega''}{2\pi} \hat{A}(\omega)\hat{B}(\omega')\hat{C}(\omega'')\e^{-\i\omega(t_1-t_2)}\e^{-\i\omega'(t_2-t_3)}\e^{-\i\omega''(t_3-t_1')} 
    \end{equation}
    Exchanging the order of integration and highlighting  $t_2$ and $t_3$
    \begin{align}
      \hat{G}_{\omega}^{+-} &= \lim_{t_1\rightarrow t_1'} \int \frac{\D\omega}{2\pi}  \frac{\D\omega'}{2\pi}  \frac{\D\omega''}{2\pi}\hat{A}(\omega)\hat{B}(\omega')\hat{C}(\omega'')  \nonumber\\
                      & \qquad \times \int \D t_2 \D t_3 \e^{-\i\omega t_1} e^{-\i t_2(\omega-\omega')} e^{-\i t_3(\omega'-\omega'')} \e^{-\i\omega'' t_1'} \nonumber\\
                      &=\lim_{t_1\rightarrow t_1'} \int \frac{\D\omega}{2\pi}  \hat{A}(\omega)\hat{B}(\omega)\hat{C}(\omega) \e^{-\i\omega (t_1-t_1')}
    \end{align} 
    Using the property that integrals of the form $\int \D t e^{-\i t(\omega-\omega')}=\delta_{\omega,\omega'}$. With an abuse of notation, in the following, we shall use the simplified form:
    \begin{equation}
      \hat{D}^{+-}= \hat{A}^{R}\hat{B}^{R}\hat{C}^{+-} + \hat{A}^{R}\hat{B}^{+-}\hat{C}^{A} + \hat{A}^{+-}\hat{B}^{A}\hat{C}^{A}        
    \end{equation} 
    also for its time counterpart.

    \subsection{Keldysh equation}\label{sec_keldysh}
    Keldysh has shown in his seminal paper how to express non-equilibrium Green functions in term of equilibrium ones. However, his derivation is quite cumbersome because the power of the Langreth' theorem was not available at that time. For this reason, we derive Keldysh formula expressing non-equilibrium Green functions in terms of equilibrium ones following the simpler approach of Ref.~\cite{jauho_book}, based on Langreth' theorem on the contour Dyson equation~\eqref{eq_dyson_g+-}:
    \begin{equation}
      \hat{G}_{1}^{+-}= \hat{g}_{0}^{+-} + \hat{g}_{0}^{R}\hat{\Sigma}^{R} \hat{G}^{+-} + \hat{g}_{0}^{R}\hat{\Sigma}^{+-} \hat{G}^{A} + \hat{g}_{0}^{R}\hat{\Sigma}^{A} \hat{G}^{A}
    \end{equation}
    Iterating once, i.e. replacing $\hat{G}^{+-}$ by itself, we obtain:
    \begin{equation}
      \hat{G}_{2}^{+-}= \hat{g}_{0}^{+-} + \hat{g}_{0}^{R}\hat{\Sigma}^{R} \left( \hat{g}_{0}^{R}\hat{\Sigma}^{R} \hat{G}^{+-} + \hat{g}_{0}^{R}\hat{\Sigma}^{+-} \hat{G}^{A} + \hat{g}^{+-}\hat{\Sigma}^{A} \hat{G}^{A} \right) + \hat{g}^{+-}\hat{\Sigma}^{+-} \hat{G}^{A} + \hat{g}^{+-}\hat{\Sigma}^{A} \hat{G}^{A}
    \end{equation} 
    which can be written as
    \begin{equation}
      \hat{G}_{2}^{+-}= (\identite + \hat{g}_{0}^{R}\hat{\Sigma}^{R})  \hat{g}_{0}^{+-} (\identite + \hat{\Sigma}^{A}\hat{G}^{A}) + (\hat{g}_{0}^{R}+ \hat{g}_{0}^{R}\hat{\Sigma}^{R}\hat{g}_{0}^{R}) \hat{\Sigma}^{+-} \hat{G}^{A} + \hat{g}_{0}^{R}\hat{\Sigma}^{R}\hat{g}_{0}^{R}\hat{\Sigma}^{R}\hat{G}^{+-}
    \end{equation}
    Iterating once again
    \begin{multline}
      \hat{G}_{3}^{+-}= (\identite + {\hat{g}_{0}^{R}\hat{\Sigma}^{R}+\hat{g}_{0}^{R}\hat{\Sigma}^{R}\hat{g}_{0}^{R}\hat{\Sigma}^{R}})  \hat{g}_{0}^{+-} (\identite + \hat{\Sigma}^{A}\hat{G}^{A}) + \\
      (\hat{g}_{0}^{R}+ \hat{g}_{0}^{R}\hat{\Sigma}^{R}\hat{g}_{0}^{R}+ \hat{g}_{0}^{R}\hat{\Sigma}^{R}\hat{g}_{0}^{R}\hat{g}_{0}^{R}\hat{\Sigma}^{R}) \hat{\Sigma}^{+-} \hat{G}^{A}  + \hat{g}_{0}^{R}\hat{\Sigma}^{R}\hat{g}_{0}^{R}\hat{\Sigma}^{R}\hat{g}_{0}^{R}\hat{\Sigma}^{R}\hat{G}^{+-}
    \end{multline}
    This leads to the infinite order
    \begin{equation}\label{eq_keldysh}\color{red}\fbox{$
      \hat{G}^{+-}= (\identite + \hat{G}^{R}\hat{\Sigma}^{R})  \hat{g}_{0}^{+-} (\identite + \hat{\Sigma}^{A}\hat{G}^{A}) + \hat{G}^{R} \hat{\Sigma}^{+-} \hat{G}^{A}$}
    \end{equation}
    provided that series converges, i.e.:
    \begin{equation}
      \lim_{n\rightarrow\infty} \left(  \hat{g}_{0}^{R} \hat{\Sigma}^{R} \right)^n \hat{G}^{+-} = 0
    \end{equation}
    In the original Keldysh' paper, this result was written for an other Green function: $\hat{F}=\hat{G}^{+-}+\hat{G}^{-+}$ (eq. 50 of Ref.~\cite{keldysh}). The result~\eqref{eq_keldysh} is the key result at the basis of our expression for the BEEM current~\eqref{eq_BEEM_current}.
    
    \FloatBarrier

\chapter{Ballistic Electron Emission Microscopy}\label{chapt_BEES}
  \lettrine[lines=3, lhang=0.35, loversize=0.77, findent=2em,nindent=-0.7em,slope=-1em]{B}{allistic Electron Emission Microscopy} (BEEM) is a tech-nique derived from the Scanning Tunneling Microscope (STM) that was proposed before the advent of spintronics by Kaiser and Bell in 1988~\cite{kaiser_bell1-PhysRevLett.60.1406,kaiser_bell2-PhysRevLett.61.2368}.
It was initially devoted to the characterization of electronic proper-ties of buried metal/semiconductor interfaces: the so-called ``Schottky barrier'' that appears by putting a metal in close contact with a semi-conductor.

\begin{figure}[!tbp]
  \centering
  \subbottom[\label{fig_BEEM}]{%
    \includegraphics[width=0.48\linewidth]{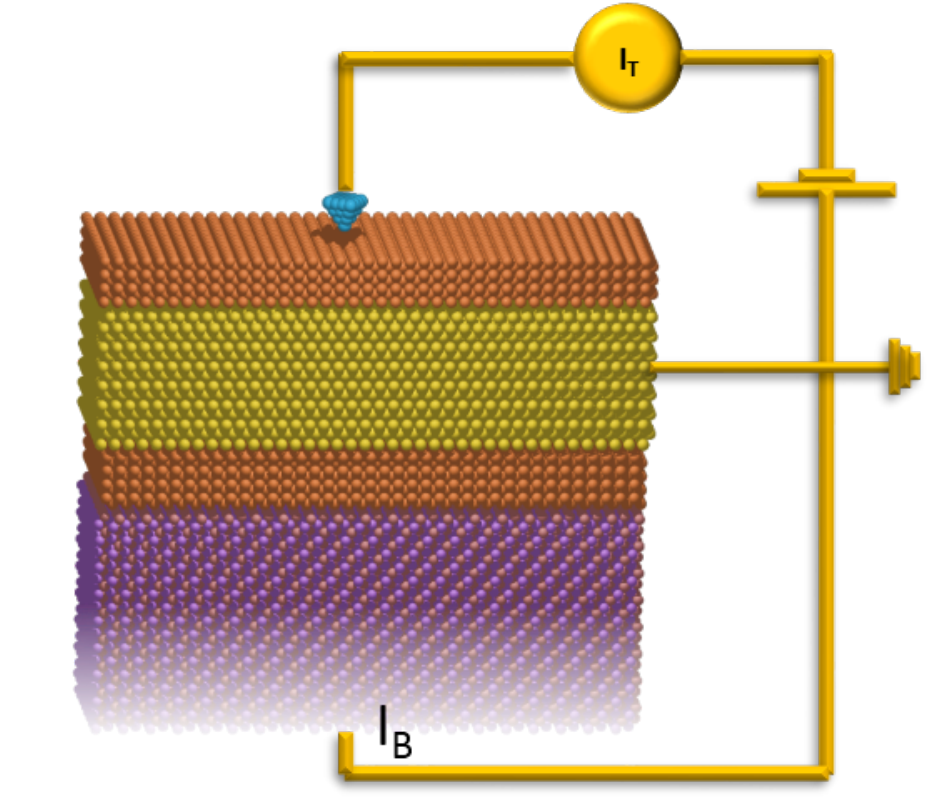}} \quad
  \subbottom[\label{fig_BEEM_pot}]{%
    \includegraphics[width=0.48\linewidth]{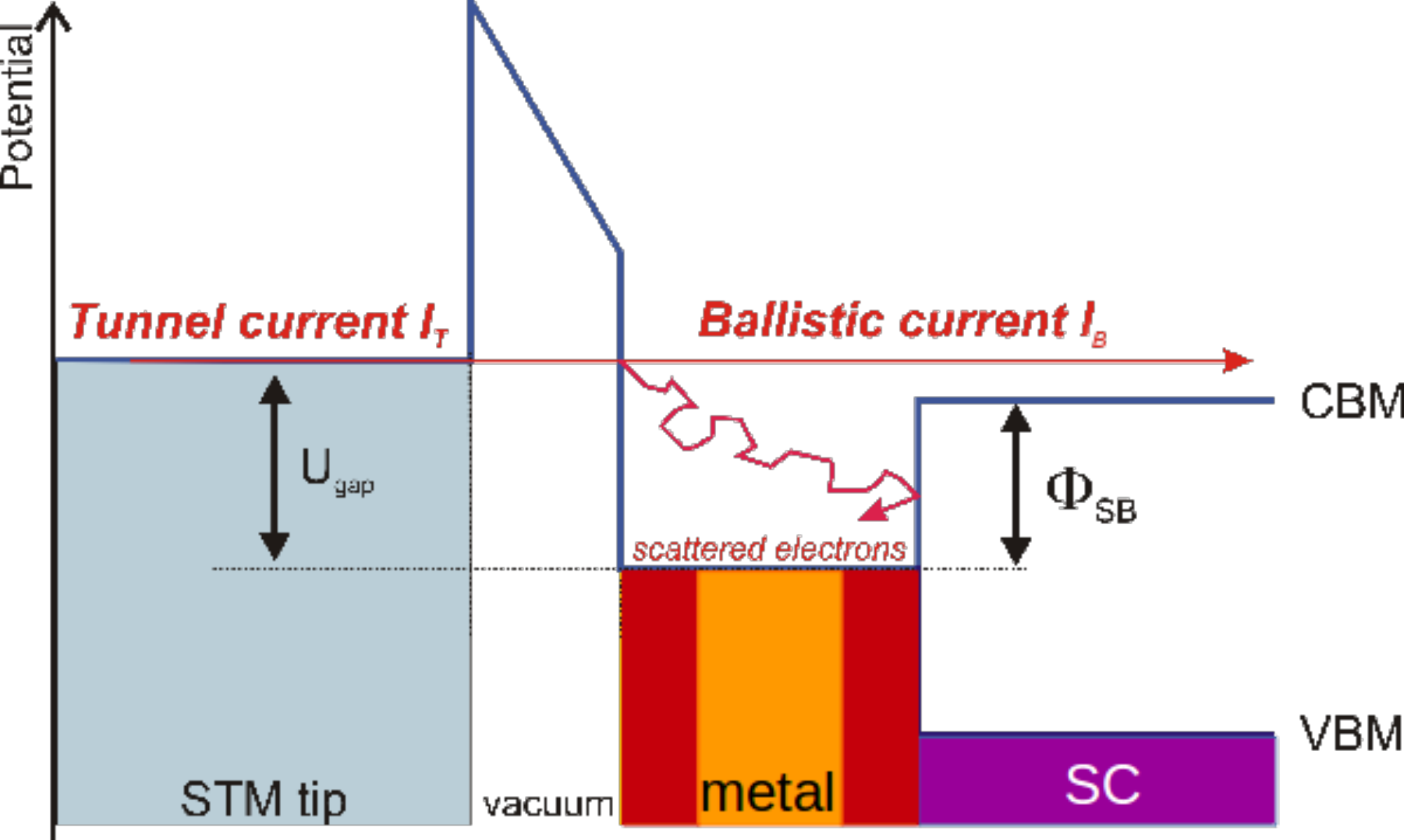}}
  \caption{\subcaptionref{fig_BEEM} Ballistic Electron Emission Microscope and \subcaptionref{fig_BEEM_pot} potential representation. Electrons are injected from a STM tip by tunnel effect, if their energy is higher than the height of the Schottky barrier (at the metal/semi-conductor interface) they are collected at the back of the Semi-conductor. Otherwise, all electrons are thermalized in the metallic films and evacuated by a contact at the surface of the metallic film linked to the mass.\label{fig_BEEM_principe}}
\end{figure}

This microscopy technique takes advantage of this Schottky barrier. As shown in figure~\ref{fig_BEEM_principe}, a current $I_t$ is injected in the metal (typically some nA) with a STM tip.
A tiny part of the injected electron current propagates elastically through the metal slab. 
If the energy of these electrons is higher than the height of the Schottky barrier (of the order of 1 eV) they can cross the metal/semi-conductor interface and be collected (BEEM current $I_\mathrm{B}\sim pA$). 
It should be underlined that, as clear from the ratio $\frac{I_t}{I_B}\sim 10^3$, most of the electrons injected do not pass the Schottky barrier and are thermalized in any case. There is only a small fraction of electrons that behaves elastically, and both experimental and theoretical studies are focused on these electrons, which are called ``ballistic electrons'' for historical reasons.

\begin{figure}[!h]
  \centering
    \includegraphics[width=0.90\linewidth]{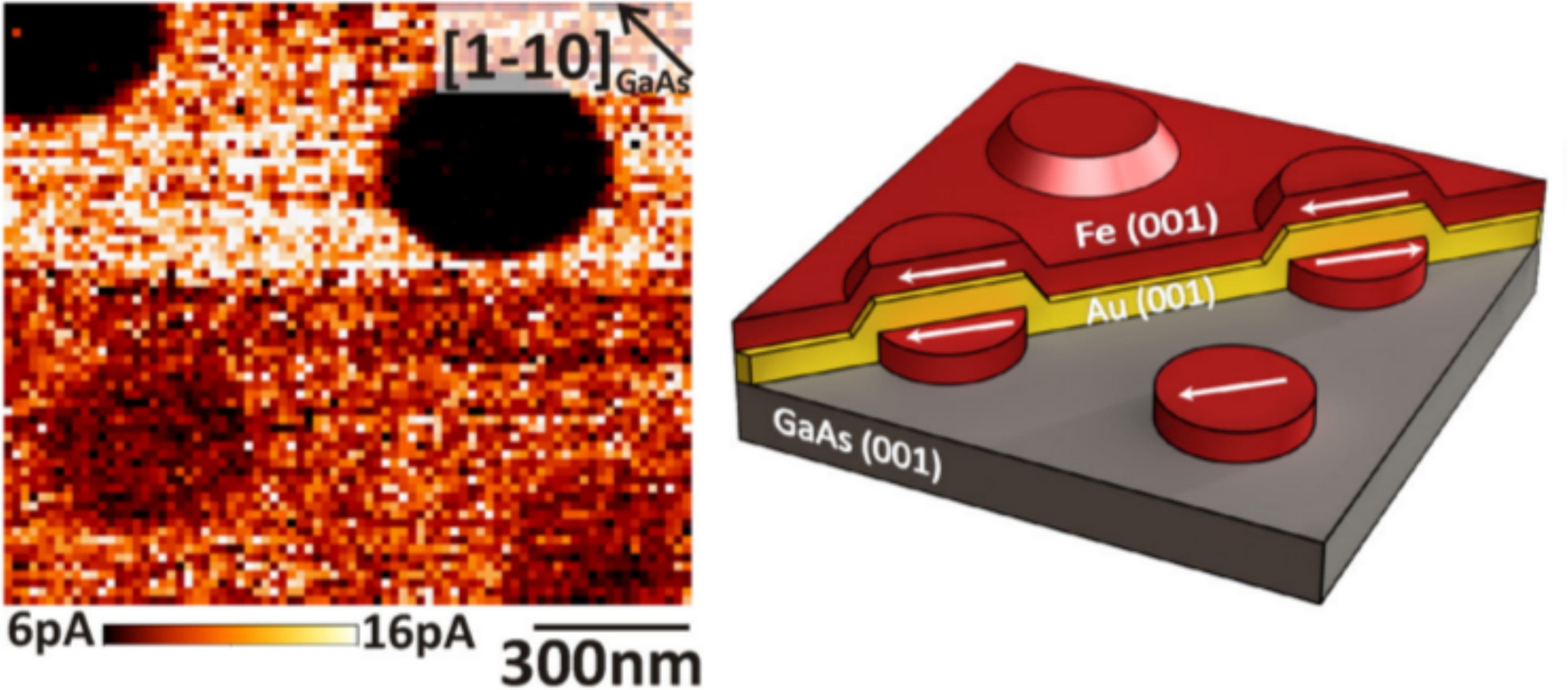}
    \caption{Image of a buried structure (Fe(1nm)/Au(6nm)/Fe(1nm)/GaAs(001)) made using the BEEM imaging mode ($U_{gap}=1.5$ V), from Ref.~\cite{these_marie}. The dark regions are low current areas, while bright regions are high current area.}\label{fig_BEEM_picture}
\end{figure}

As the height of the Schottky barrier $\phi_{SB}$ depends on the metal/semi-conductor interface, measuring the BEEM current $I_B$ allows to picture buried structures at the interface. The BEEM  allows in this way, by scanning along the (say, $xy$) surface, to obtain an image that is a cartography of the transparency of the interface to electrons at a given energy, as shown in Fig.~\ref{fig_BEEM_picture}.

\begin{figure}[!tbp]
  \centering
    \includegraphics[width=0.48\linewidth]{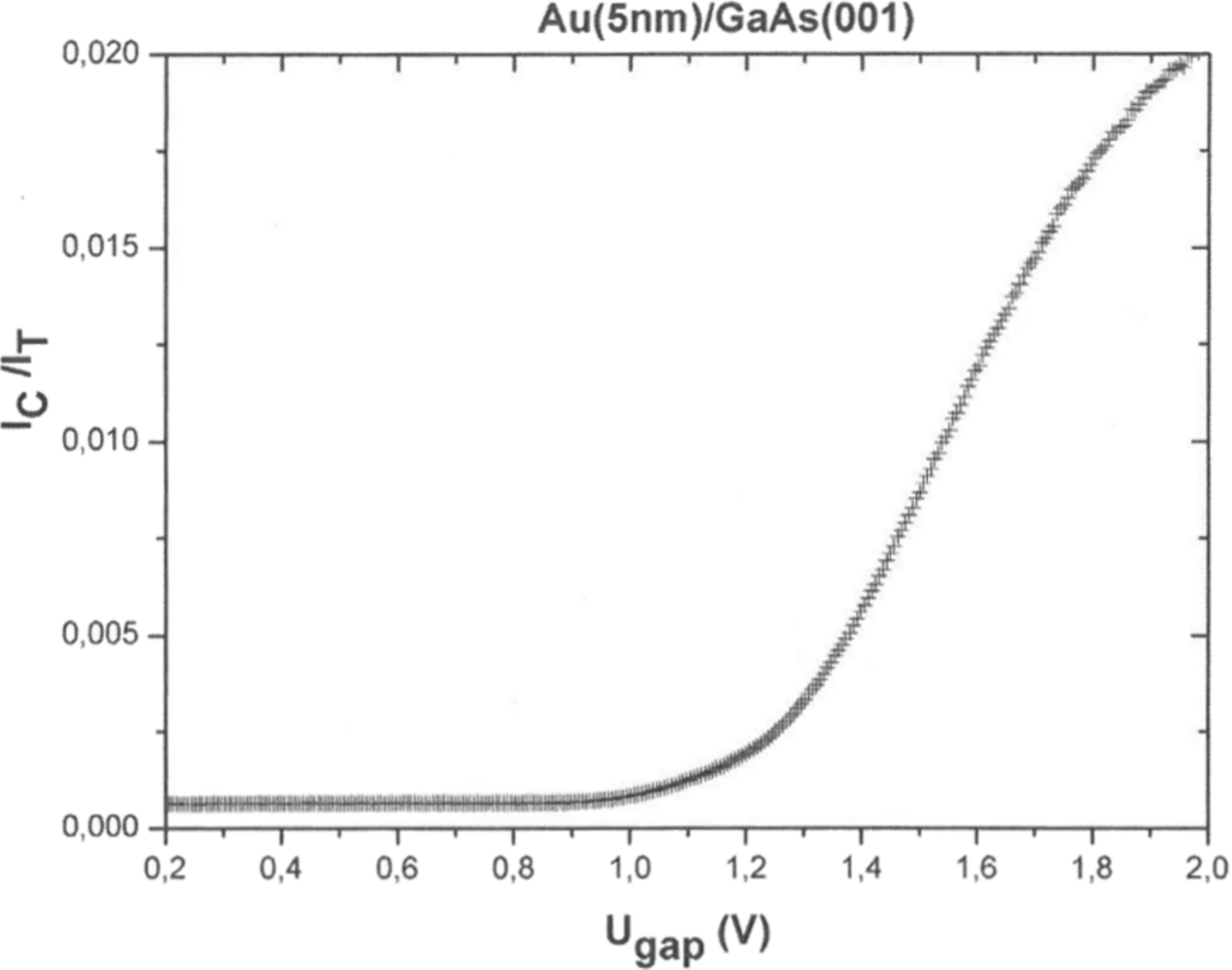}
    \caption{Ballistic Electron Emission Spectroscopy performed on Au(5nm)/GaAs(001). The BEEM current $I_B(U_{gap})$ starts to increase above a given threshold: the Schottky barrier. In this mode, the position $xy$ of the tip remains constant, as well as the tunneling current through a feedback loop which varies the altitude of the tip with respect to the surface. }\label{fig_BEES_AuGaAs}
\end{figure}

Another possibility is to keep unchanged the $xy$-position of the tip on the surface and record the evolution of the BEEM current $I_B$ with respect to the bias $U_\mathrm{gap}$ (Fig.~\ref{fig_BEES_AuGaAs}). In this mode, the tunneling current remains constant through a feedback loop which varies the altitude of the tip with respect to the surface. Amongst other things, this mode gives a direct measure of the Schottky barrier. It is the mode that we shall analyze in this manuscript.

  \section{Free-electron-like models for BEEM current}\label{sec_free_electron}
  \subsection{Kaiser \& Bell and Ludeke \& Prietsch model}
    In order to give a good description of a BEEM experiment, one has to describe several physical process for the injected electrons (see Fig.~\ref{fig_BEEM_principe}): 
    \begin{enumerate}
    \item the tunnel injection from the STM tip to the sample
    \item the propagation of the electrons within the metal
    \item their transmission inside the semiconductor to be detected as BEEM current.
    \end{enumerate}
    The first model, proposed by Kaiser and Bell~\cite{kaiser_bell2-PhysRevLett.61.2368}, has been developed in order to extract the height of the Schottky barrier from BEEM experiment. In their approach, they consider that the BEEM current is the flux of ballistic electrons (free-electrons) that can enter the semiconductor:
    \begin{equation}
      I_B = e\int \frac{\D \vec{k}}{(2\pi)^3}\ \vec{v}(\vec{k})\cdot\vec{n}\ \left[ F^{\mathrm{tip}}\left(E-U_\mathrm{gap},T\right) - F^M\left(E,T\right)\right] \ T\left(E,\vec{k}\right)\ \theta\left(E-\phi_{SB}\right)
    \end{equation} 
    with the velocity of electrons $\vec{v}(\vec{k})$, the normal vector to the interface $\vec{n}$, $F$ the Fermi distributions in the tip and in the metal, the probability $T$ for electrons to cross the junction  and the Heaviside function $\theta$ which gives the allowed energy condition for electrons to cross the interface. 
    
    Defining $v_z$ and $k_z$ the components of the group velocity and wave-vector of the electrons normal to the interface, and by considering the temperature $T=0$ K, the equation becomes
    \begin{equation}\label{eq_I_kb}
      I_B =  \frac{e}{(2\pi)^3}  \int_{E_{BC}^{SC}}^{eU_\mathrm{gap}} \D E^{\mathrm{tip}} \theta\left(E-\phi_{SB}\right) \int \D\vec{k}_{\sslash}T\left( E,\vec{k}_{\sslash} \right)  \kpara 
    \end{equation} 
    using the change of variable  $v_z=\frac{\D E}{\D k_z}$.
    
    Several hypotheses are required to simplify this equation. First, the metal/semiconductor interface is supposed abrupt and the component parallel to the interface of the wave vector $\kpara$ is conserved
    \begin{equation}
      \kpara^{M} = \kpara^{SC}=\kpara
    \end{equation} 
    The electrons are supposed to be free particles whose energy is given by
    \begin{equation}\label{eq_free_electron}
      E=\frac{\hbar^2k^2}{2m}
    \end{equation} 
    The transmission coefficient $T$ is supposed to be energy independent $T(E,\kpara)=T(\kpara)$. Using all those approximations on Eq.~\eqref{eq_I_kb} gives~\cite{Prietsch}:
    \begin{equation}\label{eq_I_kb_final}
      I_B =  \frac{e}{(2\pi)^3}  \int_{E_{BC}^{SC}}^{eU_\mathrm{gap}} \D E^{\mathrm{tip}} \int \D\vec{k}_{\sslash}  T(\kpara) \kpara 
    \end{equation}
    After integration, it gives a BEEM current proportional to the square of the electron energy, above the Schottky barrier:
    \begin{equation}\label{eq_powerlaw_KB}
      I_B \propto  (eU_\mathrm{gap}-\phi_{SB})^2
    \end{equation} 
    
    However, such a formula was considered not to be accurate enough by Ludeke and Prietsch \cite{ludeke,Prietsch} who improved this free-electron-like model by supposing
    that the transmission coefficient $T$ depends on the energy. Then, the current becomes:
    \begin{equation}
      I_B \propto \int_{E_{BC}^{SC}}^{eU_\mathrm{gap}} \D E^{\mathrm{tip}}  \int \D \vec{k}_{\sslash}\  F(E^{\mathrm{tip}}-U_\mathrm{gap},T)\ D(E^{\mathrm{tip}},U_\mathrm{gap})\ \e^{\frac{d}{\lambda(E^{\mathrm{tip})}}}\ \vec{k}_{\sslash} \ T(E^{\mathrm{tip}},\vec{k}_{\sslash})
    \end{equation} 
    Here again $F$ is the Fermi-Dirac distribution of the electrons of the tip at temperature $T$, $D$ is the probability for those electrons to tunnel from the tip to the metal and $T$ is the probability for electrons to cross the metal/semiconductor interface. The exponential term represents the attenuation of the electrons with respect to their mean free-path $\lambda$ and the thickness $d$ of the metallic slab. If we consider again that $\vec{k}_{\sslash}$ is conserved at the interface and that the electron energy is close to the bottom of the conduction-band of the semiconductor, then, it is possible to write the transmission coefficient $T$ as~\cite{Prietsch}:
    \begin{equation}
      T \propto \sqrt{\frac{2m^*(E-\phi_{SB})}{\hbar^2}-\vec{k}_{\sslash}^{2}}
    \end{equation} 
    where $m^*$ is the effective mass at the bottom of the conduction band.
    Moreover, if we consider that the temperature $T=0$ K and that $\lambda$ and $D$ are constant in the considered energy range, the BEEM current becomes:
    \begin{align}
      I_B &\propto \int_{E_{BC}^{SC}}^{eU_\mathrm{gap}} \D E^{\mathrm{tip}} (E-\phi_{SB})^{3/2} \nonumber \\
      I_B &\propto  (eU_\mathrm{gap}-\phi_{SB})^{5/2} \label{eq_powerlaw_LP}
    \end{align}
    At higher energy, other mechanisms of diffusion must be taken into account in order to fit the experimental curves. However, most of the time, BEEM experiments are performed near the Schottky barrier and it is sufficient to use this simple power law to fit experimental data and find the height of the Schottky barrier.

  \subsection{Transmission at the metal/semiconductor interface}\label{sec_acceptance_cone}
    
\begin{figure}[bt]
      \centering
    \includegraphics[width=0.48\linewidth]{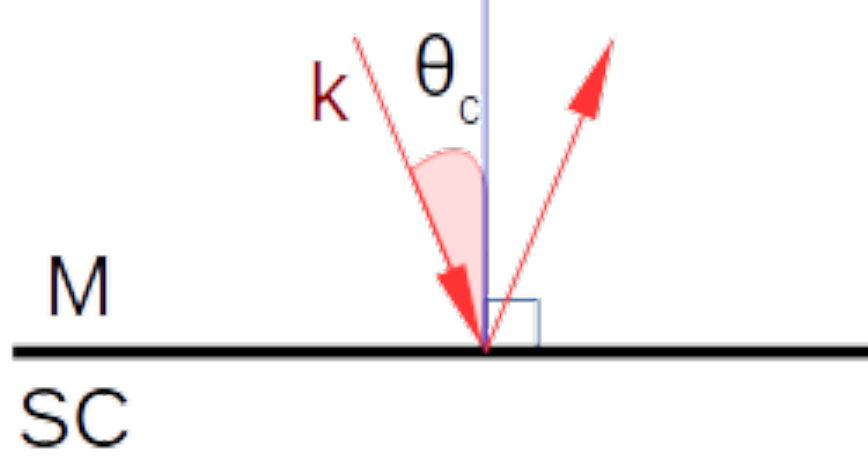}
      \caption{\label{fig_theta_c}In free-electron model, electrons that reach the junction can be reflected or refracted in the same way as light. By analogy with geometrical optics, a critical angle, above which electrons are reflected, exists.} 
    \end{figure}

    When we consider that the interface is abrupt and that the parallel component of the wave vector to the interface is conserved, we can draw an analogy with geometrical optics:  electrons that reach the junction can be reflected or refracted, in the same way as the light. A critical angle should then exists, above which, electrons cannot enter the semiconductor (see Fig.\ref{fig_theta_c}). 
    This calculation was performed by Kaiser and Bell \cite{kaiser_bell1-PhysRevLett.60.1406,kaiser_bell2-PhysRevLett.61.2368} using the energy conservation at the interface:
    \begin{equation}\label{eq_resolution}
     \theta_c = \arcsin \left[ {\frac{m^* (eU_{\mathrm{gap}}-\phi_{SB})}{m (eU_{\mathrm{gap}}+E_{F})}} \right]^{1/2}
    \end{equation}
    For instance, it gives a critical angle equal to 2.2\degree\ for Au/GaAs ($m^*=0.067m$ for $\Gamma$ valley, $\phi_{SB}=0.86$eV, $U_{\mathrm{gap}}=1$V, $E_F=5.53$eV). Using this angle we find the lateral resolution $\Delta x = 2d \tan \theta_c = 7$\AA. As such a calculation is basically dependent on the free-electron energy dispersion, Eq.~\eqref{eq_free_electron}, its conclusion is valid also for the Ludeke and Prietsch model, that assumes the same free-electron energy dispersion.

\section{Some key experimental results}
   This section gives a brief overview of some key experimental results which cannot be explained by a free-electron approach. We first present some results obtained by the Surfaces and Interfaces team of the Materials and Nanosciences department of the Physical Institute of Rennes and how they fit experimental data in order to find the height of the Schottky barrier. Then we present older experimental results for Au/Si. The latter is the historical reason which motivates the research of a better model, as we shall see.

  \subsection{Au(110)/GaAs(001)}\label{subsec_AuGaAs}
    Figure~\ref{fig_BEES_AuGaAs} represents Ballistic Electron Emission Spectroscopy (sometimes abbreviated BEES) curve for Au(110)/GaAs(001)~\cite{sophie-these} and the band structure of gallium arsenide obtained through Density Functional Theory calculations and Abinit code, within the Local Density Approximation~\cite{abinit}. 
    
\begin{figure}[bt]
      \centering
      \subbottom[\label{fig_BEES_fit1}]{%
	\includegraphics[width=0.48\linewidth]{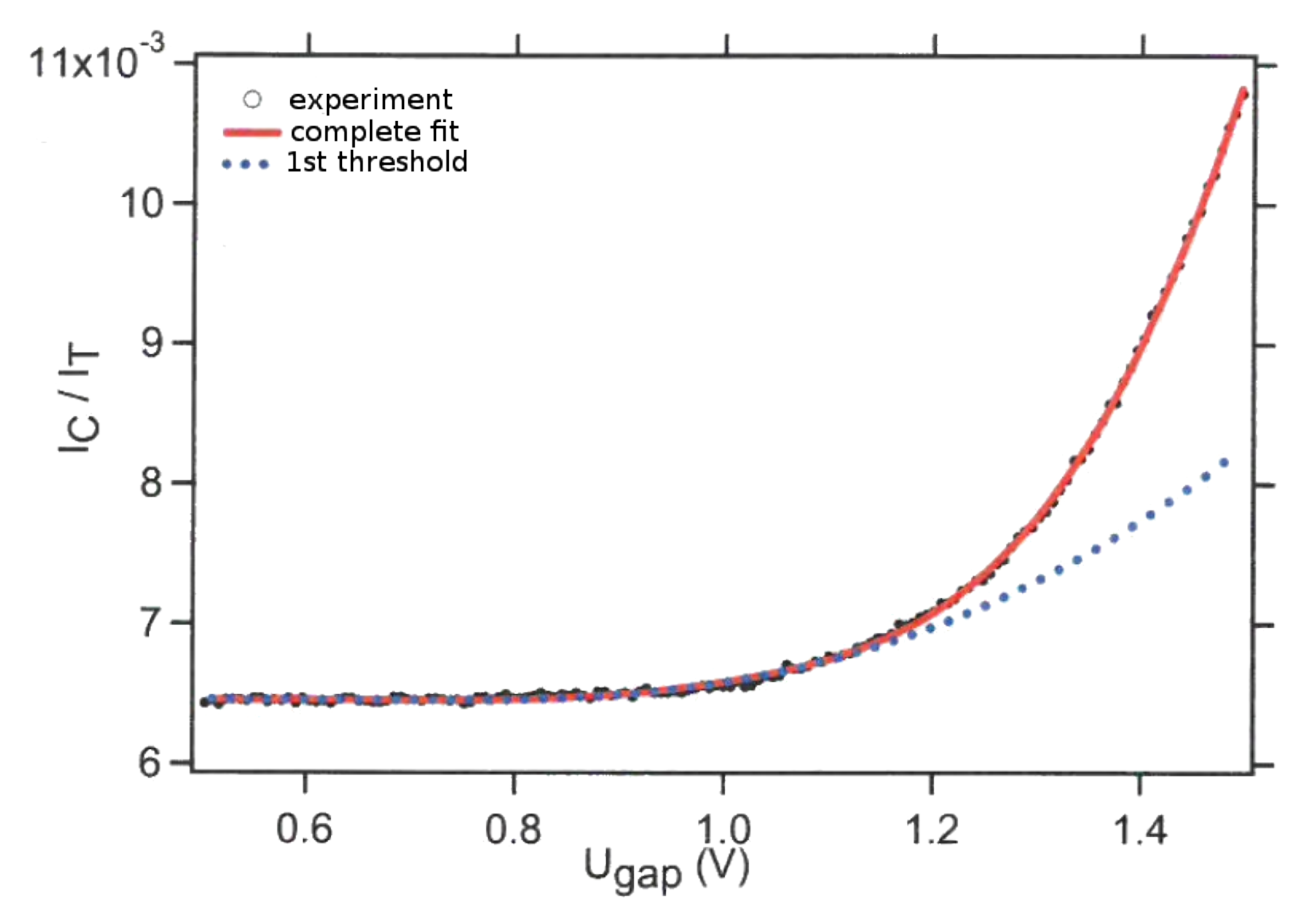}}
      \subbottom[\label{fig_BEES_fit3}]{%
	\includegraphics[width=0.48\linewidth]{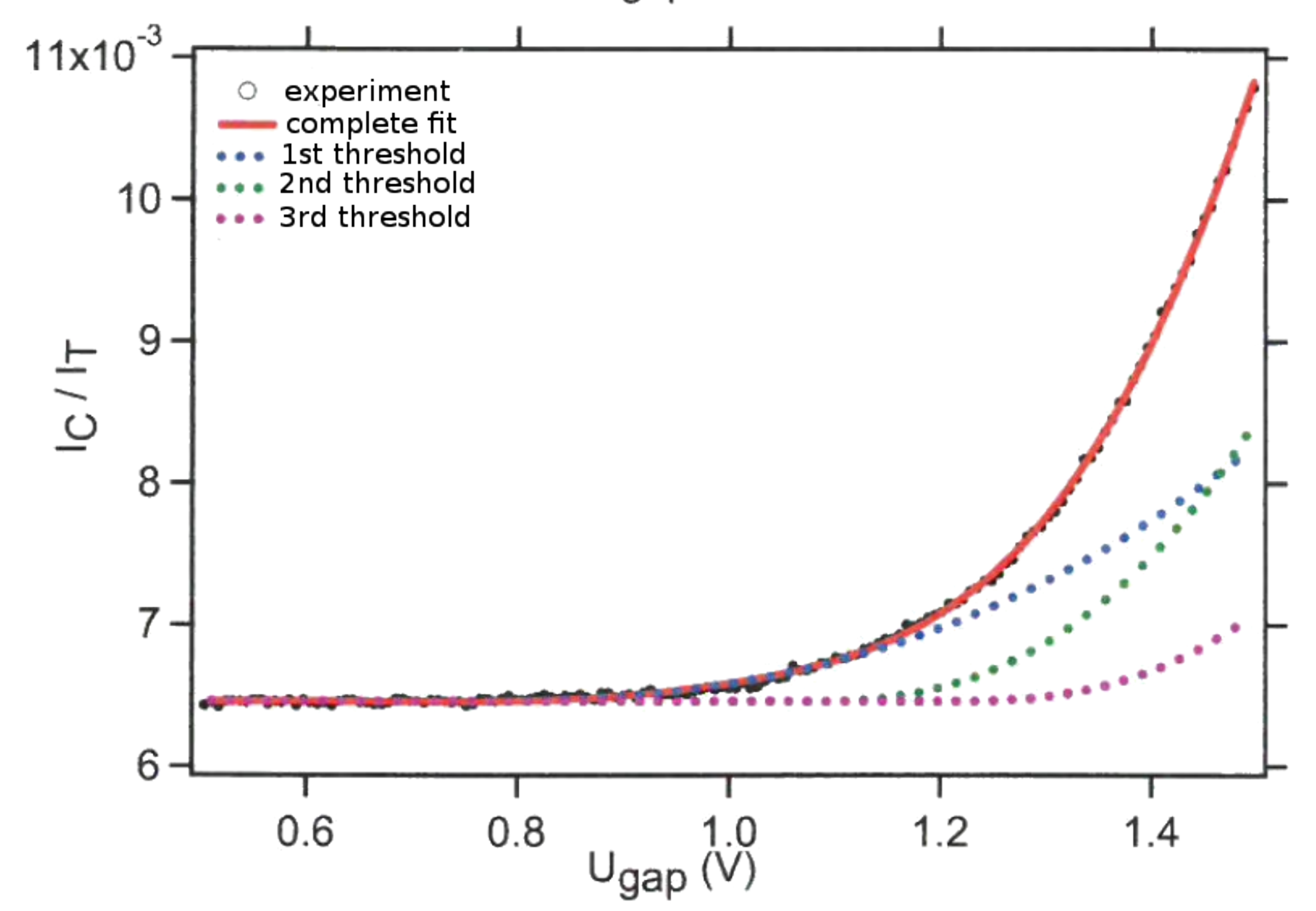}}\\
      \subbottom[\label{GaAs_bs}]{\includegraphics[width=0.5\linewidth]{BEEM/pictures/GaAs_bs}}
      \caption{\label{fig_BEES_fit} \subcaptionref{fig_BEES_fit1} and \subcaptionref{fig_BEES_fit3}: Fit of experimental spectroscopy curve \cite{sophie-these}, using respectively 1 and 3 thresholds (Eq.\eqref{eq_3threshold}): $\phi_{\Gamma}=\varepsilon_F+0.75$ eV, $\phi_{L}=\phi_{\Gamma}+0.33$ eV and $\phi_{X}=\phi_{\Gamma}+0.48$ eV.  \subcaptionref{GaAs_bs}: band structure of GaAs obtained through DFT/LDA calculations. The 3 thresholds of the fit correspond to the 3 minima, $\Gamma$, $L$ and $X$ of the conduction band of GaAs.} 
    \end{figure}

    The BEEM spectroscopy curve of Au(110)/GaAs(001) has been fitted (Fig.~\ref{fig_BEES_fit}) using the above power law \ref{eq_powerlaw_LP} of Ludeke and Prietsch (which works better than Kaiser and Bell Eq.~\eqref{eq_powerlaw_KB}): 
    \begin{align}
      \frac{I_B}{I_t} =a_0 + a_1(E-\phi_{\Gamma})^{5/2} 
    \end{align}
    The parameter $\phi_{\Gamma}$ gives the height of the Schottky barrier (the minimum of the conduction band of the semiconductor at point $\Gamma$). Ref.~\cite{sophie-these} gives a Schottky barrier $\phi_{SB}=0.81eV$. However as we can see in Fig.~\ref{fig_BEES_fit1}, even if the fit gives a correct value of the Schottky barrier, it cannot describe the experimental data at higher energy. In order to do that, we have to consider electron injection in other valleys of the conduction band of the semiconductor by including other thresholds in the ballistic-electron current formula:
    \begin{align}\label{eq_3threshold}
      \frac{I_B}{I_t} = a_0+ a_1(E-\phi_{\Gamma})^{5/2} + a_2(E-\phi_{L})^{5/2} +a_3(E-\phi_{X})^{5/2} 
    \end{align}
    Here we consider that electrons can be injected in $\Gamma$, $X$ and $L$ valleys, depending on their energy. Indeed, by fitting experimental curves, Sophie Guézo \emph{et al.}~\cite{sophie-these} found  $\phi_{\Gamma}=0.75$ eV, $\phi_{L}=\phi_{\Gamma}+0.33$ eV and $\phi_{X}=\phi_{\Gamma}+0.48$ eV (Fig.\ref{fig_BEES_fit3}), which is in good agreement with the GaAs band-structure calculations (Fig.\ref{GaAs_bs}).
    In other terms:
    \begin{itemize}
     \item at $\phi_{\Gamma}=0.75$ eV, electrons can be injected in the minimum of the conduction band of GaAs, the $\Gamma$ valley. Below, injection is impossible because there is no available density of states within the semiconductor. This energy is the height of the Schottky barrier.
     \item at $\phi_{L}=\phi_{\Gamma}+0.33$ eV, another valley is accessible for electrons: the $L$ valley. 
     \item at $\phi_{X}=\phi_{\Gamma}+0.48$ eV some other accessible density of states are accessible via the $X$ valley. 
    \end{itemize}
    Depending on the energy, the electrons can cross the interface through up to three channels which explain the need of three different thresholds in the experimental fit of Fig.~\ref{fig_BEES_AuGaAs}. Moreover, for one of these channels, the L-valley, the wave-vector $\kpara$ is different from 0. As the electrons in the tip are mainly injected with $\kpara=0$, this is in direct contradiction with the assumption that electrons behave like free particles. This phenomenological approach works quite well to fit experimental data, but it cannot make predictions. We need to use another theoretical approach which is not phenomenological, as proposed by F. J. Garcia-Vidal \emph{et al.}~\cite{Garcia-Vidal-PhysRevLett.76.807}, to explain the data of Au(111)/Si(111) versus Au(111)/Si(001), as shown below.

%

  \subsection{Au(111)/Si(111) and Au(111)/Si(001)}
\begin{figure}[bt]
      \centering
      \subbottom[\label{subfig_AuSi111}]{%
	\includegraphics[width=0.48\linewidth]{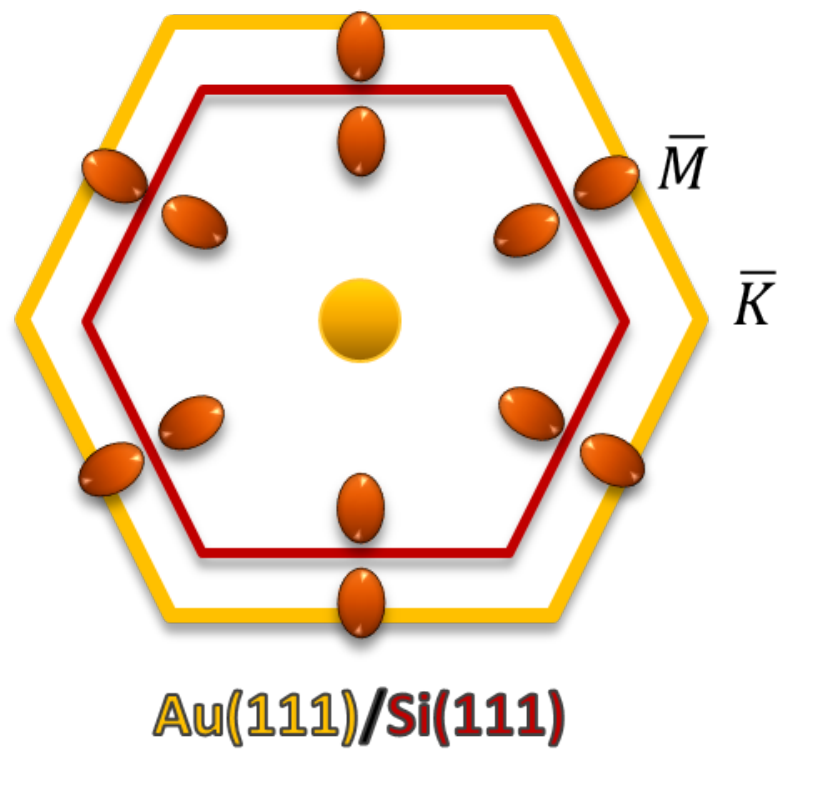}}
      \subbottom[\label{subfig_AuSi001}]{%
	\includegraphics[width=0.425\linewidth]{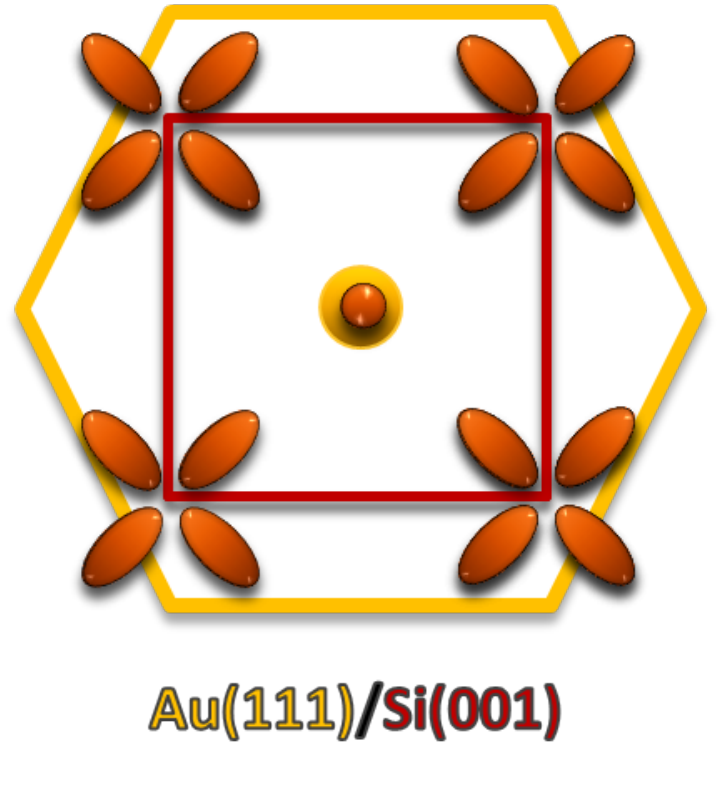}}\\
      \subbottom[\label{subfig_Si_iso_e}]{%
	\includegraphics[width=0.35\linewidth]{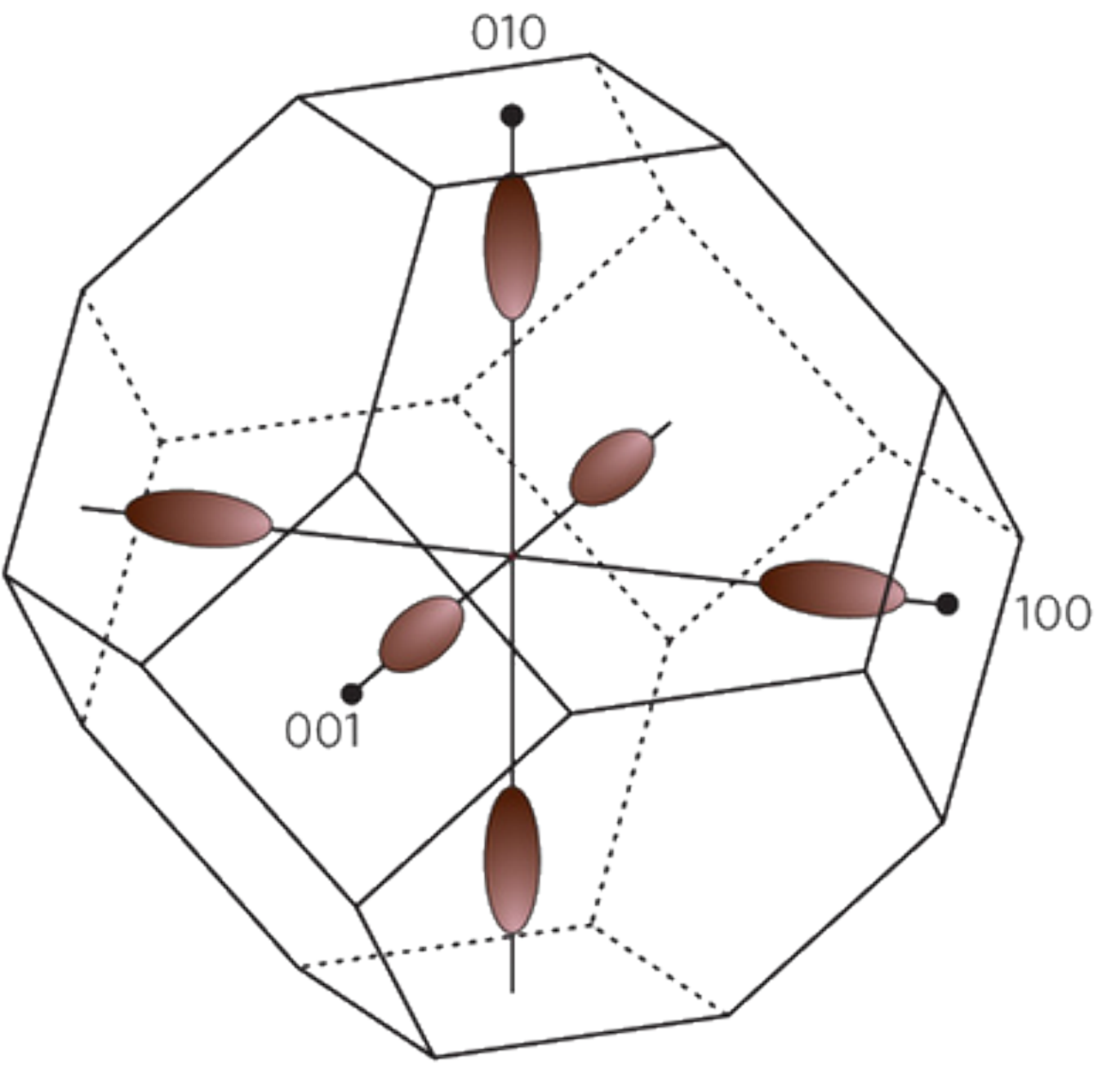}}
      \caption{\label{fig_Au_Si_free_el} \subcaptionref{subfig_AuSi111} Au(111)/Si(001) and  \subcaptionref{subfig_AuSi001} Au(111)/Si(111). The yellow hexagone represents the projection of the gold FCC-Brillouin zone in the (111) direction. The small yellow circle represents the ballistic electrons. In order to propagate inside the semi-conductor, ballistic electrons have to match available density of states in the semi-conductor. Here it is represented by the red ellipses that are the projections of the constant-energy ellipsoids of Si \subcaptionref{subfig_Si_iso_e} in the (111) and the (001) directions. For Au(111)/Si(111), pure ballistic-electrons cannot cross the metal/semi-conductor interface, unlike the Au(111)/Si(001) system.} 
    \end{figure}

    The Au(111)/Si(111) and Au(111)/Si(001) systems also demonstrate the failure of free-electron models to describe accurately experimental data: in Ref.~\cite{Milliken-AuSi_PhysRevB.46.12826} it was observed that the BEEM current is almost the same for both orientation of the silicon.
    However, by looking at the band structure of silicon (see for instance Ref.~\cite{Papa-handbook}) we can see that there are no available states inside the acceptance cone defined by Eq.~\eqref{eq_resolution} for Si(111), unlike the Si(001) direction. It means that in a ballistic free-electron hypothesis with $\kpara$ conservation at the metal/semiconductor interface, \emph{i.e.} $\kpara\sim 0$ also in the semiconductor slab,  the BEEM current should be very different for those two systems (Fig.~\ref{fig_Au_Si_free_el}). 
    As in Subsec.~\ref{subsec_AuGaAs}, this finding points towards a strong limitation of the free-electron energy dispersion hypothesis at the basis of both Kaiser/Bell and Ludeke/Prietsch models, or of the $\kpara$ conservation at the metal/semiconductor interface.
    
    Therefore all possible explanations can be divided in two classes: the first class \cite{Ludeke-kpara_non_cons_PhysRevLett.71.1760,Schowalter-PhysRevB.43.9308,bauer-PhysRevLett.71.149} explained these results by questioning the elastic diffusion of electrons at interfaces and supposing non-conservation of $\vec{k}_\sslash$ at interfaces due to roughness and defects. The opposite point of view has been proposed by Fernando Flores' and Pedro L. de Andres' groups~\cite{Garcia-Vidal-PhysRevLett.76.807}. In their model, electrons follow the band structure of the metal in which they propagate and this leads to big deviation from the small cone of Eq.~\eqref{eq_resolution}. This theory proved successful~\cite{Pedro-ProgSurfS_66_2001,Garcia-Vidal-PhysRevLett.76.807,Reuter-PhysRevB.58.14036,Pedro-phys_scripta_T66_277} when coupled to the almost ideally layered structure of presently available surfaces~\cite{sophie-PhysRevB.81.085319}. It is detailed in the next section and this is also the one we used in this thesis.

    \section{Band-structure-like models}
    
    In the case of metallic films, electrons should feel the periodic potential of the material in which they propagate already after few layers~\cite{Reuter-PhysRevB.58.14036}. By elastic scattering, they choose some preferential directions of propagation and for the same reason, lack of allowed density of states can prohibit some reciprocal-space directions, as in the usual propagation of Bloch electrons. In 1996, F. J. Garcia-Vidal, P.L. de Andres and F. Flores have proposed a model where electrons propagate within the metal by taking into account of the band structure of the material in which they propagate. They describe the system as a metallic slab $M$ in interaction $I$ with a STM tip $T$:
    \begin{equation}
     H = H_T + H_M + H_I
    \end{equation}  
    
    Starting from this hypothesis it is possible to use two methods in order to calculate the BEEM current: the original one, developed by F. Flores, is based on Keldysh approach and is described in Sec.~\ref{sec_non_eq_calc}. A simpler, approximated approach, based on a direct equilibrium calculation that we have developed in this thesis, is presented in Sec.~\ref{sec_eq_calc}.

    \subsection{Non-equilibrium calculations}\label{sec_non_eq_calc}
      In their approach, F. J. Garcia-Vidal \emph{et al.} have used the non-equilibrium perturbation theory (Sec.~\ref{sec_nept_gf}) in order to express the BEEM current (Sec.~\ref{section_beem_current}). The key quantities of their approach is the Keldysh Green function (Sec.~\ref{sec_keldysh}) which can in principle be calculated through several methods. For instance, we can use Density Functional Theory, as described in \cite{siesta-NEGF_DFT_PhysRevB.65.165401} and implemented in the SIESTA code, by replacing the usual input electron density by the non-equilibrium one. The electron density follows from the non-equilibrium Green function:
      \begin{equation}
       n(x)= -iG^{+-}(x=x',t=t') = \int \frac{\D \varepsilon}{2\i\pi} G^{+-}(x=x',\varepsilon)
      \end{equation} 
      Using the Keldysh equation, the only required ingredients are therefore the retarded (advanced) Green function and the self energy. Depending on the system, those quantities can be more or less easily obtained. The self consistent loop is then:
      $$\mbox{initial }n(x) \Rightarrow \mbox{SIESTA} \Rightarrow \psi_{KS}(x) \Rightarrow \mbox{NEGF} \Rightarrow \mbox{new } n(x)$$
      However this kind of method might become very expensive in term of calculation time for large systems (<20 layers, with 9 orbitals per layer) and requires expertise to use it properly.

      In the following we shall use instead a tight-binding approach expressed within the second quantization formalism. This is the approach proposed by F. J. Garcia-Vidal \emph{et al.}. They have re-expressed the current in terms of equilibrium Green functions (that we know how to calculate) and hopping matrices which are calculated iteratively in a very efficient procedure (see Chap.~\ref{chapt_NEPT} and \ref{chapt_results}). 
      We also decided to use the tight-binding parametrization for two main reasons. First, bthe Hubbard-$U$ parameter can be relatively simply included for future work (see Chap.~\ref{chapt_perspectives}). Second, it provides a simpler way to play with the parameter at the interfaces that can better respond to the experimentalists need.
      We have hence started a collaboration with Fernando Flores and Pedro de Andres of the Universidad Aut\'onoma of Madrid.

\begin{figure}[!htb]
	\centering
	  \includegraphics[width=0.90\linewidth]{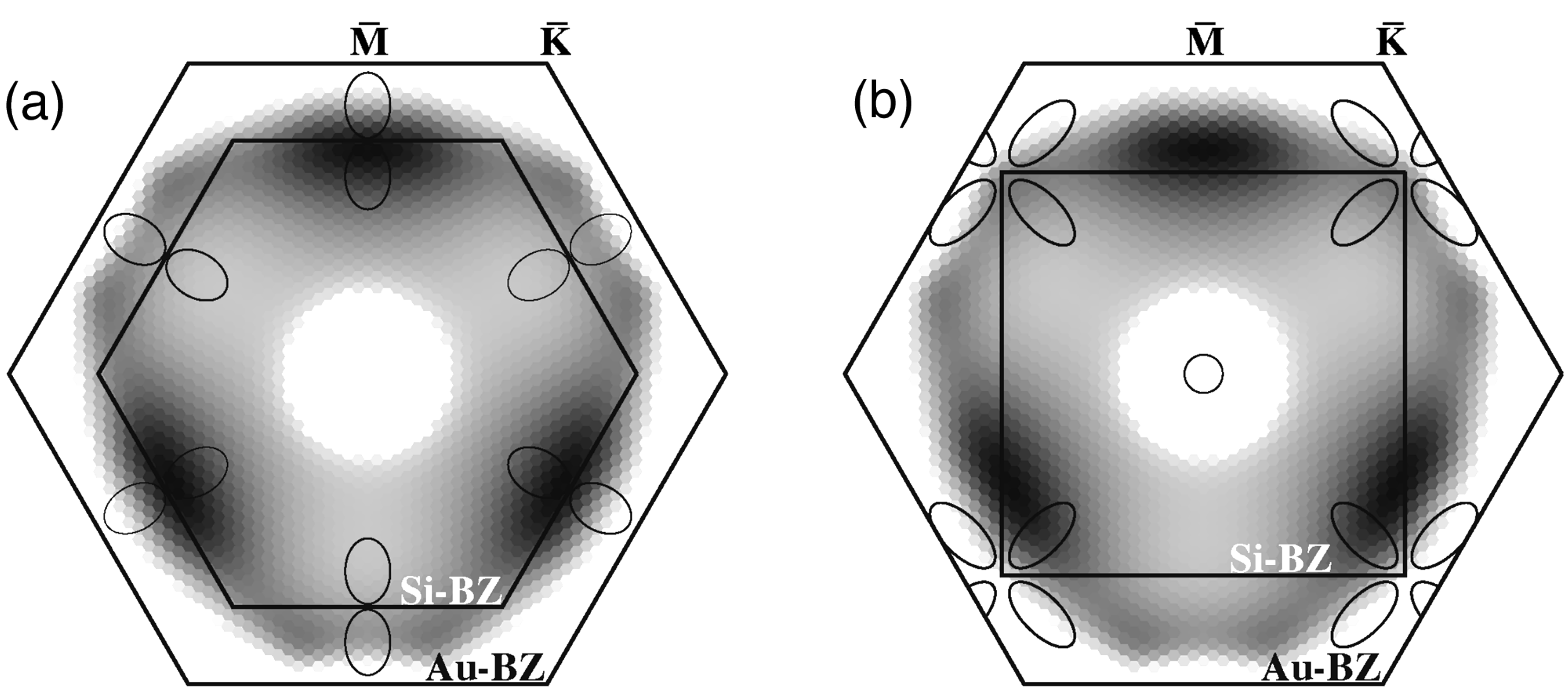}\\ $\phantom{a}$ \\
	  \includegraphics[width=0.4\linewidth]{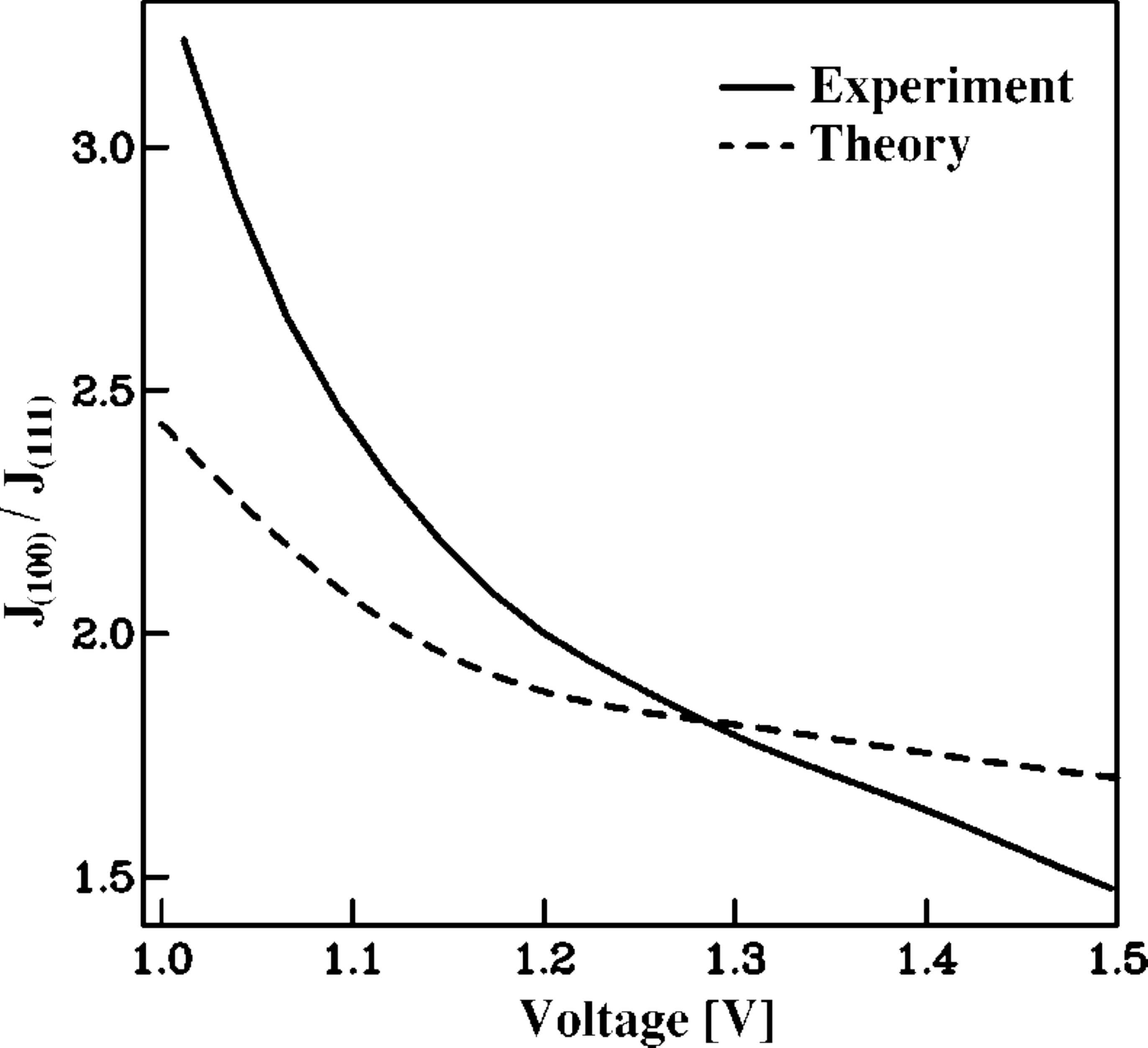} \\ (c)
	  \caption{(a) Au(111)/Si(111) and (b) Au(111)/Si(001) from \cite{Reuter-PhysRevB.58.14036}. High elastic-current is in black and the available DOS in Si is represented by ellipses. Unlike the free-electron model, a BEEM current is expected for the (111) direction of Si. The BEEM current is the overlap between the elastic current and the available DOS. (c) Reuter \emph{et al.} have calculated the ratio of the BEEM current for the two orientations. Despite the strong approximation of a transmission coefficient equal to 1, the result is qualitatively similar to experimental results. }\label{fig_Au_Si_reuter}
      \end{figure}

      The Madrid's group has shown the following results for Au/Si (Fig.~\ref{fig_Au_Si_reuter}):
      \begin{itemize}
       \item There is no propagation at all in the direction (111) of the gold slab. This is in direct contradiction with the free-electron-like model. Even if the electron are injected with a given $k_{\sslash}$, they lose this memory when they propagate inside the gold metal due to its band structure. 
       \item The high-current region matches the available density of states inside the silicon for both orientation (100) and (111). This explains the similar experimental observations in Ref.~\cite{Milliken-AuSi_PhysRevB.46.12826}.
       \item The obtained resolution is compatible with the experimental nanometric resolution due to the focusing properties of forward elastic scattering.
      \end{itemize}
      In conclusion they explain experimental results in a purely elastic limit, without the necessity to invoke any further scattering process at the interface. 
      This was confirmed as stated above by the high-quality epitaxial interfaces grown at the Surfaces and Interfaces group of IPR, where scattering processes at the interface are not expected~\cite{sophie-PhysRevB.81.085319}.
      
      However the results of Garcia-Vidal \emph{et al.} concerned only semi-infinite structures and were tested only on FCC gold, for which a tight-binding description in terms of just first-neighbors is technically possible. This is however not possible for BCC Fe, that is a key element of spintronics: in section \ref{sec_towards_spintronics_results} we show why a tight-binding description of iron with nearest neighbors does not work. For this reason, in Chap.~\ref{chapt_NEPT} we shall extend the Keldysh formalism so as to handle both a finite-layer system and second and third-nearest neighbors.

    \subsection{Equilibrium Calculation}\label{sec_eq_calc}
      In spite of the successful approach of F. Flores' group based on Keldysh formalism by taking into account the band structure, we asked ourselves whether the key ingredient for this success was just the band structure and whether one could avoid, in a first approximation, a full non-equilibrium calculation, that is quite heavy (see Chap.~\ref{chapt_NEPT}). Moreover, a simpler-band structure, equilibrium calculation would be much more intuitive for the whole experimentalist community. The calculation in this case could be done as follows: if $\rho$ electrons per unit volume all move with velocity $v_{\vec{k}}$  the current density is:
      \begin{align}\label{eq_j_eq}
       j_{\kpara} &=-\rho ev_{{\kpara}} \nonumber
       \\&= \sum_n \int \D \vec{k}_{\mathrm{epitaxy}} \vec{\nabla}\varepsilon_{\vec{k}}^{n} \cdot \delta(\varepsilon^n - \varepsilon_{\vec{k}}^{n})
      \end{align} 
      where the velocity is given by the gradient of energies and where $ \delta(\varepsilon^n - \varepsilon_{\vec{k}}^{n})$ is the non-integrated density of states  at $\vec{k}$.
      Of course, at equilibrium, the overall current is zero. The trick used here is to evaluate Eq.~\eqref{eq_j_eq} only along the epitaxy direction, \emph{i.e.}, the real direction of propagation of the current. This will never provide us with the absolute value of the current, but allows a relative analysis in $\kpara$ (parallel to interfaces, which are perpendicular to the epitaxy direction), that is what is demanded for the experiments. The results of this approach are shown in Sec.~\ref{sec_eq_calc_results}. Though approximated, its simplicity should be compared with the heavy artillery of Chap.~\ref{chapt_NEPT}.
  \section{Towards spintronics}\label{sec_towards_spintronics}
      \begin{figure}[b]
        \centering
	   \subbottom[\label{fig_GMR}]{\includegraphics[width=0.8\linewidth]{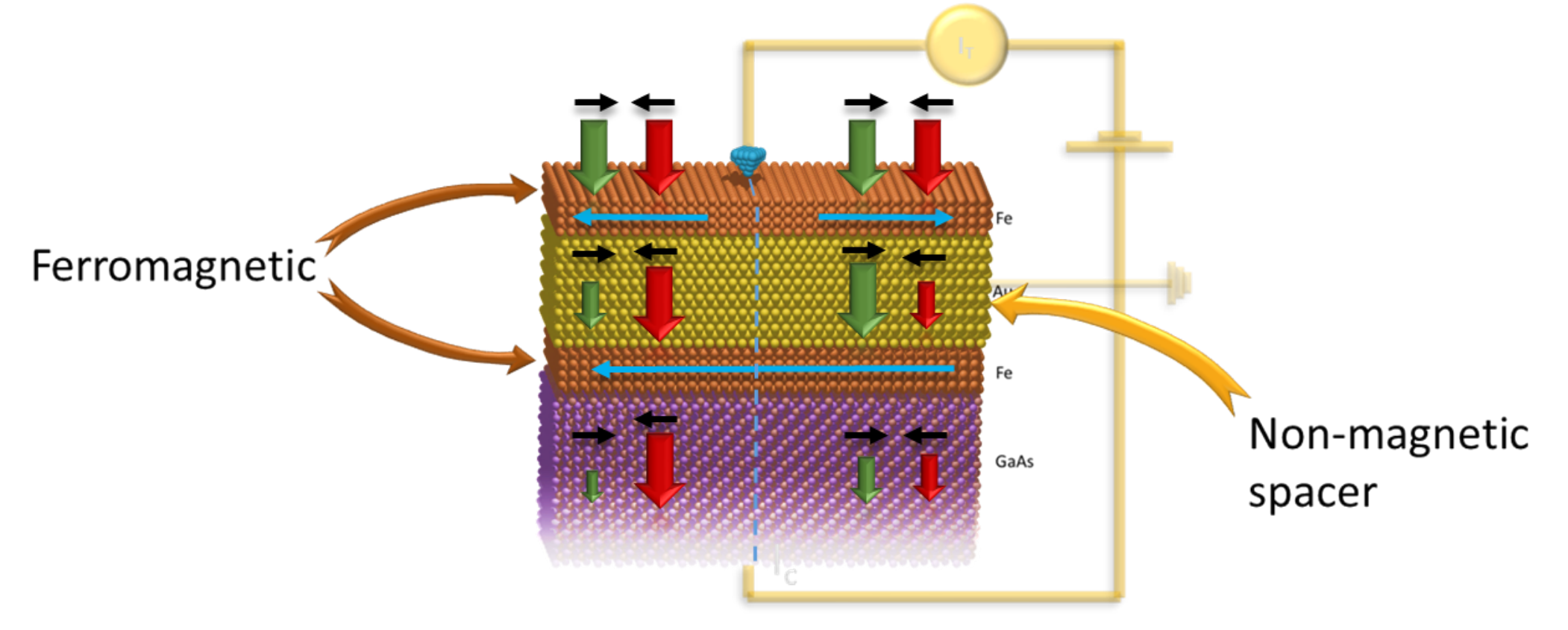}}
	   \subbottom[\label{fig_exp_spin_valve}]{\includegraphics[width=0.15\linewidth]{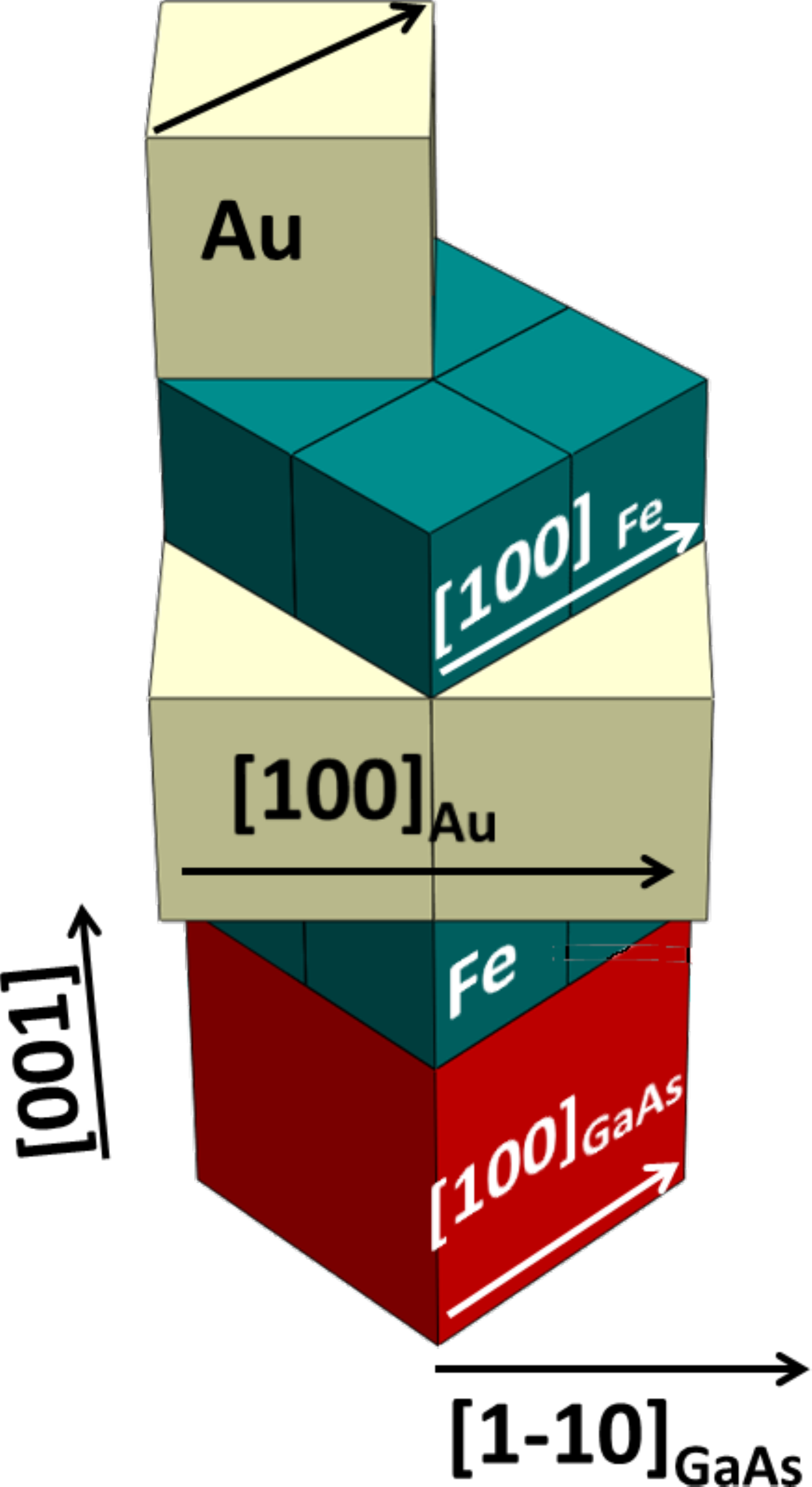}}
          \caption{\subcaptionref{fig_GMR} MagnetoResistance (GMR) device. The BEEM current depends on alignment of the ferromagnetic electrodes, because of the different transmittance for parallel or anti-parallel alignment. \subcaptionref{fig_exp_spin_valve} Structure of the Fe/Au/Fe/GaAs spinvalve studied by M. Hervé {et al.} \cite{Marie-apl_spin_filtering,Marie-jap113_quantitative_magnetic_imaging}. The top film of gold is used to avoid oxydation of the iron electrode.}\label{fig_GMR_spinvalve}          
      \end{figure}

    \subsection{Fe/GaAs[100]}
      This system has been extensively study here at Rennes~\cite{these_marie,sophie-these}. The group has shown than they can growth \emph{in-situ} Fe on GaAs by Molecular Beam Epitaxy (MBE) with an interface of a very good quality. The growth is ``cube on cube'' of four cells of iron (cell paramater=2.87\AA) on one cell of Gallium arsenide (cell parameter=5.65\AA) with the relation Fe(001)[100]//GaAs(001)[100] and a misfit of 1.4\%. 

    \subsection{Fe/Au/Fe/GaAs[001], a spin-valve}
      The Fe/Au/Fe/GaAs(100) spin-valve  has been studied by Marie Hervé during her Ph.D \cite{these_marie}. The gold growth on the iron following the epitaxial orientation: 
      \begin{center}
        Au(001)[100]//Fe(001)[110] 
      \end{center}
      with a misfit of 0.3\%. The last film of iron growth with the same orientation: 
      \begin{center}
             Fe(001)[110]//Au(001)[100]. 
      \end{center}
      In other words, the system is from bottom to top (Fig.~\ref{fig_exp_spin_valve}): 1 zinc-blend cell of gallium arsenide, 4 body-centered-cubic cells of iron in the same direction as GaAs~; 1 face-centered-cubic cell of gold with a rotation of 45\degree\ with respect to the BCC cell of iron ; then a new BCC cell of iron with a in-plane rotation of 45\degree\ with respect to the FCC cell. In practice there is a last slab of gold, a cap, in order to avoid oxidation of the iron film. 
      
      In order to simplify the structure for future calculation (as further detailed in section~\ref{sec_spinvalve_results1} and \ref{sec_towards_spintronics_results}), the 45\degree\ rotating FCC-cell can be seen as a tetragonal-centered cell with the same cell parameter as iron (2.87\AA) in the horizontal plane, and with the usual gold parameter in the vertical plane (4.08\AA).

      The aim of this study was to obtain the largest giant magneto resistance as possible. The magneto-current ($MC$) is defined as the relative variation of the BEEM current between parallel and anti-parallel magnetization of the two ferromagnetic slab (here, the iron):
      \begin{equation}\label{eq_GMR}
       MC = \frac{J_\mathrm{B}^{\mathrm{P}} - J_\mathrm{B}^{\mathrm{AP}}}{J_\mathrm{B}^{\mathrm{AP}}}
      \end{equation}
      Notice that the commonly used formula for relative variations applied to this spin-valve would have rather led to $MC = \frac{J_\mathrm{B}^{\mathrm{P}} - J_\mathrm{B}^{\mathrm{AP}}}{J_\mathrm{B}^{\mathrm{P}}}$. However, probably because of ``psychological'' reasons the form which is used in the literature is the one that gives higher ratio, \emph{i.e.} Eq.~\eqref{eq_GMR}.

\begin{figure}[htb]
        \centering
          \subbottom[\label{fig_BEES_MC_GaAs}]{\includegraphics[width=0.41\linewidth, trim=0 0 0 45, clip]{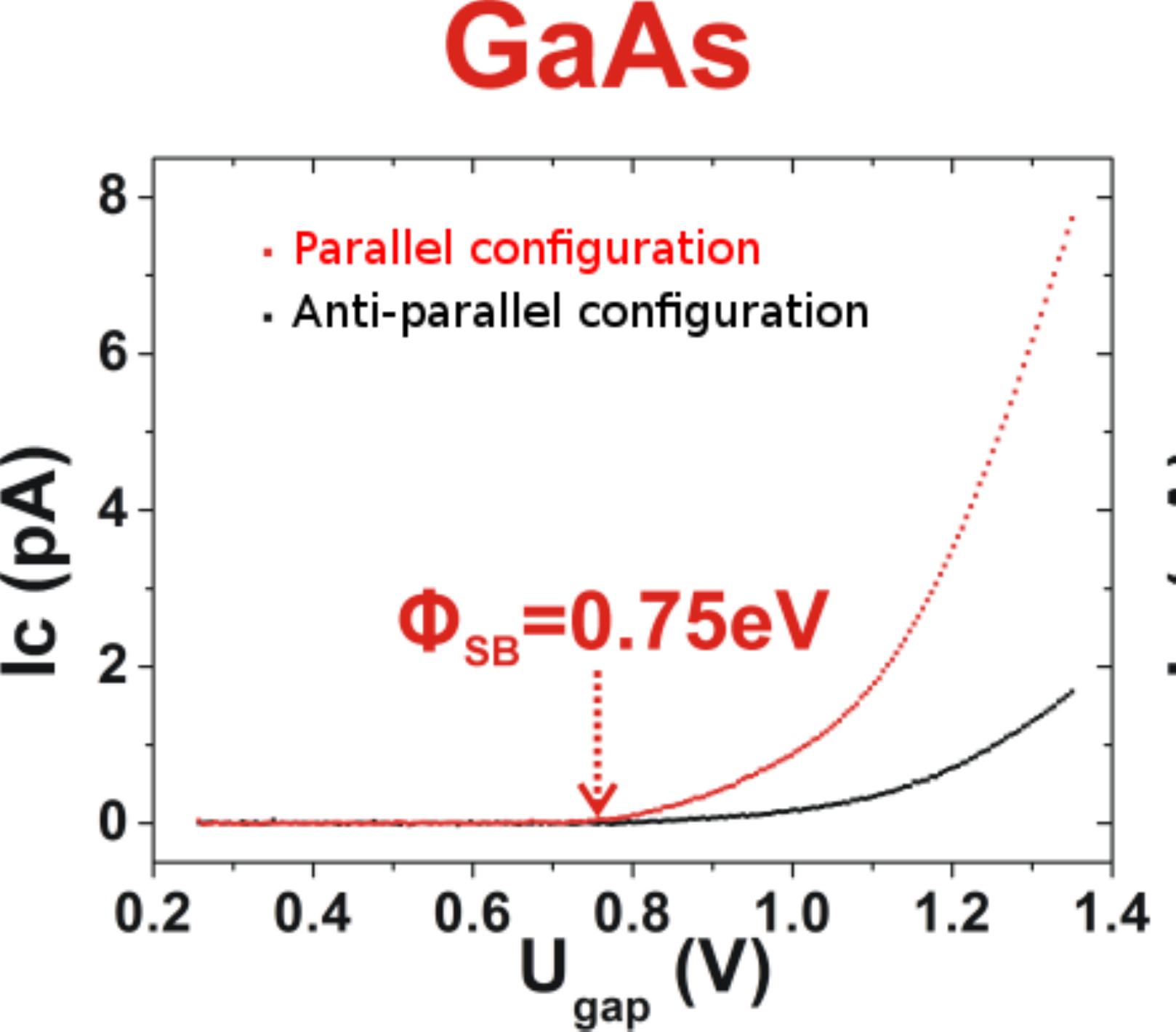}}
          \subbottom[\label{fig_MC_GaAs}]{\includegraphics[width=0.41\linewidth]{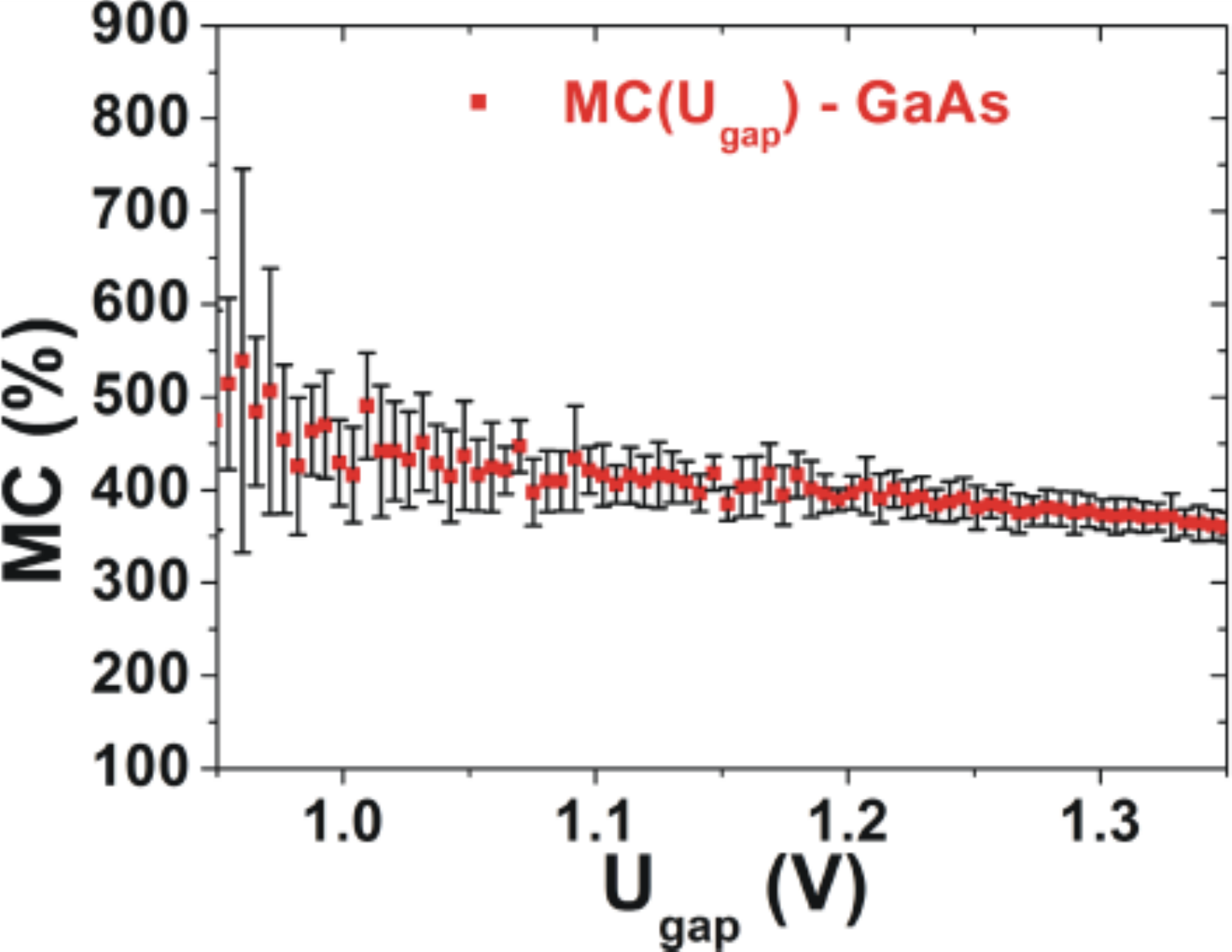}}
          \caption{\subcaptionref{fig_BEES_MC_GaAs} Room temperature Ballistic Electron Magnetic Spectroscopy perform on the Fe/Au/Fe/GaAs spinvalve, by M. Hervé \emph{et al.}\cite{Marie-apl_spin_filtering} and  \subcaptionref{fig_MC_GaAs} resulting magneto-current. The MC seems to increase near the Schottky barrier.}
          {\label{fig_exp_gmr}}
      \end{figure}

      Hervé and co-workers managed to obtain a magneto-current of 400\% at room temperature, as shown in Fig.~\ref{fig_exp_gmr}. 
      Moreover, they showed that the magneto-current was almost independent of the thickness of the iron slab, which suggests that these are interface filtering effects.
      This is also reinforced by the dependence on energy of the magneto-current, which reaches 500\% near the Schottky barrier, as we shall see in Sec.~\ref{sec_eq_calc_results}. 
      
\chapter{Non-equilibrium perturbation-theory applied to BEEM}\label{chapt_NEPT}
  \lettrine[lines=3, lhang=0.35, loversize=0.77, findent=2em,nindent=-0.5em,slope=-1em]{T}{his chapter} presents the expression of the BEEM current based on non-equilibrium Green functions. After defining the system and the notations, we derive the current formula which describes the propagation of electrons inside a layered structure. As we shall see, the key quantities of the final expression are the equilibrium retarded and advanced Green functions of the sample and the density of states of the tip.
Before performing in chapter \ref{chapt_NEPT}, the calculation of these equilibrium Green functions for our finite-slab case (thin films),  we present in Sec.~\ref{sec_decimation}, for future comparison, the decimation method due to F.~Guinea \emph{et al.}~\cite{flores-decimation}. The decimation is based on Dyson equation and designed to find the Green function of a semi-infinite homogeneous structure (a surface followed by an infinity of identical layers).
This approach is the one coded in the program BEEM v2.1~\cite{reuter-beem_v2.1} and used in Ref.~\cite{Reuter-PhysRevB.58.14036}. 
However, by construction, the decimation cannot describe the propagation of electrons inside thin films, in particular when they are not homogeneous (multi-materials). For such 
finite systems, it is necessary to evaluate the Green functions, layer by layer. Such an approach is described in Sec.~\ref{sec_modelization_finite}. 
In particular, we shall use two different calculation procedures, each with its pros and cons: the equation of motion, in section \ref{sec_finite_through_eom}, and the perturbation expansion (Dyson equation) in section \ref{sec_pert_method}. The first method allows to obtain a straightforward iteration procedure for the evaluation of the $n$-layer Green function (Sec.~\ref{sec_iterative_procedure}). However, such a procedure works well only for nearest-neighbor hopping and we could not generalize it to the case of second and third nearest-neighbors, needed to describe iron. To this aim, Dyson equation in perturbation theory, though quite cumbersome (Sec.~\ref{sec_pert_method_2nd_neighbor}), is more appropriate.

\section{BEEM current within Keldysh formalism}\label{section_beem_current}
      \begin{figure}[!tb]
	\centering
	  \includegraphics[width=0.90\linewidth]{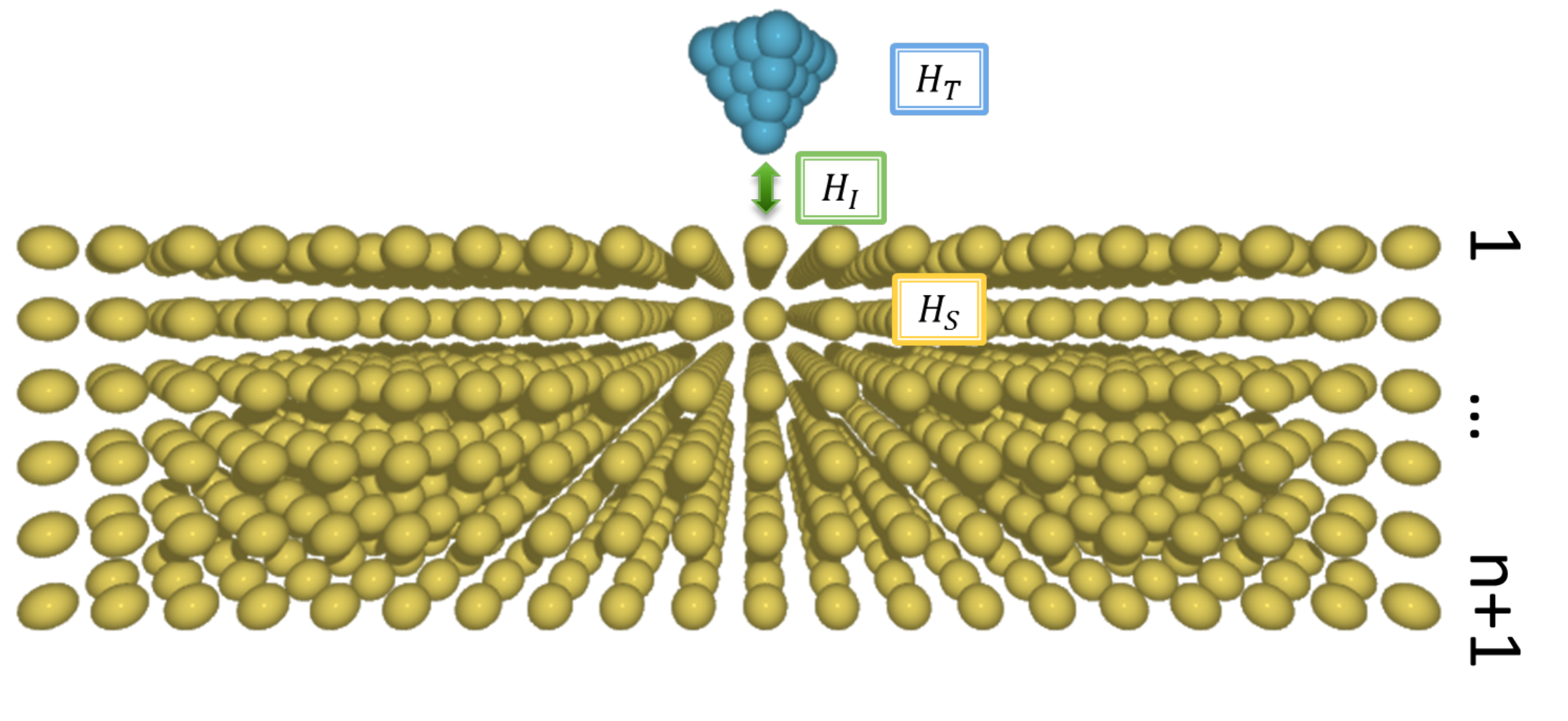}
	  \caption{Representation of our BEEM system: the STM tip is in interaction with a sample made of a finite number of layers $n+1$ in $z$ direction, but infinite in the $(x,y)$ plane (the horizontal plane). The STM tip and the sample are each at equilibrium (described by $H_T$ and $H_S$) yet the chemical potential of the tip, $\mu_T$, being higher than the one of the sample, $\mu_S$, as the tip is at the highest electric-potential. The full Hamiltonian is given in Eq.~\eqref{eq_hamiltonian_keldysh}. We want to calculate the elastic current at layer $n+1$, as described by Eq.~\eqref{eq_current_second_q}}\label{fig_beem_schematics}
      \end{figure}

  In Chapter~\ref{chapt_BEES}, we have seen that in order to give a good description of a BEEM experiment, one has to describe several physical process for the injected electrons: 
  \begin{enumerate}
    \item the tunnel injection from the STM tip to the sample
    \item the propagation of the electrons within the metal
    \item their transmission (and in principle propagation, though this is never treated in the literature) inside the semi-conductor to be detected as BEEM current.
  \end{enumerate}
  As current propagation is, by definition, a non-equilibrium process (though we limit ourselves to the stationary case), a natural way to find an expression for the BEEM current, between two layers $l$ and $m$, is to use non-equilibrium perturbation theory. 
  The framework, illustrated by Fig.~\ref{fig_beem_schematics}, is the following~\cite{Caroli-JPhysC4_1971,Garcia-Vidal-PhysRevLett.76.807,Pedro-ProgSurfS_66_2001}: we first consider the STM tip and the sample as separate objects non-interacting one another, each at equilibrium ($H_T$ and $H_S$ below). Yet, the two chemical potentials are different, that of the tip being higher (this mimics the fact that the tip is at a higher potential than the metal). 
  We then switch the interaction on ($H_I$ below) by allowing electrons to hop from the tip to the sample. This term is treated as a perturbation and is expressed as hopping matrices $t_{\alpha,i}$ that link tunneling active atoms in the tip, $\alpha$, with the corresponding ones in the sample, $i$. In what follows, Greek letters concern the tip and Latin letters the sample. The full Hamiltonian is
  \begin{important}{align}
      H   = &H_T+H_S+H_I \nonumber
  \\& H_T = \sum_{\alpha} \varepsilon_{\alpha} n_{\alpha} + \sum_{\alpha,\beta} t_{\alpha,\beta} c^{\dagger}_{\alpha} c_{\beta} + \mathrm{h.c.}\nonumber
  \\ &H_S = \sum_{i} \varepsilon_{i} n_{i} + \sum_{i,j} t_{i,j} c^{\dagger}_{i} c_{j} + \mathrm{h.c.}\nonumber
  \\& H_I = \sum_{\alpha,i}t_{\alpha,i} c^{\dagger}_{\alpha} c_{j} + \mathrm{h.c.} \label{eq_hamiltonian_keldysh}
  \end{important}
  where the time dependence on operators has been dropped in order to lighten the notation. It is interesting to note at this point that, in the original work of Caroli \emph{et al.}~\cite{Caroli-JPhysC4_1971,Caroli-JPhysC1972} from which the above procedure is borrowed, the authors underlined the fact that ``one might raise a major objection to the above procedure'' as ``the DC bias is first established, and only later the coupling between the barrier and the electrode''. In their system, a metal/insulator/metal junction, indeed the physical realization did not follow the above procedure: the metal/insulator/metal junction already existed and then, the bias was switched on. This situation is not properly described by Eq.~(\ref{eq_hamiltonian_keldysh}). It is however useful to underline that in our BEEM case, the physical process really corresponds to the procedure described by Eq.~(\ref{eq_hamiltonian_keldysh}): a DC bias is first established between the sample $H_S$ and the tip $H_T$, and only later (term 
described 
by $H_I$) the STM tip is brought near the sample.

  In our case, the perturbation $t_{\alpha,i}$ is instantaneous and real. 
  $t_{\alpha,i}$ is real because we work with a real orbital basis and this follows from Eq.~\eqref{eq_hopping_def}.
  We can also consider it to be instantaneous because it is a tunneling process, whose time duration is usually negligible.\footnote{We should remember that, however, time duration of tunneling processes are not properly defined within the orthodox quantum mechanics formalism.}
  If the perturbation is instantaneous, then the retarded Green function is zero because of the Heaviside function in the definition:
  \begin{equation}
    \lim_{t'\rightarrow t^+} G^{R}(t,t') = \lim_{t'\rightarrow t^+} -\i \theta(t-t')\Braket{\psi_0 | \left\{ \annihilation{i}(t), \creation{j}(t') \right\} |\psi_0} = 0
  \end{equation}
  Therefore, $H_I$ does not connect the two branches of the contour C (fig.~\ref{fig_keldysh_contour}), and the lesser Green function identifies to the usual causal Green function:\footnote{In the reference \cite{Caroli-JPhysC4_1971}, Caroli \emph{et al.} used the Keldysh definition for the retarded and advanced Green functions, which is not the usual one (used here): $G^R_{\mathrm{Keldysh}}=G^A_{\mathrm{literature}}$. In their second paper \cite{Caroli-JPhysC1972}, Caroli \emph{et al.} used the usual definition.}
  \begin{equation}\label{eq_G++_G+-}
    G^{++}(\mathit{1,1'}) = G^{+-}(\mathit{1,1'}) + G^{R}(\mathit{1,1'})
  \end{equation}  

    \begin{figure}[!hbt]
    \centering
      \includegraphics[width=0.90\linewidth]{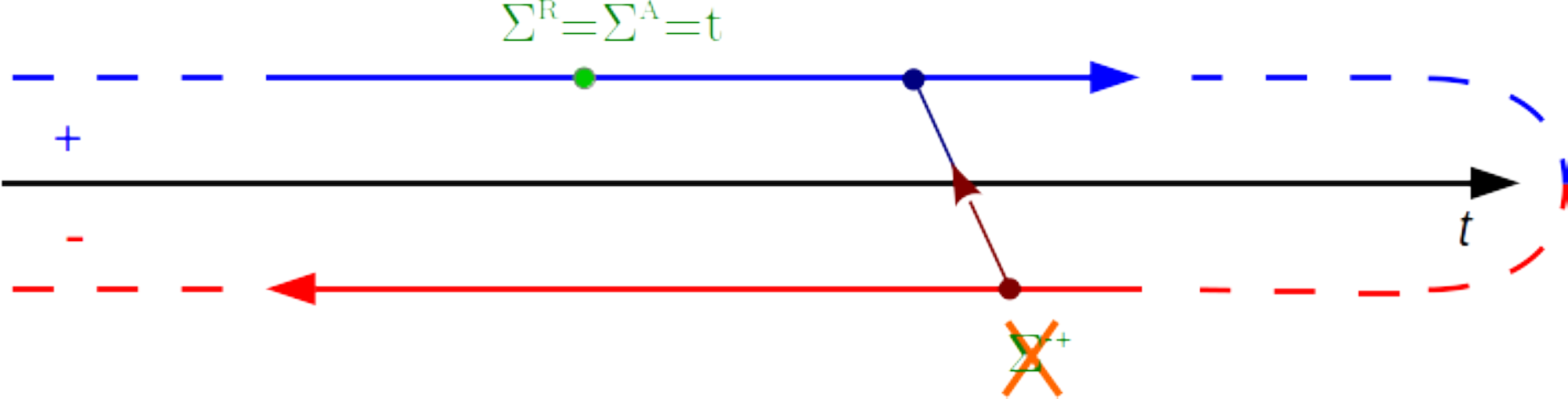}
      \caption{The perturbation induced by the STM tip is instantenous.}\label{fig_instantaneity}
  \end{figure}
  
  This instantaneity and the fact that the perturbation is $t_{\alpha,i}$ and real imply that the self-energies in Eq.~\eqref{eq_keldysh}
  $
    G^{+-}= (\identite + G^{R}\Sigma^{R})  g_{0}^{+-} (\identite + \Sigma^{A}G^{A}) + G^{R} \Sigma^{+-} G^{A}
  $
  are (Fig. \ref{fig_instantaneity}):
  \begin{empheq}[box=\fbox]{align}\label{eq_instant1}
      \Sigma^R_{\alpha,i} &=t_{\alpha,i} = t_{i,\alpha} = \Sigma^A_{\alpha,i}
    \\ \Sigma^{+-}_{\alpha,i} &= 0\label{eq_instant2}
  \end{empheq}
  Hence, the Keldysh Green function through Langreth theorem is:
  \begin{equation}\label{keldysh_formula_BEEM}\color{red}\fbox{$
    G^{+-} = \left( G t G \right)^{+-} = G^R t G^{+-} + G^{+-} t G^A $}
  \end{equation}

  We need to write the expression of the BEEM current in terms of the non-equili\-brium Green functions. In one dimension, the electron current $J_{i,i+1}$ from point $i$ to point $i+1$ (suppose $i$ above $i+1$) is defined as the sum of all electrons hopping from the sites above the site $i$ to the sites below the site $i+1$ ($i$ and $i+1$ included). To this, we have to subtract the sum of all electrons hopping from the sites below the site $i+1$ to the sites above site $i$ (again, $i$ and $i+1$ included). 
  
  Using creation and annihilation operators of the second quantization, the current operator can be written as~\cite{Caroli-JPhysC4_1971}:
  \begin{equation}\label{eq_current_caroli}
    J_{i,i+1} = \sum_{\substack{l\geq i+1 \\ m\leq i}} A_{l,m}\hat{c}^{\dagger}_{l}(t+0^+)\hat{c}_{m}(t) - \sum_{\substack{l\leq i \\ m\geq i+1}} A_{l,m}\hat{c}^{\dagger}_{l}(t+0^+)\hat{c}_{m}(t)
  \end{equation}
  The first term of the right-hand side of \ref{eq_current_caroli} destroys an electron on $m$ above the site $i$ at time $t$ and creates it on $l$ below the site $i+1$ at a later time $t+0^+$. 
  The second term, instead, destroys an electron $m$ below the site $i$ at time $t$ and creates it on $l$ above the site $i+1$ at a later time $t+0^+$. 
  The coefficients $A_{l,m}$ measure the amplitude of the process.
  The Heisenberg picture is assumed here and the small quantity $0+$ is required mathematically in order to ensure analytical properties of Green functions.
  
  From now on, in order to describe the BEEM current of epitaxial thin films, we replace the site indices $i,i+1$ with layer indices. Moreover, as in this work we limit the analysis to the study of $\kpara$-filtering at metallic interfaces (see Chap.~\ref{chapt_results}), for a metallic film made of $n+1$ layers, the current we are looking for, is the current from above towards the last layer $n+1$ (Fig.~\ref{fig_beem_schematics}. The evaluation of the BEEM current is therefore performed as in Ref.~\cite{Reuter-PhysRevB.58.14036}, by projecting the metal current density on the available density of states of the semi-conductor. 
  In this way, the double sum in Eq.~\eqref{eq_current_caroli} reduce to a simple sum:
  \begin{align}\label{eq_current_second_q}
   \hat{J}_{n+1} = \sum_{m\leq n} A_{n+1,m}\hat{c}^{\dagger}_{n+1}(t+0^+)\hat{c}_{m}(t) - \sum_{m\leq n } A_{m,n+1}\hat{c}^{\dagger}_{m}(t+0^+)\hat{c}_{n+1}(t)
  \end{align}
  In practice, this sum runs over the first and second nearest layers only.\footnote{In BCC structure, one has to consider the second and third neighbors in order to reproduce well the band structure. However, the third neighbors reside in the second nearest layer (as described in Sec.~\ref{sec_ham_sub}). That is why the sum runs over the first and second nearest layers only.}

  In order to determine the arbitrary constant $A_{n+1,m}$ we impose the time evolution of the occupation number operator $\hat{n}_{l}=\hat{c}^{\dagger}_l(t+0^+) \hat{c}_l(t)$, around any site $l$, under stationary conditions:
  \begin{align}
   \hat{J}_{n+1} &= -e \braket{\partial_t \hat{n}_{n+1}} \nonumber
   \\            &= -\frac{\i e}{\hbar} \braket{[\hat{n}_{n+1},H]}
  \end{align}  
  According to the derivation in appendix \ref{appendix_commutator} and equation (\ref{equation_of_motion_n}) the current reads
  \begin{align}\label{eq_Jtemp}
   J_{n+1} &= \frac{\i e}{\hbar}  \sum_{{m}} \left( t_{n+1,m} \braket{c^{\dagger}_{n+1}c_{m}} - t_{m,n+1}^{*} \braket{c^{\dagger}_{m}c_{n+1}}\right)\nonumber
   \\     &= \frac{e}{\hbar}        \sum_{{m}}  t_{n+1,m}\left[ G^{+-}_{m,n+1}(t+0^+)           -  G^{+-}_{n+1,m}(t+0^+)            \right]
  \end{align}
  using $t_{m,n+1}^{*}= t_{n+1,m}$ and defining the non-equilibrium lesser Green function (cf Eq.~\eqref{eq_negf}):
  \begin{align}
   G^{+-}_{m,n+1}(t+0^+) &= \i \Braket{0^{\mathrm{neq}} | \hat{c}^{\dagger}_{n+1} \hat{c}_m | 0^{\mathrm{neq}} } \\
   G^{+-}_{n+1,m}(t+0^+) &= \i \Braket{0^{\mathrm{neq}} | \hat{c}^{\dagger}_m \hat{c}_{n+1} | 0^{\mathrm{neq}} }
  \end{align}  
  Notice that in Ref.~\cite{Caroli-JPhysC4_1971}, it is the causal Green function (defined as $G^{++}$ here) that is used, but as shown in Eq.~\eqref{eq_G++_G+-}, both Green functions can be used as the perturbation is instantaneous.
  
   As the BEEM is a non equilibrium but stationary process, it is more convenient to Fourier transform and work in the energy domain (while the time domain is better suited to study transitory regimes):
   \begin{equation}\label{eq_TF_gpm}
     G^{+-}_{m,n+1}(t-t') = \int_{-\infty}^{+\infty}\frac{\D \omega}{2\pi} G^{+-}_{m,n+1}(\omega) \e^{-\i\omega(t-t')}
   \end{equation}
   As we are interested in the limit $t'\rightarrow t+0^+$, Eq.~\eqref{eq_TF_gpm} becomes
   \begin{align}
    G^{+-}_{m,n+1}(t+0^{+}) = \int_{-\infty}^{+\infty}  \frac{\D\omega}{2\pi} G^{+-}_{m,n+1}(\omega)\e^{-\i\omega\eta^+}
   \end{align}
   where $\eta$ is a positive infinitesimal part needed for a proper contour integral in the complex $\omega$-plane. 
   From Eq.~\eqref{eq_Jtemp}, this leads to
   \begin{equation}\label{current_E_general}\color{red}
    \boxed{J_{n+1}= \frac{e}{\hbar} \int \frac{\D E}{2\pi} \Tr\ \left[ \hat{t}_{n+1,m}\left(   G^{+-}_{m,n+1}(E)-  G^{+-}_{n+1,m}(E) \right)\right] }
   \end{equation} 
   In Sec.~\ref{sec_keldysh}, we have seen how to calculate non-equilibrium Green functions, using a perturbation expansion based on Langreth theorem. For STM and BEEM currents, the above definitions Eq. \eqref{eq_instant1} and \eqref{eq_instant2}  allow us to write the Dyson equation (\ref{eq_dyson_g+-}) as:
  \begin{align}\label{eq_G+-mn+1}
      G^{+-}_{m,n+1} &= g^{+-}_{m,n+1} + \sum_{\alpha,i} \left( g_{m,i} t_{i,\alpha} G_{\alpha,n+1} \right)^{+-} \nonumber \\
                     &= g^{+-}_{m,n+1} + \sum_{\alpha,i} \left(  g_{m,i}^R t_{i,\alpha} G_{\alpha,n+1}^{+-} + g_{m,i}^{+-} t_{i,\alpha} G_{\alpha,n+1}^A \right)
  \end{align}
  We need the expression of the new non-equilibrium Green function:
  \begin{align}\label{eq_G+-alphan+1}
   G^{+-}_{\alpha,n+1} &= \sum_{\beta,j} \left( g_{\alpha,\beta} t_{\beta,j} G_{j,n+1} \right)^{+-} \nonumber \\
                       &= \sum_{\beta,j} \left( g_{\alpha,\beta}^R t_{\beta,j} G_{j,n+1}^{+-} + g_{\alpha,\beta}^{+-} t_{\beta,j} G_{j,n+1}^{A} \right)
  \end{align}
  Once again we need another non-equilibrium Green function:
  \begin{align}\label{eq_G+-in+1}
   G^{+-}_{j,n+1} &= g^{+-}_{j,n+1} + \sum_{\gamma,k} \left( g_{j,k} t_{k,\gamma} G_{\gamma,n+1} \right)^{+-} \nonumber \\
                  &= g^{+-}_{j,n+1} + \sum_{\gamma,k} \left( g_{j,k}^R t_{k,\gamma} G_{\gamma,n+1}^{+-} + g_{j,k}^{+-} t_{k,\gamma} G_{\gamma,n+1}^{A} \right)
  \end{align} 
  We notice that we have as many non-equilibrium Green functions as hopping terms from the tip to the sample. As the distance between tip and sample in BEEM varies around 5\AA, we will consider only hopping from the last atom of the tip, to the first layer of the sample, \emph{i.e.} $t_{\alpha,i}=t_{0,1}$ and vice-versa, so that we can now close the system by writing:
  \begin{align}
    G^{+-}_{m,n+1} &= g^{+-}_{m,n+1} +  g_{m,1}^R t_{1,0} G_{0,n+1}^{+-} + g_{m,1}^{+-} t_{1,0} G_{0,n+1}^A     \\
    G^{+-}_{0,n+1} &= g_{0,0}^R t_{0,1} G_{1,n+1}^{+-} + g_{0,0}^{+-} t_{0,1} G_{1,n+1}^{A} \\
    G^{+-}_{1,n+1} &=g^{+-}_{j,n+1} +  g_{1,1}^R t_{1,0} G_{0,n+1}^{+-} + g_{1,1}^{+-} t_{1,0} G_{0,n+1}^{A} 
  \end{align}
  And we obtain
  \begin{align}
   G^{+-}_{0,n+1} &= \underbrace{\left( \identite - g_{0,0}^R t_{0,1} g_{1,1}^R t_{1,0}  \right)^{-1}}_{D_{0,0}^R} 
                          \left(\begin{array}{l} g_{0,0}^R    t_{0,1} g_{1,1}^{+-} t_{1,0} G_{0,n+1}^A \\
                               + g_{0,0}^{R} t_{0,1} g_{1,n+1}^{+-} \\
                               + g_{0,0}^{+-} t_{0,1} G_{1,n+1}^{A} \end{array}\right)
  \end{align}
  Re-injected in Eq. \eqref{eq_G+-mn+1} we obtain:
  \begin{align}
   G^{+-}_{m,n+1} &= g^{+-}_{m,n+1} \\
      &\phantom{=} +   g_{m,1}^R t_{1,0} D_{0,0}^R g_{0,0}^R    t_{0,1} g_{1,1}^{+-} t_{1,0} G_{0,n+1}^A \\
      &\phantom{=} +            g_{m,1}^R t_{1,0} D_{0,0}^R g_{0,0}^{R} t_{0,1} g_{1,n+1}^{+-} \\
      &\phantom{=} +            g_{m,1}^R t_{1,0} D_{0,0}^R g_{0,0}^{+-} t_{0,1} G_{1,n+1}^{A} \\
      &\phantom{=} +                       g_{m,1}^{+-} t_{1,0} G_{0,n+1}^A 
  \end{align}  
  That can be simplified using
  \begin{align}
    g_{m,1}^R t_{1,0} D_{0,0}^R g_{0,0}^{R} & = g_{m,1}^R t_{1,0} G_{0,0}^{R} = G_{m,0}^{R}    
  \end{align}
  and
  \begin{align}
   g_{m,1}^R t_{1,0} D_{0,0}^R &= g_{m,1}^R t_{1,0} \left( \identite - g_{0,0}^R t_{0,1} g_{1,1}^R t_{1,0}  \right)^{-1} \\
                                              &= g_{m,1}^R \left( t_{1,0}^{-1} - g_{0,0}^R t_{0,1} g_{1,1}^R t_{1,0} t_{1,0}^{-1} \right)^{-1} \\
                                              &= g_{m,1}^R \left( t_{1,0}^{-1} - t_{1,0}^{-1} t_{1,0} g_{0,0}^R t_{0,1} g_{1,1}^R \right)^{-1} \\
                                              &= g_{m,1}^R \left( \identite - g_{0,0}^R t_{0,1} g_{1,1}^R t_{1,0}  \right)^{-1} t_{1,0} \\
                                              &= g_{m,1}^R \left( \identite - t_{1,0} g_{0,0}^R t_{0,1} g_{1,1}^R  \right)^{-1} t_{1,0} \\
                                              &= G_{m,1}^R t_{1,0}
  \end{align}
  So finally,  $G^{+-}_{m,n+1}$ is:
  \begin{align}
   G^{+-}_{m,n+1} &= g^{+-}_{m,n+1} \\
      &\phantom{=} +   G_{m,0}^R t_{0,1} g_{1,1}^{+-} t_{1,0} G_{0,n+1}^A \\
      &\phantom{=} +                         G_{m,0}^{R} t_{0,1} g_{1,n+1}^{+-} \\
      &\phantom{=} +   G_{m,1}^R t_{1,0} g_{0,0}^{+-} t_{0,1} G_{1,n+1}^{A} \\
      &\phantom{=} +                       g_{m,1}^{+-} t_{1,0} G_{0,n+1}^A 
  \end{align}
  $G^{+-}_{n+1,m}$ is obtained by switching indices $m$ and $n+1$.

\begin{figure}[!bt]
    \centering
      \includegraphics[width=0.50\linewidth]{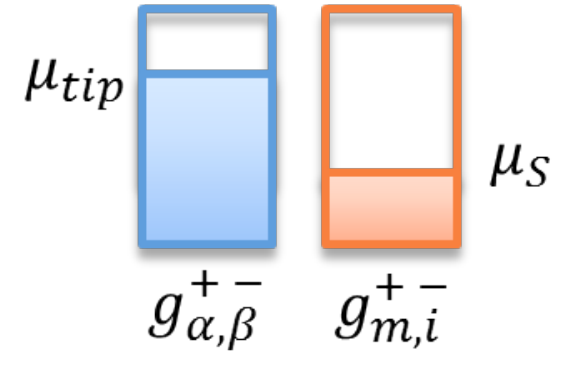}
      \caption{The Keldysh Green functions for the uncoupled system (\emph{i.e.} for the tip alone and for the sample alone) are linked to the chemical potentials as shown in Eq.~\eqref{eq_g+-_chemical}.}\label{fig_chemical_pot}
  \end{figure}

  The Keldysh Green function $g^{+-}$ of the uncoupled system (and hence equilibrium) can be expressed as retarded and advanced equilibrium Green functions and the Fermi distribution ($f_T(E)$ for the tip and $f_S(E)$ for the sample):
  \begin{align}\label{eq_III19}
    g^{+-}_{1,1}          &= f_S (g^{A}_{1,1}-g^{R}_{1,1}) \\
    g^{+-}_{0,0} &= f_T (g^{A}_{0,0}-g^{R}_{0,0}) \label{eq_III20}
  \end{align}
  Equations \eqref{eq_III19} and \eqref{eq_III20} are a consequence of the explicit expressions of $g^{+-}$, $g^{R}$ and $g^{A}$ for uncoupled systems (Fig.~\ref{fig_chemical_pot}), that are:
  \begin{align}
    g^{+-}_{m,1}(t-t') &= \i \Braket{\creation{m}(t')\annihilation{i}(t)} \nonumber
    \\                 &= \i f\ \e^{-\i \varepsilon_{i}(t-t')} \label{eq_g+-_chemical}
    \\ g_{m,1}^R{(t-t')} &= -\i\theta(t-t') \Braket{\left\{\annihilation{i}(t),\creation{m}(t')\right\}} \nonumber
    \\                      &= -\i\theta(t-t')f\ \e^{-\i \varepsilon_{i}(t-t')} 
    \\ g_{m,1}^A{(t-t')} &= \i\theta(t'-t) \Braket{\left\{\annihilation{i}(t),\creation{m}(t')\right\}} \nonumber
    \\                      &= \i\theta(t'-t)f\ \e^{-\i \varepsilon_{i}(t-t')} 
  \end{align}

  Those equations lead to:
  \begin{equation}\label{g+--g+-}
    G^{+-}_{m,n+1}-G^{+-}_{n+1,m} = g^{+-}_{m,n+1} - g^{+-}_{n+1,m} + f_T U_T + f_S U_S
  \end{equation}
  where
  \begin{align}
      U_T =  \big[ & G^R_{m,1} t_{1,0} (g^{A}_{0,0}-g^{R}_{0,0}) t_{0,1} G^A_{1,n+1}  \nonumber
  \\                                    -& G^R_{n+1,i} t_{1,0} (g^{A}_{0,0}-g^{R}_{0,0}) t_{0,1} G^A_{1,m} \big]
  \end{align} 
  concerns the tip and where
  \begin{align}
    U_S =  \left[\begin{array}{l}  G^R_{m,0}   t_{0,1} (g^{A}_{1,n+1}-g^{R}_{1,n+1}) 
                                               +(g^{A}_{m,1}-g^{R}_{m,1}) t_{1,0} G^A_{0,n+1}  
    \\                                           -G^R_{n+1,0} t_{0,1} (g^{A}_{1,m}-g^{R}_{1,m}) 
                                               - (g^{A}_{n+1,i}-g^{R}_{n+1,i}) t_{1,0} G^A_{0,m}  
    \\                                           +G_{m,0}^R t_{0,1} g_{1,1}^{+-} t_{1,0} G_{0,n+1}^A 
                                               -G_{n+1,0}^R t_{0,1} g_{1,1}^{+-} t_{1,0} G_{0,m}^A  \end{array}\right]
  \end{align}
  concerns the sample. We now, consider each term separately.

\subsubsection*{$\blacksquare$ $g^{+-}_{m,n+1} - g^{+-}_{n+1,m}$}
  The first term $g^{+-}_{m,n+1} - g^{+-}_{n+1,m}$ in Eq.~\eqref{g+--g+-} corresponds to the current between $l$ and $m$ inside the metal in absence of coupling between the tip and the sample (equilibrium situation), i.e. zero.
  \begin{equation}\fbox{$
    g^{+-}_{m,n+1} - g^{+-}_{n+1,m}=0$}
  \end{equation} 

\subsubsection*{$\blacksquare$ $U_T$}
  Let's consider first the term associated with the tip $U_T$. Using $G^A_{m,1}=\left( G^R_{1,m} \right)^\dagger$ and $t_{1,0}=t_{0,1}^\dagger$ one can write
  \begin{align}
    G^R_{m,1} t_{1,0} (g^{A}_{0,0}-g^{R}_{0,0}) t_{0,1} G^A_{1,n+1} = - \left[ G^R_{n+1,1} t_{1,0} (g^{A}_{0,0}-g^{R}_{0,0}) t_{0,1} G^A_{1,m} \right]^\dagger
  \end{align}
  It allows to write $U_T$ as a real part of a matrix:
  \begin{align}
      U_T &= 2 \Re  \big[  G^R_{m,1} t_{1,0} (g^{A}_{0,0}-g^{R}_{0,0}) t_{0,1} G^A_{1,n+1} \big] \nonumber\\
          &= 2 \Re  \big[  G^R_{m,1} t_{1,0} g^{A}_{0,0} t_{0,1} G^A_{1,n+1} -  G^R_{m,1} t_{1,0} g^{R}_{0,0} t_{0,1} G^A_{1,n+1} \big] \nonumber \\
        \boxedalign{U_T  &= 2 \Re  \big[ G^R_{m,1} t_{1,0} G^{A}_{0,n+1} -  G^R_{m,0} t_{0,1} G^A_{1,n+1}\big]}
  \end{align}  

\subsubsection*{$\blacksquare$ $U_S$}
  $U_S$ also can be expressed as a real part of a matrix using same arguments:
    \begin{align}
    U_S &=  \left[\begin{array}{lcl}  G^R_{m,0} t_{0,1} (g^{A}_{1,n+1}-g^{R}_{1,n+1})                &-& G^R_{n+1,0} t_{0,1} (g^{A}_{1,m}-g^{R}_{1,m}) \\
                                  + (g^{A}_{m,1}-g^{R}_{m,1}) t_{1,0} G^A_{0,n+1}                    &-& (g^{A}_{n+1,i}-g^{R}_{n+1,i}) t_{1,0} G^A_{0,m}  
    \\                            + G_{m,0}^R t_{0,1} (g^{A}_{1,1}-g^{R}_{1,1}) t_{1,0} G_{0,n+1}^A  &-&G_{n+1,0}^R t_{0,1} (g^{A}_{1,1}-g^{R}_{1,1}) t_{1,0} G_{0,m}^A  \end{array}\right]
    \\ \label{eq_Us_interm}
     &= 2 \Re  \left[\begin{array}{lcl} \phantom{+}
                                        G^R_{m,0} t_{0,1} g^{A}_{1,n+1} &-& G^R_{m,0} t_{0,1} g^{R}_{1,n+1} \\
                                       +g^{A}_{m,1} t_{1,0} G^A_{0,n+1} &-& g^{R}_{m,1} t_{1,0} G^A_{0,n+1} \\
                                       +G_{m,0}^R t_{0,1} g^{A}_{1,1} t_{1,0} G_{0,n+1}^A &-& G_{m,0}^R t_{0,1} g^{R}_{1,1} t_{1,0} G_{0,n+1}^A 
                                       \end{array}\right]                                   
  \end{align}
  here, we can use the fact that:
  \begin{align}
   G_{m,0}^R t_{0,1} {\color{red}g^{A}_{1,1} t_{1,0} G_{0,n+1}^A} &=  G_{m,0}^R t_{0,1} {\color{red}\left( G_{1,n+1}^A - g_{1,n+1}^A \right)} \\
   {\color{blue}G_{m,0}^R t_{0,1} g^{R}_{1,1}} t_{1,0} G_{0,n+1}^A &={\color{blue} \left( G_{m,1}^R - g_{m,1}^R \right)}t_{1,0} G_{0,n+1}^A
  \end{align}
  That allows to simplify Eq. \eqref{eq_Us_interm}:
  \begin{align}
   U_S &= 2 \Re  \left[\begin{array}{l} \phantom{+}
                                         {\color{red}G^R_{m,0} t_{0,1} g^{A}_{1,n+1}} {\color{orange}- G^R_{m,0} t_{0,1} g^{R}_{1,n+1}} \\
                                       +{\color{orange} g^{A}_{m,1} t_{1,0} G^A_{0,n+1}} {\color{blue}- g^{R}_{m,1} t_{1,0} G^A_{0,n+1}} \\
                                       + G_{m,0}^R t_{0,1} G_{1,n+1}^A   { \color{red}-G_{m,0}^R t_{0,1} g_{1,n+1}^A }  \\
                                       - G_{m,1}^R t_{1,0} G_{0,n+1}^A   {\color{blue}+ g_{m,1}^R t_{1,0} G_{0,n+1}^A} \end{array}\right]
  \end{align}
  The two terms in blue, as well as the two terms in red, obviously cancel. The two terms in orange also cancel as we are evaluating the real part of the difference between two complex conjugates $\Re \left( {\color{orange} g^{A}_{m,1} t_{1,0} G^A_{0,n+1} - G^R_{m,0} t_{0,1} g^{R}_{1,n+1}} \right)=0$.

%

  Finally, $U_S$ is:
  \begin{align}
  \boxedalign{  U_S &= 2 \Re  \left[  G_{m,0}^R t_{0,1} G_{1,n+1}^A   
                                                            - G_{m,1}^R t_{1,0} G_{0,n+1}^A  \right]}
  \end{align}

\subsubsection*{$\blacksquare$ Current formula}
  At this point, $G^{+-}_{m,n+1}-G^{+-}_{n+1,m}$ is given by the exact formula
  \begin{align}
    G^{+-}_{m,n+1}-G^{+-}_{n+1,m} &= f_T U_T + f_S U_S \nonumber\\
                                  &= 2 \Re  \left(f_T-f_S\right) \left[G^R_{m,1} t_{1,0} G^{A}_{0,n+1} -  G^R_{m,0} t_{0,1} G^A_{1,n+1}\right]
  \end{align}
  Expanding $ G^{A}_{0,n+1}$ and $G^R_{m,0}$ we have:
  \begin{align}
   G^{+-}_{m,n+1}-G^{+-}_{n+1,m} &= 2 \Re  \left(f_T-f_S\right) \left[G^R_{m,1} t_{1,0} g^{A}_{0,0} t_{0,1} G^{A}_{1,n+1}
                                                                                    -G^R_{m,1} t_{1,0} g^{R}_{0,0} t_{0,1} G^{A}_{1,n+1} \right] \nonumber\\
                                 &=  2 \Re  \left(f_T-f_S\right) \left[G^R_{m,1} t_{1,0}(g^{A}_{0,0}- g^{R}_{0,0}) t_{0,1} G^{A}_{1,n+1} \right]
  \end{align}
  As the matrix density of states of the tip (Eq.~\eqref{eq_dos_from_Gr}) is given by $\rho_{0,0}=\frac{1}{\pi}\Im g^{A}_{0,0}=-\frac{1}{\pi}\Im g^{R}_{0,0}$ we have finally
  \begin{align}
   G^{+-}_{m,n+1}-G^{+-}_{n+1,m} &=  2 \Re  \left(f_T-f_S\right) \left[G^R_{m,1} t_{1,0}(2\i\pi\rho_{0,0}) t_{0,1} G^{A}_{1,n+1} \right] \nonumber\\
                                 &= 4\pi \left(f_T-f_S\right) \Im   \left[G^R_{m,1} t_{1,0} \rho_{0,0} t_{0,1} G^{A}_{1,n+1} \right]
  \end{align}

   Hence, the current is:
  \begin{equation}\label{eq_current_general}
   J_{n+1} = \frac{4e\pi}{\hbar} \left(f_T-f_S\right) \Im \int_{0}^{\infty} \frac{\D \omega}{2\pi} \Tr \sum_m  t_{n+1,m} G^R_{m,1} t_{1,0} \rho_{0,0} t_{0,1} G^{A}_{1,n+1} 
  \end{equation} 
  This equation is the generalization of the one in Ref.~\cite{Reuter-PhysRevB.58.14036}, where the approximation described below were assumed since the beginning.
  We can further simplify Eq.~\eqref{eq_current_general} with the following assumptions that apply in our case: in BEEM, a typical distance between the sample and the tip is $\sim$5\AA. Assuming that the coupling matrices $t_{0,1}$ are much smaller than hopping matrices in the metal (as in tunneling condition), one can work in the lowest-order perturbation theory and not ``renormalize'' the Green function. In other words, the denominators defined above are $D^{R,A}=\identite$. This leads to the simpler expression
  \begin{align}
  G^{+-}_{m,n+1}-G^{+-}_{n+1,m} &=4 \pi (f_T-f_S) \Im  g^{R}_{m,1} t_{1,0} \rho_{0,0} t_{0,1} g^{A}_{1,n+1}
  \end{align}
  where capital $G$ are now small $g$.
  Finally, assuming zero temperature, the current is expressed as an integral over a window of energies ranging from the Schottky barrier height $\phi_{SB}$ up to the applied voltage:
  \begin{important}{align}\color{red}
  J_{n+1} = \frac{4e\pi}{\hbar}\Im \int_{\phi_{SB}}^{eV} \frac{\D E}{2\pi} \Tr \sum_m   t_{n+1,m} g^{R}_{m,1} t_{1,0} \rho_{0,0} t_{0,1} g^{A}_{1,n+1}\label{eq_BEEM_current}
  \end{important}
  The current is now expressed with equilibrium Green functions, and DOS, of isolated systems, quantities which can be calculated with the usual equilibrium formalism, as detailed in the sections below. 
  With this equation, two of the three physical processes (the tunnel injection from the STM tip to the sample, the propagation of the electrons within the metal) can be described by calculating the equilibrium retarded and advanced Green functions. The third physical process (transmission and propagation inside the semi-conductor to be detected as BEEM current) will be dealt with in Chap.~\ref{chapt_perspectives}.

  \section{Modeling of a semi-infinite slab}\label{sec_decimation}
  In order to highlight analogies and differences with the finite-slab case that is the main object of this work, it is useful to introduce here the case of a semi-infinite slab, treated through the so-called decimation procedure with nearest-neighbor hopping by F. Flores group~\cite{flores-decimation,Pedro-phys_scripta_T66_277,lannoo-decimation}.

  This decimation procedure is a way to obtained $G_{1,2^n}$ (the Green function from layer 1 to layer 2 to the power of $n$) through a set of Dyson equations, faster than by iterating layer by layer, because, at each iteration of Dyson equations, the number of layers is doubled (\emph{i.e.}, we move from a $n$-layer system to a $2n$-layer system, instead of a $n+1$-layer system of the layer-by-layer procedure).
  It can in principle be used to evaluate exactly any homogeneous system with a number of layers $2^n$.
  When the system becomes thick enough (semi-infinite), any propagation from one surface (layer 1) to the other (layer $2^n$), expressed by $G_{1,2^n}$, vanishes and some simplification occurs~\cite{Pedro-phys_scripta_T66_277}. 
  We would like to remind, however, that though the decimation technique was used in Refs.~\cite{flores-decimation,Pedro-phys_scripta_T66_277,lannoo-decimation} with this ``semi-infinite approximation'', it is otherwise an exact procedure.
  Here for example, we present the calculation for $G_{1,4}$ of a four layers slab with nearest neighbor hopping. Our starting point is the one-layer Green function. We suppose to have two different layers: A, with Green function $\greenone{1,1}$ and B, with Green function $\greenone{2,2}$.
  
  The Dyson expansion of the Green functions of two planes ``AB'' in interaction is:
  \begin{align}
    \green{1,2}{2} &= \greenone{1,1} t_{1,2} \greenone{2,2} + \greenone{1,1} t_{1,2} \greenone{2,2} t_{2,1} \greenone{1,1} t_{1,2} \greenone{2,2} + \dots \nonumber
    \\             &= \green{1,1}{2} t_{1,2} \greenone{2,2}
    \\ \green{1,1}{2} &= \greenone{1,1} +\greenone{1,1} t_{1,2} \greenone{2,2} t_{2,1} \greenone{1,1} + \dots \nonumber
    \\  	      &= \greenone{1,1}  + \greenone{1,1} t_{1,2} \greenone{2,2} t_{2,1}  \green{1,1}{2}
  \end{align}
  Solving the system, we find
  \begin{align}\label{G112}
      \green{1,1}{2} &= \left(\identite -\greenone{1,1} t_{1,2} \greenone{2,2} t_{2,1} \right)^{-1} \greenone{1,1} 
    \\\green{1,2}{2} &= \left(\identite -\greenone{1,1} t_{1,2} \greenone{2,2} t_{2,1} \right)^{-1} \greenone{1,1} t_{1,2} \greenone{2,2} \label{G122}
  \end{align}
  and also, by switching indexes:
  \begin{align}\label{G222}
      \green{2,2}{2} &= \left(\identite -\greenone{2,2} t_{2,1} \greenone{1,1} t_{1,2} \right)^{-1} \greenone{2,2} 
    \\\green{2,1}{2} &= \left(\identite -\greenone{2,2} t_{2,1} \greenone{1,1} t_{1,2} \right)^{-1} \greenone{2,2} t_{2,1} \greenone{1,1} \label{G212}
  \end{align}
  Here the exponent $(n)$ means ``a slab made of n layers'' and the subscript represents the layer's number (as usual). A small $g$ is for unperturbed monolayer (redundant with exponent (1)). Note that as every quantity is a matrix (which is labeled with a hat), the multiplication are thus non-commutative. 
  
  Consider now 2 layers ``AB'' in contact with two other layers ``AB''. This means that the perturbation is the hopping $t_{2,3}$ between layer 2 and layer 3. The Green functions $\green{1,2}{2}$ and $\green{1,1}{2}$ are the zeroth order Green function starting from which we have to write the ``full'' (n=4) Green functions $\green{1,1}{4}$, $\green{4,1}{4}$, $\green{2,1}{4}$ and $\green{3,1}{4}$. From Dyson equation, we have:
  \begin{align}
       \green{1,1}{4} &= \green{1,1}{2} + \green{1,2}{2} t_{2,3} \green{3,1}{4}
    \\ \green{3,1}{4} &= G_{33}^{(2)} t_{3,2} \green{2,1}{4} \label{eq_G314}
    \\ \green{2,1}{4} &= G_{21}^{(2)} + \green{2,2}{2} t_{2,3} \green{3,1}{4}
    \\ \green{4,1}{4} &= G_{43}^{(2)} t_{3,2} \green{2,1}{4}
  \end{align}
  using the equalities ``layer 1 = layer 3'' and ``layer 2 = layer 4'', it leads to
  \begin{align}
       \green{4,1}{4} &= G_{21}^{(2)} t_{3,2} \green{2,1}{4} 
    \\ \green{2,1}{4} &= G_{21}^{(2)} + \green{2,2}{2} t_{2,3} \underbrace{ G_{33}^{(2)} t_{3,2} \green{2,1}{4}}_{\green{3,1}{4}}
    \\ \green{1,1}{4} &= \green{1,1}{2} + \green{1,2}{2} t_{2,3} \green{3,1}{4}
  \end{align}
  Using $ \green{2,1}{4}=(\identite- \green{2,2}{2} t_{2,3} \green{1,1}{2} t_{3,2})^{-1} G_{21}^{(2)}$ (this equation is the formally same as eq.(\ref{G222}) by replacing unperturbed Green functions by perturbed Green functions) we finally get:
  \begin{align}
    \green{4,1}{4} &=  G_{21}^{(2)} t_{3,2} (\identite- \green{2,2}{2} t_{2,3} \green{1,1}{2} t_{3,2})^{-1} G_{21}^{(2)}
  \end{align}
  by expressing $ G_{ij}^{(2)}$ in terms of $ g_{ij}^{(1)}$ (equations (\ref{G112}) to (\ref{G212})) we obtain:
  \begin{align}
    \green{4,1}{4} =& \underbrace{(\identite- \greenone{2,2} t_{2,1} \greenone{1,1} t_{1,2})^{-1} \greenone{2,2} t_{2,1} \greenone{1,1}}_{G_{21}^{(2)}} t_{3,2} \cdot \nonumber
    \\            &\left[\identite- \underbrace{\left(\identite -\greenone{2,2} t_{2,1} \greenone{1,1} t_{1,2} \right)^{-1} \greenone{2,2}}_{\green{2,2}{2}}  t_{2,3} 
                                    \underbrace{\left(\identite -\greenone{1,1} t_{1,2} \greenone{2,2} t_{2,1} \right)^{-1} \greenone{1,1}}_{\green{1,1}{2}}  t_{3,2} \right]^{-1} \cdot \nonumber
    \\            &\underbrace{\left(\identite -\greenone{2,2} t_{2,1} \greenone{1,1} t_{1,2} \right)^{-1} \greenone{2,2}t_{2,1} \greenone{1,1} }_{G_{21}^{(2)}}
  \end{align}
  This Green function, in turn, becomes the new zeroth order from which we can write the ``full'' (n=8) Green functions, and so forth. It is clear that is a very effective way to calculate propagators. Moreover, as stated above, it is an exact derivation. However, it only describes slabs of $2\time2^n$ layers: it does not give the Green functions of any number of layers, and, especially, of any composition. This is the reason that has led to its dismissal in the present work. We are interested in studying hepitaxial thin films of the kind: Fe(1nm)/Au(2.6nm)/Fe(1.2nm)/GaAs. We need therefore a method allowing to deal with each single layer separately. This is shown in section \ref{sec_modelization_finite}.
  
\begin{figure}
    \centering
    \subbottom[\label{subfig_transm_mat}]{
      \includegraphics[width=0.3\linewidth]{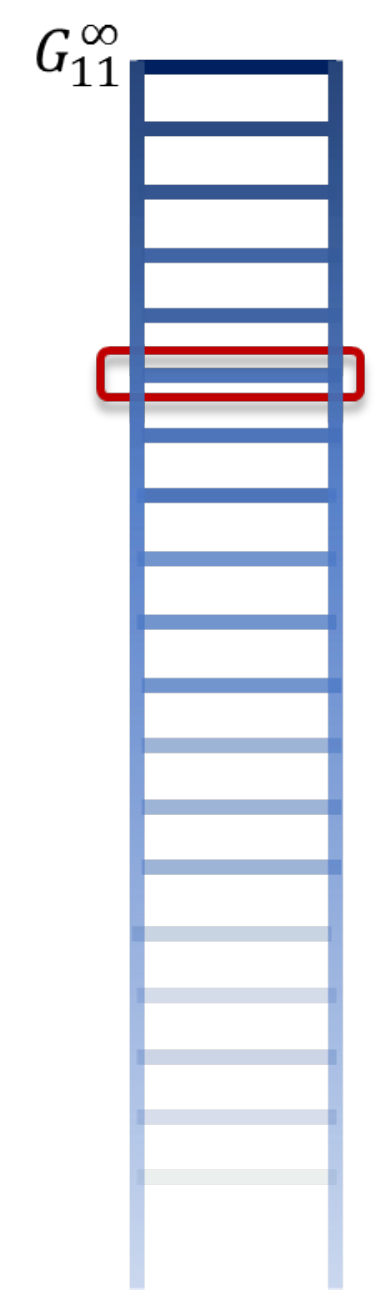}} \qquad\qquad
    \subbottom[\label{subfig_finite}]{
      \includegraphics[width=0.25\linewidth]{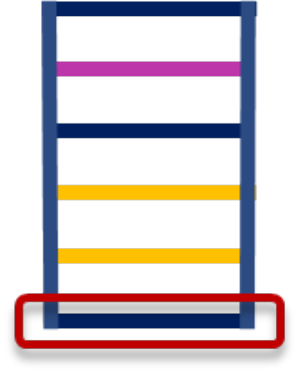}}
    \caption{Semi-infinite vs finite procedures. \subcaptionref{subfig_transm_mat} BEEM v2.1 \cite{reuter-beem_v2.1} uses the decimation procedure combined with a transfer-matrix approach in order to calculate the elastic current inside a semi-infinite slab that is, by construction, layer periodic. \subcaptionref{subfig_finite} Calculate Green functions through a layer-by-layer procedure allows to describe thin heterostructures.}
    \label{fig_inf_vs_finite}
  \end{figure}

  Before finishing this subsection on the decimation technique, however, it is useful for technical reasons to linger on the following point.  
  In the ancient code BEEM v2.1, K. Reuter had implemented a transfer matrix method which allowed to calculate propagator at a given layer $m$, $\green{m,1}{2n}$, inside a semi-infinite structure, starting from the exact Green function $\green{1,1}{2n}$. The equation was written as follows:
  \begin{equation}\label{eq_form_trans_mat}
    \green{m,1}{2n} = (\hat{M}^{n})^{m-1} \green{1,1}{2n}
  \end{equation}
  where $\hat{M}^{(n)}= \green{1,1}{n} t_{n+1,n}$ is called the transfer matrix. 
  \begin{demo}
    This idea comes from the generalization of Eq.~\eqref{eq_G314} that can be written as:
    \begin{align}
      G_{n+1,1}^{(2n)} &= G_{n+1,n+1}^{(n)} t_{n+1,n} G_{n,1}^{(2n)} \nonumber
      \\               &= \underbrace{G_{1,1}^{(n)} t_{n+1,n}}_{\hat{M}^{(n)}} G_{n,1}^{(2n)}
    \end{align}
    where we have used $\green{1,1}{n}=\green{n+1,n+1}{n}$. The meaning of this last equation is clear: the matrix $\hat{M}^{(n)}$ allows to propagate from layer $n$ to layer $n+1$ of the $2n$-layer slab. We want to see if this matrix can be used to propagate from layer 1, to any layers, as expressed in Eq.~\eqref{eq_form_trans_mat}.
    
    From Dyson equation we have the following expression for the propagator from layer 1 to layer 2 of a $2n$-layer slab:
    \begin{equation}\label{eq_G21n}
      \green{2,1}{2n} = \green{2,1}{n} + \green{2,n}{n} t_{n,n+1} \underbrace{\green{1,1}{n} t_{n+1,n}}_{\hat{M}^{(n)}} \green{n,1}{2n}
    \end{equation}
    where we have used again $\green{1,1}{n}=\green{n+1,n+1}{n}$. We recognize $\hat{M}^{(n)}=\green{1,1}{n} t_{n+1,n}$.
    In order to have an equation of the form of Eq.~\eqref{eq_form_trans_mat}, the following equality has to be satisfied:
    \begin{equation}\label{eq_III66}
      \hat{M}^{(n)} \green{1,1}{2n} = \eqref{eq_G21n}
    \end{equation}
    The exact expression for $\green{1,1}{2n}$, from Dyson equation, is however:
    \begin{equation}
      \green{1,1}{2n}=\green{1,1}{n} + \green{1,n}{n} t_{n,n+1} \hat{M}^{(n)} \green{n,1}{2n}
    \end{equation} 
    Therefore, Eq.~\eqref{eq_III66} is satisfied only if the following equation is satisfied:
    \begin{equation}
      \hat{M}^{(n)} \green{1,1}{n} + \hat{M}^{(n)} \green{1,n}{n} t_{n,n+1} \hat{M}^{(n)} \green{n,1}{2n} = \green{2,1}{n} + \green{2,n}{n} t_{n,n+1} {\green{1,1}{n} t_{n+1,n}} \green{n,1}{2n}
    \end{equation}
    The latter equality is true only if:
    \begin{equation}\left\{\begin{array}{rl}
      \hat{M}^{(n)} \green{1,1}{n} & = \green{2,1}{n} \\
      \green{2,n}{n}   & = \hat{M}^{(n)} \green{1,n}{n} \end{array}\right.\label{eq_condition_transfer_mat}
    \end{equation}
    This is however generally not so, because we should have at the same time:
    \begin{equation}\left\{\begin{array}{rl}
      \green{2,1}{n} &= \hat{M}^{(n/2)} \green{1,1}{n}  \\
      \green{2,n}{n} & = \hat{M}^{(n/2)} \green{1,n}{n} \end{array}\right.\label{eq_condition_transfer_mat2}
    \end{equation}
    It turns out that Eqs.~\eqref{eq_condition_transfer_mat} and \eqref{eq_condition_transfer_mat2} can be both satisfied only for a semi-infinite system, where we have $\green{1,1}{n}=\green{1,1}{2n}$ and then $\hat{M}^{(n/2)}=\hat{M}^{(n)}$. 
    In that case only, we can use the transfer matrix $M^{n}$ in order to find any Green functions $\green{m,1}{2n}$, using
    \begin{equation}\label{eq_transfer_matrix}
      \green{m,1}{2n} = (\hat{M}^{(n)})^{m-1} \green{1,1}{2n}
    \end{equation}
  \end{demo}
  In conclusion, the transfer matrix approach is a very useful tool to get the current at any layer inside a semi infinite slab (Fig.~\ref{subfig_transm_mat}). However, it is based on the assumption that $\hat{M}^{(n)}=\hat{M}^{(n/2)}$ is valid only for the semi-infinite slab. The method has therefore not be designed to describe thin films. For that, we need another procedure to calculate the propagators, as described in the next section.
  \section{Modeling of a finite structure}\label{sec_modelization_finite}

Besides the decimation technique, it is possible to find the retarded and advanced Green functions needed for Eq.~\eqref{eq_BEEM_current}
through an exact layer-by-layer procedure. The advantage of this procedure, though less efficient, is that it becomes possible to describe structures of low thickness, as well as structures that are not layer periodic (Fig.~\ref{subfig_finite}), like spinvalves.
  
As stated in the introduction to this chapter, we shall employ two different methods for this layer-by-layer calculation: the equations of motion (Sec.~\ref{sec_finite_through_eom}) and the perturbation expansion (Sec.~\ref{sec_pert_method}).

  \subsection{Layer-by-layer equation of motion}\label{sec_finite_through_eom}

    It is quite easy to solve the equation of motion for few layers  but it very quickly becomes tedious even if straightforward (already for more than 5 layers). That is why we have found an iterative procedure to find the $n$-layer Green function. For that, we have solved the Green functions $G_{1,n}$ for slabs of $n=2$, $n=3$ and $n=4$ and have established a formula which gives the Green function of $n$-layers starting from 1 layer and iterating up to the $n$ layers. For simplicity, we describe the derivation considering only nearest-neighbors interactions. It should be reminded that by using the method of equation of motion, we do not need to pass through the interaction representation: all operators in this subsection are Heisenberg operators and their time-evolution is governed by Eq.~\eqref{eq_heisenberg_eq_motion}, as shown in section \ref{sec_schrodinger_heisenberg_picture}

    \subsubsection{Few layer procedure}
    \subsubsubsection*{$\bullet$ Case of two layers}
    The Hamiltonian is:
      \begin{equation}\label{eq_ham_2layers}
        H =  \sum_{\kpara} \left(\varepsilon^{(1)}_{\kpara} \ccdagger{1\kpara}{1\kpara}+ \varepsilon^{(2)}_{\kpara} \ccdagger{2\kpara}{2\kpara} + t_{12} \ccdagger{1\kpara}{2\kpara} + t_{21} \ccdagger{2\kpara}{1\kpara}\right)          
      \end{equation}
      As shown in section~\ref{sec_resol_through_eom_chapt1}, the Green functions can be calculated using the formula
      \begin{equation}
       (\hbar\omega- \varepsilon) G_{ij\sigma}(\omega) = \delta_{i,j} + \sum_m t_{im} G_{mj}(\omega)
      \end{equation}
      Specializing it to the Hamiltonian~\eqref{eq_ham_2layers}, we get:
      \begin{align}
         (\hbar\omega- \varepsilon^{(1)}) \green{1,2}{2}(\omega) &= t_{1,2}\green{2,2}{2}(\omega)
       \\(\hbar\omega- \varepsilon^{(2)}) \green{2,2}{2}(\omega) &= 1 + t_{2,1}\green{1,2}{2}(\omega)
      \end{align}
      Note that every quantities are matrices and hence do not commute. 
      
      Defining $\left( \greenone{i,i}\right)^{-1} = (\hbar\omega- \varepsilon^{(i)})$, and dropping the $\omega$ dependence, the solutions are:
      \begin{align}
      	\green{1,2}{1} &= \greenone{1,1} t_{1,2}\left[ \left( \greenone{2,2} \right)^{-1}-t_{2,1}\greenone{1,1}t_{1,2} \right]^{-1} \label{eq_G12_2layers}\\
        \green{2,2}{2} &= \left[ \left( \greenone{2,2} \right)^{-1}-t_{2,1}\greenone{1,1}t_{1,2} \right]^{-1} \label{eq_G22_2layers}
      \end{align}
      Remind that we want to find a general formula which allows to find any $G_{1,n}^{(n)}$, then, we have to find a similar set of equations for 3 layers.     

      \begin{remark}
	In the case of scalar quantities, it is possible to find the poles and the spectral weight by identifying
	\begin{align}
	  \green{1,2}{2}(\omega) &= \frac{\hbar\omega-\varepsilon_2
					}{(\hbar\omega)^2-(\varepsilon_1+\varepsilon_2)\hbar\omega+(\varepsilon_1\varepsilon_2-t_{12}t_{21})
					  }       
	\end{align}
	with
	\begin{align}
	  \green{1,2}{2}(\omega) &= \frac{A_1}{\hbar\omega-E_1} + \frac{A_2}{\hbar\omega-E_2} \nonumber
	  \\                 &= \frac{(A_1+A_2)\hbar\omega-(A_1E_2+A_2E_1)}{(\hbar\omega)^2-(E_1+E_2)\hbar\omega+E_1E_2}
	\end{align}
	because the denominator is quadratic. The two poles $E_1$ and $E_2$ and their weights $A_1$ and $A_2$ are
	\begin{align}
	  E_1 &= \frac{1}{2}(\varepsilon_1+\varepsilon_2) - \frac{1}{2} \sqrt{(\varepsilon_1-\varepsilon_2)^2+4t_{12}t_{21}}
	\\E_2 &= \frac{1}{2}(\varepsilon_1+\varepsilon_2) + \frac{1}{2} \sqrt{(\varepsilon_1-\varepsilon_2)^2+4t_{12}t_{21}}
	\\A_1 &= \frac{1}{2}\left( 1+\frac{1}{\sqrt{1+\frac{4t_{12}t_{21}}{(\varepsilon_1-\varepsilon_2)^2} }} \right)
	\\A_2 &= \frac{1}{2}\left( 1-\frac{1}{\sqrt{1+\frac{4t_{12}t_{21}}{(\varepsilon_1-\varepsilon_2)^2} }} \right)
	\end{align}
	For matrix hopping terms $t_{i,j}$, it is not possible to find an analytical expression in that way, and we must use a numerical approach. 
      \end{remark}

    \subsubsubsection*{$\bullet$ Case of three layers}
      
      For three layers, the system of equations of motion to solve after Fourier transform is:
      \begin{align}
           (\hbar\omega-\varepsilon^{(3)}) \green{3,3}{3}(\omega) &= 1 + t_{3,2}\green{2,3}{3}(\omega)
        \\ (\hbar\omega-\varepsilon^{(2)}) \green{2,3}{3}(\omega) &= t_{2,1}\green{1,3}{3}(\omega) + t_{2,3}\green{3,3}{3}(\omega)
        \\ (\hbar\omega-\varepsilon^{(1)}) \green{1,3}{3}(\omega) &= t_{1,3}\green{3,3}{3}(t-t')
      \end{align}
      and with some algebra
      \begin{align}\label{eq_G13_3layers}
	\green{1,3}{3} &= \greenone{1,1} t_{1,2} \green{2,3}{3} \\
	\green{2,3}{3} &= \left[\left(\greenone{2,2}\right)^{-1} - t_{2,1} \greenone{1,1} t_{1,2}\right]^{-1} t_{2,3} \green{3,3}{3} \\
	\green{3,3}{3} &= \Bigg\{ \left( \green{3,3}{1} \right)^{-1} - t_{3,2} \underbrace{\left[ \left( \greenone{2,2} \right)^{-1} - t_{2,1} \greenone{1,1} t_{1,2} \right]^{-1}}_{\green{2,2}{2}} t_{2,3} \Bigg\}^{-1} \label{eq_G33_3layers}
      \end{align}
      where we recognize $\green{2,2}{2}$ in the expression of $\green{3,3}{3}$. An iterative procedure begin to emerge: we moved to four layers, in order to confirm it.

    \subsubsubsection*{$\bullet$ Case of four layers}
      Working with the same method for the four layers configuration, we obtain:
      \begin{align}\label{eq_G14_4layers}
	\green{1,4}{4} &= \greenone{1,1} t_{1,2} \green{2,4}{4} \\
	\green{2,4}{4} &= \left[\left(\greenone{2,2}\right)^{-1} - t_{2,1} \greenone{1,1} t_{1,2}\right]^{-1} t_{2,3} \green{3,4}{4} \\
	\green{3,4}{4} &= \left\{ \left( \green{3,3}{1} \right)^{-1} - t_{3,2} \left[  \left( \greenone{2,2} \right)^{-1} - t_{2,1} \greenone{1,1} t_{1,2} \right]^{-1} t_{2,3} \right\}^{-1} t_{3,4} \green{4,4}{4}\\
	\green{4,4}{4} &= \Bigg( \green{4,4}{1}- t_{4,3} 
	  \underbrace{\Bigg\{ \left( \green{3,3}{1} \right)^{-1} - t_{3,2} \underbrace{\left[ \left( \greenone{2,2} \right)^{-1} - t_{2,1} \greenone{1,1} t_{1,2} \right]^{-1}}_{\green{2,2}{2}} t_{2,3} \Bigg\}^{-1}}_{\green{3,3}{3}}
	  t_{3,4} \Bigg)^{-1} \label{eq_G44_4layers}
      \end{align}
      It is then clear that we can express $\green{4,4}{4}$ in terms of $\green{3,3}{3}$ which is itself expressed in term of $\green{2,2}{2}$, and finally find $\green{1,4}{4}$.

    \subsubsection{Iterative procedure}\label{sec_iterative_procedure}
      From the previous equations~\eqref{eq_G13_3layers} to \eqref{eq_G33_3layers}, of the three-layer case, and \eqref{eq_G14_4layers} to \eqref{eq_G44_4layers}, of the four-layer cases, it is possible to deduce the following iterative formulas:
      \begin{equation}\label{eq_iterative_G1np1np1}
       	\green{1,n+1}{n+1} = \prod_{i=1}^{n+1} G_{i,i}^{(i)} t_{i,i+1}
      \end{equation} 
      \begin{equation*}
	\begin{array}{rl}
	\mbox{with } & \left\{\begin{array}{rl}
	                    \greenone{1,1} &= \left(\hbar\omega - \varepsilon^{(1)} \right)^{-1} \\
	                    G_{i+1,i+1}^{(i+1)} &= \left[\left( g_{i+1,i+1}^{(1)} \right)^{-1}- t_{i+1,i} G_{i,i}^{(i)}  t_{i,i+1}\right]^{-1} 	\\
	                    t_{n+1,n+2} &= \identite
	                   \end{array}\right.
	\end{array}
      \end{equation*}
      As we wrote Eq.~\eqref{eq_iterative_G1np1np1} with a sum up to $n+1$, we have to specify the special condition $t_{n+1,n+2} = \identite$ since the layer $n+2$ does not exist.
      We see from these equations, that as long as we know both the hopping from one layer to the next one, and the Hamiltonian of the isolated layers, then we can find the advanced Green function $G_{n+1,1}^{(A,n+1)} = \left[\green{1,n+1}{R,n+1}\right]^{\dagger}$ of Eq.~\eqref{eq_BEEM_current}. 
      
      In order to calculate the retarded Green function $\green{1,n}{R,n+1}$ required by Eq.~\eqref{eq_BEEM_current} we can proceed in the same way as for $G_{1,i}^{i+1}$. We obtain a generalization  similar to Eq.~\eqref{eq_iterative_G1np1np1}:
      \begin{equation}\label{eq_iterative_G1nnp1}
	\begin{array}{rl}
	\green{1,n}{n+1}   &= \prod_{i=1}^{n} G_{i,i}^{(i)} t_{i,i+1} \\ \\
	\mbox{with } & \left\{\begin{array}{rl}
	                    \greenone{1,1} &= \left(\hbar\omega - \varepsilon^{(1)} \right)^{-1} \\
	                    G_{i+1,i+1}^{(i+1)} &= \left[\left( g_{i+1,i+1}^{(1)} \right)^{-1}- t_{i+1,i} G_{i,i}^{(i)}  t_{i,i+1}\right]^{-1} \\ 
	                    t_{n,n+1} & = \identite
	                   \end{array}\right.
	\end{array}
      \end{equation}       
      This time, as the sum runs up to $n$ layers, $t_{n,n+1}$ does not exist and must be set equal to the identity.

    \subsubsection{Effective hopping}\label{sec_effective_hopping}
      It is possible to write the above iterative procedure in another form using a transfer-matrix approach similar to the one defined in the decimation procedure (Sec.~\ref{sec_decimation}). 
      The advantage to proceed as follows is that it allows a direct comparison with the equations obtained through Dyson equation. Besides, it is more intuitive, as we shall see.
      
      We can start, as above, from the two-layer case. Equations~\eqref{eq_G12_2layers} and \eqref{eq_G22_2layers} can be rewritten as:
      \begin{align}\label{eq_g12_Teff}
	\green{1,2}{2} &= \greenone{1,1} t_{1,2}\left[  \left(\identite - \greenone{2,2} t_{2,1}\greenone{1,1}t_{1,2} \right) \left( \greenone{2,2} \right)^{-1}\right]^{-1} \nonumber \\
                       &= \greenone{1,1} \underbrace{t_{1,2}\left[\identite -\greenone{2,2} t_{2,1}\greenone{1,1}t_{1,2} \right]^{-1}}_{T_{1,2}^{eff}} \greenone{2,2} \\
	\green{2,2}{2} &=  \left(\identite -\greenone{2,2}t_{2,1}\greenone{1,1}t_{1,2}  \right)^{-1} \greenone{2,2} 
      \end{align}      
      where $T_{1,2}^{eff}=t_{1,2}\left[\identite -t_{2,1}\greenone{1,1}t_{1,2} \greenone{2,2}\right]^{-1}$ is the effective hopping which links layers 1 and 2 one to another, by taking into account their interaction through the denominator (energy-dependent renormalization). 
      \\*
      
      We move to three layers, in order to see if we can find a similar effective-hopping $T_{2,3}^{eff}$. Developing Eq.~\eqref{eq_G13_3layers}, we find:
      \begin{multline}
       \green{1,3}{3} =\\ \underbrace{\greenone{1,1} t_{1,2} \left[ \left( \greenone{2,2}\right)^{-1} - t_{2,1}\greenone{1,1}t_{1,2}  \right]^{-1}}_{\green{1,2}{2} {\scriptsize \mbox{( Eq.~\eqref{eq_G12_2layers})}}}
       t_{2,3} \Bigg\{ \left( \greenone{3,3}\right)^{-1} - t_{3,2} 
           \underbrace{\left[ \left( \greenone{2,2}\right)^{-1} - t_{2,1}\greenone{1,1}t_{1,2}  \right]^{-1}}_{\green{2,2}{2} {\mbox{\scriptsize ( Eq.~\eqref{eq_G22_2layers})}}}
           t_{2,3} \Bigg\}^{-1} \\
            = \green{1,2}{2} t_{2,3}  \Bigg\{ \left( \greenone{3,3}\right)^{-1} - t_{3,2} \green{2,2}{2} t_{2,3} \Bigg\}^{-1}
      \end{multline}      
      Using the same factorization as above, we have:
      \begin{equation}
       \green{1,3}{3} = \green{1,2}{2} 
                        \underbrace{t_{2,3}  \Bigg\{ \identite - \greenone{3,3} t_{3,2} \green{2,2}{2} t_{2,3} \Bigg\}^{-1}}_{T_{2,3}^{eff}}
                        \greenone{3,3}
      \end{equation} 
      From that, we see that the effective hopping obeys the iterative expression
      \begin{equation}\label{eq_effective_hopping}
       T_{i,i+1}^{eff} =  t_{i,i+1} \left[\identite -  g_{i+1,i+1}^{1} t_{i+1,i} \green{i,i}{i} t_{i,i+1} \right]^{-1}
      \end{equation} 
      and the Green function can be obtained through
      \begin{equation}\label{eq_G1n_Teff}
       \green{1,i+1}{i+1} = \green{1,i}{i} T_{i,i+1}^{eff} \green{i+1,i+1}{1}
      \end{equation}
      We can choose either equation~\eqref{eq_iterative_G1np1np1} or equation~\eqref{eq_G1n_Teff} in order to find the Green functions. The latter has the advantage of being more intuitive: electrons jumps from a slab of $n$ layers to an isolated layer $n+1$ with a probability $t_{n,n+1}$, and this process is renormalized due to the interaction between the slab and the isolated layer.

      Both methods converge very fast, so that, choosing one or the other is not critical.      
      In conclusion, we have found iterative equations which give the propagator for any number of layers considering only first-neighbor interactions by solving the equation of motion. We could use the same procedure for second and third neighbor hoppings. However, their derivation becomes much more complicated and it is really impractical to find an iterative formula (especially because we deal with matrices). To this purpose, the perturbation approach is better suited as detailed in the next section.

\subsection{Layer-by-layer perturbation expansion}\label{sec_pert_method}
      
  In the above subsection, we have seen how to obtain the Green functions of $n$ layers through the exact derivation of equation of motion. Here, we first show that perturbation approach and equation of motion give the same results in the case of nearest-neighbor hopping. Then, in subsection~\ref{sec_pert_method_2nd_neighbor}, we extend the expression of Green functions including second and third-nearest-neighbor hopping.
  
  The aim is to find an iterative formula which gives the Green function of $n+1$ interacting layers as a function of the Green functions of $n$ layers and of the isolated $n+1$ layer. In order to do that, we consider that the perturbation is the added layer (see section \ref{sec_hban} for a more precise definition of the terms):
  \begin{align}\label{eq_hamiltonian_expansion}
    H_0 &= \overbrace{\sum_{\kpara} \sum_{i=1}^{n} \varepsilon^{(i)}_{\kpara} \ccdagger{i\kpara}{i\kpara} + \sum_{\kpara}\sum_{i\neq j}^{n} t_{ij\kpara}\ccdagger{i\kpara}{j\kpara}}^{H_n} 
    \quad+ \overbrace{\sum_{\kpara} \ccdagger{n+1\kpara}{n+1\kpara}}^{H_{n+1}} \\
    H_I &=\underbrace{ \sum_{\kpara} t_{n,n+1,\kpara} \ccdagger{n,\kpara}{n+1,\kpara}}_{H^{(n,n+1)}_I} + \underbrace{ \sum_{\kpara} t_{n-1,n+1,\kpara} \ccdagger{n-1,\kpara}{n+1,\kpara}}_{H^{(n-1,n+1)}_I}
  \end{align}
  In the next subsection, we consider the case where $H^{(n-1,n+1)}_I=0$, \emph{i.e.}, only nearest-neighbor hopping, in order to compare with the results of section \ref{sec_finite_through_eom}. From now on, we shall drop the $\kpara$ dependence for simplicity.

  \subsubsection{Nearest-layer hopping}\label{sec_pert_method_1st_neighbor}  

        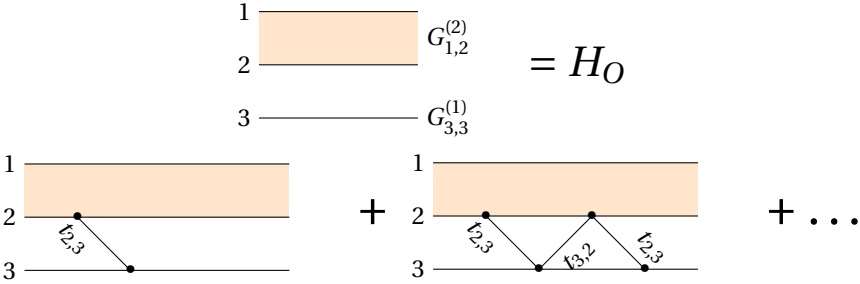
\begin{figure}[!bt]\begin{center}
	\begin{tikzpicture}[scale=0.7]
	  \draw (0,0) node[left]{3} -- (3,0) node[right]{$G^{(1)}_{3,3}$} ;
	  \fill [orange, opacity=0.2] (0,2)  -- (3,2)  -- (3,1)  -- (0,1)  -- cycle;
	  \draw( 0,2) node[left]{1} -- (3,2) node[right]{};
	  \draw (3,1) node[above right]{$G^{(2)}_{1,2}$} -- (0,1) node[left]{2};
	  \draw (6,1) node {{\huge $=H_O$}};
	\end{tikzpicture}
	\\*
	
	\begin{minipage}{.4\linewidth}
	  \begin{tikzpicture}[scale=0.7]
	    \draw (0,0) node[left]{3} -- (5,0) node[right]{} ;
	    \fill [orange, opacity=0.2] (0,2) -- (5,2) -- (5,1) -- (0,1) -- cycle;
	    \draw (0,2) node[left]{1} -- (5,2) node[right]{};
	    \draw (5,1) node[right]{} -- (0,1) node[left]{2} ;
	    \draw (8,0.5) node[left]{} ;
	    \draw (1,1) node{$\bullet$} -- (2,0) node[midway, sloped, below left]{$t_{2,3}$} node{$\bullet$};
	  \end{tikzpicture}
	\end{minipage}{\Huge +\ }
	\begin{minipage}{.4\linewidth}
	  \begin{tikzpicture}[scale=0.7]
	    \draw (0,0) node[left]{3} -- (5,0) node[right]{} ;
	    \fill [orange, opacity=0.2]  (0,2) -- (5,2) -- (5,1) -- (0,1) -- cycle;
	    \draw (0,2) node[left]{1} -- (5,2) node[right]{};
	    \draw (5,1) node[right]{} -- (0,1) node[left]{2} ;
	    \draw (8,0.5) node[left]{} ;
	    \draw (1,1) node{$\bullet$} -- (2,0)  node[midway, sloped, below left]{$t_{2,3}$} node{$\bullet$} -- (3,1)  node[midway, sloped, below]{$t_{3,2}$} node{$\bullet$} -- (4,0)  node[midway, sloped, above right]{$t_{2,3}$} node{$\bullet$};
	  \end{tikzpicture}
	\end{minipage}{\Huge+\ \dots}\caption{Diagrammatic representation of perturbation expansion for a slab of two layers put in contact with a third isolated-layer through $t_{2,3}$ (and $t_{3,2}$).}\label{fig_diagram_pert}
       \end{center}
    \end{figure}
    If we consider only  nearest-neighbor interaction, a system of, \emph{e.g.}, two layers interacting with one layer can be represented with diagrams, as pictured in Fig.~\ref{fig_diagram_pert}
    According to Dyson equation (cf.~Sec.~\ref{sec_pert_expansion}) the perturbation expansion is then
    \begin{align}
      \green{1,3}{3} &= \green{1,2}{2} t_{2,3} \green{3,3}{1} + \green{1,2}{2} t_{2,3} \green{3,3}{1} t_{3,2} \green{2,2}{2} t_{2,3} \green{3,3}{1} + \dots \nonumber
      \\             &= \green{1,2}{2} t_{2,3} \green{3,3}{3}
    \end{align}
    This result can be generalized in a straightforward way as:
    \begin{align}\label{eq_G1np1_1st}
     \green{1,n+1}{n+1} &= \green{1,n}{n} t_{n,n+1} \green{n+1,n+1}{1} + \green{1,n}{n} t_{n,n+1} \green{n+1,n+1}{1} t_{n+1,n} \green{n,n}{n} t_{n,n+1} \green{n+1,n+1}{1}\nonumber \\
                        &= \green{1,n}{n} t_{n,n+1} \green{n+1,n+1}{n+1}
    \end{align}
    From this we see that in order to obtain $\green{1,n+1}{n+1}$ another Green function is required: $\green{n+1,n+1}{n+1}$. Its Dyson equation is:
    \begin{align}
     \green{n+1,n+1}{n+1} &= \green{n+1,n+1}{1} +\green{n+1,n+1}{1} t_{n+1,n} \green{n,n}{n} t_{n,n+1} \green{n+1,n+1}{1} + \dots \nonumber\\
                          &= \green{n+1,n+1}{1}  + \green{n+1,n+1}{1} t_{n+1,n} \green{n,n}{n} t_{n,n+1}  \green{n+1,n+1}{n+1} 
    \end{align} 
    This equation can be easily solved as:
    \begin{equation}
      \green{n+1,n+1}{n+1} = \left[ \identite - \green{n+1,n+1}{1} t_{n+1,n} \green{n,n}{n} t_{n,n+1}  \right]^{-1} \green{n+1,n+1}{1}
    \end{equation} 
    Reinjected in Eq.~\eqref{eq_G1np1_1st}, it gives
    \begin{equation}
     \green{1,n+1}{n+1} = \green{1,n}{n} \underbrace{t_{n,n+1}  \left[ \identite - \green{n+1,n+1}{1} t_{n+1,n} \green{n,n}{n} t_{n,n+1}  \right]^{-1}}_{T_{n,n+1}^{eff}} \green{n+1,n+1}{1}
    \end{equation} 
    where $T_{n,n+1}^{eff}=t_{n,n+1}  \left[ \identite - \green{n+1,n+1}{1} t_{n+1,n} \green{n,n}{n} t_{n,n+1}  \right]^{-1}$ is the same effective hopping as defined in Sec.~\ref{sec_effective_hopping} (Eq.~\eqref{eq_effective_hopping}). The denominator contains the surface Green function  $\green{n,n}{n}$ of the previous iteration. Unlike the transfer matrix of decimation for a semi-infinite slab, the effective hopping has to be recalculated each time a layer is added because of the finite nature of the slab. This is the main difference between our finite system and the semi-infinite of Refs.~\cite{Garcia-Vidal-PhysRevLett.76.807,Reuter-PhysRevB.58.14036,Pedro-ProgSurfS_66_2001}. 
    
    Considering only nearest-neighbor hopping is sufficient to reproduce quite well (see Chap.~\ref{chapt_results}) the band structure of face centered cubic crystals (like gold, silver or Nickel). Nevertheless, for body centered structure (like iron), we have to consider second and third nearest-neighbors (see Chap.~\ref{chapt_results}). As we shall see below, the derivation becomes far more tedious.

\subsubsection{Second-nearest-layer hopping}\label{sec_pert_method_2nd_neighbor}

    In the BEEM current formula, Eq.~\eqref{eq_BEEM_current}, 3 Green functions are required in order to calculate the current at layer $n$ of a $n$-layer slab: $G_{1,n-1}^{R(n)}$,  $G_{n,1}^{A(n)}=\left(G_{1,n}^{R(n)}\right)^\dagger$ and $G_{1,n-2}^{R(n)}$. All these three Green functions can be obtained through an iterative procedure: we have to express the $G_{i,j}^{(n+1)}$ in terms of $G_{i,j}^{(n)}$.
    
    Note that for body-centered structures, like iron, that requires third-neighbor parameters to fit well the band structure, the third nearest-neighbors are located in (110) positions. Consequently, by considering all hopping terms between layer $i$ and layer $i\pm 2$, the third-neighbor interactions are completely taken into account and there is no need to consider hopping terms between layers $i$ and $i\pm 3$.
    
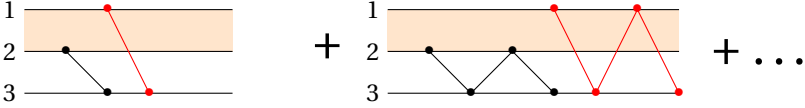
\begin{figure}[!bt]\begin{center}
	\begin{minipage}{.35\linewidth}
	  \begin{tikzpicture}[scale=0.55]
	    \draw (0,0) node[left]{3} -- (5,0) node[right]{} ;
	    \fill [orange, opacity=0.2] (0,2) -- (5,2) -- (5,1) -- (0,1) -- cycle;
	    \draw (0,2) node[left]{1} -- (5,2) node[right]{};
	    \draw (5,1) node[right]{} -- (0,1) node[left]{2} ;
	    \draw (1,1) node{$\bullet$} -- (2,0)  node{$\bullet$};
	    \draw (2,2)[color=red] node{$\bullet$} -- (3,0)  node{$\bullet$};
	  \end{tikzpicture}
	\end{minipage}{\Huge +\ }
	\begin{minipage}{.5\linewidth}
	  \begin{tikzpicture}[scale=0.55]
	    \draw (0,0) node[left]{3} -- (7,0) node[right]{} ;
	    \fill [orange, opacity=0.2] (0,2) -- (7,2) -- (7,1) -- (0,1) -- cycle;
	    \draw (0,2) node[left]{1} -- (7,2) node[right]{};
	    \draw (7,1) node[right]{} -- (0,1) node[left]{2} ;	    
	    \draw (1,1) node{$\bullet$} -- (2,0)   node{$\bullet$} -- (3,1)  node{$\bullet$} -- (4,0)   node{$\bullet$};
	    \draw[color=red] (4,2) node{$\bullet$} -- (5,0)   node{$\bullet$} -- (6,2)  node{$\bullet$} -- (7,0)  node{$\bullet$};
	    \draw (9,1) node {\Huge+\ \dots};
	  \end{tikzpicture}
	\end{minipage}\caption{Diagrammatic representation of perturbation expansion for a slab of two layers put in contact with a third isolated-layer through $t_{2,3}$ and $t_{1,3}$.}\label{fig_diagram_pert_2nd}
       \end{center}
    \end{figure}

    The starting point of the derivation is to use a Dyson expansion of a $n$-layer slab that is put in contact with an isolated layer (labeled $n+1$) through nearest and second nearest-neighbor hopping. This means that we consider both $H_I^{(n,n+1)}$ and $H_I^{(n-1,n+1)}$ in Eq.~\eqref{eq_hamiltonian_expansion}. Diagrammatically this is shown in Fig.~\ref{fig_diagram_pert_2nd}

    \subsubsubsection{${\color{red}\green{1,n+1}{n+1}}$}\label{sec_G1np1np1}

    From perturbation theory (Dyson equation), we get:
    \begin{align}
      \green{1,n+1}{n+1} = &   \green{1,n}{n} t_{n,n+1} \green{n+1,n+1}{1} + \green{1,n-1}{n} t_{n-1,n+1} \green{n+1,n+1}{1}  \nonumber
      \\                   & + \green{1,n\phantom{-1}}{n} t_{n,n+1\phantom{-1}} \green{n+1,n+1}{1} t_{n+1,n\phantom{-1}}   \green{n,n\phantom{-1-1}}{n}     t_{n,n+1\phantom{-1}}   \green{n+1,n+1}{1} \nonumber
      \\                   & + \green{1,n\phantom{-1}}{n} t_{n,n+1\phantom{-1}} \green{n+1,n+1}{1} t_{n+1,n\phantom{-1}}   \green{n,n-1\phantom{-1}}{n}     t_{n-1,n+1}             \green{n+1,n+1}{1} \nonumber
      \\                   & + \green{1,n\phantom{-1}}{n} t_{n,n+1\phantom{-1}} \green{n+1,n+1}{1} t_{n+1,n-1}             \green{n-1,n-1}{n}               t_{n-1,n+1}             \green{n+1,n+1}{1} \nonumber
      \\                   & + \green{1,n\phantom{-1}}{n} t_{n,n+1\phantom{-1}} \green{n+1,n+1}{1} t_{n+1,n-1}             \green{n-1,n\phantom{-1}}{n}     t_{n,n+1\phantom{-1}}   \green{n+1,n+1}{1} \nonumber
      \\                   & + \green{1,n-1}{n} t_{n-1,n+1} \green{n+1,n+1}{1} t_{n+1,n\phantom{-1}}   \green{n,n\phantom{-1-1}}{n}     t_{n,n+1\phantom{-1}}   \green{n+1,n+1}{1} \nonumber
      \\                   & + \green{1,n-1}{n} t_{n-1,n+1} \green{n+1,n+1}{1} t_{n+1,n\phantom{-1}}   \green{n,n-1\phantom{-1}}{n}     t_{n-1,n+1}             \green{n+1,n+1}{1} \nonumber
      \\                   & + \green{1,n-1}{n} t_{n-1,n+1} \green{n+1,n+1}{1} t_{n+1,n-1}             \green{n-1,n-1}{n}               t_{n-1,n+1}             \green{n+1,n+1}{1} \nonumber
      \\                   & + \green{1,n-1}{n} t_{n-1,n+1} \green{n+1,n+1}{1} t_{n+1,n-1}             \green{n-1,n\phantom{-1}}{n}     t_{n,n+1\phantom{-1}}   \green{n+1,n+1}{1} \nonumber
      \\                   & + \dots
    \end{align}
    We can highlight a new Green function ${\color{green}\green{n+1,n+1}{n+1}}$ by factorizing:
    \begin{align}\label{eq_G1np1np1_temp}
      {\color{red}\green{1,n+1}{n+1}} = & \phantom{+}\ \ \green{1,n\phantom{-1}}{n} t_{n,n+1\phantom{-1}} \green{n+1,n+1}{1} 
			      \left[ \begin{array}{l}
						\identite + t_{n+1,n\phantom{-1}}   \green{n,n\phantom{-1-1}}{n}     t_{n,n+1\phantom{-1}}   \green{n+1,n+1}{1}
					      \\\phantom{\identite}+ t_{n+1,n\phantom{-1}}   \green{n,n-1\phantom{-1}}{n}     t_{n-1,n+1}             \green{n+1,n+1}{1}
					      \\\phantom{\identite}+ t_{n+1,n-1}             \green{n-1,n-1}{n}     t_{n-1,n+1}             \green{n+1,n+1}{1}
					      \\\phantom{\identite}+ t_{n+1,n-1}             \green{n-1,n\phantom{-1}}{n}               t_{n,n+1\phantom{-1}}   \green{n+1,n+1}{1} +\dots
					      \end{array}
			      \right] \nonumber
    \\                    & + \green{1,n-1}{n} t_{n-1,n+1} \underbrace{\green{n+1,n+1}{1}
    \left[ \begin{array}{l}
						\identite+           t_{n+1,n\phantom{-1}}   \green{n,n\phantom{-1-1}}{n}     t_{n,n+1\phantom{-1}}   \green{n+1,n+1}{1}
					      \\\phantom{\identite}+ t_{n+1,n\phantom{-1}}   \green{n,n-1\phantom{-1}}{n}     t_{n-1,n+1}             \green{n+1,n+1}{1}
					      \\\phantom{\identite}+ t_{n+1,n-1}             \green{n-1,n-1}{n}               t_{n-1,n+1}             \green{n+1,n+1}{1}
					      \\\phantom{\identite}+ t_{n+1,n-1}             \green{n-1,n\phantom{-1}}{n}     t_{n,n+1\phantom{-1}}   \green{n+1,n+1}{1} +\dots
					      \end{array}
			      \right]}_{\green{n+1,n+1}{n+1}}
    \end{align}
    \begin{equation}\label{eq_G1np1np1}\fbox{$
      {\color{red}{\color{red}\green{1,n+1}{n+1}}} = [{\color{red}\green{1,n}{n}}t_{n,n+1} + {\color{blue}\green{1,n-1}{n}}t_{n-1,n+1}] {\color{green}\green{n+1,n+1}{n+1}} $}
    \end{equation}
    Where ${\color{red}\green{1,n}{n}}$ is the previous iteration and where ${\color{green}\green{n+1,n+1}{n+1}}$ and  ${\color{blue}\green{1,n-1}{n}}$ are calculated in sec.~\ref{sec_Gnp1np1np1} and sec.~\ref{sec_G1nnp1}.

  \subsubsubsection{${\color{green}\green{n+1,n+1}{n+1}}$}\label{sec_Gnp1np1np1}  
    \begin{align}
      \green{n+1,n+1}{n+1} = & \green{n+1,n+1}{1} \nonumber
      \\                     & + \green{n+1,n+1}{1} t_{n+1,n\phantom{-1}} \green{n,n\phantom{-1-1}}{n} t_{n,n+1\phantom{-1}} \green{n+1,n+1}{1} \nonumber
      \\                     & + \green{n+1,n+1}{1} t_{n+1,n\phantom{-1}} \green{n,n-1\phantom{-1}}{n} t_{n-1,n+1}           \green{n+1,n+1}{1} \nonumber
      \\                     & + \green{n+1,n+1}{1} t_{n+1,n-1}           \green{n-1,n-1}{n}           t_{n-1,n+1}           \green{n+1,n+1}{1} \nonumber
      \\                     & + \green{n+1,n+1}{1} t_{n+1,n-1}           \green{n-1,n\phantom{-1}}{n} t_{n,n+1\phantom{-1}} \green{n+1,n+1}{1} \nonumber
      \\                     & + \dots
    \end{align}
    Where we recognize the last two factors of Eq.~\eqref{eq_G1np1np1_temp}. At infinite order it gives:
    \begin{equation}
      \green{n+1,n+1}{n+1} =\green{n+1,n+1}{1} + \green{n+1,n+1}{1}\left[ \begin{array}{l}
									        \phantom{+} t_{n+1,n\phantom{-1}} \green{n,n\phantom{-1-1}}{n} t_{n,n+1\phantom{-1}}
									   \\   + t_{n+1,n\phantom{-1}} \green{n,n-1\phantom{-1}}{n} t_{n-1,n+1}          
									   \\   + t_{n+1,n-1}           \green{n-1,n-1}{n}           t_{n-1,n+1}           
									   \\   + t_{n+1,n-1}           \green{n-1,n\phantom{-1}}{n} t_{n,n+1\phantom{-1}} 
									    \end{array}\right]
						  \green{n+1,n+1}{n+1}
    \end{equation}
    Finally:
    \begin{equation}\label{Gnp1np1np1}\fcolorbox{red}{white}{$
      {\color{green}\green{n+1,n+1}{n+1}} = \left[ \identite - \green{n+1,n+1}{1}\left( \begin{array}{l}
									        \phantom{+} t_{n+1,n\phantom{-1}} {\color{green}\green{n,n\phantom{-1-1}}{n}} t_{n,n+1\phantom{-1}}
									   \\   + t_{n+1,n\phantom{-1}} {\color{cyan}\green{n,n-1\phantom{-1}}{n}} t_{n-1,n+1}          
									   \\   + t_{n+1,n-1}           {\color{orange}\green{n-1,n-1}{n}}           t_{n-1,n+1}           
									   \\   + t_{n+1,n-1}           {\color{cyan}\green{n-1,n\phantom{-1}}{n}} t_{n,n+1\phantom{-1}} 
									    \end{array}\right) \right]^{-1} \green{n+1,n+1}{1}$}
    \end{equation}
    ${\color{green}\green{n,n}{n}}$ is the previous iteration, ${\color{cyan}\green{n,n-1}{n}}$ and $ {\color{cyan}\green{n-1,n}{n}}$ are derived below in sec.~\ref{sec_Gnp1nnp1} and ${\color{orange}\green{n-1,n-1}{n}}$ is derived in sec.~\ref{sec_Gnnnp1_Gnm1nnp1}.

  \subsubsubsection{${\color{blue}\green{1,n}{n+1}}$}\label{sec_G1nnp1}  
    Eq.~\eqref{eq_G1np1np1} shows that ${\color{blue}\green{1,n-1}{n}}$ is required to get ${\color{red}\green{1,n+1}{n+1}}$. Hence, we need to find a way to calculate it iteratively, i.e. to rewrite this Green function for the $n\rightarrow n+1$ case:
    \begin{align}\label{eq_G1nnp1_temp1}
      \green{1,n}{n+1} = & \green{1,n}{n}           + \green{1,n\phantom{-1}}{n} t_{n,n+1\phantom{-1}} \green{n+1,n+1}{1} t_{n+1,n\phantom{-1}} \green{n,n}{n} \nonumber
      \\                 & \phantom{\green{1,n}{n}} + \green{1,n\phantom{-1}}{n} t_{n,n+1\phantom{-1}} \green{n+1,n+1}{1} t_{n+1,n-1}           \green{n-1,n}{n} \nonumber
      \\                 & \phantom{\green{1,n}{n}} + \green{1,n-1}{n}           t_{n-1,n+1}           \green{n+1,n+1}{1} t_{n+1,n-1}           \green{n-1,n}{n} \nonumber
      \\                 & \phantom{\green{1,n}{n}} + \green{1,n-1}{n}           t_{n-1,n+1}           \green{n+1,n+1}{1} t_{n+1,n\phantom{-1}} \green{n,n}{n} \nonumber
      \\                 & + \dots
    \end{align}
    As in Eq.~\eqref{eq_G1np1np1}, it can be rewritten in a simplest form by factorizing and highlighting $\green{n,n}{n+1}$ and $\green{n-1,n}{n+1}$ (sec.~\ref{sec_Gnnnp1_Gnm1nnp1}):
    \begin{multline}\label{eq_G1nnp1_temp}
      \green{1,n}{n+1} =  \green{1,n}{n} + \\ \left[ \green{1,n}{n} t_{n,n+1} + \green{1,n-1}{n} t_{n-1,n+1} \right] \green{n+1,n+1}{1}
                                            \left[ t_{n+1,n-1} \underbrace{\left(\green{n-1,n}{n} + \dots\right)}_{\green{n-1,n}{n+1}}
                                                  +t_{n+1,n  } \underbrace{\left(\green{n,n}{n}   + \dots\right)}_{\green{n,n}{n+1}} \right]
    \end{multline}
    \begin{multline}\label{eq_G1nnp1}
     {\color{blue}\green{1,n}{n+1}} = {\color{red}\green{1,n}{n}} + \\ \left[ {\color{red}\green{1,n}{n}} t_{n,n+1}      + {\color{blue}\green{1,n-1}{n}} t_{n-1,n+1} \right] \green{n+1,n+1}{1}
                                            \left[ t_{n+1,n-1}{\color{Fuchsia}\green{n-1,n}{n+1}} + t_{n+1,n} {\color{orange}\green{n,n}{n+1}}   \right]
    \end{multline} 
    Here, another set of Green functions is required: ${\color{blue}\green{1,n-1}{n}}$ is obtained from the previous iteration of this formula and ${\color{red}\green{1,n}{n}}$ is obtained from the previous iteration of Eq.~\eqref{eq_G1np1np1}. ${\color{Fuchsia}\green{n-1,n}{n+1}}$ and ${\color{orange}\green{n,n}{n+1}}$ are obtained at the current iteration from equations \eqref{eq_Gnm1nnp1} and \eqref{eq_Gnnnp1}.

  \subsubsubsection{${\color{cyan}\green{n+1,n}{n+1}}$ and ${\color{cyan}\green{n,n+1}{n+1}}$}\label{sec_Gnp1nnp1}
    \begin{align}
      \green{n+1,n}{n+1} &= \green{n+1,n+1}{1} t_{n+1,n} \green{n,n}{n} + \green{n+1,n+1}{1} t_{n+1,n-1} \green{n-1,n}{n} + \dots \\
      \green{n,n+1}{n+1} &= \green{n,n}{n} t_{n,n+1} \green{n+1,n+1}{1} + \green{n,n-1}{n} t_{n-1,n+1} \green{n+1,n+1}{1} + \dots
    \end{align}
    Considering the next orders they become:
    \begin{equation}\fcolorbox{red}{white}{$
      {\color{cyan}\green{n+1,n}{n+1}} = \green{n+1,n+1}{1} t_{n+1,n} {\color{orange}\green{n,n}{n+1}} + \green{n+1,n+1}{1} t_{n+1,n-1} {\color{Fuchsia}\green{n-1,n}{n+1}} $}
    \end{equation}
    and
    \begin{equation}\label{eq_Gnnp1np1}\fcolorbox{red}{white}{$
       {\color{cyan}\green{n,n+1}{n+1}} =  {\color{orange}\green{n,n}{n+1}}  t_{n,n+1} \green{n+1,n+1}{1} + {\color{Fuchsia}\green{n,n-1}{n+1}} t_{n-1,n+1} \green{n+1,n+1}{1} $}
    \end{equation} 
    Where, again, ${\color{orange}\green{n,n}{n+1}}$ from eq~\eqref{eq_Gnnnp1} and  ${\color{Fuchsia}\green{n-1,n}{n+1}}$ from Eq.~\eqref{eq_Gnm1nnp1} are required.

  \subsubsubsection{ Closure of the system: ${\color{orange}\green{n,n}{n+1}}$ and ${\color{Fuchsia}\green{n-1,n}{n+1}}$}\label{sec_Gnnnp1_Gnm1nnp1}
    $\green{n,n}{n+1}$ has the same Dyson expansion as $\green{1,n}{n+1}$ (Eq.~\eqref{eq_G1nnp1_temp1}) except for the first label of each term:
    \begin{align}
      \green{n,n}{n+1} = & \green{n,n}{n}           + \green{n,n\phantom{-1}}{n} t_{n,n+1\phantom{-1}} \green{n+1,n+1}{1} t_{n+1,n\phantom{-1}} \green{n,n}{n} \nonumber
      \\                 & \phantom{\green{n,n}{n}} + \green{n,n\phantom{-1}}{n} t_{n,n+1\phantom{-1}} \green{n+1,n+1}{1} t_{n+1,n-1}           \green{n-1,n}{n} \nonumber
      \\                 & \phantom{\green{n,n}{n}} + \green{n,n-1}{n}           t_{n-1,n+1}           \green{n+1,n+1}{1} t_{n+1,n-1}           \green{n-1,n}{n} \nonumber
      \\                 & \phantom{\green{n,n}{n}} + \green{n,n-1}{n}           t_{n-1,n+1}           \green{n+1,n+1}{1} t_{n+1,n\phantom{-1}} \green{n,n}{n} \nonumber
      \\                 & + \dots
    \end{align}
    and its factorized form is
   \begin{align}
     \green{n,n}{n+1} =& \green{n,n}{n}           + \left[ \green{n,n}{n} t_{n,n+1} + \green{n,n-1}{n} t_{n-1,n+1} \right] \green{n+1,n+1}{1} t_{n+1,n-1}\green{n-1,n}{n+1} \nonumber
     \\                & \phantom{\green{n,n}{n}} + \left[ \green{n,n}{n} t_{n,n+1} + \green{n,n-1}{n} t_{n-1,n+1} \right] \green{n+1,n+1}{1} t_{n+1,n}  \green{n,n}{n+1}   
    \end{align}
    Regrouping $\green{n,n}{n+1}$:
    \begin{multline}\label{eq_Gnnnp1_temp}
      \left[ \identite - \left( \green{n,n}{n} t_{n,n+1} + \green{n,n-1}{n} t_{n-1,n+1} \right) \green{n+1,n+1}{1}t_{n+1,n} \right] \green{n,n}{n+1} =
                         \\ \green{n,n}{n} + \left[ \green{n,n}{n} t_{n,n+1} + \green{n,n-1}{n} t_{n-1,n+1} \right] \green{n+1,n+1}{1} t_{n+1,n-1}\green{n-1,n}{n+1}
    \end{multline}
    \\*

    Regarding $\green{n-1,n}{n+1}$, the Dyson expansion is, again, the same as $\green{1,n}{n+1}$ (Eq.~\eqref{eq_G1nnp1_temp1}) except for the first label of each term:
    \begin{align}
      \green{n-1,n}{n+1} = & \green{n-1,n}{n}           + \green{n-1,n\phantom{-1}}{n} t_{n,n+1\phantom{-1}} \green{n+1,n+1}{1} t_{n+1,n\phantom{-1}} \green{n,n}{n} \nonumber
      \\                   & \phantom{\green{n-1,n}{n}} + \green{n-1,n\phantom{-1}}{n} t_{n,n+1\phantom{-1}} \green{n+1,n+1}{1} t_{n+1,n-1}           \green{n-1,n}{n} \nonumber
      \\                   & \phantom{\green{n-1,n}{n}} + \green{n-1,n-1}{n}           t_{n-1,n+1}           \green{n+1,n+1}{1} t_{n+1,n-1}           \green{n-1,n}{n} \nonumber
      \\                   & \phantom{\green{n-1,n}{n}} + \green{n-1,n-1}{n}           t_{n-1,n+1}           \green{n+1,n+1}{1} t_{n+1,n\phantom{-1}} \green{n,n}{n} \nonumber
      \\                   & + \dots
    \end{align}
    Hence:
    \begin{align}
      \green{n-1,n}{n+1} =& \green{n-1,n}{n}           + \left[ \green{n-1,n}{n} t_{n,n+1} + \green{n-1,n-1}{n} t_{n-1,n+1} \right] \green{n+1,n+1}{1} t_{n+1,n-1}\green{n-1,n}{n+1} \nonumber
      \\                  & \phantom{\green{n-1,n}{n}} + \left[ \green{n-1,n}{n} t_{n,n+1} + \green{n-1,n-1}{n} t_{n-1,n+1} \right] \green{n+1,n+1}{1} t_{n+1,n\phantom{-1}}  \green{n,n}{n+1}   
    \end{align}
    The equation for $\green{n,n-1}{n+1}$ is the same but it should be read from right to left:
    \begin{align}
      \green{n,n-1}{n+1} =& \green{n,n-1}{n}           + \green{n,n-1}{n+1}  t_{n-1,n+1} \green{n+1,n+1}{1} \left[t_{n+1,n} \green{n,n-1}{n} + t_{n+1,n-1} \green{n-1,n-1}{n} \right]  \nonumber
      \\                  & \phantom{\green{n-1,n}{n}} + \green{n,n\phantom{-1}}{n+1} t_{n,n+1\phantom{-1}} \green{n+1,n+1}{1} \left[t_{n+1,n} \green{n,n-1}{n} + t_{n+1,n-1} \green{n-1,n-1}{n} \right] 
    \end{align}
    \begin{multline}\label{eq_Gnm1nnp1_temp}
      \left[ \identite   - \left( \green{n-1,n}{n} t_{n,n+1} + \green{n-1,n-1}{n} t_{n-1,n+1} \right) \green{n+1,n+1}{1} t_{n+1,n-1} \right] \green{n-1,n}{n+1} =
     \\ \green{n-1,n}{n} + \left[ \green{n-1,n}{n} t_{n,n+1} + \green{n-1,n-1}{n} t_{n-1,n+1} \right] \green{n+1,n+1}{1} t_{n+1,n} \green{n,n}{n+1}
    \end{multline}

    By solving the system of Eq.~\eqref{eq_Gnnnp1_temp} and \eqref{eq_Gnm1nnp1_temp} the whole system of equations can be closed:
    \begin{align}\label{eq_Gnnnp1}
      {\color{orange}\green{n,n}{n+1}} & = \Bigg\{ \identite - \big( {\color{green}\green{n,n}{n}} t_{n,n+1} + {\color{cyan}\green{n,n-1}{n}} t_{n-1,n+1} \big) \green{n+1,n+1}{1} \nonumber
      \\ & \phantom{= 1 }\times \bigg[  \identite - t_{n+1,n-1} \Big( \identite + \big( {\color{cyan}\green{n-1,n}{n}} t_{n,n+1} + {\color{orange}\green{n-1,n-1}{n}} t_{n-1,n+1} \big) \green{n+1,n+1}{1} t_{n+1,n-1} \Big)^{-1} \nonumber
      \\ & \phantom{= 1\times 2} \times \big( {\color{cyan}\green{n-1,n}{n}} t_{n,n+1} + {\color{orange}\green{n-1,n-1}{n}} t_{n-1,n+1} \big) \green{n+1,n+1}{1} \bigg] t_{n+1,n}  \Bigg\}^{-1} \nonumber
      \\ & \times \Bigg\{ {\color{green}\green{n,n}{n}} + \Big( {\color{green}\green{n,n}{n}} t_{n,n+1} + {\color{cyan}\green{n,n-1}{n}} t_{n-1,n+1} \Big) \green{n+1,n+1}{1}t_{n+1,n-1} \nonumber
      \\ & \phantom{= 1}\times \bigg[ \identite - \Big( {\color{cyan}\green{n-1,n}{n}} t_{n,n+1} + {\color{orange}\green{n-1,n-1}{n}} t_{n-1,n+1} \Big)  \green{n+1,n+1}{1}t_{n+1,n-1} \bigg]^{-1} {\color{cyan}\green{n-1,n}{n}} \Bigg\}
    \end{align}
  and
  \begin{align}\label{eq_Gnm1nnp1}
    {\color{Fuchsia}\green{n-1,n}{n+1}} & = \Bigg\{ \identite - \big( {\color{cyan}\green{n-1,n}{n}} t_{n,n+1} + {\color{orange}\green{n-1,n-1}{n}} t_{n-1,n+1} \big) \green{n+1,n+1}{1} \nonumber
    \\ & \phantom{= 1 }\times \bigg[  \identite - t_{n+1,n} \Big( \identite + \big( {\color{green}\green{n,n}{n}} t_{n,n+1} + {\color{cyan}\green{n,n-1}{n}} t_{n-1,n+1} \big) \green{n+1,n+1}{1} t_{n+1,n} \Big)^{-1} \nonumber
    \\ & \phantom{= 1\times 2} \times \big( {\color{green}\green{n,n}{n}} t_{n,n+1} + {\color{cyan}\green{n,n-1}{n}} t_{n-1,n+1} \big) \green{n+1,n+1}{1} \bigg] t_{n+1,n-1}  \Bigg\}^{-1} \nonumber
    \\ & \times \Bigg\{ {\color{cyan}{\color{cyan}\green{n-1,n}{n}}} + \Big( {\color{cyan}{\color{cyan}\green{n-1,n}{n}}} t_{n,n+1} + {\color{orange}\green{n-1,n-1}{n}} t_{n-1,n+1} \Big) \green{n+1,n+1}{1}t_{n+1,n} \nonumber
    \\ & \phantom{= 1}\times \bigg[ \identite - \Big( {\color{green}\green{n,n}{n}} t_{n,n+1} + {\color{cyan}\green{n,n-1}{n}} t_{n-1,n+1} \Big)  \green{n+1,n+1}{1}t_{n+1,n} \bigg]^{-1} {\color{green}\green{n,n}{n}}\Bigg\}
  \end{align}
  \begin{align}\label{eq_Gnnm1np1}
    {\color{Fuchsia}\green{n,n-1}{n+1}} & = \Bigg\{ {\color{green}\green{n,n}{n}} \bigg[ \identite - t_{n,n+1}  \green{n+1,n+1}{1} \Big( t_{n+1,n-1} {\color{cyan}\green{n-1,n}{n}} +   t_{n+1,n} {\color{green}\green{n,n}{n}} \Big)  \bigg]^{-1} \nonumber 
    \\ & \phantom{= 1 }\times t_{n,n+1}  \green{n+1,n+1}{1} \Big( t_{n+1,n-1} {\color{orange}\green{n-1,n-1}{n}} + t_{n+1,n} {\color{cyan}\green{n,n-1}{n}}  \Big) + {\color{cyan}\green{n-1,n}{n}} \nonumber \Bigg\} \nonumber
    \\ & \times 
    \Bigg\{ \identite - t_{n-1,n+1} \bigg[ \green{n+1,n+1}{1} \big( t_{n+1,n-1} {\color{cyan}\green{n-1,n}{n}} +   t_{n+1,n} {\color{green}\green{n,n}{n}} \big) \nonumber
    \\& \phantom{= 1\times} \times \Big( \identite + t_{n,n+1} \green{n+1,n+1}{1} \big( t_{n+1,n-1} {\color{cyan}\green{n-1,n}{n}} +   t_{n+1,n} {\color{green}\green{n,n}{n}} \big) \Big)^{-1}     
    t_{n,n+1}\bigg] \nonumber
    \\  & \phantom{= 1 }\times \green{n+1,n+1}{1} \Big( t_{n+1,n-1} {\color{orange}\green{n-1,n-1}{n}} + t_{n+1,n} {\color{cyan}\green{n,n-1}{n}}  \Big) \Bigg\}
  \end{align}

  \subsubsubsection{${\color{Sepia}\green{1,n-1}{n+1}}$}
    At this point, we have two of the three Green functions which are present in the BEEM current equation \eqref{eq_BEEM_current}:  $G_{1,n-1}^{R(n)}$ and  $G_{n,1}^{A(n)}=\left(  G_{1,n}^{R(n)}\right)^\dagger$. The last Green function is $G_{1,n-2}^{R(n)}$, whose Dyson equation writes:
    \begin{align}
	\green{1,n-1}{n+1} = & \green{1,n-1}{n}           + \green{1,n\phantom{-1}}{n} t_{n,n+1\phantom{-1}} \green{n+1,n+1}{1} t_{n+1,n\phantom{-1}} \green{n,n-1}{n} \nonumber
	\\                   & \phantom{\green{1,n-1}{n}} + \green{1,n\phantom{-1}}{n} t_{n,n+1\phantom{-1}} \green{n+1,n+1}{1} t_{n+1,n-1}           \green{n-1,n-1}{n} \nonumber
	\\                   & \phantom{\green{1,n-1}{n}} + \green{1,n-1}{n}           t_{n-1,n+1}           \green{n+1,n+1}{1} t_{n+1,n-1}           \green{n-1,n-1}{n} \nonumber
	\\                   & \phantom{\green{1,n-1}{n}} + \green{1,n-1}{n}           t_{n-1,n+1}           \green{n+1,n+1}{1} t_{n+1,n\phantom{-1}} \green{n,n-1}{n} \nonumber
	\\                   & + \dots
      \end{align}
      As in Eq.~\eqref{eq_G1np1np1}, it can be rewritten in a simpler form by factorizing and highlighting $\green{n,n-1}{n+1}$ and $\green{n-1,n-1}{n+1}$ (sec.~\ref{sec_Gnnnp1_Gnm1nnp1} and sec.~\ref{sec_Gnm1nm1np1}):
      \begin{multline}\label{eq_G1nm1np1}
      {\color{Sepia}\green{1,n-1}{n+1}} = {\color{blue}\green{1,n-1}{n}} +  \\ \left[ {\color{blue}\green{1,n-1}{n}} t_{n-1,n+1}      + {\color{red}\green{1,n}{n}} t_{n,n+1} \right] \green{n+1,n+1}{1}
					      \left[ t_{n+1,n-1}{\color{Goldenrod}\green{n-1,n-1}{n+1}} + t_{n+1,n} {\color{Fuchsia}\green{n,n-1}{n+1}}   \right]
      \end{multline} 
      $\green{1,n-1}{n}$, $\green{1,n}{n}$ and $\green{n,n-1}{n+1}$ are already calculated above. We only need to calculate $\green{n-1,n-1}{n+1}$.

    \subsubsubsection{${\color{Goldenrod}\green{n-1,n-1}{n+1}}$}\label{sec_Gnm1nm1np1}
      Following the same procedure as above, the Dyson expansion can be written as:
      \begin{multline}
      \green{n-1,n-1}{n+1} = \green{n-1,n-1}{n} + \\ \left[ \green{n-1,n-1}{n} t_{n-1,n+1}      + \green{n-1,n}{n} t_{n,n+1} \right] \green{n+1,n+1}{1}
					  \left[ t_{n+1,n-1}\green{n-1,n-1}{n+1} + t_{n+1,n} \green{n,n-1}{n+1}   \right] 
      \end{multline}
      Finally,
      \begin{align}\label{eq_Gnm1nm1np1}
      {\color{Goldenrod}\green{n-1,n-1}{n+1}}  &= \left[ \identite - \left(  {\color{orange}\green{n-1,n-1}{n}} t_{n-1,n+1} +{\color{cyan}\green{n-1,n}{n}} t_{n,n+1} \right) \green{n+1,n+1}{1} t_{n+1,n-1}\right]^{-1}
       \\               &\phantom{= } \times \left(  {\color{orange}\green{n-1,n-1}{n}} t_{n-1,n+1} +{\color{cyan}\green{n-1,n}{n}} t_{n,n+1} \right) \green{n+1,n+1}{1}  t_{n+1,n} {\color{Fuchsia}\green{n,n-1}{n+1}}
      \end{align}

  \subsubsubsection{Example: How to obtain $\green{1,4}{4}$}\label{sec_eg_G14}
    Given the complexity of the whole calculation scheme, we provide here a specific example for the case with $n+1=4$ layers. In what follows, we enumerate the different steps in reverse order with respect to the code, but we keep the numeration of the code that performs the procedure in the opposite order (from step 6 to step 1). The reason is that the order from step 6 to step 1 is more appropriate to explain whereas the opposite order from step 1 to step 6 is needed to do the calculations.
    
    Starting from a slab of 3 layers ($n=3$), let us calculate $\green{1,4}{4}$:
    {\renewcommand{\theenumi}{step \arabic{enumi}}
    \begin{etaremune}[leftmargin=32pt]
      \item \label{enum_G144}
	      In order to calculate
		\begin{equation}\label{eq_G144}
		  \green{1,4}{4} = [\green{1,3}{3}t_{3,4} + \green{1,2}{3}t_{2,4}] \green{4,4}{4}
		\end{equation}
	      For the iteration with $n+1=4$, we need from the previous iteration ($n+1=3$): $\green{1,3}{3}$ (\ref{enum_G144}) and $\green{1,2}{3}$ (\ref{enum_G134}). And we also need $\green{4,4}{4}$ (\ref{enum_G444}) from the current iteration ($n+1=4$).
      \item \label{enum_G444}
	      For
		\begin{equation}
		  \green{4,4}{4} = \left[ \identite - \green{4,4}{1}\left( \begin{array}{l}
											    \phantom{+} t_{4,3} \green{3,3}{3} t_{3,4}
										      \\   +            t_{4,3} \green{3,2}{3} t_{2,4}          
										      \\   +            t_{4,2} \green{2,2}{3} t_{2,4}           
										      \\   +            t_{4,2} \green{2,3}{3} t_{3,4} 
											\end{array}\right) \right]^{-1} \green{4,4}{1}
		\end{equation}
	      only Green functions of the previous iteration are required: $\green{3,3}{3}$ (\ref{enum_G444}), $\green{3,2}{3}$ and $\green{2,3}{3}$ (\ref{enum_G434}), and $\green{2,2}{3}$ (\ref{enum_G334}).
      \item \label{enum_G134}
	      The Green function
		  \begin{equation}\label{eq_G134}
		  \green{1,3}{4} = \green{1,3}{3} +  \left[ \green{1,3}{3} t_{3,4}      + \green{1,2}{3} t_{2,4} \right] \green{4,4}{1}
							  \left[ t_{4,2}\green{2,3}{4} + t_{4,3} \green{3,3}{4}   \right]
		  \end{equation}
	      can be obtained with $\green{1,3}{3}$ (\ref{enum_G144}), $\green{1,2}{3}$ (\ref{enum_G134}) of the previous iteration and with $\green{2,3}{4}$ (\ref{enum_G234}) and $\green{3,3}{4}$ (\ref{enum_G334}) of the current iteration.
      \item \label{enum_G434}
	      Here,
		\begin{equation}
		  \green{4,3}{4} = \green{4,4}{1} t_{4,3} \green{3,3}{4} + \green{4,4}{1} t_{4,2} \green{2,3}{4}
		\end{equation}
	      requires $\green{2,3}{4}$ (\ref{enum_G234}) and $\green{3,3}{4}$ (\ref{enum_G334}).
      \item \label{enum_G234}
	      For
		\begin{align}
		  \green{2,3}{4} & = \Bigg\{ \identite - \big( \green{2,3}{3} t_{3,4} + \green{2,2}{3} t_{2,4} \big) \green{4,4}{1} \nonumber
		  \\ & \phantom{= 1 }\times \bigg[  \identite - t_{4,3} \Big( \identite - \big( \green{3,3}{3} t_{3,4} + \green{3,2}{3} t_{2,4} \big) \green{4,4}{1} t_{4,3} \Big)^{-1} \nonumber
		  \\ & \phantom{= 1\times 2} \times \big( \green{3,3}{3} t_{3,4} + \green{3,2}{3} t_{2,4} \big) \green{4,4}{1} \bigg] t_{4,2}  \Bigg\}^{-1} \nonumber
		  \\ & \times \Bigg\{ \green{2,3}{3} + \Big( \green{2,3}{3} t_{3,4} + \green{2,2}{3} t_{2,4} \Big) \green{4,4}{1}t_{4,3} \nonumber
		  \\ & \phantom{= 1}\times \bigg[ \identite - \Big( \green{3,3}{3} t_{3,4} + \green{3,2}{3} t_{2,4} \Big)  \green{4,4}{1}t_{4,3} \bigg]^{-1} \green{3,3}{3}\Bigg\}
		\end{align}
	      only Green functions of the previous iteration are needed: $\green{3,2}{3}$ and $\green{2,3}{3}$ (\ref{enum_G434}), $\green{2,2}{3}$ (\ref{enum_G334}) and $\green{3,3}{3}$ (\ref{enum_G444}).
      \item \label{enum_G334}
	      Finally,
		\begin{align}
		  \green{3,3}{4} & = \Bigg\{ \identite - \big( \green{3,3}{3} t_{3,4} + \green{3,2}{3} t_{2,4} \big) \green{4,4}{1} \nonumber
		  \\ & \phantom{= 1 }\times \bigg[  \identite - t_{4,2} \Big( \identite - \big( \green{2,3}{3} t_{3,4} + \green{2,2}{3} t_{2,4} \big) \green{4,4}{1} t_{4,2} \Big)^{-1} \nonumber
		  \\ & \phantom{= 1\times 2} \times \big( \green{2,3}{3} t_{3,4} + \green{2,2}{3} t_{2,4} \big) \green{4,4}{1} \bigg] t_{4,3}  \Bigg\}^{-1} \nonumber
		  \\ & \times \Bigg\{ \green{3,3}{3} + \Big( \green{3,3}{3} t_{3,4} + \green{3,2}{3} t_{2,4} \Big) \green{4,4}{1}t_{4,2} \nonumber
		  \\ & \phantom{= 1}\times \bigg[ \identite - \Big( \green{2,3}{3} t_{3,4} + \green{2,2}{3} t_{2,4} \Big)  \green{4,4}{1}t_{4,2} \bigg]^{-1} \green{2,3}{3} \Bigg\}
		\end{align}
	      is obtained with the same Green functions of the previous iteration: $\green{3,2}{3}$ and $\green{2,3}{3}\dagger$ (\ref{enum_G434}), $\green{2,2}{3}$ (\ref{enum_G334}) and $\green{3,3}{3}$ (\ref{enum_G444}).             

    \end{etaremune}}
    As stated above, considering next-nearest layer interactions, an additional Green function is required: $G_{1,n-2}^{R(n)}$. In our $n=3$ case such a Green function $\green{1,2}{4}$ is obtained as follow:    

{\renewcommand{\theenumi}{step \alph{enumi}}
\renewcommand{\labelenumi}{\theenumi.}
      \begin{etaremune}
        \item \label{G124}
             \begin{equation}\label{eq_G124}
           \green{1,2}{4} = \green{1,2}{3} +  \left[ \green{1,2}{3} t_{2,4} + \green{1,3}{3} t_{3,4} \right] \green{4,4}{1}
                                                \left[ t_{4,2}\green{2,2}{4} + t_{4,3} \green{3,2}{4}   \right]
            \end{equation} 
            Equation \eqref{eq_G124} is obtained using $\green{1,2}{3}$, $\green{1,3}{3}$ and $\green{3,2}{4}$ from \ref{enum_G134} and \ref{enum_G144} of the previous iteration and from \ref{enum_G234} and \ref{enum_G224}  of the current iteration. 
        \item \label{enum_G224}
            In order to close Eq.~\eqref{eq_G124} we need to evaluate also:
            \begin{align}
              \green{2,2}{4}  &= \left[ \identite - \left(  \green{2,2}{3} t_{2,4} +\green{2,3}{3} t_{3,4} \right) \green{4,4}{1} t_{4,2}\right]^{-1} \nonumber
              \\               &\phantom{= } \times \left(  \green{2,2}{3} t_{2,4} +\green{2,3}{3} t_{3,4} \right)  \green{4,4}{1}  t_{4,3} \green{3,2}{4}
            \end{align}
            For this,we need $ \green{2,2}{3}$ and $\green{2,3}{3}$ from \ref{enum_G334}and \ref{enum_G434} of the previous iteration and $\green{3,2}{4}$ from \ref{enum_G234} of the current iteration.
      \end{etaremune}}

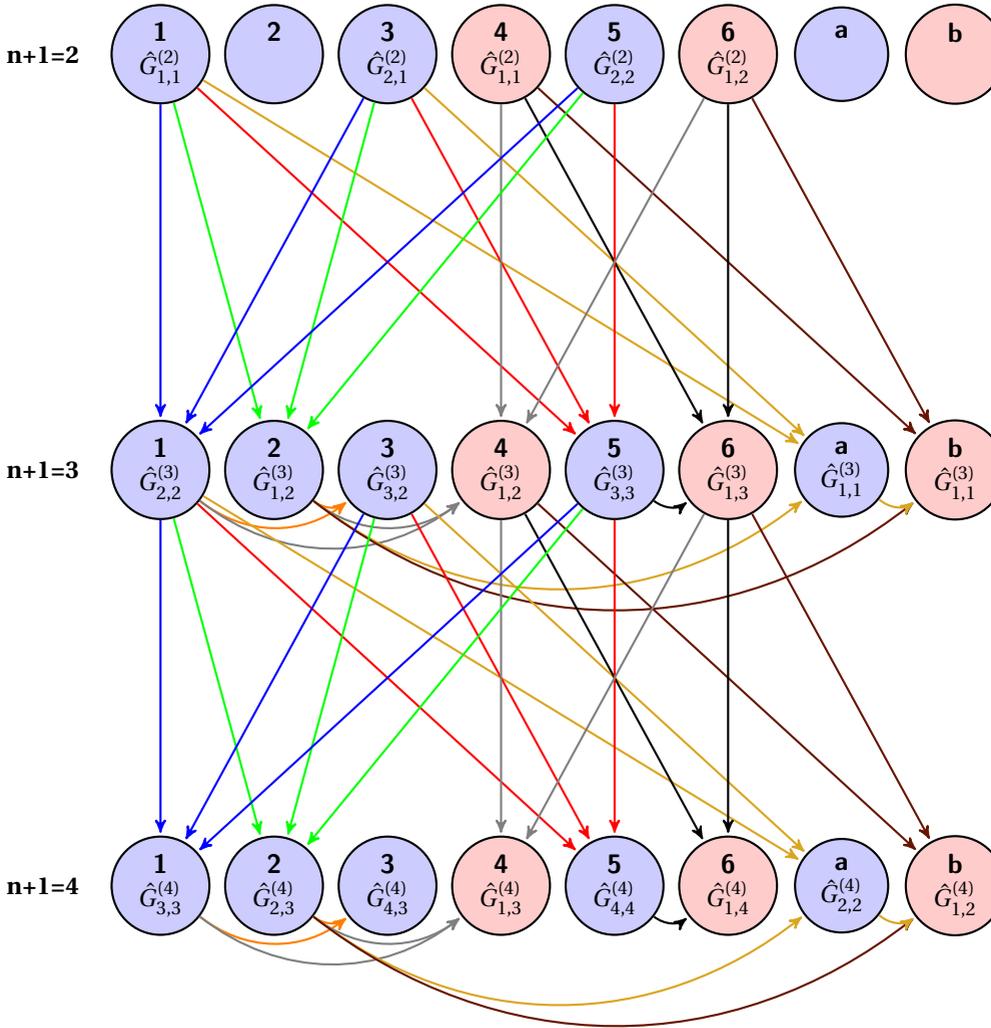
\begin{figure}[!hbt]
  \begin{tikzpicture}[
		      ->,>=stealth',shorten >=1pt,auto,node distance=1.5cm, thick,
		      main node/.style={circle,fill=blue!20,draw, font=\sffamily\bfseries},
		      green node/.style={circle,fill=red!20,draw, font=\sffamily\bfseries}
		      ]
      
    \node[main node] (A1) { \shortstack{1 \\ $\green{1,1}{2}$}};
    \node[main node] (A2) [right of=A1] {\shortstack{2 \\ \phantom{$\green{1,2}{2}$}}};
    \node[main node] (A3) [right of=A2] {\shortstack{3 \\ $\green{2,1}{2}$}};
    \node[green node] (A4) [right of=A3] {\shortstack{4 \\ $\green{1,1}{2}$}};
    \node[main node] (A5) [right of=A4] {\shortstack{5 \\ $\green{2,2}{2}$}};
    \node[green node] (A6) [right of=A5] {\shortstack{6 \\ $\green{1,2}{2}$}};
    \node[main node] (Aa) [right of=A6] {\shortstack{a \\  \phantom{$\green{0,0}{2}$}}};
    \node[green node] (Ab) [right of=Aa] {\shortstack{b \\  \phantom{$\green{1,1}{2}$}}};
    \node[left =0.3cm of A1] {\textbf{{n+1=2}}};
    
    \node[main node] (B1) [below =4.15cm of A1] {\shortstack{1 \\ $\green{2,2}{3}$}};
    \node[main node] (B2) [right of=B1] {\shortstack{2 \\ $\green{1,2}{3}$}};
    \node[main node] (B3) [right of=B2] {\shortstack{3 \\ $\green{3,2}{3}$}};
    \node[green node] (B4) [right of=B3] {\shortstack{4 \\ $\green{1,2}{3}$}};
    \node[main node] (B5) [right of=B4] {\shortstack{5 \\ $\green{3,3}{3}$}};
    \node[green node] (B6) [right of=B5] {\shortstack{6 \\ $\green{1,3}{3}$}};
    \node[main node] (Ba) [right of=B6] {\shortstack{a \\ $\green{1,1}{3}$}};
    \node[green node] (Bb) [right of=Ba] {\shortstack{b \\ $\green{1,1}{3}$}};
    \node[left =0.3cm of B1] {\textbf{{n+1=3}}};

    \node[main node] (C1) [below =4.15cm of B1] {\shortstack{1 \\ $\green{3,3}{4}$}};
    \node[main node] (C2) [right of=C1] {\shortstack{2 \\ $\green{2,3}{4}$}};
    \node[main node] (C3) [right of=C2] {\shortstack{3 \\ $\green{4,3}{4}$}};
    \node[green node] (C4) [right of=C3] {\shortstack{4 \\ $\green{1,3}{4}$}};
    \node[main node] (C5) [right of=C4] {\shortstack{5 \\ $\green{4,4}{4}$}};
    \node[green node] (C6) [right of=C5] {\shortstack{6 \\ $\green{1,4}{4}$}};  
    \node[main node] (Ca) [right of=C6] {\shortstack{a \\ $\green{2,2}{4}$}};
    \node[green node] (Cb) [right of=Ca] {\shortstack{b \\ $\green{1,2}{4}$}};
    \node[left =0.3cm of C1] {\textbf{{n+1=4}}};
    
    \path[every node/.style={font=\sffamily\small, fill=white,inner sep=1pt}]
	(A1) edge[blue]  (B1)
	     edge[green] (B2)
	     edge[red]   (B5)
	     edge[Goldenrod] (Ba)
	(A3) edge[blue]  (B1)
	     edge[green] (B2)
	     edge[red]   (B5)
             edge[Goldenrod] (Ba)
	(A4) edge[gray]  (B4)
	     edge[black] (B6)
             edge[Sepia] (Bb)
	(A5) edge[blue]  (B1)
	     edge[green] (B2)
	     edge[red]   (B5)
	(A6) edge[gray]  (B4)
	     edge[black] (B6)
             edge[Sepia] (Bb)
	(B1) edge[bend right=38, orange] (B3)
	     edge[bend right=38, gray]   (B4)
	(B2) edge[bend right=38, orange] (B3)
	     edge[bend right=38, gray]   (B4)
             edge[bend right=38, Goldenrod] (Ba)
             edge[bend right=38, Sepia] (Bb)
	(B5) edge[bend right=38, black]  (B6)
	(Ba) edge[bend right=38, Goldenrod] (Bb)
	(B1) edge[blue]  (C1)
	     edge[green] (C2)
	     edge[red]   (C5)
             edge[Goldenrod] (Ca)
	(B3) edge[blue]  (C1)
	     edge[green] (C2)
	     edge[red]   (C5)
             edge[Goldenrod] (Ca)
	(B4) edge[gray]  (C4)
	     edge[black] (C6)
             edge[Sepia] (Cb)
	(B5) edge[blue]  (C1)
	     edge[green] (C2)
	     edge[red]   (C5)
	(B6) edge[gray]  (C4)
             edge[Sepia] (Cb)
	     edge[black] (C6)
	(C1) edge[bend right=38, orange] (C3)
	     edge[bend right=38, gray]   (C4)
	(C2) edge[bend right=38, orange] (C3)
	     edge[bend right=38, gray]   (C4)
             edge[bend right=38, Goldenrod] (Ca)
             edge[bend right=38, Sepia] (Cb)
	(C5) edge[bend right=38, black]  (C6)
        (Ca) edge[bend right=38, Goldenrod] (Cb)
	;
  \end{tikzpicture}
  \caption{Flow chart of the algorithm used to calculate the Green functions $\hat{G}_{1,3}^{R,(4)}$, $\hat{G}_{1,2}^{R,(4)}$ and $\hat{G}_{4,1}^{A,(4)}$ (pink circles). The label 1, 2, 3, 4, 5 and 6 are the same as those used in sec~\ref{sec_eg_G14}. This is also the same order used in the code to calculate the various Green functions. At the first iteration (n=1, we are looking for n+1=2), there is no next-nearest layers: $\green{i,j}{2}$ is only expressed in term of $\green{i,j}{1}$. At the second iteration there are next-nearest-layer interactions, but still one degenerate case: step~\ref{enum_G134} = step~\ref{enum_G234}. At the third iteration, each Green function is different (at least all indexes are). Equations  \eqref{eq_G144},  \eqref{eq_G134} and  \eqref{eq_G124} give the retarded Green function and the advanced Green function of the current formula (eq.~(\ref{eq_BEEM_current})).}\label{fig_GF_flow_chart}
\end{figure}

    In conclusion, in order to apply the current formula \eqref{eq_BEEM_current} in this 4-layer case, three Green functions are needed and have been evaluated with the above steps:
    \begin{align*}
     \hat{G}_{1,3}^{R,(4)} \mbox{ Eq. }\eqref{eq_G134} &\ ; &  \hat{G}_{1,2}^{R,(4)} \mbox{ Eq. }\eqref{eq_G124} 
     &\ ;&  \hat{G}_{4,1}^{A,(4)} \eqref{eq_G124}=\left( \hat{G}_{1,4}^{R,(4)} \right)^{\dagger} \mbox{ Eq. } \eqref{eq_G144}
    \end{align*}
    Figure.~\ref{fig_GF_flow_chart} offers a better visualization of the algorithm. 
    
\chapter{BEEM program}\label{chapt_BEEM_prog}
  \lettrine[lines=3, lhang=0.35, loversize=0.77, findent=3em,nindent=-0.5em,slope=-1em]{A}{s stated} in the previous chapter, K. Reuter and P. de Andres from the Universidad Autonoma de Madrid  have developed a Fortran code called BEEM v2.1 which calculates the BEEM current using the decimation approach for semi-infinite structures.
At that time, the decimation was by far the best choice for thick slabs, because of the $2^n$-algorithm ($n$=number of layers) allowing a fast convergence (see Sec.~\ref{sec_decimation}). However, today, developments in computing make it possible to implement an exact calculation of the Green functions layer by layer in order to describe a finite system, even for a large number of layers. The layer by layer approach has the advantage of allowing a better analysis of present experiments, where the number of layers can be controlled at the level of monolayer. Moreover, it allows to change the layer character (eg, Au/Fe/Au/Fe/Fe\dots) whereas a description by means of decimation require a homogeneous system.

To this aim, I have created a new code, based on the layer-by-layer approach, named BEEM v3.0, after BEEM v2.1, as some subroutines have been borrowed from it. We start by presenting the flow chart of the code and then we present some key parts.

\section{Flow chart}
  The aim of the code is to perform numerically the calculations described in chapter \ref{chapt_NEPT}. In order to achieve this, I have organized the code according to the flow chart described in Fig.~\ref{fig_code_flowchart}. The main parts of the code are:
  \begin{enumerate}
    \item The input files are read. The main input file contains the name of all the others input files, including 2-center parameters and atomic positions for each slab. For instance, for a Fe/Au structure, there are 3 input files (in addition to the main input files and the k point input files): one for Fe, one for Au and one for the interface FeAu.
    \item A loop over the slab is performed
	  \begin{enumerate}
            \item The two subroutines det\_neighbors and det\_matrixelements build the hopping matrices for each atom.
	    \item A loop over $\kpara$ is called
		  \begin{enumerate}
		    \item det\_tightham and arrange\_tightham build the hamiltonian matrices (on site hamiltonian and hopping matrices in $k$ space)
		    \item Green functions are calculated (through the procedure described in Sec.~\ref{sec_pert_method_2nd_neighbor}).
		    \item The current is calculated (through Eq.~\eqref{eq_BEEM_current}).
		    \item The density of states is calculated if there is a loop over energy (controlled in the main input files) by taking the imaginary part of the propagator.
		  \end{enumerate}
          \end{enumerate}
  \end{enumerate}
  The action of each of the main subroutines is detailed in the next sections.
  \insertfigure{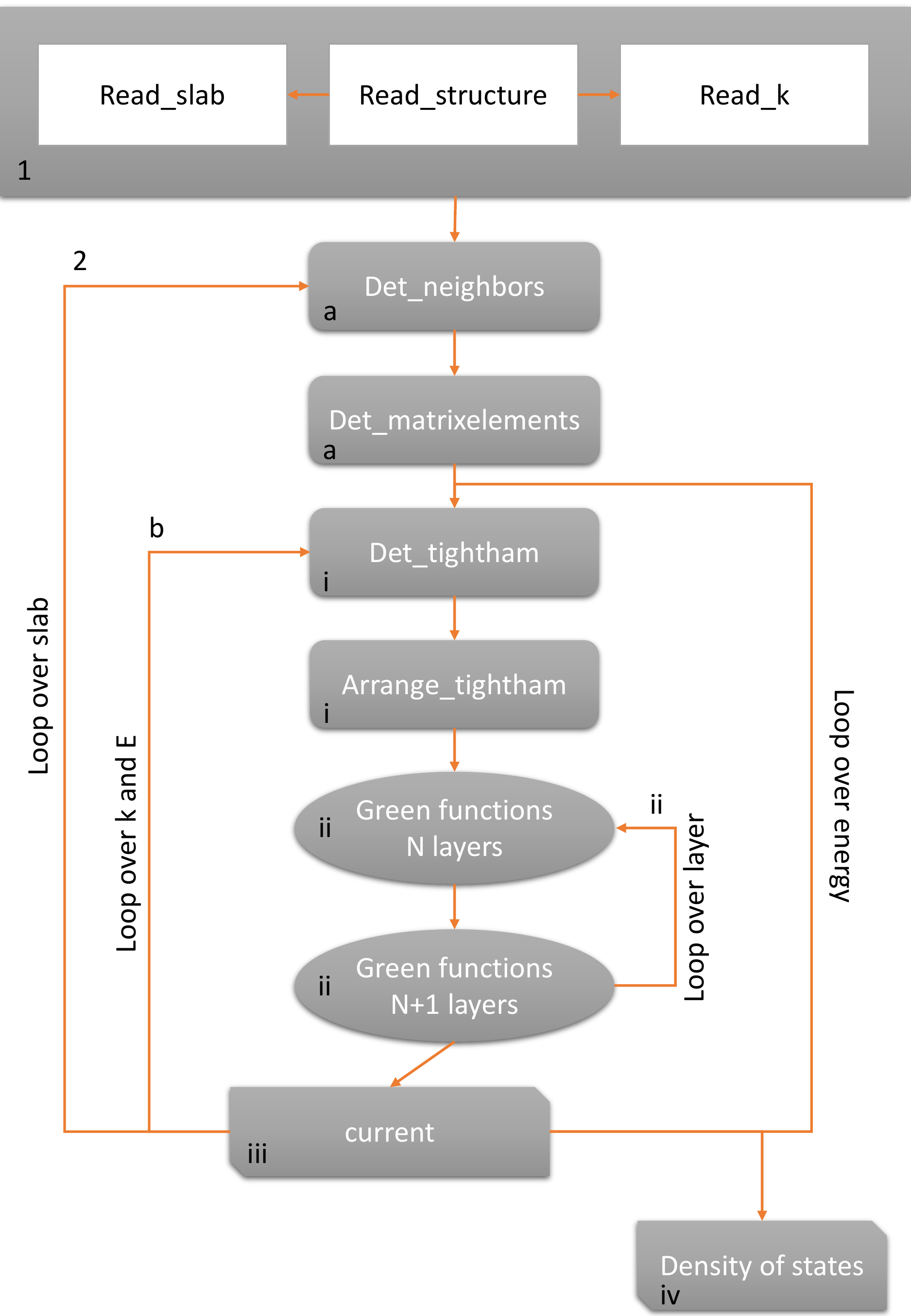}{Flow chart of BEEM v3.0.}{\label{fig_code_flowchart}}

\section{Execution of the code and input files}

  \subsection{Execution, input and output}
    When executing the code, we need first of all to specify the structure for which we want to evaluate the current. The code opens all the required input files automatically. For instance, the command
    \begin{verbatim}
     ./BEEM.exe structure
    \end{verbatim}
    opens the input file \emph{structure.in}  (for instance \emph{FeAuFe.in}, \emph{Au111\_13layers.in} etc\dots) and creates the output files \emph{structure.log} (that contains the log of the execution), \emph{structure\_current.dat} (that contains the current) and \emph{structure\_dos.dat} (that contains the density of states).

  \subsection{Input files}
    The code requires three input files. The first, \emph{structure.in}, is given  with the above execution-command. It contains the name of the two other input-files: the one used to build the Hamiltonian and the one which contains the $k$-point grid. These three input files are described below.

    \subsubsection{The main input-file \emph{structure.in} (Listing \ref{code_main_input})}
      The main input-file contains the structure and the parameters that can be controlled experimentally. An example of such a file is reported in  listing~\ref{code_main_input}. The main parameters that can be controlled are:
      \begin{description}
      \item[line 3] the number of slabs in the structure. i.e. the number of different materials + the interfaces. eg: for a FeAu structure there are 3 different kinds of slabs. Fe, FeAu interface and Au.
      \item[lines 4 to 6] name of the files which contain the data required to build the Hamiltonian (they are described in subsection~\ref{sec_ham_input}); the number of layers in each slab is also given here.
      \item[line 7] name of the file which contains the k-points where the current has to be calculated (grid in the 2D Brillouin zone), described in subsection~\ref{sec_k_input}.
      \item[lines 9 to 12] repeat the unit cell in (x,y) directions in order to find the different neighbors when the Hamiltonian is built. Here, $(x,y)$ refer to the in-layer coordinates, as detailed in
      section~\ref{sec_ham_sub}.
      \item[line 13] plane up to which neighbors have to be searched
      \item[lines 15 to 23] parameters of the tip, for future use. At the moment the tip is considered as placed at the coordinate origin in the plane $(x,y)$. See chapter~\ref{chapt_perspectives} for further remarks.
      \item[lines 25 to 27] set minimal energy equal to maximal energy in order to calculate the BEEM current at a given energy. For DOS calculation choose a range of energy and a step in eV.
      \item[line 29] mean free path for Green functions = $\eta$, the damping parameter. A typical value is $k_B T$ at room temperature, \emph{i.e.} about 0.025 eV.
      \end{description}

\begin{lstlisting}[numbers=left,numberstyle=\tiny,float=htbb,frame=lines,basicstyle=\tiny\ttfamily,captionpos=b,caption={Main input file.},label=code_main_input,breaklines=true]
=============================== experiment data ================================
----- global structure and input files Fe/Au --------
 3                         ! nb of slab + interfaces. eg: FeAu= 3
	Fe.in     10       ! slab 1: ham filename, number of layers
	Fe_Au.in   2       ! slab 2: the interface, 2 layers (2nd nn hopping)
	Au.in      9       ! slab 3: ham filename, number of layers
        kpoint_BCC001.in   ! name of the k-point input file
----- Data on plane, where current is to be calculated
 -2                               ! n1a    Plane goes in x-direction        I4
  2                               ! n1b    from n1a to n1b (index integer)  I4
 -2                               ! n2a    Plane goes in y-direction        I4
  2                               ! n2b    from n2a to n2b (index integer)  I4
  2                               ! Max. plane-distance of neighbours       I4
----- tip data ------     ! NOT USED FOR NOW except the nb of orbitals
   9                               ! No. of tip-orbitals  (max=9)           I4
  0.000                            ! Tip fermi-level in eV (sample=0.00)  F7.3
  1.000                            ! STM bias in eV                       F7.3
  3.000                            ! Pot. barrier for tunneling in eV     F7.3
  0.000000  0.000000  -5.00000     ! Tip position in A (xyz)             3F7.3
  5.100                            ! Max. tunneling-dist. in A            F7.3
   1                               ! No. of max. allowed t-atoms (arrays)   I4
   1                               ! No. of atom-types/layers involved      I4
  0.000                            ! Temperature in K                     F7.3
----- Data on energy-integral interval
  1.000                            ! Min. energy in eV                    F7.3
  1.000                            ! Max. energy in eV                    F7.3
  0.100                            ! Increment in loop in eV              F7.3
----- Data for Green function calculation  
  0.025                            ! Mean free path for G-functions       F8.4	
\end{lstlisting}

    \subsubsection{The Hamiltonian input-file (Listing \ref{ham_input})}\label{sec_ham_input}
      The Hamiltonian input-file contains the data required to build the Hamiltonian whose inversion is needed to evaluate the $\greenone{i,i}$ Green functions. Pre-existent data-bases are set for non-expert users, for most common structures.
      \begin{description}
       \item[lines 3 to 5] 2D lattice parameter in \AA{} and vectors of the 2D unit-cell (in reduce coordinates). We must choose a 2D unit cell in order to have one atom per cell. A BCC cell can be used. However, for a FCC, the requirement of one atom per cell in the layer is not respected. For instance, for gold on iron, the FCC cell is 45\degree\ rotated with respect to the BCC cell. In that case, The FCC cell can be described with a tetragonal centered whose in-plane parameter is the same as BCC, and whose $z$ parameter (epitaxy direction) is the one of the usual FCC cell. The tetragonal cell is used in the Listing \ref{ham_input} (gold).
       \item[line 6] number of atoms in the 3D unit-cell. See line 11 for more details.
       \item[line 7] number of different chemical species. 
       \item[line 8] dimension of the Hamiltonian = number of orbitals $\times$ number of atoms in the unit cell.
       \item[line 9] maximum number of allowed neighbors. Used to dimension arrays.
       \item[line 10] maximal value for the azimuthal quantum number $l$. eg: 2 for $d$-orbitals.
       \item[line 11] position of atoms in the unit cell (reduced coordinates) + on-site energies (with $\varepsilon_d = \frac{\varepsilon_{eg}+\varepsilon_{t2g}}{2}$ as they are very close). Warning, there is a trick here in order to simplify the code when it calculates the hopping matrices: as it needs to calculate hopping matrices up to second-nearest plane (i.e. up to third nearest-neighbors), we must add two layers to the usual unit cell, as it is done here for the FCC cell of Au(001). The reason for that is illustrated in Fig.~\ref{fig_stacking} and explained in Sec.~\ref{sec_extract_H}. You shall find other examples in the database, and it is also documented in the Hamiltonian subroutines.
       \item[lines 16-52] two-center parameters for nearest, second-nearest and third-nearest neighbors. The first data-line (eg line 18) is for hopping from atom of type i to type j. Warning,  if there is more than one type of atom, for instance Fe=1 and Au=2, all the hopping combinations have to be specified: 1 with 1, 2 with 2, 1 with 2 and 2 with 1. For the next 9 lines, the two first digits are the azimuthal quantum number $l$ and ``$-1 -1$'' ends a data set.
       \item[line 53] minimal and maximal radii of the shell where first, second and third neighbors $j$ of atom $i$ are searched. Warning, again,  we have to specify all possible combinations for $i$ and $j$ when there is more than one type of atoms, as above.
      \end{description}

\begin{lstlisting}[numbers=left,numberstyle=\tiny,float=hbt,frame=lines,basicstyle=\tiny\ttfamily,captionpos=b,caption={Hamiltonian input file.},label=ham_input,breaklines=true]
------------------ Input File for Tight-Binding Calculation ------------------
------------------------- Data is for Au(001) Layer --------------------------
   2.87000                         ! Lattice parameter in A              F10.5
   1.00000   0.00000   0.00000     ! Unit_vector_1 (in units of a)      3F10.5
   0.00000   1.00000   0.00000     ! Unit_vector_2                      3F10.5
  4                                ! No. of atoms in unit-cell              I3
  1                                ! No. of diff. chemical species          I3
 36                                ! Dimension of H-matrix (atoms*orb)      I4
 12                                ! No. of max. allowed neighbours (arrays)I4
  2  2  2  2                       ! lmax for each atom in unit-cell       XI3
       x/y/z in units of a     chem  es       ep        ed    3F10.5,I3,3F10.5
   0.50000   0.50000   0.00000  1   0.32926   10.08141  -3.82279  ! Au 
   0.00000   0.00000   0.70711  1   0.32926   10.08141  -3.82279  ! Au    
   0.50000   0.50000   1.41422  1   0.32926   10.08141  -3.82279  ! Au 
   0.00000   0.00000   2.12133  1   0.32926   10.08141  -3.82279  ! Au 
Parametrized interactions between atoms in lattice (all in eV):
first neighbor
 1  1                               !  Data  between  type  i&j   2I3
  0  0  -0.90886                      !  sssigma  2I3,3F10.5    
  0  1   1.32261                      !  spsigma      
  0  2  -0.64246                      !  sdsigma      
  1  0  -1.32261                      !  pssigma      
  1  1   2.43079  -0.22381            !  ppsigma,pppi    
  1  2  -0.87063   0.25796            !  pdsigma,pdpi    
  2  0  -0.64246                      !  dssigma      
  2  1   0.87063  -0.25796            !  dpsigma,dppi    
  2  2  -0.67634   0.35701  -0.06218  !  ddsigma,ddpi,dddelta  
-1 -1                               ! End of data i&j             2I3
second neighbor
 1  1                               !  Data  between  type  i&j   2I3
  0  0   0.03769                      !  sssigma  2I3,3F10.5    
  0  1   0.03551                      !  spsigma      
  0  2  -0.10667                      !  sdsigma      
  1  0  -0.03551                      !  pssigma      
  1  1   0.50436  -0.13946            !  ppsigma,pppi    
  1  2  -0.10368   0.06395            !  pdsigma,pdpi    
  2  0  -0.10667                      !  dssigma      
  2  1   0.10368  -0.06395            !  dpsigma,dppi    
  2  2  -0.04150   0.03265  -0.00776  !  ddsigma,ddpi,dddelta  
-1 -1                               ! End of data i&j             2I3
third neighbor
 1  1                               ! Data between type i&j       2I3
  0  0   0.00000                      !  sssigma  2I3,3F10.5  
  0  1  -0.00000                      !  spsigma    
  0  2  -0.00000                      !  sdsigma    
  1  0   0.00000                      !  pssigma    
  1  1  -0.00000   0.00000            !  ppsigma,pppi  
  1  2  -0.00000  -0.00000            !  pdsigma,pdpi  
  2  0  -0.00000                      !  dssigma    
  2  1   0.00000   0.00000            !  dpsigma,dppi  
  2  2   0.00000   0.00000  -0.00000  !  ddsigma,ddpi,dddelta
-1 -1                               ! End of data i&j             2I3
Maximum interaction radius between atoms i & j in lattice (in A):
first neighbor
  1  1   0.10000  2.90000             ! i,j,rmin,rmax(i,j)      2I3,F10.5
second neighbor
  1  1   2.90100  4.10000             ! i,j,rmin,rmax(i,j)      2I3,F10.5
third neighbor
  1  1   4.11000  4.12000             ! i,j,rmin,rmax(i,j)      2I3,F10.5  
------------------------------------------------------------------------------

\end{lstlisting}
      

      \FloatBarrier

    \subsubsection{The k-point input-file}\label{sec_k_input}
      This input file contains a grid of k-points which belong to the 2D Brillouin-zone. Such a grid can be created by my own code for a rectangular set or can be imported from another program for general, non-orthogonal sets. For example, the hexagonal grid needed for Au(111) was imported from R. Ramirez (CSIC, UAM) code within BEEM v2.1. The first line of the file must be the number of k-points to be read (for dimensioning).

\section{Building the hopping matrices and in-layer Hamiltonian}\label{sec_ham_sub}

  As seen in the chapter~\ref{chapt_NEPT} the only required ingredients to calculate the current is the retarded and advanced Green functions of isolated layers and the matrices which describe the hopping from one plane to another. Therefore, we have to build the Hamiltonian of a slab, which is achieved in two steps.
  
  \subsection{The tight binding matrix}
    After reading the input files, two subroutines are used in order to build the tight binding matrix: \emph{det\_neighbor.f90} and \emph{det\_matrixelements.f90}.
    
    The first one determines, for each atom in the unit cell, all its first, second and third neighbors within a given radius (between rmin and rmax), as specified after line 53 of the  Hamiltonian input  file. Then, using as input the atomic and neighbor positions, as well as the parametrized interactions (both from Hamiltonian input file), \emph{det\_matrixelements.f90} subroutine determines all interatomic matrix-elements needed for the tight-binding Hamiltonian. Rotation matrices allowing to determine the overlap for the interatomic transition-matrix elements are included from the older BEEM v2.1 code: they are encoded in \emph{rot\_coord.f}. This subroutine and those that are called by it allow the correct rotation of the spherical harmonics using the tensor algebra of the Wigner matrices~\cite[Chap. 4 \& 5]{varshalovich}.
    
    The results are square matrices of dimension ``\emph{orb}'' which describe the probability to jump from an orbital at site $i$ to another one at site $j$. There are as many matrices as neighbors for each atom of the unit cell. For structures made of different materials, \emph{e.g.} Fe/Au, the program calculates these hopping matrices for Fe, Au and the interface FeAu. 
    Once these matrices are calculated, it is possible to build the matrix elements of the Hamiltonian.

  \subsection{The Hamiltonian matrix \emph{hban}}\label{sec_hban}
    We identify a given atom by the label $i_x,i_y,i_z$ and its neighbors by $i_x+\delta_x, i_y+\delta_y, i_z+\delta_z$. In this way, the tight-binding Hamiltonian of Eq.~\ref{sec_tight_binding_model} can be written as:
    \begin{align}\label{eq_hban}
      H = \sum_{i_x,i_y,i_z} \sum_{\delta_x,\delta_y,\delta_z} \sum_{m,m'} t^{(m,m')}_{\!\!\!\begin{array}{l}
                                                                                        \scriptstyle(i_x, i_y, i_z),\\ \scriptstyle(i_x+\delta_x, i_y+\delta_y, i_z+\delta_z)
                                                                                      \end{array}}
        \hat{c}_{i_x,i_y,i_z}^{\dagger(m)} \hat{c}_{i_x+\delta_x, i_y+\delta_y, i_z+\delta_z}^{(m')} + \textrm{h.c.}
    \end{align}
    Where the terms $$ t^{(m,m')}_{\!\!\!\begin{array}{l} \scriptstyle(i_x, i_y, i_z),\\ \scriptstyle(i_x+\delta_x, i_y+\delta_y, i_z+\delta_z)\end{array}} \hat{c}_{i_x,i_y,i_z}^{\dagger(m)}  \hat{c}_{i_x+\delta_x, i_y+\delta_y, i_z+\delta_z}^{(m')}$$ destroy an orbital $m$ at $\vec{i}+\vec{\delta}$ to create another one $m'$ at $\vec{i}$ with an amplitude $t$. These $t$ are the ones obtained through the \emph{det\_matrixelements.f90} subroutine.
    
    As the system is finite in the $z$ direction, it is not possible to perform a Fourier transform in this direction. 
    Hence, we split the $xy$ plane, where the Fourier transform can be performed, and the $z$-component, where it cannot. This implies rewriting the Hamiltonian \eqref{eq_hban} as the sum of two terms: $H=\sum_{i_z}H_{i_z,i_z} + \sum_{i_z,\delta_z\neq 0} H_{i_z,\delta_z}$, where $H_{i_z,i_z}$ corresponds to the layer-to-layer hopping (in the $z$-direction). In formulæ
    \begin{align}\label{H_ii}
      H_{i_z,i_z} &= \sum_{i_x,i_y} \sum_{\delta_x,\delta_y} \sum_{m,m'} t^{(m,m')}_{\delta_x,\delta_y,i_z}
        \hat{c}_{i_x,i_y,i_z}^{\dagger(m)} \hat{c}_{i_x+\delta_x, i_y+\delta_y, i_z}^{(m')} + \textrm{h.c.} \\
      H_{i_z,\delta_z} &= \sum_{i_x,i_y} \sum_{\delta_x,\delta_y} \sum_{m,m'} t^{(m,m')}_{\delta_x,\delta_y,\delta_z}
        \hat{c}_{i_x,i_y,i_z+\delta_z}^{\dagger(m)} \hat{c}_{i_x+\delta_x, i_y+\delta_y, i_z}^{(m')} + \textrm{h.c.} \label{Hiz_deltaz}   
    \end{align}
    where $t^{(m,m')}_{\delta_x,\delta_y}$ is the in-layer hopping term for layer $i_z$ which is independent of $i_x$ and $i_y$ because of translation invariance.
    
    As we saw above, the slab periodicity in the $(x,y)$-plane allows us to perform the following Fourier transform of the ladder operators:
    \begin{align}
      \hat{c}_{i_x,i_y,i_z}^{\dagger(m)} &= \frac{1}{\sqrt{N}} \sum_{k_x,k_y} \hat{c}_{k_x,k_y,i_z}^{\dagger(m)} e^{\i(k_x i_x + k_y i_y)} \\
      \hat{c}_{i_x+\delta_x,i_y+\delta_y,i_z}^{(m)} &= \frac{1}{\sqrt{N}} \sum_{k_x,k_y} \hat{c}_{k_x,k_y,i_z}^{(m)} e^{-\i[k_x (i_x+\delta_x) + k_y (i_y+\delta_y)]}
    \end{align}
    and analogously for the Hermitian conjugated terms $\hat{c}_{i_x,i_y,i_z}^{(m)}=\left[\hat{c}_{i_x,i_y,i_z}^{\dagger(m)}\right]^*$. Re-inject-ing it in Eq.~\eqref{H_ii} we find
    \begin{align}
      H_{i_z,i_z} =& \sum_{i_x,i_y}\sum_{\delta_x,\delta_y} \sum_{m,m'} \varepsilon^{(m,m')}_{\delta_x,\delta_y,i_z} \sum_{k_x,k_y}\sum_{k'_x,k'_y} 
                            \frac{1}{N} \e^{\i(k_x i_x + k_y i_y)} \e^{-\i[k'_x (i_x+\delta_x) + k'_y (i_y+\delta_y)]}
      \nonumber\\            & \cdot \hat{c}_{k_x,k_y,i_z}^{\dagger(m)}  \hat{c}_{k'_x,k'_y,i_z}^{(m')}
      \nonumber\\           =& \sum_{\delta_x,\delta_y} \sum_{m,m'} t^{(m,m')}_{\delta_x,\delta_y,i_z} \sum_{k_x,k_y}\sum_{k'_x,k'_y} 
                    \underbrace{\frac{1}{N}  \sum_{i_x,i_y} \e^{-\i[(k'_x-k_x)i_x + (k'_y-k_y)i_y]}}_{\delta_{\kpara\kpara'} } 
      \nonumber\\   &\cdot \hat{c}_{k_x,k_y,i_z}^{\dagger(m)}  \hat{c}_{k'_x,k'_y,i_z}^{(m')} 
                            \e^{-\i(k'_x\delta_x + k'_y \delta_y)}
    \nonumber\\ H_{i_z,i_z}(\kpara)=& \sum_{\kpara} \sum_{m,m'} \varepsilon^{(m',m)}_{\kpara,i_z} \hat{c}_{\kpara,i_z}^{\dagger(m)}\hat{c}_{\kpara,i_z}^{(m')}
    \end{align}
    where $\varepsilon^{(m',m)}_{\kpara,i_z}$ is the matrix energy over the orbital indexes $(m,m')$ for layer $i_z$.
    The same can be done for the off-layer terms, i.e. the hopping from one layer to another. These can be expressed as:
    \begin{equation}
       H_{i_z,i_z+\delta_z}(\kpara) =  \sum_{\kpara} \sum_{m,m'} \varepsilon^{(m,m')}_{\kpara,i_z,i_z+\delta_z} 
                                      \hat{c}_{\kpara,i_z}^{\dagger(m)}  \hat{c}_{\kpara,i_z+\delta_z}^{(m')} + \mathrm{h.c.} 
    \end{equation}
    with $\varepsilon^{(m,m')}_{\kpara,i_z,i_z+\delta_z}= \sum_{\delta_x,\delta_y} \sum_{m,m'} t^{(m,m')}_{\delta_x,\delta_y,i_z,i_z+\delta_z} \e^{\i(k_x \delta_x + k_y \delta_y)}$ the hopping matrices in $\kpara$ space from layer $i_z$ to layer $i_z+\delta_z$.
    Let us now consider, as an example, the case of three interacting layers. It is useful, for iterative purposes, to write the Hamiltonian in the following matrix form:
    \begin{align}
      \tilde{H} = \left[\begin{array}{ccc}
                  H_{11} & H_{12} & H_{13}\\
                  H_{21} & H_{22} & H_{23}\\
                  H_{31} & H_{32} & H_{33}
                \end{array}
          \right]
    \end{align}
    The Hamiltonian here, is just an example which describes three interacting layers. The diagonal part, of this block-Hamiltonian is the Hamiltonian of the isolated layer $i_z$ (for $i_z =1$, 2 or 3).    
    The off-diagonal part of the block-Hamiltonian corresponds instead to the layer-to-layer hoppings (Eq.~\eqref{Hiz_deltaz}). For example, $H_{1,2}$ and $H_{2,1}$ are respectively the hopping matrices from layer 1 to 2 and from layer 2 to 1:
    \begin{align}\label{H_ij}
      H_{1,2} = \sum_{i_x,i_y} \sum_{\delta_x,\delta_y} \sum_{m,m'} t^{(m,m')}_{\!\!\!\begin{array}{l}
                                                                                        \scriptstyle(i_x, i_y, 1),\\ \scriptstyle(i_x+\delta_x, i_y+\delta_y, 2)
                                                                                      \end{array}}
    \hat{c}_{i_x,i_y,1}^{\dagger(m)} \hat{c}_{i_x+\delta_x, i_y+\delta_y, 2}^{(m')} + \textrm{h.c.}
    \end{align}
    And similarly for $H_{13}$ and $H_{31}$. As in this thesis we work with a local orbital basis of $s$, $p$, and $d$ orbitals, each of the sub-Hamiltonians ($H_{1,1}$,$H_{2,1}$ etc\dots ) is a $9\times9$ matrix.

    After Fourier transform, if we consider the hopping from layer 1 to layer 2, eq.~(\ref{H_ij}) becomes:
    \begin{align}
      H_{1,2}(\kpara)=& \sum_{\kpara} \sum_{m,m'} \varepsilon^{(m,m')}_{\kpara,1,2} 
                                      \hat{c}_{\kpara,1}^{\dagger(m)}  \hat{c}_{\kpara,2}^{(m')} + \mathrm{h.c.} 
    \end{align}
    whose matrix representation is
    \begin{flalign*}
      & H_{1,2}(\kpara)= & &
    \end{flalign*} 
    {\small
    \begin{align}\label{H_ij2}
       \bordermatrix{~ & \ket{s} & \ket{p_x} & \ket{p_y} & \ket{p_z} & \ket{d_{xy}} & \ket{d_{yz}} & \ket{d_{xz}} & \ket{d_{x^2-y^2}} & \ket{d_{z^2}} \cr
                                                  \hfill \bra{s}           & H_{1,2}^{(s,s)}(\kpara)   & H_{1,2}^{(s,p_x)}(\kpara) & \dots & & & & & & \cr
                                                  \hfill\bra{p_x}         &  H_{1,2}^{(p_x,s)}(\kpara) & \ddots& & & & & & & \cr
                                                  \hfill \bra{p_y}         & \vdots & & & & & & & & \cr
                                                  \hfill \bra{p_z}         & & & & & & & & & \cr
                                                  \hfill \bra{d_{xy}}        & & & & & & & & & \cr
                                                  \hfill \bra{d_{yz}}        & & & & & & & & & \cr
                                                  \hfill\bra{d_{xz}}        & & & & & & & & & \cr
                                                  \hfill\bra{d_{x^2-y^2}} & & & & & & & & & \cr
                                                  \hfill\bra{d_{z^2}}      & & & & & & & & & \cr}
    \end{align}}%
    For instance with $m=\ket{s}$ and $m'=\ket{p_x}$ we have:
    \begin{align}
       H_{1,2}^{(s,p_x)}(\kpara) = \Braket{s|H_{1,2}(\kpara)|p_x}= &\sum_{\delta_x,\delta_y} t^{(s,p_x)}_{\delta_x,\delta_y,\delta_z} e^{-\i(k_x\delta_x + k_y \delta_y)} 
      \nonumber\\                                   &+\sum_{\delta_x,\delta_y} t^{(p_x,s)}_{\delta_x,\delta_y,\delta_z} e^{+\i(k_x\delta_x + k_y \delta_y)} 
    \end{align}
    
    This 2D Fourier transform is performed in the \emph{det\_dettightham.f90} subroutine, which is called within the k-point loop. It requires the atomic position and the hopping matrix for each neighbor.

  \subsection{Extracting the hopping matrices and the in-layer matrices from \emph{hban}}\label{sec_extract_H}
    In the previous version of the BEEM program (v2.1) the full matrix was used in order to calculate the Green functions. However, it could be troublesome for finite systems because if the Hamiltonian describes a slab made of 3 layers, it propagates 3 layers by 3 layers and we have to deal with dimensioning problem of matrices. To avoid those complications, \emph{arrange\_tightham.f90} extracts the required matrices: in-layer Hamiltonians $H_{i_z,i_z}$, nearest-layer hopping $H_{i_z,i_z+1}$ and next-nearest-layer hopping $H_{i_z,i_z+2}$. From now on, we shall call $H_{i_z,i_z+1}$ and  $H_{i_z,i_z+2}$ as $t_{i_z,i_z+1}$ and  $t_{i_z,i_z+2}$ respectively. As shown in chapter~\ref{chapt_NEPT} second and third-nearest-neighbor interactions both imply that electrons jump from one layer to the next-nearest one, for a BCC lattice, like for iron. 
    
\begin{figure}[!hbtp]
    \centering 
    \subbottom[\label{subfig_fcc_unit}]{\tdplotsetmaincoords{70}{10}
    \begin{tikzpicture}[tdplot_main_coords, scale = 3]

    \draw[thin] (0,0,0) -- (0,0,1) -- (1,0,1) -- (1,0,0) -- cycle;
    \draw[thin] (1,0,1) -- (1,1,1) -- (1,1,0)-- (1,0,0) -- cycle ;
    \draw[thin] (0,0,1) -- (0,1,1) -- (1,1,1);
    \draw[thin,dashed] (0,0,0) -- (0,1,0) -- (0,1,1);
    \draw[thin,dashed] (0,1,0) -- (1,1,0);

    \tdplottransformmainscreen{0}{0.5}{0.5}
    \shadedraw[tdplot_screen_coords, ball color = red] (\tdplotresx,\tdplotresy) circle (0.04);

    \tdplottransformmainscreen{0.5}{0.5}{0}
    \shadedraw[tdplot_screen_coords, ball color = red] (\tdplotresx,\tdplotresy) circle (0.04);

    \tdplottransformmainscreen{0.5}{1}{0.5}
    \shadedraw[tdplot_screen_coords, ball color = red] (\tdplotresx,\tdplotresy) circle (0.04);

    \tdplottransformmainscreen{0}{1}{0}
    \shadedraw[tdplot_screen_coords, ball color = blue] (\tdplotresx,\tdplotresy) circle (0.04);  
    
    \tdplottransformmainscreen{0}{1}{1}
    \shadedraw[tdplot_screen_coords, ball color = blue] (\tdplotresx,\tdplotresy) circle (0.04);


    \tdplottransformmainscreen{1}{0}{0}
    \shadedraw[tdplot_screen_coords, ball color = blue] (\tdplotresx,\tdplotresy) circle (0.04);

    \tdplottransformmainscreen{1}{1}{0}
    \shadedraw[tdplot_screen_coords, ball color = blue] (\tdplotresx,\tdplotresy) circle (0.04);


    \tdplottransformmainscreen{0}{0}{1}
    \shadedraw[tdplot_screen_coords, ball color = blue] (\tdplotresx,\tdplotresy) circle (0.04);

    \tdplottransformmainscreen{1}{0.5}{0.5}
    \shadedraw[tdplot_screen_coords, ball color = red] (\tdplotresx,\tdplotresy) circle (0.04);

    \tdplottransformmainscreen{0.5}{0.5}{1}
    \shadedraw[tdplot_screen_coords, ball color = red] (\tdplotresx,\tdplotresy) circle (0.04);

    \path[draw, color=green, fill=green!15, opacity=0.7] (1,1,1) -- (1.5,0.5,1) -- (1.5,0,1.5) -- (1,0.5,1.5) -- cycle;
    \path[draw, color=green, fill=green!40, opacity=0.7] (0,0,0) -- (0,-0.5,0.5) --  (1.0,0.5,1.5) -- (1.0,1,1) -- cycle;
    \path[draw, color=green, fill=green!20, opacity=0.7] (0.5,-0.5,0) -- (0,0,0) -- (1,1,1) -- (1.5,0.5,1) -- cycle;
    \draw[thin,color=green] (0,-0.5,0.5) --  (0.5,-1,0.5) -- (0.5,-0.5,0);
    \draw[thin,color=green] (0.5,-1,0.5) -- (1.5,0,1.5);

    \draw[thin,color=green] (1,1,1) -- (1.5,0.5,1) -- (1.5,0,1.5) -- (1,0.5,1.5) -- cycle;
    \draw[thin,color=green] (1,1,1) -- (1.5,0.5,1) -- (1.5,0,1.5) -- (1,0.5,1.5) -- cycle;

    \tdplottransformmainscreen{0}{0}{0}
    \shadedraw[tdplot_screen_coords, ball color = blue] (\tdplotresx,\tdplotresy) circle (0.04) node[below right] {$A_1$};
    
    \draw[dotted,color=green!60!black] (1,0,1) -- (1.25,0.5,1.25)  node {$\bullet$};
    \tdplottransformmainscreen{1}{0}{1}
    \shadedraw[tdplot_screen_coords, ball color = blue] (\tdplotresx,\tdplotresy) circle (0.04) node[below right] {$C_1$};

    \draw[dotted,color=green!60!black] (0.5,0,0.5) -- (1.125,0.75,1.125) node {$\bullet$};
    \tdplottransformmainscreen{0.5}{0}{0.5}
    \shadedraw[tdplot_screen_coords, ball color = red] (\tdplotresx,\tdplotresy) circle (0.04) node[below right] {$B_1$};

    \tdplottransformmainscreen{1}{1}{1}
    \shadedraw[tdplot_screen_coords, ball color = blue] (\tdplotresx,\tdplotresy) circle (0.04)node[left] {$A_2$};

    \draw[->,>=stealth,color=orange!60!black,thick] (0,0,0) -- (0.5,-0.5,0)node[below left]{x};
    \draw[->,>=stealth,color=orange!60!black,thick] (0,0,0) -- (0,-0.5,0.5)node[above left]{y};
    \draw[->,>=stealth,color=orange!60!black,thick] (0,0,0) -- (1,1,1)node[above left]{z};    
    
    \end{tikzpicture}}
%
%
%
%
%
%
%
%
  \subbottom[\label{subfig_hoppings}]{\tdplotsetmaincoords{70}{10}
  \begin{tikzpicture}[tdplot_main_coords, scale = 3]

  \draw[thin] (1,0,0) -- (1.5,0.87,0) -- (0.5,0.87,0) ;
  \draw[thin] (0,0,-1.73) -- (1,0,-1.73) -- (1.5,0.87,-1.73) -- (0.5,0.87,-1.73) -- cycle;
  \draw[thin] (0,0,0) -- (0,0,-1.73);
  \draw[thin] (1,0,0) -- (1,0,-1.73);
  \draw[thin] (1.5,0.87,0) -- (1.5,0.87,-1.73);
  \draw[thin] (0.5,0.87,0) -- (0.5,0.87,-1.73);

  \coordinate (A1) at (0,0,0);
  \coordinate (B1) at (0.5,0.29,-0.58);
  \coordinate (C1) at (1,0.58,-1.15);
  \coordinate (A2) at (0,0,-1.73);
  \coordinate (B2) at (0.5,0.29,-2.31);
  
  \draw[->,>=stealth,color=orange!60!black,thick] (A1) -- (1,0,0) node[above] {(1,0,0)};
  \draw[->,>=stealth,color=orange!60!black,thick] (A1) -- (0.5,0.87,0) node[above] {(0.5,0.87,0)};
  
  \draw[thin, dotted] (0.5,0.29,-0.58) -- (0.5,0.29,-1.73) node {$\bullet$};
  \draw[thin, dotted] (1,0.58,-1.15) -- (1,0.58,-1.73) node {$\bullet$};
  
  \tdplottransformmainscreen{0}{0}{0}
  \shadedraw[tdplot_screen_coords, ball color = red] (\tdplotresx,\tdplotresy) circle (0.04) node[below right] {$A_1$};
    
  \tdplottransformmainscreen{0.5}{0.29}{-0.58}
  \shadedraw[tdplot_screen_coords, ball color = green] (\tdplotresx,\tdplotresy) circle (0.04) node[below right] {$B_1$};
  
  \tdplottransformmainscreen{1}{0.58}{-1.15}
  \shadedraw[tdplot_screen_coords, ball color = blue] (\tdplotresx,\tdplotresy) circle (0.04) node[below right] {$C_1$};
  
  \tdplottransformmainscreen{0}{0}{-1.73}
  \shadedraw[tdplot_screen_coords, ball color = red] (\tdplotresx,\tdplotresy) circle (0.04) node[below right] {$A_2$};    
  
  \tdplottransformmainscreen{0.5}{0.29}{-2.31}
  \shadedraw[tdplot_screen_coords, ball color = green] (\tdplotresx,\tdplotresy) circle (0.04) node[below right] {$B_2$};

  \draw[->,>=stealth,color=green] (A1) to[bend right](B1);
  \draw[->,>=stealth,color=blue]  (B1) to[bend right](C1);
  \draw[->,>=stealth,color=red]   (C1) to[bend right](A2);

  \node[color=green] at (0.25,0.15,-0.29) {$t_{1,2}$};
  \node[color=blue]  at (0.75,0.45,-0.90) {$t_{2,3}$};
  \node[color=red] at (0.4,0.20,-1.15) {$t_{3,4}$};

  \node[color=blue] at (0.2,0.19,-0.65) {$t_{1,3}$};
  \node[color=red] at (0.15,0.1,-1.15) {$t_{2,4}$};
  \node[color=green] at (0.7,0.45,-1.65) {$t_{3,5}$};
  \node[color=green] at (0,0,-2.15) {$t_{4,5}=t_{1,2}$};

  \draw[->,>=stealth,color=blue]  (A1) to[bend right](C1);
  \draw[->,>=stealth,color=red]   (B1) to[bend right](A2);
  \draw[->,>=stealth,color=green] (C1) to[bend right](B2);
  \draw[->,>=stealth,color=green] (A2) to[bend right](B2);

  \end{tikzpicture}}\caption{\label{fig_stacking} \subcaptionref{subfig_fcc_unit} Example of a ABCABC\dots stacking in the (111) direction of the FCC Brillouin zone. This is for instance the structure of Au(111). \subcaptionref{subfig_hoppings} At the first iteration in the calculation of the Green functions (2 layers in contact), electrons can jump from $A_1$ to $B_1$ only. At the second iteration, 2 layers in contact with a third, they can jump from $B_1$ to $C_1$ and from $A_1$ to $C_1$, and so forth.}
\end{figure}
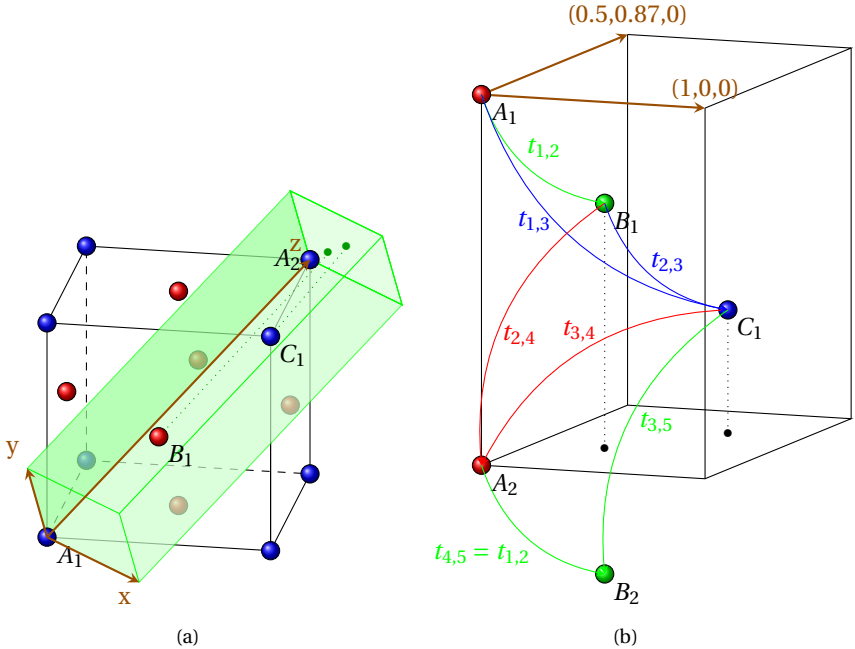

    This is the origin of the tricky part of the \emph{line 11} of the hamiltonian input-file (see sec.~\ref{sec_ham_input}). Consider now a FCC structure in (111) direction. In that case, the structure is a stacking $A_1B_1C_1A_2B_2C_2A_3B_3$\dots and the following hopping are required: $t_{A_1,B_1}$, $t_{B_1,C_1}$, $t_{C_1,A_2}$ and $t_{A_1,C_1}$, $t_{B_1,A_2}$, $t_{C_1,B_2}$ as illustrated in Fig.~\ref{fig_stacking}. In order force the program to compute the hopping matrices $t_{C_1,A_2}$ and $t_{B_1,A_2}$, $t_{C_1,B_2}$ we need to give the positions of atom $A_2$ and $B_2$.
    
    Because of the iterative procedure described in sec.~\ref{sec_pert_method_2nd_neighbor}, these matrices are stored as arrays in the following way:
    \begin{enumerate}
      \item First, electrons propagate from $A_{1}$ to $B_{1}$ (cf. Fig.~\ref{fig_stacking}) $\rightarrow$ $t_{A_1,B_1} =\mathrm{Tij1(1)}$
      \item Then electrons propagate from $B_{1}$ to $C_{1}$ and from $A_{1}$ to $C_{1}$ $\rightarrow$ $t_{B_1,C_1} =\mathrm{Tij1(2)}$ and $t_{A_1,C_1} =\mathrm{Tij2(2)}$
      \item Then electrons propagate from $C_{1}$ to $A_{2}$ and from $B_{1}$ to $A_{2}$ $\rightarrow$ $t_{C_1,A_2} = \mathrm{Tij1(3)}$ and $t_{B_1,A_2} =\mathrm{Tij2(3)}$
      \item Finally electrons propagate from $A_{2}$ to $B_{2}$ and from $C_{1}$ to $B_{2}$ $\rightarrow$ $t_{A_2,B_2} =\mathrm{Tij1(1)}$ and $t_{C_1,B_2} = \mathrm{Tij2(1)}$
    \end{enumerate}
    The $Tij1$ is for nearest-layer hopping and $Tij2$ is for next-nearest ones.    
    To summarize, given a periodicity in $z$ direction, we have to add as an input at least two more layers, as it is done in the Hamiltonian input-file above (listing \ref{ham_input}). In any case, it is already done in the database of Hamiltonian input-files provided with the code.
    
    Once all of those matrices are initiated, the algorithm described in sec.~\ref{sec_pert_method_2nd_neighbor} is used to get the Green functions for the BEEM current.
    
\section{Calculating the propagators and the current}
  The propagators are calculated as described in Sec.~\ref{sec_pert_method_2nd_neighbor} and illustrated the Fig.~\ref{fig_GF_flow_chart} that is here reproduced for clarity.
  
\begin{figure}[!hbt]
  \begin{tikzpicture}[
		      ->,>=stealth',shorten >=1pt,auto,node distance=1.5cm, thick,
		      main node/.style={circle,fill=blue!20,draw, font=\sffamily\bfseries},
		      green node/.style={circle,fill=red!20,draw, font=\sffamily\bfseries}
		      ]
      
    \node[main node] (A1) { \shortstack{1 \\ $\green{1,1}{2}$}};
    \node[main node] (A2) [right of=A1] {\shortstack{2 \\ \phantom{$\green{1,2}{2}$}}};
    \node[main node] (A3) [right of=A2] {\shortstack{3 \\ $\green{2,1}{2}$}};
    \node[green node] (A4) [right of=A3] {\shortstack{4 \\ $\green{1,1}{2}$}};
    \node[main node] (A5) [right of=A4] {\shortstack{5 \\ $\green{2,2}{2}$}};
    \node[green node] (A6) [right of=A5] {\shortstack{6 \\ $\green{1,2}{2}$}};
    \node[main node] (Aa) [right of=A6] {\shortstack{a \\  \phantom{$\green{0,0}{2}$}}};
    \node[green node] (Ab) [right of=Aa] {\shortstack{b \\  \phantom{$\green{1,1}{2}$}}};
    \node[right =0.3cm of Ab] {\textbf{{n+1=2}}};
    
    \node[main node] (B1) [below =4.15cm of A1] {\shortstack{1 \\ $\green{2,2}{3}$}};
    \node[main node] (B2) [right of=B1] {\shortstack{2 \\ $\green{1,2}{3}$}};
    \node[main node] (B3) [right of=B2] {\shortstack{3 \\ $\green{3,2}{3}$}};
    \node[green node] (B4) [right of=B3] {\shortstack{4 \\ $\green{1,2}{3}$}};
    \node[main node] (B5) [right of=B4] {\shortstack{5 \\ $\green{3,3}{3}$}};
    \node[green node] (B6) [right of=B5] {\shortstack{6 \\ $\green{1,3}{3}$}};
    \node[main node] (Ba) [right of=B6] {\shortstack{a \\ $\green{1,1}{3}$}};
    \node[green node] (Bb) [right of=Ba] {\shortstack{b \\ $\green{1,1}{3}$}};
    \node[right =0.3cm of Bb] {\textbf{{n+1=3}}};

    \node[main node] (C1) [below =4.15cm of B1] {\shortstack{1 \\ $\green{3,3}{4}$}};
    \node[main node] (C2) [right of=C1] {\shortstack{2 \\ $\green{2,3}{4}$}};
    \node[main node] (C3) [right of=C2] {\shortstack{3 \\ $\green{4,3}{4}$}};
    \node[green node] (C4) [right of=C3] {\shortstack{4 \\ $\green{1,3}{4}$}};
    \node[main node] (C5) [right of=C4] {\shortstack{5 \\ $\green{4,4}{4}$}};
    \node[green node] (C6) [right of=C5] {\shortstack{6 \\ $\green{1,4}{4}$}};  
    \node[main node] (Ca) [right of=C6] {\shortstack{a \\ $\green{2,2}{4}$}};
    \node[green node] (Cb) [right of=Ca] {\shortstack{b \\ $\green{1,2}{4}$}};
    \node[right =0.3cm of Cb] {\textbf{{n+1=4}}};
    
    \path[every node/.style={font=\sffamily\small, fill=white,inner sep=1pt}]
	(A1) edge[blue]  (B1)
	     edge[green] (B2)
	     edge[red]   (B5)
	     edge[Goldenrod] (Ba)
	(A3) edge[blue]  (B1)
	     edge[green] (B2)
	     edge[red]   (B5)
             edge[Goldenrod] (Ba)
	(A4) edge[gray]  (B4)
	     edge[black] (B6)
             edge[Sepia] (Bb)
	(A5) edge[blue]  (B1)
	     edge[green] (B2)
	     edge[red]   (B5)
	(A6) edge[gray]  (B4)
	     edge[black] (B6)
             edge[Sepia] (Bb)
	(B1) edge[bend right=38, orange] (B3)
	     edge[bend right=38, gray]   (B4)
	(B2) edge[bend right=38, orange] (B3)
	     edge[bend right=38, gray]   (B4)
             edge[bend right=38, Goldenrod] (Ba)
             edge[bend right=38, Sepia] (Bb)
	(B5) edge[bend right=38, black]  (B6)
	(Ba) edge[bend right=38, Goldenrod] (Bb)
	(B1) edge[blue]  (C1)
	     edge[green] (C2)
	     edge[red]   (C5)
             edge[Goldenrod] (Ca)
	(B3) edge[blue]  (C1)
	     edge[green] (C2)
	     edge[red]   (C5)
             edge[Goldenrod] (Ca)
	(B4) edge[gray]  (C4)
	     edge[black] (C6)
             edge[Sepia] (Cb)
	(B5) edge[blue]  (C1)
	     edge[green] (C2)
	     edge[red]   (C5)
	(B6) edge[gray]  (C4)
             edge[Sepia] (Cb)
	     edge[black] (C6)
	(C1) edge[bend right=38, orange] (C3)
	     edge[bend right=38, gray]   (C4)
	(C2) edge[bend right=38, orange] (C3)
	     edge[bend right=38, gray]   (C4)
             edge[bend right=38, Goldenrod] (Ca)
             edge[bend right=38, Sepia] (Cb)
	(C5) edge[bend right=38, black]  (C6)
        (Ca) edge[bend right=38, Goldenrod] (Cb)
	;
  \end{tikzpicture}
  \caption{Flow chart of the algorithm used to calculate the Green functions $\hat{G}_{1,3}^{R,(4)}$, $\hat{G}_{1,2}^{R,(4)}$ and $\hat{G}_{4,1}^{A,(4)}$ (pink circles). The label 1, 2, 3, 4, 5 and 6 are the same as those used in sec~\ref{sec_eg_G14}. This is also the same order used in the code to calculate the various Green functions. At the first iteration (n=1, we are looking for n+1=2), there is no next-nearest layers: $\green{i,j}{2}$ is only expressed in term of $\green{i,j}{1}$. At the second iteration there are next-nearest-layer interactions, but still one degenerate case: step~\ref{enum_G134} = step~\ref{enum_G234}. At the third iteration, each Green function is different (at least all indexes are). Equations  \eqref{eq_G144},  \eqref{eq_G134} and  \eqref{eq_G124} give the retarded Green function and the advanced Green function of the current formula (eq.~(\ref{eq_BEEM_current})).}\label{fig_GF_flow_chart_bis}
\end{figure}

  For the first iteration ($n=1$), only the  Green function of the first isolated layer is different from zero:
  \begin{center}
  \begin{lstlisting}%[basicstyle=\tiny\ttfamily, keywordstyle=\color{blue},otherkeywords={.AND.},stringstyle=\color{red},commentstyle=\color{gray}]
  
    IF(islab==1 .AND. ilayer==1) THEN
       Gnnm1  (:,:) = (0.d0,0.d0)
       Gnm1n  (:,:) = (0.d0,0.d0)
       Gnm1nm1(:,:) = (0.d0,0.d0)
       G1nm1  (:,:) = (0.d0,0.d0)
       Gnn    (:,:) = gii(:,:,1)
       G1n    (:,:) = gii(:,:,1)
    END IF
  \end{lstlisting}    
  \end{center}
  In the program we have used the notations:
  \begin{align*}
       \mathrm{Gnnm1}   &= \green{n,n-1}{n}   \\
       \mathrm{Gnm1n}   &= \green{n-1,n}{n}   \\
       \mathrm{Gnm1nm1} &= \green{n-1,n-1}{n}   \\
      \mathrm {G1nm1}   &= \green{1,n-1}{n}   \\
      \mathrm {Gnn}     &= \green{n,n}{n}   \\
       \mathrm{G1n}     &= \green{1,n}{n}   \\
       \mathrm{gii(:,:,1)} &= \green{1,1}{1}  
  \end{align*}
  Of course, with just one layer, all Green functions with $n-1$ label are identically zero. Then, we move to the second iteration and get:
  \begin{lstlisting}%[language=fortran,basicstyle=\tiny\ttfamily, keywordstyle=\color{blue},otherkeywords={.AND.},stringstyle=\color{red},commentstyle=\color{gray}]
  
    Gnm1nm1  = Gnnnp1       ! step 1
    Gnnm1    = Gnp1nnp1     ! step 3
    Gnm1n    = Gnnp1np1     ! step 3
    G1nm1    = G1nnp1       ! step 4
    Gnn      = Gnp1np1np1   ! step 5
    G1n      = G1np1np1     ! step 6
    G1nm2    = G1nm1np1     ! step 8
  \end{lstlisting}
  Where ``step i'' refers to the steps in figure \ref{fig_GF_flow_chart}.
  At the end of the iteration up to layer $n$, $\mathrm{G1nm2} = G_{1,n-2}^{R,(n)}$, $\mathrm{G1nm1} = G_{1,n-1}^{R,(n)}$ and  $\mathrm{GAn1}=\left[G_{1,n}^{R(n)}\right]^{\dagger}$ are used to calculate the current, according to the equation \eqref{eq_BEEM_current}

  Finally, the current is calculated for any k-points, and written in the output file \emph{structure\_current.dat}. If a loop over the energy has been asked, then the DOS is calculated and stored in  \emph{structure\_dos.dat}. Examples are given in chapter~\ref{chapt_results}.
  
    \FloatBarrier

\chapter{Results and discussion}\label{chapt_results}
  \lettrine[lines=3, lhang=0.35, loversize=0.77, findent=2.5em,nindent=-0.5em,slope=-1em]{I}{n the previous chapters}, we have introduced Ballistic Electron Emission Miscroscopy and the different ways to model the elastic scattering of electrons with increasing number of layers, and their pro-pagation. As shown above, the tight-binding approach is probably the best intuitive method to deal with hopping from one layer to another and it allows a simple way to ``play'' with the parameters at the interfaces, that can better respond to the experimentalist needs.
For these reasons we decided to use a tight-binding approach, instead of the non-equilibrium Green-function Density Functional Theory (NEGF-DFT), for which moreover, the required time of calculation for our systems could have been very expensive. 

    \begin{table}\fontsize{7.8}{13}\selectfont
         \begin{equation*}\rowcolors{1}{MidnightBlue!10}{}\begin{array}{ll}
	 E_{s,s} &= V_{ss\sigma}
      \\ E_{s,x} &= l V_{sp\sigma}
      \\ E_{x,x} &= l^2 V_{pp\sigma} + (1 - l^2) V_{pp\pi}
      \\ E_{x,y} &= l m V_{pp\sigma} - l m V_{pp\pi}
      \\ E_{x,z} &= l n V_{pp\sigma} - l n V_{pp\pi}
      \\ E_{s,xy} &= \sqrt{3} l m V_{sd\sigma}
      \\ E_{s,x^2-y^2} &= \frac{\sqrt{3}}{2} (l^2 - m^2) V_{sd\sigma}
      \\ E_{s,3z^2-r^2} &= [n^2 - (l^2 + m^2) / 2] V_{sd\sigma}
      \\ E_{x,xy} &= \sqrt{3} l^2 m V_{pd\sigma} + m (1 - 2 l^2) V_{pd\pi}
      \\ E_{x,yz} &= \sqrt{3} l m n V_{pd\sigma} - 2 l m n V_{pd\pi}
      \\ E_{x,zx} &= \sqrt{3} l^2 n V_{pd\sigma} + n (1 - 2 l^2) V_{pd\pi}
      \\ E_{x,x^2-y^2} &= \frac{\sqrt{3}}{2} l (l^2 - m^2) V_{pd\sigma} + l (1 - l^2 + m^2) V_{pd\pi}
      \\ E_{y,x^2-y^2} &= \frac{\sqrt{3}}{2} m(l^2 - m^2) V_{pd\sigma} - m (1 + l^2 - m ^2) V_{pd\pi}
      \\ E_{z,x^2-y^2} &= \frac{\sqrt{3}}{2} n(l^2 - m^2) V_{pd\sigma} - n(l^2 - m^2) V_{pd\pi}
      \\ E_{x,3z^2-r^2} &= l[n^2 - (l^2 + m^2)/2]V_{pd\sigma} - \sqrt{3} l n^2 V_{pd\pi}
      \\ E_{y,3z^2-r^2} &= m [n^2 - (l^2 + m^2) / 2] V_{pd\sigma} - \sqrt{3} m n^2 V_{pd\pi}
      \\ E_{z,3z^2-r^2} &= n [n^2 - (l^2 + m^2) / 2] V_{pd\sigma} + \sqrt{3} n (l^2 + m^2) V_{pd\pi}
      \\ E_{xy,xy} &= 3 l^2 m^2 V_{dd\sigma} + (l^2 + m^2 - 4 l^2 m^2) V_{dd\pi} + (n^2 + l^2 m^2) V_{dd\delta}
      \\ E_{xy,yz} &= 3 l m^2 nV_{dd\sigma} + l n (1 - 4 m^2) V_{dd\pi} + l n (m^2 - 1) V_{dd\delta}
      \\ E_{xy,zx} &= 3 l^2 m n V_{dd\sigma} + m n (1 - 4 l^2) V_{dd\pi} + m n (l^2 - 1) V_{dd\delta}
      \\ E_{xy,x^2-y^2} &= \frac{3}{2} l m (l^2 - m^2) V_{dd\sigma} + 2 l m (m^2 - l^2) V_{dd\pi} + l m (l^2 - m^2) / 2 V_{dd\delta}
      \\ E_{yz,x^2-y^2} &= \frac{3}{2} m n (l^2 - m^2) V_{dd\sigma} - m n [1 + 2(l^2 - m^2)] V_{dd\pi} + m n [1 + (l^2 - m^2) / 2] V_{dd\delta}
      \\ E_{zx,x^2-y^2} &= \frac{3}{2} n l (l^2 - m^2) V_{dd\sigma} + n l [1 - 2(l^2 - m^2)] V_{dd\pi} - n l [1 - (l^2 - m^2) / 2] V_{dd\delta}
      \\ E_{xy,3z^2-r^2} &= \sqrt{3} \left[ l m (n^2 - (l^2 + m^2) / 2) V_{dd\sigma} - 2 l m n^2 V_{dd\pi} + l m (1 + n^2) / 2 V_{dd\delta} \right]
      \\ E_{yz,3z^2-r^2} &= \sqrt{3} \left[ m n (n^2 - (l^2 + m^2) / 2) V_{dd\sigma} + m n (l^2 + m^2 - n^2) V_{dd\pi} - m n (l^2 + m^2) / 2 V_{dd\delta} \right]
      \\ E_{zx,3z^2-r^2} &= \sqrt{3} \left[ l n (n^2 - (l^2 + m^2) / 2) V_{dd\sigma} + l n (l^2 + m^2 - n^2) V_{dd\pi} - l n (l^2 + m^2) / 2 V_{dd\delta} \right]
      \\ E_{x^2-y^2,x^2-y^2} &= \frac{3}{4} (l^2 - m^2)^2 V_{dd\sigma} + [l^2 + m^2 - (l^2 - m^2)^2] V_{dd\pi} + [n^2 + (l^2 - m^2)^2 / 4] V_{dd\delta}
      \\ E_{x^2-y^2,3z^2-r^2} &= \sqrt{3} \left[ (l^2 - m^2) [n^2 - (l^2 + m^2) / 2] V_{dd\sigma} / 2 + n^2 (m^2 - l^2) V_{dd\pi} + (1 + n^2)(l^2 - m^2) / 4 V_{dd\delta}\right]
      \\ E_{3z^2-r^2,3z^2-r^2} &= [n^2 - (l^2 + m^2) / 2]^2 V_{dd\sigma} + 3 n^2 (l^2 + m^2) V_{dd\pi} + \frac{3}{4} (l^2 + m^2)^2 V_{dd\delta}
    \end{array}\end{equation*}
    \caption{\label{tab_SK_system}\normalsize Table of interatomic matrix elements due to Slater and Koster \cite{Slater-Koster_PhysRev.94.1498}. The table expresses the matrix elements as functions of LCAO 2-center bond integrals between two orbitals, $i$ and $j$, on adjacent atoms. The bond integrals are for example the $V_{ss\sigma}$,$ V_{pp\pi}$ and $V_{dd\delta}$ for sigma, pi and delta bonds.     $l$, $m$ and $n$ are direction cosines of the atomic position $\vec{v}$: $l = \frac{\vec{v} \cdot {\hat{a}_1} }{ \left \Vert \vec{v} \right \Vert }$, $m =  \frac{\vec{v} \cdot {\hat{a}_2} }{ \left \Vert \vec{v} \right \Vert }$, $n = \frac{\vec{v} \cdot {\hat{a}_3} }{ \left \Vert \vec{v} \right \Vert }$.}
    \end{table}

Within the tight-binding approach we have tested two different models for the BEEM current: an equilibrium model and a non-equilibrium one. 
The first (see section \ref{sec_eq_calc_results} for a proper definition) has the merit to be conceptually simpler.
We have employed it in the case of bulk band structure, so that the results can be interpreted very intuitively, as we shall see below.
We shall compare this equilibrium model with the more elaborate (and conceptually more precise) model based on the non-equilibrium Green-functions formalism presented in Chap.~\ref{chapt_NEPT}. It is expected that the propagation of electrons within thin films is better described by the realistic non-equilibrium model. 
Nevertheless, it turns out that the band structure effects contained in the equilibrium calculation can explain some experimental results and allow us to make predictions that, given the present level of experimental sensitivity, are hardly distinguishable from the more correct non-equilibrium ones.

In this chapter, we first present the effectiveness of tight-binding parametrization used for our calculations (in Sec.~\ref{sec_tb_parametrization}). In section \ref{sec_eq_calc_results}, we present some results obtained with the equilibrium approach. Finally, we move to the non-equilibrium approach in section \ref{sec_quantum_approach}, before drawing our conclusions.

\section{Tight-binding parametrization}\label{sec_tb_parametrization}
  The tight-binding approach is a parametrized calculation of band-structure that allows us to reproduce electronic band-structures with a root-mean-square error of the order of $(50\pm20)$meV, as shown for example in Ref.~\cite{Papa-handbook}. We shall not linger on the basic aspects of the theory, that are well described elsewhere (see for example Ashcroft~\cite{ashcroft} for a first introduction and Papaconstantopoulos \cite{Papa-handbook} for a more advanced treatment). In this section we shall rather describe two different approaches to handle the tight-binding approximation: Papaconstantopoulos' approach, based on Slater-Koster two-center parameters, and Harrison's approach (even in the Papaconstantopoulos modified version), based on a minimal set of universal hopping parameters.

  \subsection{Papaconstantopoulos' approach}
    Dimitrios A. Papaconstantopoulos wrote the first systematic tight-binding parametrization of all metals obtained by fitting ab-initio calculations. This work was published as a ``handbook of the band structure of elemental solids'' in which ab-initio band structure-calculations have been recast on a linear combination of atomic orbitals (LCAO) basis using the Slater-Koster (SK) method~\cite{Slater-Koster_PhysRev.94.1498} as an interpolation scheme (Table~\ref{tab_SK_system}). The ab-initio calculations have been done using the augmented plane-wave method (APW) that includes the mass velocity and Darwin relativistic effects, but without spin-orbit coupling. Then, the 2- and 3-center SK parameters are obtained by a fitting procedure:\footnote{The full procedure is well described in the introduction of the handbook and therefore we do not repeat it here.} the fit is done for the six first bands on a uniform mesh of $k$ points using a least-square program, and energies at extra high symmetry points (nine bands are calculated for the latter) are fixed using analytical formulæ.        

\begin{figure}[btp]
      \centering
      \includegraphics[width=0.97\linewidth]{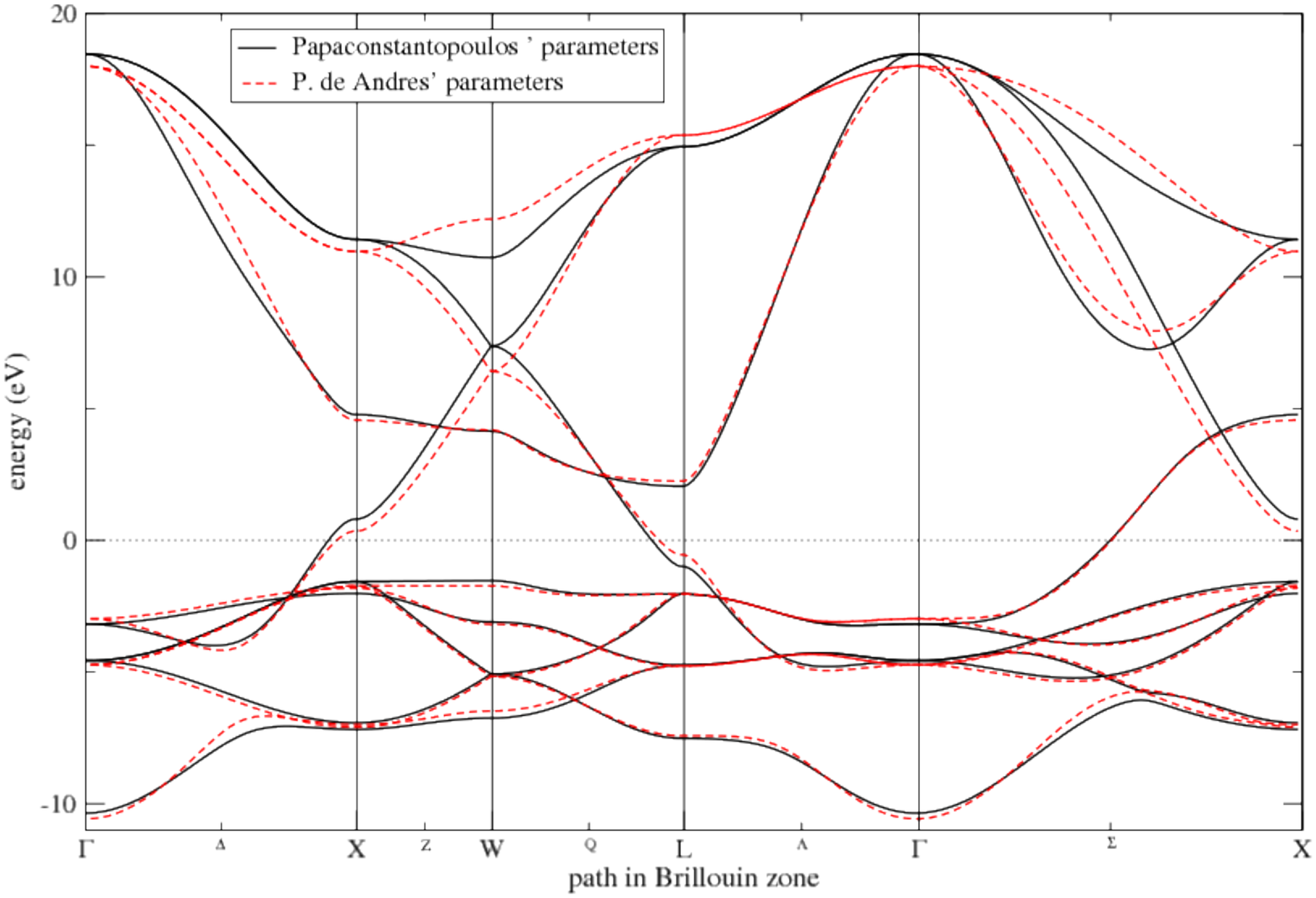}
      \caption{Band structure of gold considering only nearest-neighbor hopping from BEEM v2.1 input files (in dashed red lines) vs second nearest-neighbor hopping from \cite{Papa-handbook} (in black lines). The two band structures are in quite good agreement, in particular around the Fermi level (RMS error < 70 meV for the sixth band).}{\label{fig_Au_bs}}
    \end{figure}

\begin{figure}[btp]
      \centering
      \includegraphics[width=0.8\linewidth]{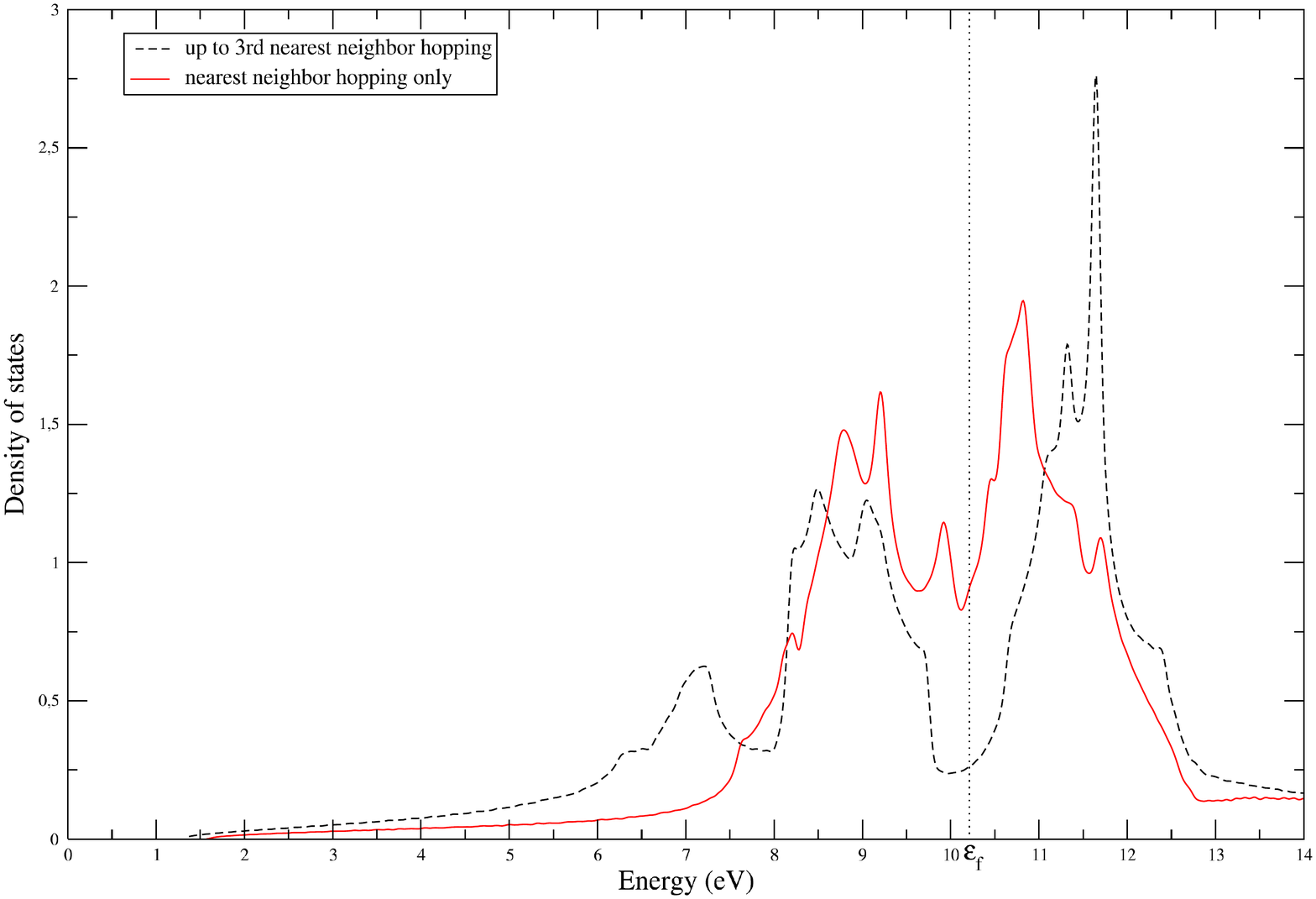}
      \caption{Density of states per unit cell of spin-down iron considering nearest-neighbor hopping only (red curve) and up to third-nearest-neighbor hopping (black dashed curve) using parameters of Papaconstantopoulos' handbook \cite{Papa-handbook}. The DOS for the calculation with nearest-neighbor hopping only is instead very different. In general, we cannot describe BCC structures considering only nearest-neighbor hopping.\label{fig_bs_Fe_1st}}
    \end{figure}

    By means of a subroutine explicitly inserted in my code BEEM v3, I have been able to reproduce some band structures of metal of interest for BEEM. Some are reproduced here: Fe (fig.~\ref{fig_Fe_Au_BS}), Au (fig.~\ref{fig_Au_bs}) and Ag (fig.~\ref{fig_silver_harrison}). All those band structures have been obtained with the so-called 3-center parameters. It is also possible to use the 2-center parameters reported in Papaconstantopoulos book by using the Fortran  ``2\_to\_center.f90'' subroutine of my code.
    The drawback of the accurate matching (root mean square errors $\sim$ 30 to 70 meV) of Papaconstantopoulos' parametrization with ab-initio calculations is the high number of tight-binding parameters required ($\sim$ 25 up to 35).
    Note that this number of parameters can be reduced for FCC crystals by considering only nearest-neighbor hopping, as shown in figure \ref{fig_Au_bs}. However, for BCC structures, we have no choice but to consider third-nearest-neighbor hopping, as shown in Fig.~\ref{fig_bs_Fe_1st}. For this reason, we have considered nearest-neighbor hopping for gold (and because we want to compare our results with the previous results using BEEM v2.1 code) and up to third-nearest-neighbor hopping for iron.
    
    In order to reduce the number of parameters, Harrison proposed an alternative parametrization of tight-binding that, though less precise (as we shall see below), drastically reduces the number of required parameters, and allows straightforward generalizations to interfaces.

  \subsection{Harrison's approach}
    Harrison has developed an elegant theory of the solid state where he seeks to explain electronic properties analytically, with a reduced number of external parameters. His starting points are the fact that Slater-Koster parameters can be expressed with power laws $\propto d^{-2}$,  $\propto d^{-4}$, or  $\propto d^{-5}$ depending on the type of orbital interactions (as shown below) and that by fitting free electron bands or band structures of germanium and silicon, he had found very close coefficients. 
    The latter remark made him assume that there might be general coefficients to describe all band structures.  
    With this approximation, he was able with, a minimal set of parameters, to reproduce the band structures of many materials.

\begin{figure}
      \centering
      \includegraphics[trim=0cm 1cm 6cm 3cm,clip,width=0.97\linewidth]{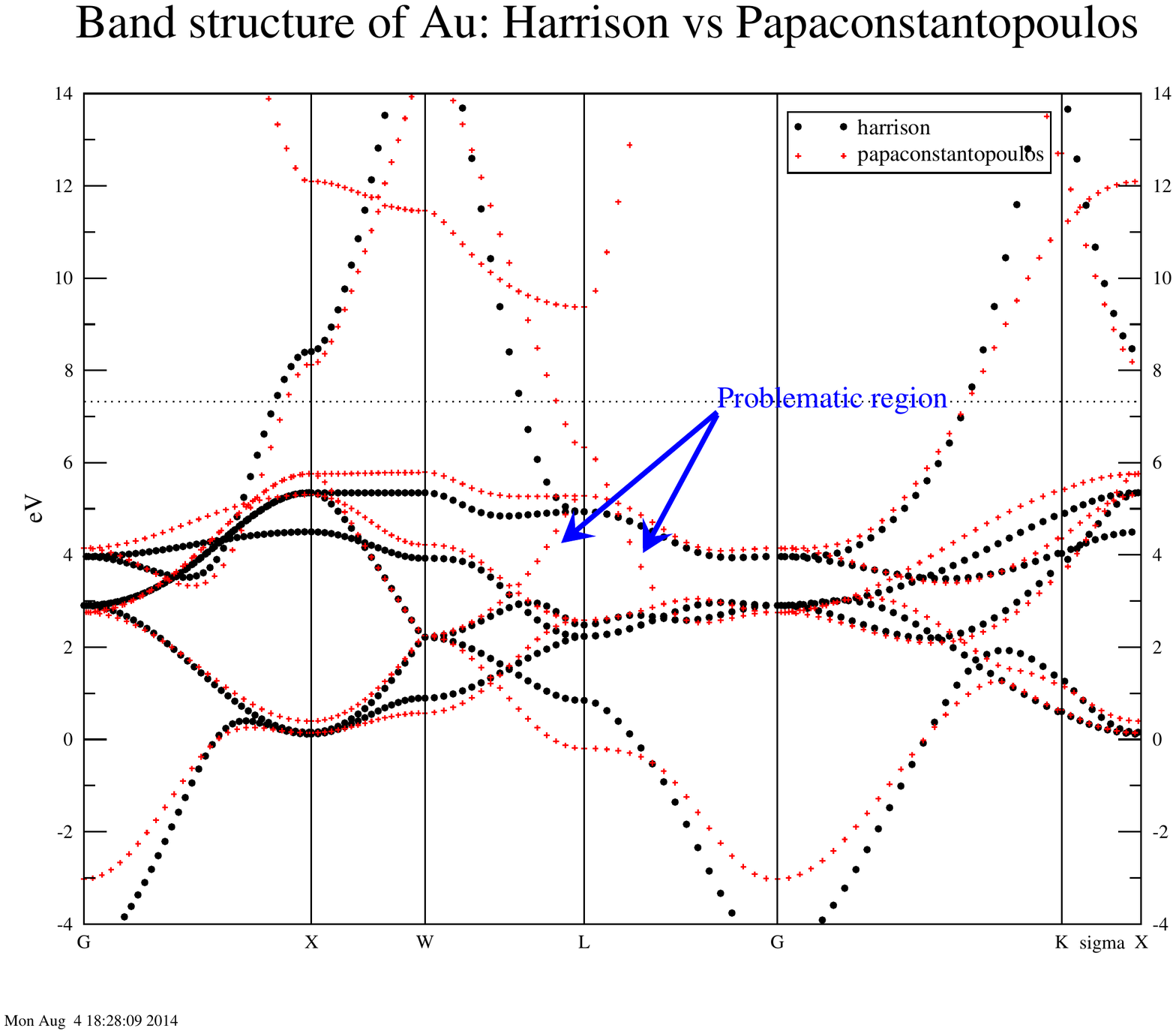}
      \caption{Band structure of gold calculated with Harrison's parameters \cite{Harrison-book_1989} (black-dashed lines) and with Papaconstantopoulos' parameters \cite{Papa-handbook} (red-dotted lines). The topology of the Harrison band-structure is very similar to the Papaconstantopoulos one. However, even if such similarities with only 3 parameters are quite impressive, some problematic regions remain, like for instance, in the 2-4eV range, around the L-point. }{\label{fig_harrison_vs_papa_Au}}
    \end{figure}

    For instance, in order to build the gold band structure of Fig.~\ref{fig_harrison_vs_papa_Au}, we can use the universal parameters:
    \begin{align}\label{eq_harrison_Vllm}
      V_{ll'm} &= \beta_{ll'm} \frac{\hbar^2}{m_e} \frac{1 }{d^2} \\
      V_{ddm}  &= \beta_{ddm}  \frac{\hbar^2}{m_e} \frac{r_d^3    }{d^5} \label{eq_harrison_Vddm}
    \end{align} 
    The denominator $d$ is the distance between the considered neighbors, $l$ is a $s$ or $p$ orbital, and $\beta$ are the constant prefactors:
    \begin{align}
	\beta_{ss\sigma}&= -1.32            &   \beta_{dd\sigma}&= -\frac{45}{\pi}\nonumber\\\nonumber
	\beta_{sp\sigma}&= \phantom{-} 1.42 &   \beta_{dd\pi}   &= \phantom{-} \frac{30}{\pi}\\\nonumber
	\beta_{pp\sigma}&= \phantom{-} 2.22 &   \beta_{dd\delta}&=  -\frac{15}{2\pi}\\
	\beta_{pp\pi}   &= -0.63            &                   &                 &          
    \end{align}
    and $\hbar^2/m=7.62$eV\AA{}$^2$. In his previous book~\cite{Harrison-book_1989}, Harrison also consider matrix elements $V_{ldm} = \beta_{ldm}  \frac{\hbar^2}{m_e} \frac{r_d^{3/2}}{d^{7/2}}$, which are here absent.
    
    $\gamma_s$ and $r_d$ are both material-dependent parameters and can be found in L. Shi and D. Papaconstantopoulos' paper \cite{Papa-Harrison_PhysRevB.70.205101}. For gold, they are:
    \begin{align*}
      \varepsilon_s &= -\phantom{1} 6.980 \mbox{ eV} \\
      \varepsilon_d &= -17.780 \mbox{ eV} \\
      r_d           &=  \phantom{-1} 1.007 \mbox{ \AA{}}
    \end{align*}
    and the distance is
    \begin{align}
        d_{\mathrm{1st}} &= {a\sqrt{2}}
      \\d_{\mathrm{2nd}} &= a
    \end{align}
    with $a=4.08$\AA{}~\cite{Papa-handbook}.

    This reduction of the problem is of course always less accurate that a full \emph{ab-initio} numerical solution. For instance, Fig.~\ref{fig_harrison_vs_papa_Au} shows the differences between Harrison's approach and Papaconstantopoulos' prametrization of APW calculations. It is interesting to note that the band structure is nicely reproduced except around the $L$ point where a band below Fermi level (within the range 2-4 eV) is completely wrong : there is a difference of more than 2.5 eV ! Moreover, the minimum energy, at $\Gamma$ point, is also several eV too low.
    We reproduced Harrison's band structures also for other materials and found similar behaviors. 

    It is useful to remind that some researchers, Andrey Umerski \emph{et al.}~\cite{umerski}, modeled tunneling magneto-resistance (TMR) using this theory for the description of the interfaces between the different materials of the structures. As electrons cross the tunneling junction with a wave vector $\kpara=0$, our results show that Harrison's approach can be used, as the band structure of the studied material is well reproduced around $\Gamma$ point. However, one has to be cautious before using Harrison's approach for any band structures.

    Interestingly, in 2004, D. Papaconstantopoulos improved Harrison's theory \cite{Papa-Harrison_PhysRevB.70.205101} by adding a $p$ on-site energy to the $s$ and $d$ on-site energies used by Harrison, by modifying the $sp$ hopping integrals with a new dimensionless parameter $\gamma_s$ and by considering the $V_{ldm} = \beta_{ldm}  \frac{\hbar^2}{m_e} \frac{r_d^{3/2}}{d^{7/2}}$ of the previous book. In spite of this extra parameter $\gamma_s$, the number of free parameters in this modified Harrison's approach is still limited. In this section, we shall see if and how this method can be used to describe multi-material hopping in order to describe the transport from, for instance, a layer of gold to a layer of iron.

    \subsubsection{Modified Harrison tight-binding parametrization}
      So, besides the parameters described above in Eqs.~\eqref{eq_harrison_Vllm} and \eqref{eq_harrison_Vddm}, we have an extra parameter and an extra matrix element:
      \begin{align}
	V_{ll'm} &= \beta_{ll'm} \frac{\hbar^2}{m_e} \frac{\gamma_s }{d^2} \\
	V_{ldm}  &= \beta_{ldm}  \frac{\hbar^2}{m_e} \frac{r_d^{3/2}}{d^{7/2}} \\
	V_{ddm}  &= \beta_{ddm}  \frac{\hbar^2}{m_e} \frac{r_d^3    }{d^5}
      \end{align} 
      The denominator $d$ is the distance between the considered neighbors, $l$ is a $s$ or $p$ orbital, and $\beta$ are the constant prefactors:
      \begin{align}
	  \beta_{ss\sigma}&= -0.90            & \beta_{sd\sigma}&= -3.12            & \beta_{dd\sigma}&= -21.22\\\nonumber
	  \beta_{sp\sigma}&= \phantom{-} 1.44 & \beta_{pd\sigma}&= -4.26            & \beta_{dd\pi}   &= \phantom{-} 12.60\\\nonumber
	  \beta_{pp\sigma}&= \phantom{-} 2.19 & \beta_{pd\pi}   &= \phantom{-} 2.08 & \beta_{dd\delta}&=  -\phantom{0}2.29\\\nonumber
	  \beta_{pp\pi}   &= -0.03            &                 &                   &                 &          
      \end{align}
      $\gamma_s$ and $r_d$ are both material-dependent parameters and a complete table for the most of the elements usually employed in material science can be found in L. Shi and D. Papaconstantopoulos' paper \cite{Papa-Harrison_PhysRevB.70.205101}.
      
      With only three equations, ten universal constants and two material-dependent parameters  $\gamma_s$ and $r_d$ plus the distance $d$ between the two atoms, the two center parameters of Slater \& Koster can be found in order to build the band structure.

    \subsubsection{Silver band structure}
      To ensure our results we have reproduced the band structure for silver in Fig.~\ref{fig_silver_harrison}, as in Ref.~\cite{Papa-Harrison_PhysRevB.70.205101}: we consider first and second nearest-neighbors in a 4.064\AA{} FCC. It should be reminded that Papaconstantopoulos's correction of Harrison's parameters in Ref.~\cite{Papa-Harrison_PhysRevB.70.205101} aimed at fitting six $s-d$ bands of the transition metals by keeping the same set of the Harrison $\beta$ parameters for all of them, in order to achieve similar universality as that of Harrison. This is why, the 7th-9th bands are not fitted very well (RMS error of 2eV, similarly in the Papaconstantopoulos' book the RMS errors are the greater for those bands $\sim$ 70 meV), particularly around the X point. However, the result is good enough for the region which interests us, i.e., 1 eV above Fermi level\footnote{The RMS error of the sixth band is 0.47eV that is quite high. However, at 1ev, the differences in energy are $\sim$ 40 up to 100 meV.}, and for the resolution of the BEEM.
      
      \insertfigure{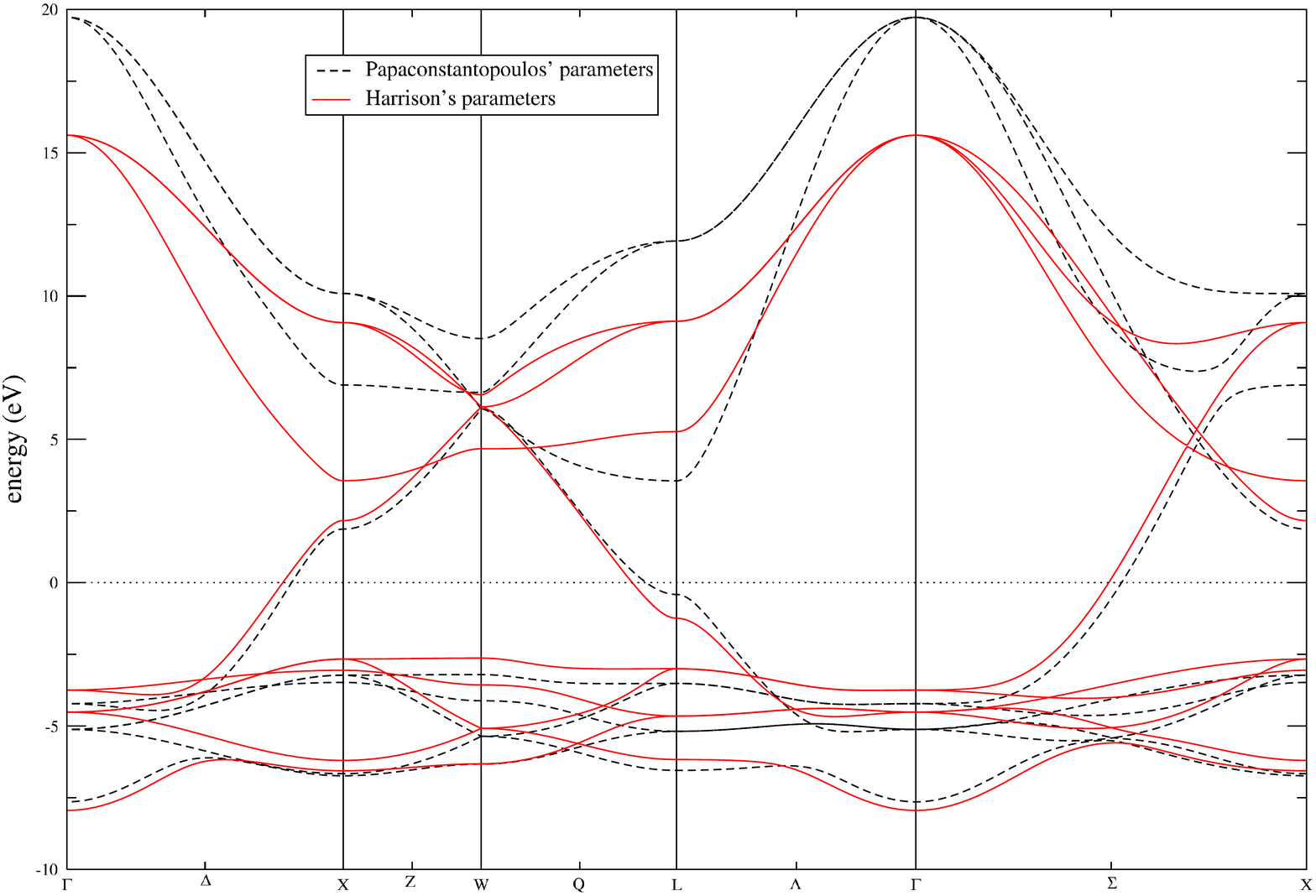}{Band structure of silver calculated with the modified Harrison's theory \cite{Papa-Harrison_PhysRevB.70.205101}. The dotted line is the APW parametrized band structure due to Papaconstantopoulos \cite{Papa-handbook}. Their matching is excellent at low energy. At high energy, their matching is worst. However, the comparisons at high energy should be avoided because the RMS error of the high energy bands of Papaconstantopoulos' parametrization with respect to the APW calculations is quite high: 70 meV for the 6th band of Silver, against less than 25 meV for 1st to 5th bands. In any case, for the BEEM current, we are interested in the 0-1.5 eV energy range above the Fermi level (set to 0 in this figure).}{\label{fig_silver_harrison}}

    \subsubsection{Multi-material parametrization}
      The description of the BEEM current through a spinvalve, say Fe/Au/Fe, requires the hopping parametrization of two different atomic species, in this case Fe and Au. In the Harrison theory, it can be shown~\cite[Chap. 17]{Harrison-book_2004} that the hopping from a material to a different one can be obtained by taking the geometric mean of some power of their $\gamma_s$ and $r_d$ parameters. The reason why we have to keep this formula is that the general coupling between atomic states of any angular-momentum quantum numbers is~\cite{wills_harrison-PhysRevB.28.4363}:
      \begin{equation}
        V_{ll'm} = \beta_{ll'm} \frac{\hbar^2 \sqrt{r_{l}^{2l-1} r_{l'}^{2l'-1}} }{md^{l+l'+1}}
      \end{equation}       
      For instance to propagate from iron to gold one has to take
      \begin{equation}
	\left(r_d^{FeAu}\right)^{3}  = \sqrt{\left(r_d^{Fe}\right)^3 \left(r_d^{Au}\right)^3}
      \end{equation}
      and keep the usual on-site parameters for each element. The distance between the two atoms if not known experimentally should be found with DFT total energy calculations, or used as an extra free parameter. In this way, it is straightforward to model any type of interface.

\section{Equilibrium evaluation of BEEM current}\label{sec_eq_calc_results}
  In Sec.~\ref{sec_free_electron}, we have introduced the first modeling of BEEM current, by Kaiser and Bell, in terms of a free-electron propagation. This picture was invalidated by the work of Garcia-Vidal \emph{et al.}~\cite{Garcia-Vidal-PhysRevLett.76.807} who showed that BEEM electrons are instead elastically scattered by the periodic potential of the crystal. In order to model this behavior, they used a fully quantum out-of-equilibrium approach based on Keldysh formalism, like the one described in Chap.~\ref{chapt_NEPT} and in the following Sec.~\ref{sec_quantum_approach}. However, it is interesting to compare this out-of-equilibrium approach with a simpler equilibrium calculation. We should specify what we mean by ``equilibrium'' calculation of the current, as by definition, there is no net current at equilibrium. 
  

  In the semi-classical theory of metal, the current density at point $\vec{r}$ and time $t$ is:
  \begin{align}
    \vec{j}(\vec{r},t) &= q\rho(\vec{r},t)\vec{v}_g(\vec{r},t)
  \end{align}
  if $\rho$ electrons per unit volume, of charge $q$, all move with velocity $\vec{v}_g(\vec{r},t)$.

  As the group velocity of electrons is proportional to the reciprocal-space gradient of their energies, after space and time Fourier transform, we obtain:
  \begin{equation}\label{eq_sc_current}
    \vec{j}_{\kpara} \propto \sum_{n={\textrm{band index}}} \int \vec{\nabla}\!_{\vec{k}}\varepsilon^n_{\vec{k}} \cdot \delta(\varepsilon^n-\varepsilon^n_{\vec{k}})\,\D \vec{k}_{z}
  \end{equation}
  where $\delta(\varepsilon^n-\varepsilon^n_{\vec{k}})$ is the density of states at energy $\varepsilon$, and $\varepsilon^n_{\vec{k}}$ the eigenvalues of the $n^{\text{th}}$ band. $\vec{k}_{z}$ is the component of the wave-vector parallel to the epitaxy direction and $\kpara$ is the component parallel to the interfaces (orthogonal to the epitaxy direction).
  
  The current is obtained in this way by bulk band-structure calculations, such as in the simple tight-binding approximation. It is clear that this integral is zero for electrons which propagate in all directions. For this current to be meaningful, we have to consider that electrons propagate only in $k_z>0$ direction. This point of view has the advantage of a simpler physical understanding, compared to the formalism of section \ref{sec_quantum_approach}. This is how out-of-equilibrium is artificially introduced.
      
  Once the current-density vector\footnote{Actually, it is rather a ``channel'' in $\kpara$-space where electrons can propagate.} is calculated for each $\vec{k}$-points, all $\vec{j}_{\vec{k}}$ have to be summed with respect to the epitaxial components of the current-density. In other words, the Brillouin zone is projected in a 2D Brillouin Zone. For instance, Fig.\ref{fig_2D_FCC_brillouin_zone} represents the2D projection of the 3D Brillouin-zone of the FCC lattice along the (001), (110) and (111) directions and of 3D Brillouin-zone of the BCC lattice along the (001) direction. Once the current is projected, it can be compared to the accessible density of states (DOS) in the semiconductor. If a high current area matches accessible DOS, a current should cross the interface, assuming that the parallel component of the wave vector is conserved at the metal/semi-conductor interface.
  
  Those calculations have been done for several materials using a tight-binding code that I have written in Fortran 90, at present not included in the full non-equilibrium code BEEM v3.

  \subsection{2D projection of 3D Brillouin zones}
\begin{figure}[!hbtp]
      \centering
        \subbottom[\label{subfig_Au001}]{\tdplotsetmaincoords{75}{100}

\begin{tikzpicture}[tdplot_main_coords, scale=1.5]
  \draw[->] (-0.1,0,0) -- (1.5,0,0);
  \draw (1.5,0,0) node[right] {$k_x$};
  \draw [->] (0,-0.5,0) -- (0,1.5,0);
  \draw (0,1.5,0) node[above] {$k_y$};
  \draw [->] (0,0,-0.04) -- (0,0,1.5);
  \draw (0,0,1.5) node[above] {$k_z$};

  \coordinate (W1x) at (1,0.5,0);	
  \coordinate (W2x) at (1,-0.5,0);	
  \coordinate (W3x) at (1,0,0.5);	
  \coordinate (W4x) at (1,0,-0.5);	
  \coordinate (W-1x) at (-1,0.5,0);	
  \coordinate (W-2x) at (-1,-0.5,0);
  \coordinate (W-3x) at (-1,0,0.5);	
  \coordinate (W-4x) at (-1,0,-0.5);
  \coordinate (W1y) at (0.5,1,0);	
  \coordinate (W2y) at (-0.5,1,0);	
  \coordinate (W3y) at (0,1,0.5);	
  \coordinate (W4y) at (0,1,-0.5);	
  \coordinate (W-1y) at (0.5,-1,0);	
  \coordinate (W-2y) at (-0.5,-1,0);
  \coordinate (W-3y) at (0,-1,0.5);	
  \coordinate (W-4y) at (0,-1,-0.5);
  \coordinate (W1z) at (0.5,0,1);	
  \coordinate (W2z) at (-0.5,0,1);	
  \coordinate (W3z) at (0,0.5,1);	
  \coordinate (W4z) at (0,-0.5,1);	
  \coordinate (W-1z) at (0.5,0,-1);	
  \coordinate (W-2z) at (-0.5,0,-1);
  \coordinate (W-3z) at (0,0.5,-1);	
  \coordinate (W-4z) at (0,-0.5,-1);

  \draw[fill=gray!50,opacity=0.4] (W1x) -- (W4x) -- (W2x) -- (W3x) -- cycle; 	
  \draw[fill=gray,opacity=0.7] (W1y) -- (W3y) -- (W2y) -- (W4y) -- cycle;	 	
  \draw (W1y) -- (W3y) -- (W2y) -- (W4y) -- cycle;								

  \draw[fill=gray!50,opacity=0.4] (W1z) -- (W3z) -- (W2z) -- (W4z) -- cycle;	
  \draw (W1z) -- (W3z) -- (W2z) -- (W4z) -- cycle;								

  \draw[fill=gray!20,opacity=0.4] (W3x) -- (W2x) --(W-1y) -- (W-3y) --(W4z) -- (W1z) -- cycle ;	
  \draw (W3x) -- (W2x) --(W-1y) -- (W-3y) --(W4z) -- (W1z) -- cycle ;							
  \draw[fill=gray!50,opacity=0.4] (W3x) -- (W1z) --(W3z) -- (W3y) --(W1y) -- (W1x) -- cycle ;	
  \draw (W3x) -- (W1z) --(W3z) -- (W3y) --(W1y) -- (W1x) node[midway, sloped, above] {K} -- cycle ;								
  \draw[fill=gray!90,opacity=0.4] (W4x) -- (W1x) --(W1y) -- (W4y) --(W-3z) -- (W-1z) -- cycle ;	
  \draw (W4x) -- (W1x) --(W1y) -- (W4y) --(W-3z) -- (W-1z) -- cycle ;							
  \draw[fill=gray!20,opacity=0.4] (W4x) -- (W2x) --(W-1y) -- (W-4y) --(W-4z) -- (W-1z) -- cycle;
  \draw (W4x) -- (W2x) --(W-1y) -- (W-4y) --(W-4z) -- (W-1z) -- cycle;							

  \draw (0,0,0) node[above right] {$\Gamma$};
  \draw (0.5,0.5,0.5) node[sloped,rotate=10] {L};

  \draw (0,0,0) -- (0.5,0.5,0) -- (1,0,0) -- (0.5,-0.5,0) -- cycle; 							
  \fill[color=red!80,opacity=0.5] (0,0,1) -- (0.5,0.5,1) -- (1,0,1) -- (0.5,-0.5,1) -- cycle;	
  \fill[color=red!40,opacity=0.5] (0.5,0.5,0) -- (0.5,0.5,0.5) --(1,0,0.5) -- (1,0,0) -- cycle;	
  \fill[color=red,opacity=0.5] (0.5,0.5,0.5) -- (0.5,0.5,1) --(1,0,1) -- (1,0,0.5) -- cycle;	
  \fill[color=red!20,opacity=0.5] (0.5,-0.5,1) -- (0.5,-0.5,0) -- (1,0,0) -- (1,0,1) -- cycle;	
  \fill[color=red!50,opacity=0.5] (0.5,-0.5,1) -- (0.5,-0.5,0.5) -- (1,0,0.5) -- (1,0,1)  -- cycle;

  \draw (W1x) -- (W3x) -- (W2x) -- (W4x) -- cycle;

  \draw (1,0,0) node[above] {X};
  \draw (W3x) node[above left] {W};
  \draw (1.1,1,0.05) node[right] {X};

\draw[dashed] (-1,-1,-1)--(-1,-1,1)--(-1,1,1)--(-1,1,-1)--cycle; 
\draw[dashed] (1,-1,-1)--(1,-1,1)--(1,1,1)--(1,1,-1)--cycle; 
\draw[dashed] (1,-1,-1) -- (-1,-1,-1); 
\draw[dashed] (1,1,-1) -- (-1,1,-1) node[midway,below, sloped]{$4\pi/a$}; 
\draw[dashed] (1,1,1) -- (-1,1,1); 
\draw[dashed] (1,-1,1) -- (-1,-1,1); 

\draw[dotted] (0,0,1) -- (0,0,2.5);				
\draw[dotted] (0.5,0.5,0.5) -- (0.5,0.5,2.5);		
\draw[dotted] (1,0,0) -- (1,0,2.5);				
\draw[dotted] (0.5,-0.5,0.5) -- (0.5,-0.5,2.5);	
\draw (0,0,2.5) node[above ,sloped] {$\bar{\Gamma}$}-- (0.5,0.5,2.5)node[right ,sloped] {$\bar{X}$} -- (1,0,2.5)node[below ,sloped] {$\bar{M}$} -- (0.5,-0.5,2.5)node[left ,sloped] {$\bar{X}$} -- cycle; 

\end{tikzpicture}}
        \subbottom[\label{subfig_Au110}]{\tdplotsetmaincoords{75}{100}

\begin{tikzpicture}[tdplot_main_coords, scale=1.5]
  \draw[->] (-0.1,0,0) -- (1.5,0,0);
  \draw (1.5,0,0) node[right] {$k_x$};
  \draw [->] (0,-0.5,0) -- (0,1.5,0);
  \draw (0,1.5,0) node[above] {$k_y$};
  \draw [->] (0,0,-0.04) -- (0,0,1.5);
  \draw (0,0,1.5) node[above] {$k_z$};

  \coordinate (W1x) at (1,0.5,0);	
  \coordinate (W2x) at (1,-0.5,0);	
  \coordinate (W3x) at (1,0,0.5);	
  \coordinate (W4x) at (1,0,-0.5);	
  \coordinate (W-1x) at (-1,0.5,0);	
  \coordinate (W-2x) at (-1,-0.5,0);
  \coordinate (W-3x) at (-1,0,0.5);	
  \coordinate (W-4x) at (-1,0,-0.5);
  \coordinate (W1y) at (0.5,1,0);	
  \coordinate (W2y) at (-0.5,1,0);	
  \coordinate (W3y) at (0,1,0.5);	
  \coordinate (W4y) at (0,1,-0.5);	
  \coordinate (W-1y) at (0.5,-1,0);	
  \coordinate (W-2y) at (-0.5,-1,0);
  \coordinate (W-3y) at (0,-1,0.5);	
  \coordinate (W-4y) at (0,-1,-0.5);
  \coordinate (W1z) at (0.5,0,1);	
  \coordinate (W2z) at (-0.5,0,1);	
  \coordinate (W3z) at (0,0.5,1);	
  \coordinate (W4z) at (0,-0.5,1);	
  \coordinate (W-1z) at (0.5,0,-1);	
  \coordinate (W-2z) at (-0.5,0,-1);
  \coordinate (W-3z) at (0,0.5,-1);	
  \coordinate (W-4z) at (0,-0.5,-1);

  \draw[fill=gray!50,opacity=0.4] (W1x) -- (W4x) -- (W2x) -- (W3x) -- cycle; 	
  \draw[fill=gray,opacity=0.7] (W1y) -- (W3y) -- (W2y) -- (W4y) -- cycle;	 	
  \draw (W1y) -- (W3y) -- (W2y) -- (W4y) -- cycle;								

  \draw[fill=gray!50,opacity=0.4] (W1z) -- (W3z) -- (W2z) -- (W4z) -- cycle;	
  \draw (W1z) -- (W3z) -- (W2z) -- (W4z) -- cycle;								

  \draw[fill=gray!20,opacity=0.4] (W3x) -- (W2x) --(W-1y) -- (W-3y) --(W4z) -- (W1z) -- cycle ;	
  \draw (W3x) -- (W2x) --(W-1y) -- (W-3y) --(W4z) -- (W1z) -- cycle ;							
  \draw[fill=gray!50,opacity=0.4] (W3x) -- (W1z) --(W3z) -- (W3y) --(W1y) -- (W1x) -- cycle ;	
  \draw (W3x) -- (W1z) --(W3z) -- (W3y) --(W1y) -- (W1x) node[midway, sloped, above] {K} -- cycle ;								
  \draw[fill=gray!90,opacity=0.4] (W4x) -- (W1x) --(W1y) -- (W4y) --(W-3z) -- (W-1z) -- cycle ;	
  \draw (W4x) -- (W1x) --(W1y) -- (W4y) --(W-3z) -- (W-1z) -- cycle ;							
  \draw[fill=gray!20,opacity=0.4] (W4x) -- (W2x) --(W-1y) -- (W-4y) --(W-4z) -- (W-1z) -- cycle;
  \draw (W4x) -- (W2x) --(W-1y) -- (W-4y) --(W-4z) -- (W-1z) -- cycle;							

  \draw (0,0,0) node[above right] {$\Gamma$};
  \draw (0.5,0.5,0.5) node[sloped,rotate=10] {L};

  \draw (0,0,0) -- (1,1,0) -- (1.5,0.5,0) -- (0.5,-0.5,0) -- cycle; 									
  \fill[color=red!70,opacity=0.5] (0,0,1) -- (1,1,1) -- (1.5,0.5,1) -- (0.5,-0.5,1) -- cycle;			
  \fill[color=red!90,opacity=0.5] (1,1,0) -- (1,1,1) --(1.5,0.5,1) -- (1.5,0.5,0) -- cycle;				
  \fill[color=red!10,opacity=0.5] (0.5,-0.5,1) -- (0.5,-0.5,0) -- (1.5,0.5,0) -- (1.5,0.5,1) -- cycle;	
  \fill[color=red!40,opacity=0.5] (0.5,-0.5,1) -- (0.5,-0.5,0.5) -- (1,0,0.5) -- (1,0,0) -- (1.5,0.5,0) -- (1.5,0.5,1) -- cycle;

  \draw (W1x) -- (W3x) -- (W2x) -- (W4x) -- cycle;

  \draw (1,0,0) node[above] {X};
  \draw (W3x) node[above left] {W};
  \draw (1.1,1,0.05) node[right] {X};

\draw[dashed] (-1,-1,-1)--(-1,-1,1)--(-1,1,1)--(-1,1,-1)--cycle; 
\draw[dashed] (1,-1,-1)--(1,-1,1)--(1,1,1)--(1,1,-1)--cycle; 
\draw[dashed] (1,-1,-1) -- (-1,-1,-1); 
\draw[dashed] (1,1,-1) -- (-1,1,-1) node[midway,below, sloped]{$4\pi/a$}; 
\draw[dashed] (1,1,1) -- (-1,1,1); 
\draw[dashed] (1,-1,1) -- (-1,-1,1); 

\draw[dotted] (1,1,0) -- (3,3,0);			
\draw[dotted] (1.5,0.5,0) -- (3.5,2.5,0);	
\draw[dotted] (1,1,1) -- (3,3,1);			
\draw[dotted] (1.5,0.5,1) -- (3.5,2.5,1);	
\draw (3,3,0) node[below ,sloped] {$\bar{\Gamma}$}-- (3.5,2.5,0)node[below ,sloped] {$\bar{X}$} -- (3.5,2.5,1)node[above ,sloped] {$\bar{A}$} -- (3,3,1)node[above ,sloped] {$\bar{\Gamma}$} -- cycle; 

\draw[dotted] (0.5,0.5,0.5) -- (3,3,0.5) node[right,sloped] {$\bar{Y}$};
\draw[dotted] (1,0,0.5) -- (3.5,2.5,0.5) node[left,sloped] {$\bar{S}$};
\end{tikzpicture}}\\
        \subbottom[\label{subfig_Au111}]{\tdplotsetmaincoords{58}{110}

\begin{tikzpicture}[tdplot_main_coords, scale=1.5]
  \draw[->] (-0.1,0,0) -- (1.5,0,0);
  \draw (1.5,0,0) node[right] {$k_x$};
  \draw [->] (0,-0.5,0) -- (0,1.5,0);
  \draw (0,1.5,0) node[above] {$k_y$};
  \draw [->] (0,0,-0.04) -- (0,0,1.5);
  \draw (0,0,1.5) node[above] {$k_z$};

  \coordinate (W1x) at (1,0.5,0);	
  \coordinate (W2x) at (1,-0.5,0);	
  \coordinate (W3x) at (1,0,0.5);	
  \coordinate (W4x) at (1,0,-0.5);	
  \coordinate (W-1x) at (-1,0.5,0);	
  \coordinate (W-2x) at (-1,-0.5,0);
  \coordinate (W-3x) at (-1,0,0.5);	
  \coordinate (W-4x) at (-1,0,-0.5);
  \coordinate (W1y) at (0.5,1,0);	
  \coordinate (W2y) at (-0.5,1,0);	
  \coordinate (W3y) at (0,1,0.5);	
  \coordinate (W4y) at (0,1,-0.5);	
  \coordinate (W-1y) at (0.5,-1,0);	
  \coordinate (W-2y) at (-0.5,-1,0);
  \coordinate (W-3y) at (0,-1,0.5);	
  \coordinate (W-4y) at (0,-1,-0.5);
  \coordinate (W1z) at (0.5,0,1);	
  \coordinate (W2z) at (-0.5,0,1);	
  \coordinate (W3z) at (0,0.5,1);	
  \coordinate (W4z) at (0,-0.5,1);	
  \coordinate (W-1z) at (0.5,0,-1);	
  \coordinate (W-2z) at (-0.5,0,-1);
  \coordinate (W-3z) at (0,0.5,-1);	
  \coordinate (W-4z) at (0,-0.5,-1);

  \draw[fill=gray!50,opacity=0.4] (W1x) -- (W4x) -- (W2x) -- (W3x) -- cycle; 	
  \draw[fill=gray,opacity=0.7] (W1y) -- (W3y) -- (W2y) -- (W4y) -- cycle;	 	
  \draw (W1y) -- (W3y) -- (W2y) -- (W4y) -- cycle;								

  \draw[fill=gray!50,opacity=0.4] (W1z) -- (W3z) -- (W2z) -- (W4z) -- cycle;	
  \draw (W1z) -- (W3z) -- (W2z) -- (W4z) -- cycle;								

  \draw[fill=gray!20,opacity=0.4] (W3x) -- (W2x) --(W-1y) -- (W-3y) --(W4z) -- (W1z) -- cycle ;	
  \draw (W3x) -- (W2x) --(W-1y) -- (W-3y) --(W4z) -- (W1z) -- cycle ;							
  \draw[fill=gray!50,opacity=0.4] (W3x) -- (W1z) --(W3z) -- (W3y) --(W1y) -- (W1x) -- cycle ;	
  \draw (W3x) -- (W1z) --(W3z) -- (W3y) --(W1y) -- (W1x) node[midway, sloped, above] {K} -- cycle ;								
  \draw[fill=gray!90,opacity=0.4] (W4x) -- (W1x) --(W1y) -- (W4y) --(W-3z) -- (W-1z) -- cycle ;	
  \draw (W4x) -- (W1x) --(W1y) -- (W4y) --(W-3z) -- (W-1z) -- cycle ;							
  \draw[fill=gray!20,opacity=0.4] (W4x) -- (W2x) --(W-1y) -- (W-4y) --(W-4z) -- (W-1z) -- cycle;
  \draw (W4x) -- (W2x) --(W-1y) -- (W-4y) --(W-4z) -- (W-1z) -- cycle;							

  \draw (0,0,0) node[above right] {$\Gamma$};
  \draw (0.5,0.5,0.5) node[sloped,rotate=10] {L};


  \draw (W1x) -- (W3x) -- (W2x) -- (W4x) -- cycle;

  \draw (1,0,0) node[above] {X};
  \draw (W3x) node[above left] {W};
  \draw (1.1,1,0.05) node[right] {X};

\draw[dashed] (-1,-1,-1)--(-1,-1,1)--(-1,1,1)--(-1,1,-1)--cycle; 
\draw[dashed] (1,-1,-1)--(1,-1,1)--(1,1,1)--(1,1,-1)--cycle; 
\draw[dashed] (1,-1,-1) -- (-1,-1,-1); 
\draw[dashed] (1,1,-1) -- (-1,1,-1) node[midway,below, sloped]{$4\pi/a$}; 
\draw[dashed] (1,1,1) -- (-1,1,1); 
\draw[dashed] (1,-1,1) -- (-1,-1,1); 



\coordinate (Lbar1m) at (0.333,0.333,-0.667);
\coordinate (Ubar1m) at (0.83,0.083,-0.417);
\coordinate (Xbar1m) at (0.83,-0.17,0.17);
\coordinate (Ubar2m) at (0.83,-0.417,0.083);
\coordinate (Lbar2m) at (0.333,-0.667,0.333);

\coordinate (L1) at (0.5,0.5,-0.5);
\coordinate (U1) at (1,0.25,-0.25);
\coordinate (U2) at (1,-0.25,0.25);
\coordinate (L2) at (0.5,-0.5,0.5);

\coordinate (Lbar1) at (0.833,0.833,-0.167);
\coordinate (Ubar1) at (1.33,0.583,0.083);
\coordinate (Xbar1) at (1.33,0.33,0.33);
\coordinate (Ubar2) at (1.33,0.083,0.583);
\coordinate (Lbar2) at (0.833,-0.167,0.833);

\coordinate (Lbar1p) at (3.833,3.833,2.83);
\coordinate (Ubar1p) at (4.33,3.583,3.083);
\coordinate (Xbar1p) at (4.33,3.33,3.33);
\coordinate (Ubar2p) at (4.33,3.083,3.583);
\coordinate (Lbar2p) at (3.833,2.83,3.833);

\fill[color=red!35,opacity=0.5] (Lbar1m) -- (L1) -- (U1) -- (Ubar1m) -- cycle;
\fill[color=red!25,opacity=0.5] (Ubar1m) -- (Ubar2m) -- (U2) -- (U1) -- cycle;
\fill[color=red!10,opacity=0.5] (Lbar2m) -- (L2) -- (U2) -- (Ubar2m) -- cycle;

\fill[color=red!85,opacity=0.5] (Lbar1) -- (L1) -- (U1) -- (Ubar1) -- cycle;
\fill[color=red!70,opacity=0.5] (Ubar1) -- (Ubar2) -- (U2) -- (U1) -- cycle;
\fill[color=red!55,opacity=0.5] (Lbar2) -- (L2) -- (U2) -- (Ubar2) -- cycle;

\fill[color=red!95,opacity=0.5]  (0.5,0.5,0.5) -- (Lbar2) -- (Ubar2) -- (Xbar1) -- (Ubar1) -- (Lbar1) -- cycle ;

\draw[dotted,color=red] (0.5,0.5,0.5) -- (3.5,3.5,3.5)node[above] {$\bar{\Gamma}$};
\draw[dotted,color=red] (Lbar1) -- (Lbar1p) node[below right] {$\bar{M}_L$};
\draw[dotted,color=red] (Lbar2) -- (Lbar2p) node[above] {$\bar{M}_L$};
\draw[dotted,color=red] (Ubar2) -- (Ubar2p) ;
\draw[dotted,color=red] (Ubar1) -- (Ubar1p) node[below ] {$\bar{K}_U$};
\draw[dotted,color=red] (Xbar1) -- (Xbar1p) node[right] {$\bar{M}_X$};

\draw (3.5,3.5,3.5) -- (Lbar2p) -- (Ubar2p) -- (Xbar1p) -- (Ubar1p) -- (Lbar1p) -- cycle  ;
\end{tikzpicture}}
        \subbottom[\label{subfig_Fe001}]{\tdplotsetmaincoords{75}{100}

\begin{tikzpicture}[tdplot_main_coords, scale=1.5]
  \draw[->] (-0.1,0,0) -- (1.5,0,0);
  \draw (1.5,0,0) node[right] {$k_x$};
  \draw [->] (0,-0.5,0) -- (0,1.5,0);
  \draw (0,1.5,0) node[above] {$k_y$};
  \draw [->] (0,0,-0.04) -- (0,0,1.5);
  \draw (0,0,1.5) node[above] {$k_z$};

  \coordinate (Hz)  at (0,0,1);	
  \coordinate (H-z) at (0,0,-1);
  \coordinate (Hx)  at (1,0,0);	
  \coordinate (H-x) at (-1,0,0);	
  \coordinate (Hy)  at (0,1,0);
  \coordinate (H-y) at (0,-1,0);

  \coordinate (Pxyz)   at (0.5,0.5,0.5);	
  \coordinate (Px-yz)  at (0.5,-0.5,0.5);
  \coordinate (P-xyz)  at (-0.5,0.5,0.5);	
  \coordinate (P-x-yz) at (-0.5,-0.5,0.5);

  \coordinate (Pxy-z)   at (0.5,0.5,-0.5);	
  \coordinate (Px-y-z)  at (0.5,-0.5,-0.5);
  \coordinate (P-xy-z)  at (-0.5,0.5,-0.5);	
  \coordinate (P-x-y-z) at (-0.5,-0.5,-0.5);

  \draw (0,0,0) node[above left] {$\Gamma$};
  \draw (0,0.5,0.5) node[rotate=45] {N};

  \draw[fill=gray!40,opacity=0.4] (Hz)  -- (Pxyz) -- (Hx) -- (Px-yz) -- cycle; 	
  \draw (Hz)node[right] {H} -- (Pxyz)node[below] {P} -- (Hx) -- (Px-yz) -- cycle; 	

  \draw[fill=gray!90,opacity=0.4] (Hz) -- (Pxyz) -- (Hy) -- (P-xyz) -- cycle; 	
  \draw (Hz) -- (Pxyz) -- (Hy) -- (P-xyz) -- cycle; 	

  \draw[fill=gray!30,opacity=0.4] (Hz) -- (Px-yz) -- (H-y) -- (P-x-yz) -- cycle; 	
  \draw (Hz) -- (Px-yz) -- (H-y) -- (P-x-yz) -- cycle; 	


  \draw[fill=gray!70,opacity=0.4] (H-z) -- (Pxy-z) -- (Hx) -- (Px-y-z) -- cycle; 	
  \draw (H-z) -- (Pxy-z) -- (Hx) -- (Px-y-z) -- cycle; 	



  \draw[fill=gray!80,opacity=0.4] (Hx) -- (Pxyz) -- (Hy) -- (Pxy-z) -- cycle; 	
  \draw (Hx) -- (Pxyz) -- (Hy) -- (Pxy-z) -- cycle; 	

  \draw[fill=gray!40,opacity=0.4] (Hx) -- (Px-yz) -- (H-y) -- (Px-y-z) -- cycle; 	
  \draw (Hx) -- (Px-yz) -- (H-y) -- (Px-y-z) -- cycle; 	

\draw[dashed] (-1,-1,-1)--(-1,-1,1)--(-1,1,1)--(-1,1,-1)--cycle; 
\draw[dashed] (1,-1,-1)--(1,-1,1)--(1,1,1)--(1,1,-1)--cycle; 
\draw[dashed] (1,-1,-1) -- (-1,-1,-1); 
\draw[dashed] (1,1,-1) -- (-1,1,-1) node[midway,below, sloped]{$4\pi/a$}; 
\draw[dashed] (1,1,1) -- (-1,1,1); 
\draw[dashed] (1,-1,1) -- (-1,-1,1); 

  \fill[color=red!60,opacity=0.5] (0,0,1) -- (0.5,0,1) -- (0.5,0.5,1) -- (0,0.5,1) -- cycle; 		
  \fill[color=red!60,opacity=0.5] (0,0.5,0) -- (0.5,0.5,0) -- (0.5,0.5,0.5) -- (0,0.5,0.5) -- cycle; 	
  \fill[color=red!90,opacity=0.5] (0,0.5,0.5) -- (0.5,0.5,0.5) -- (0.5,0.5,1) -- (0,0.5,1) -- cycle; 	

  \fill[color=red!20,opacity=0.5] (0.5,0,0) -- (0.5,0.5,0) --(0.5,0.5,0.5) -- (0.5,0,0.5) -- cycle;		
  \fill[color=red!50,opacity=0.5] (0.5,0,0.5) -- (0.5,0.5,0.5) --(0.5,0.5,1) -- (0.5,0,1) -- cycle;		

\draw[dotted] (0,0,0) -- (0,0,2.5)node[above left]{$\bar{\Gamma}$};		 	 
\draw[dotted] (0.5,0,0.5) -- (0.5,0,2.5)node[below left]{$\bar{X}$};	 
\draw[dotted] (0.5,0.5,0.5) -- (0.5,0.5,2.5)node[below right]{$\bar{M}$};
\draw[dotted] (0,0.5,0.5) -- (0,0.5,2.5)node[above right]{$\bar{X}$};	 
\draw (0,0,2.5) -- (0.5,0,2.5) -- (0.5,0.5,2.5) -- (0,0.5,2.5) -- cycle ;

\end{tikzpicture}}
      \caption{2-dimensional reduced projections of the 3D Brillouin zone of the FCC and BCC lattices. The current is calculated for each k-points within the red parallelepipeds, then summed in the direction of epitaxy. \ref{subfig_Au001}, \ref{subfig_Au110} and \ref{subfig_Au111} are the projection of the reduced FCC Brillouin-zone respectively in the (001), (110) and (111) directions. \ref{subfig_Fe001} is the projection of the reduced BCC Brillouin-zone (one quarter of the full Brillouin zone) in the (001) direction.}
      \label{fig_2D_FCC_brillouin_zone}
    \end{figure}
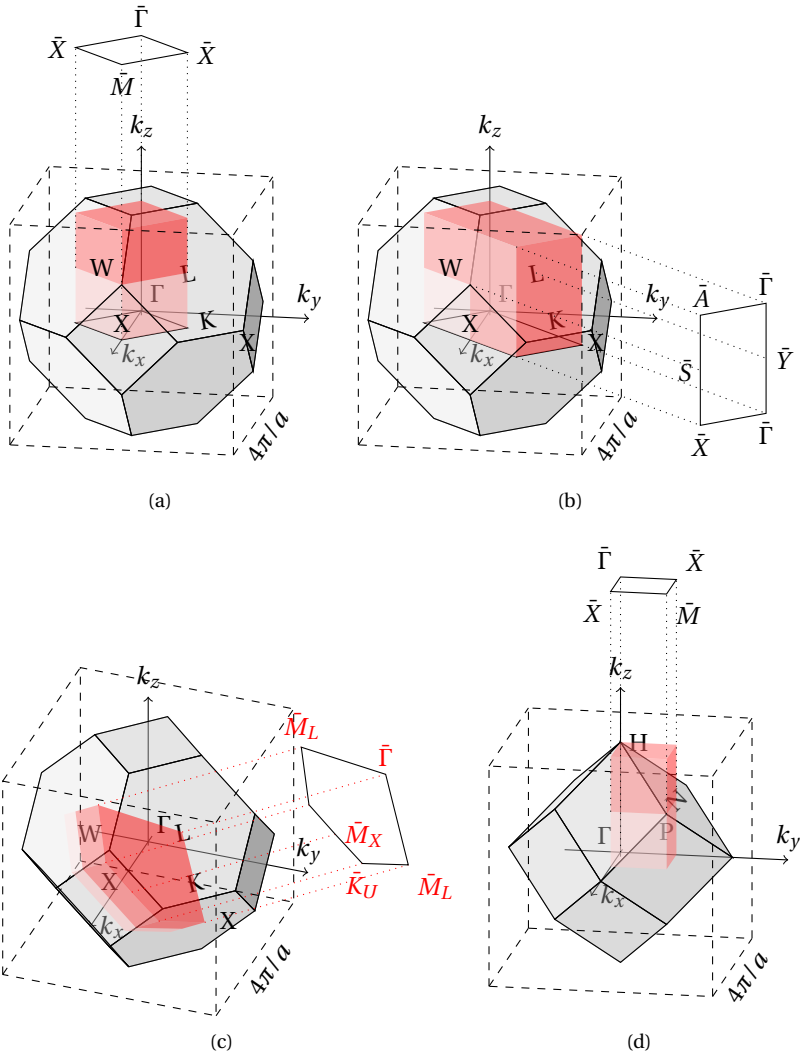

    Calculating the current is straightforward, the only cumbersome part is the summation of the current density vectors with respect to the epitaxial component. As stated above, figure~\ref{fig_2D_FCC_brillouin_zone} shows the reduced 2D Brillouin-zones for the three orientations (001), (110) and (111) of the FCC cell and for the (001) orientation of the BCC cell. The current is calculated for each $k$-points within the red polygons and then summed with respect to the epitaxial direction.
    
    In order to calculate the current we have to find the coordinates of the polygon's basis. This coordinates are given in this technical subsection, as well as the way to find them. For the following we denote the basis of the polygons with an underline, as $\underline{A}$ and the top of the polygons with an over-line, as $\overline{A}$.
    
    Consider the FCC cell, first. The basis of the polygon of Au(001) (Subfig.~\ref{subfig_Au001}) is $\underline{\Gamma}\underline{X}\underline{M}\underline{X}$ and its coordinates are    
    $$\frac{2\pi}{a}\left\{\left(0,0,0\right),\left(\frac{1}{2},\frac{1}{2},0\right),\left(1,1,0\right),\left(-\frac{1}{2},-\frac{1}{2},0\right)\right\}$$ 
    and the sum in the (001) direction runs up to $\overline{\Gamma XMX}$:
    $$\left\{\left(0,0,1\right),\left(\frac{1}{2},\frac{1}{2},1\right),\left(1,1,1\right),\left(-\frac{1}{2},-\frac{1}{2},1\right)\right\}$$
    Hence, the edges of the polygon go through 2 $L$ points and through $W$. 
      
    For Au(110) (subfigure~\ref{subfig_Au110}), the coordinates of the basis $\underline{\Gamma XA\Gamma}$ are
    $$\frac{2\pi}{a}\left\{\left(\frac{1}{2},-\frac{1}{2},\frac{1}{2}\right),\left(0,0,\frac{1}{2}\right),\left(0,0,0\right),\left(\frac{1}{2},-\frac{1}{2},0\right)\right\}$$ 
    and the sum in the (110) direction runs up to $\overline{\Gamma XA\Gamma}$ 
    $$\frac{2\pi}{a}\left\{\left(\frac{3}{2},\frac{1}{2},\frac{1}{2}\right) , \left(1,1,\frac{1}{2}\right) , \left(1,1,0\right) , \left(\frac{3}{2},\frac{1}{2},0\right)\right\}$$
    Hence, the edges of the polygon go through $X$, $W$, $L$ and $K$ points. Note that $L\in (\underline{Y}\overline{Y})$, $W\in (\underline{S}\overline{S})$, $K\in (\underline{\Gamma}\overline{\Gamma})$ and $X\in (\underline{X}\overline{X})$.
    
    The (111) direction (subfigure~\ref{subfig_Au111}) is a little trickier. It is a polygon whose basis is the third of an hexagon. The basis is $\underline{\Gamma M_1K_1K_2M_2}$, the edges go through $L_1,K_1,X,K_2,L_2$ and the top of the polygon lies on the hexagon face of the Brillouin zone $\overline{\Gamma M_1K_1K_2M_2}$. From this, we see that the reduced 2D-Brillouin-zone of the (111) direction is an hexagon with a 3-fold symmetry: starting from $\overline{M}_1$, a $2\pi/3$ rotation is required in order to find an equivalent $\overline{M}_2$ point.
    
    We start from the coordinates of the high symmetry points which are on the edges of the contour:
    \begin{align*}
     L_1(\frac{1}{2},\frac{1}{2},-\frac{1}{2}) 
     K_1(1,\frac{1}{4},0)
     K_2(1,0,\frac{1}{4})
     L_2(-\frac{1}{2},-\frac{1}{2},\frac{1}{2}) 
    \end{align*}
    and we define the vector $\vec{n}(1,1,1)$. In order to find the coordinates of the basis we need to find the intersection of the vector that goes through one of the above high symmetry points, with the basis plane. However, except for $\Gamma$ point, there is no high symmetry points on the edge of the basis. Then, it is easier to look for the intersection with the top plane of the polygon, $\overline{\Gamma MKKM}$. For instance, we want to find the intersection point $\overline{M}$ between $(\underline{M}_1L_1\overline{M}_1)$ and $(\overline{\Gamma}K'\overline{M})$. For that, we use the parametric equation:
    \begin{align}
     (\overline{M}_1)_{x} &= \overline{\Gamma}_{x} + t (\overrightarrow{\Gamma K})_{x} = L_{1,x} + t' (\vec{n})_{x}\\
     (\overline{M}_1)_{y} &= \overline{\Gamma}_{y} + t (\overrightarrow{\Gamma K})_{y} = L_{1,x} + t' (\vec{n})_{y}\\
     (\overline{M}_1)_{z} &= \overline{\Gamma}_{z} + t (\overrightarrow{\Gamma K})_{z} = L_{1,x} + t' (\vec{n})_{z}
    \end{align}
    with $\overline{\Gamma}=L=(1/2,1/2,1/2)$ and $L_1=(1/2,1/2,-1/2)$. Solving this system, we found $t=t'=1/3$ and so, $\overline{M}_1=(5/6,5/6,-1/6)$. Finding $\underline{M}$ is now easy, we only need to do a $(-1/2,-1/2,-1/2)$ translation to get $\underline{M}_1=(1/3,1/3,-2,3)$.
    Proceeding in the same way for the others point, we find:
    \begin{multline*}
      \underline{\Gamma M_1K_1K_2M_2}=\\ \frac{2\pi}{a}\left\{\left(0,0,0\right) , \left(\frac{1}{3},\frac{1}{3},-\frac{2}{3}\right) , \left(\frac{5}{6},\frac{1}{12},-\frac{5}{12}\right) , \left(\frac{5}{6},-\frac{1}{6},\frac{1}{6}\right), \left(\frac{5}{6},-\frac{5}{12},\frac{1}{12}\right) , \left(\frac{1}{3},-\frac{2}{3},\frac{1}{3}\right)\right\}     
    \end{multline*}

    The BCC (001) direction (Subfig.~\ref{subfig_Fe001}) is analogous to the FCC one. The coordinates of the basis $\underline{\Gamma X_1MX_2}$ are
    $$ \frac{2\pi}{a}\left\{\left(0,0,0\right),\left(0,\frac{1}{2},0\right),\left(\frac{1}{2},\frac{1}{2},0\right),\left(\frac{1}{2},0,0\right)\right\}$$
    and the sums runs up to 
    $$ \frac{2\pi}{a}\left\{\left(0,0,1\right),\left(0,\frac{1}{2},1\right),\left(\frac{1}{2},\frac{1}{2},1\right),\left(\frac{1}{2},0,1\right)\right\}$$
    We have limited ourself to the Fe(001) direction because in the next section we present some results only for Fe(001)/GaAs(001). Whereas for gold, we have studied Au(001)/Fe(001) and Au(111)/Si. The (110) direction has been presented above for completeness and because Au(110)/GaAs(001) has been experimentally studied  at IPR.

  \subsection{Gold: Au(001) and Au(111)}\label{sec_gold_eq}
\begin{figure}[!bt]
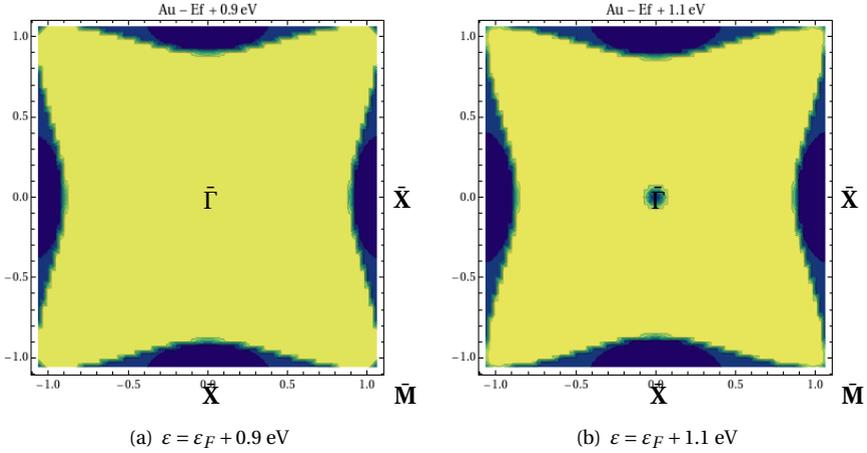

        \centering
          \subbottom[$\varepsilon=\varepsilon_F+0.9$ eV\label{subfig_09}]{\input{results_and_discussion/pictures/Au_map09.tex}}
          \subbottom[$\varepsilon=\varepsilon_F+1.1$ eV\label{subfig_11}]{\input{results_and_discussion/pictures/Au_map11.tex}}          \caption{{Current-density vector projected on the 2D Brillouin zone at \subcaptionref{subfig_09} Fermi level +0.9 eV and \subcaptionref{subfig_11} +1.1 eV for Au(001). The distribution is quite isotropic but above 1.1eV, a low-current zone appears at $\bar{\Gamma}$. If gold is grow on a material which has available DOS only around $\bar{\Gamma}$, then, the BEEM current should decrease above $\varepsilon_F+1.1$eV. }}
        \label{fig_Au_001_map}
      \end{figure}

      Figure~\ref{fig_Au_001_map} shows the mapping of the current for Au(001), at different energies:
      it is quite isotropic with a neck that appears at $\bar{\Gamma}$ for $\varepsilon=\varepsilon_F+1.1$ eV.  
      
      These results suggest that if we had another material, on which gold can grow epitaxially along the (001) direction, for which there are available states only around $k_{\sslash}=0$ ($\bar{\Gamma}$ point), then the BEEM current should decrease above 1.1 eV, due to the absence of propagation of electrons at this energy and for $\kpara=0$. This theoretical suggestion has been fully implemented in the Fe/Au/Fe spin-valve described in section \ref{sec_spinvalve_results1}

      Of course, we should now find a way to counter-check the validity of the equilibrium approach described by Eq.~\eqref{eq_sc_current}. On way to do it, is to look at the angular distribution of the current intensity, in order to compare our simple model with the results of reference \cite{Garcia-Vidal-PhysRevLett.76.807}, based on the non-equilibrium Keldysh Green-functions. 
      
\begin{figure}[!hbt]
        \begin{center}
          \subbottom[\label{subfig_isocurves}]{\includegraphics[scale=0.65]{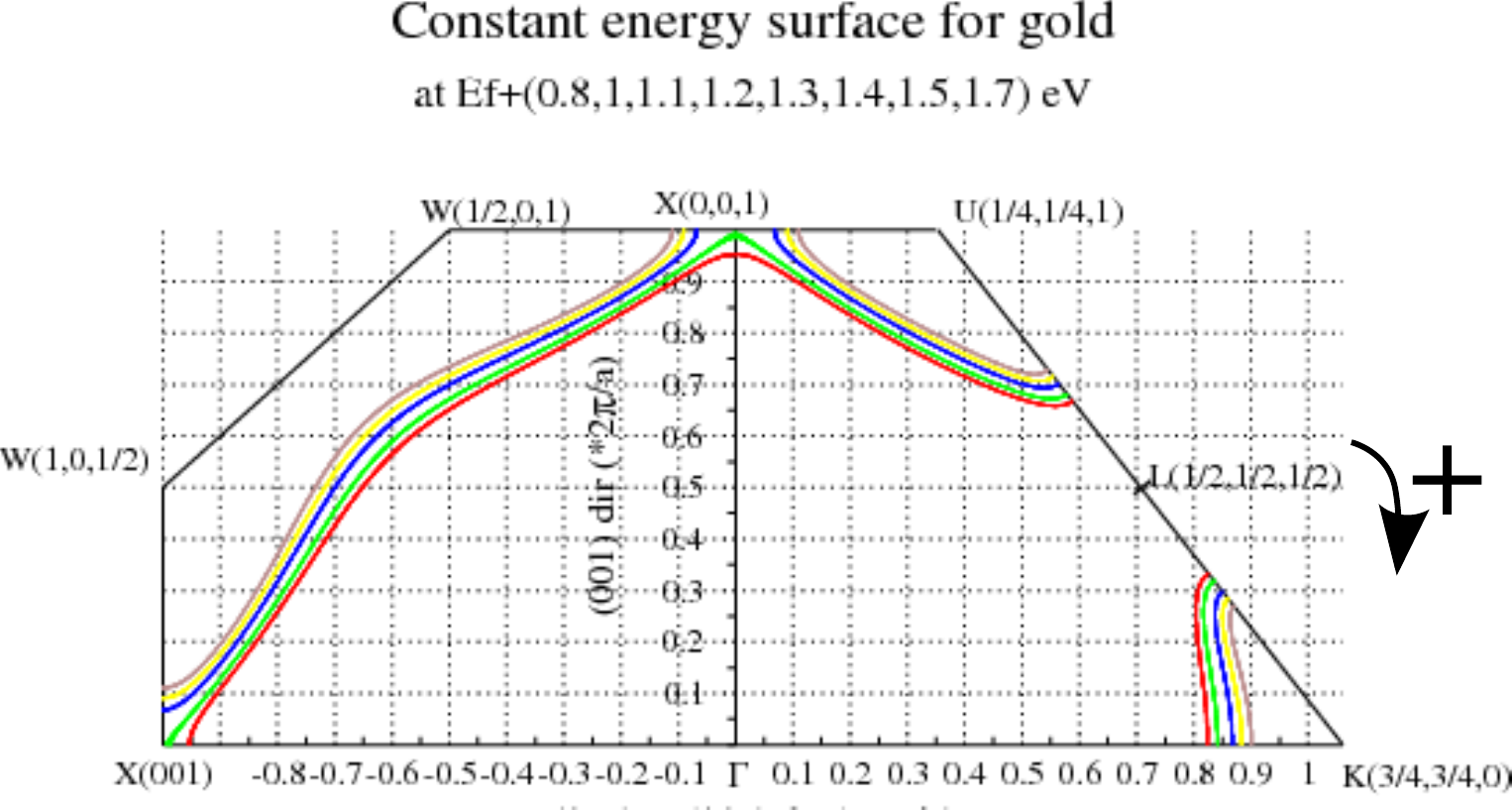}}\\
          \subbottom[\label{subfig_gradient}]{\includegraphics[width=0.8\linewidth]{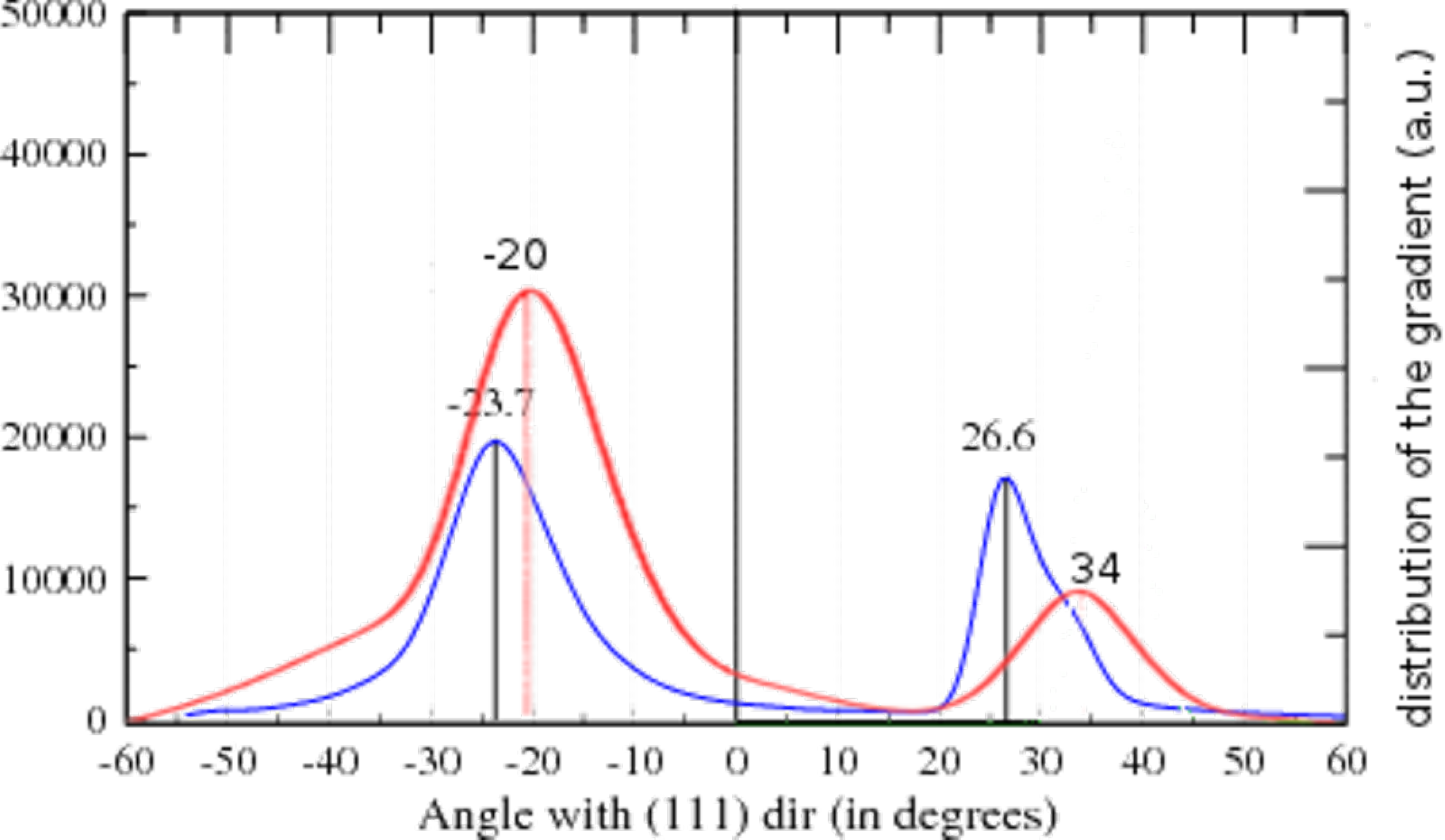}}\\
          \caption{\subcaptionref{subfig_isocurves} Iso-energetic curves and \subcaptionref{subfig_gradient} (blue line) angular distribution of the gradient for $\varepsilon=1.3$ eV with respect to (111) direction ($\Gamma$-L, the right part of \subcaptionref{subfig_isocurves}) of the FCC Brillouin zone af gold (cf. Fig.~\ref{fig_2D_FCC_brillouin_zone}). The more the iso-energetic bands are flat, the higher is the current. The red curve, extracted from  the non-equilibrium calculation of Ref.~\cite{Garcia-Vidal-PhysRevLett.76.807}, is in quite good agreement with our equilibrium results: in both cases the current peaks lie at similar angles with respect to the $\Gamma$ point in the planes shown in \subcaptionref{subfig_isocurves}. Experimentally, the difference of 7.4\degree\ cannot be be seen due to roughness at interfaces and non zero temperature.}
        \label{angular_distribution}
        \end{center}
      \end{figure}

      Figure~\ref{angular_distribution} represents constant energy curves from 0.8, to 1.7 eV above the Fermi level (Subfig.~\ref{subfig_isocurves}) in the $\Gamma KLUXWWX\Gamma$ plane of the FCC Brillouin zone (Subfig.~\ref{fig_2D_FCC_brillouin_zone}), and the angular distribution of the gradient of the $\varepsilon_F+1.3$ eV curve, with respect to the (111) direction (Subfig.~\ref{subfig_gradient}). As the current is proportional to the gradient, the peaks in the gradient distribution correspond to high current direction. That means that most of the electrons propagate at -24\degree\ and 27\degree\ with respect to the (111) direction. This behavior is qualitatively similar, within the experimental sensitivity, to what has been obtained by K. Reuter \emph{et al.} in Ref.~\cite{Reuter-PhysRevB.58.14036} (this is represented by the red curve in \ref{subfig_gradient}).
      
      This result is a strong indication that equilibrium calculations like those detailed here, can find their place in several realistic descriptions of BEEM currents.      


  \subsection{Fe(001)/GaAs(001)}

      The calculation procedure is the same as seen above for gold, but this time the accessible DOS in GaAs is superposed to the 2D Brillouin-zone current distribution, both for spin up and spin down electrons. Conservation of $k_\sslash$ implies that a current crosses the interface only if, at a given energy, k-states impinging from the metal have a corresponding empty DOS with the same $k_\sslash$. From these premises we can expect a modulation of spectral weight between parallel and anti-parallel states of the spin valve throughout the Brillouin zone, due to the spin-up/spin-down asymmetries in the band structure.
      
\begin{figure}[!hbtp]
      \begin{center}
      \subbottom[\label{subfig_Fe_down_map}$\varepsilon_\downarrow=\varepsilon_F+1.1$ Ev]{\input{results_and_discussion/pictures/Fe_down_map.tex}}
      \subbottom[\label{subfig_Fe_up_map}$\varepsilon_\uparrow=\varepsilon_F+1.1$ Ev]{\input{results_and_discussion/pictures/Fe_up_map.tex}}\\
      \subbottom[\label{fig_GaAs_bs}]{\includegraphics[width=0.6\linewidth]{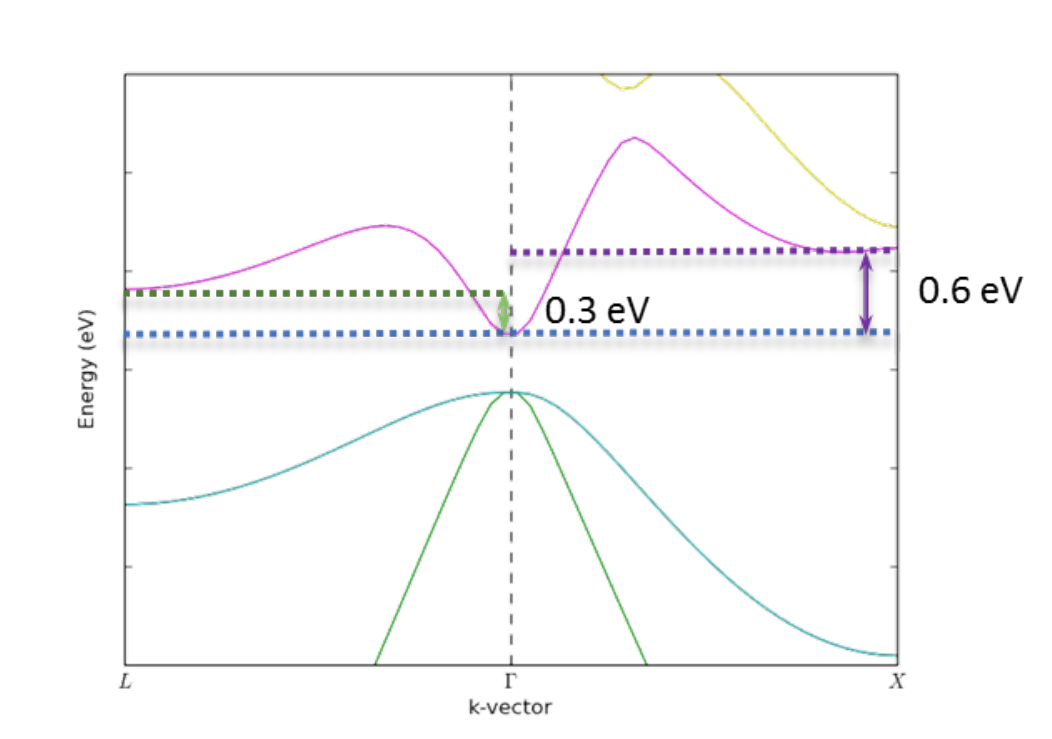}}
      \caption{The current-density vector is projected on the Fe(001) 2D Brillouin-zone for \subcaptionref{subfig_Fe_down_map} spin-down and \subcaptionref{subfig_Fe_up_map} spin-up electrons ($\varepsilon=E_F+1.1$eV). Yellow color is high current area and blue color is low color area. The available density of states in 2D Brillouin zone for GaAs(001) is represented by red disks. They correspond to the three valleys that are accessible at $\Gamma$ ($\phi_{\Gamma}=\varepsilon_F+0.75$ eV), $L$ ($\phi_{L}=\phi_{\Gamma}+0.33$ eV) and $X$ ($\phi_{X}=\phi_{\Gamma}+0.48$ eV) points in \subcaptionref{fig_GaAs_bs} the band structure of GaAs (obtained through DFT/LDA calculations using ABINIT~\cite{abinit} code). For the spin-down electrons, no electron can be injected in the L valley of GaAs (that projects in 2D to the point $\bar{X}_{GaAs}$), unlike the spin-up electrons. }
      \label{Fe_sur_GaAs}
      \end{center}
      \end{figure}

      This feature is shown in figure \ref{Fe_sur_GaAs}: the current distribution of the spin-up electrons is quite homogeneous whereas the the current distribution of spin-down electrons shows strong relative variations. Suppose that we can control the available density of states within the semiconductor around $\kpara=0$: by increasing the surface of the available DOS, we expect to increase the BEEM current. For spin-up electrons, as the current distribution is homogeneous, the BEEM current should increase linearly with the surface. On the contrary, for spin down electrons, this variation depends on the current distribution, not only the size of the available DOS area. However,  we a priori expect that the BEEM current is governed by the majority spin (i.e. spin-up electrons), whose mean free-path is higher than for the minority spin (i.e. spin-down electrons). 
      In other words, increasing the area of the available density of states should lead to a almost linear variation of the BEEM current with respect to the surface, the small non-linear part being caused by minority spin.

      For instance, we have projected the available density of states in GaAs on these current distribution: spin down-electrons can only propagate to the $\bar{\Gamma}$ point, but spin-up electrons can also cross the interface through the $\bar{X}$ point. 
      
      This is not the case for spin-up electrons (Fig.~\ref{subfig_Fe_up_map}), where a density of states is available at $\bar{X}_{GaAs}$.
      Moreover, for spin-up electrons, the current projected onto the L valley ($\bar{X}$ of GaAs) has more or less the same intensity as at $\bar{\Gamma}$ point.

  \subsection{Towards spintronics: Fe/Au/Fe/GaAs, the equilibrium approach}\label{sec_spinvalve_results1}
    In section \ref{sec_towards_spintronics} we have introduced the 
    $$\mbox{Fe(001)[100]/Au(001)[110]/Fe(001)[100]/GaAs(001)}$$ 
    spin-valve that is studied at IPR, as a Giant Magneto-Resistance (GMR) device (Fig. \ref{fig_GMR}):
    \begin{itemize}
      \item For antiparallel alignment of the ferromagnetic electrodes, the BEEM current $I_B$ is low.
      \item For parallel alignment of the ferromagnetic electrodes, the BEEM current $I_B$ is high. (up to 500\% of $I_B$ in anti-parallel configuration)
    \end{itemize}
    In this section, we shall see how the available density of states of the different materials can filter the propagation of electrons and how we can increase or decrease the BEEM current by changing the semiconductor (Subsec.~\ref{sec_bs_filtering}). Then we show how the BEEM current can vary due to wave-function filtering (Subsec.~\ref{sec_wf_sym_filtering}).

    \subsubsection{Band structure ($\kpara$) filtering}\label{sec_bs_filtering}  
      We have seen above that a polarized layer of iron leads to the polarization of the current. What happens now, if a gold layer is stacked between two ferromagnetic electrodes ? 
      From section \ref{sec_gold_eq}, we can see that the ``neck'' that appears at $\bar{\Gamma}$ at $\varepsilon=1.1$ eV in the band structure (Figs.~\ref{fig_Au_001_map} and \ref{subfig_isocurves}) acts as a filter: for $\kpara=0$, above 1 eV, no electron can enter the gold slab and hence the BEEM current should be zero.

      To check the relevance of this gap opening in gold, we compare its size with the valley opening in GaAs since, also for small values of $k_\sslash$ around 0, electrons are injected in $\Gamma$ valley of GaAs. 
      
      The valley opening could be calculated using a free-electron model, as we deal with small energy variations, and the effective mass of the semiconductor $E(\kpara)=\hbar^2\kpara^2 / 2m^*$. For GaAs, the effective mass of the $\Gamma$ valley is $m^*_{\Gamma}=0.067 m_0$ \cite{mass_reduite-vurgaftman}. For Al$_{0.4}$Ga$_{0.6}$As, the effective masses of the $\Gamma$ and $X$ valleys are respectively $m^*_{\Gamma}=0.086$ $ m_0$ and $m^*_{\Gamma}=0.226 m_0$ \cite{mass_reduite-vurgaftman}.
      
\begin{figure}[!hbtp]
      \begin{center}
      \subbottom[\label{subfig_XG_valleys}]{\includegraphics[width=0.38\linewidth]{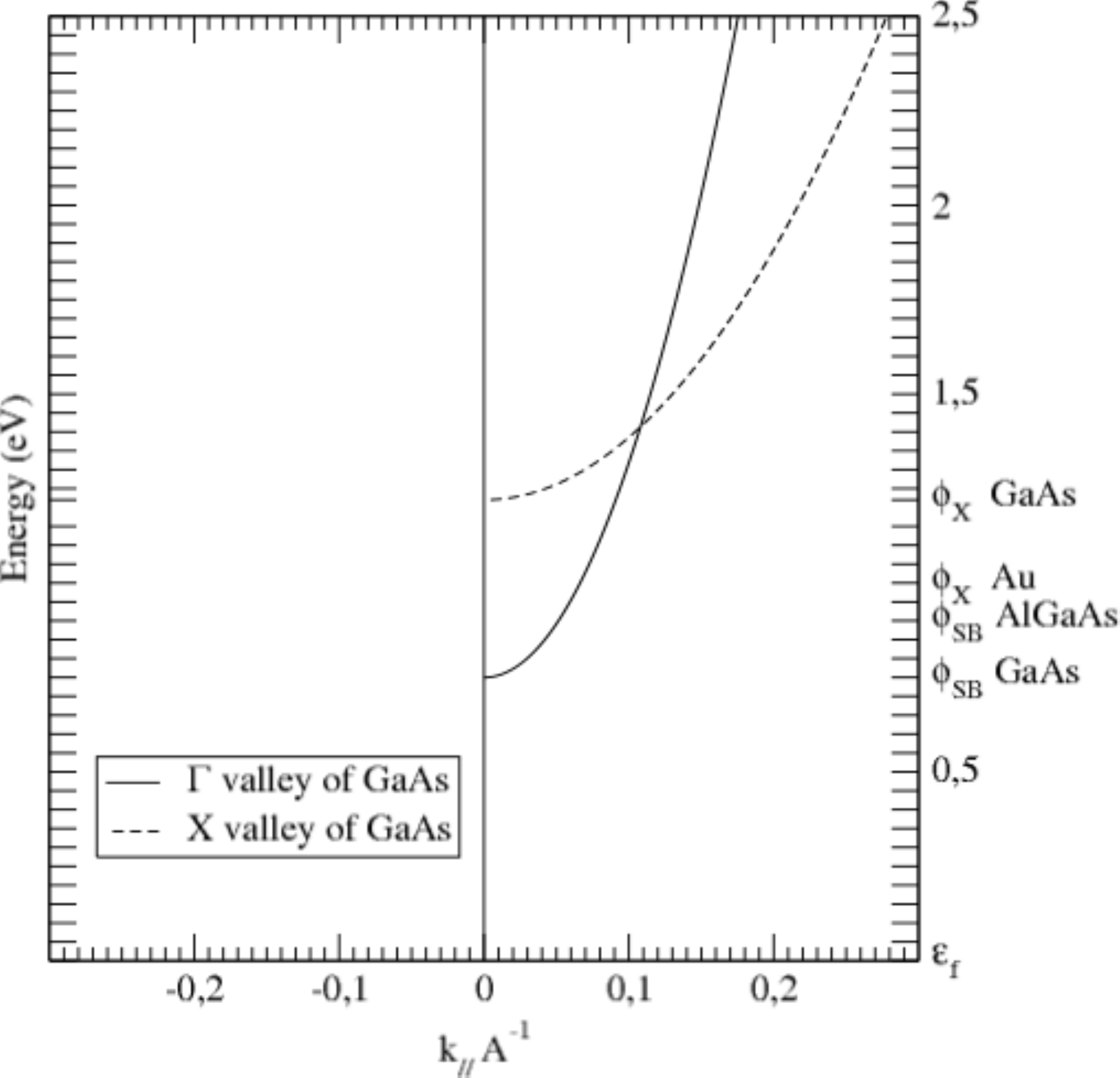}}
      \subbottom[\label{subfig_gold_neck}]{\includegraphics[width=0.38\linewidth]{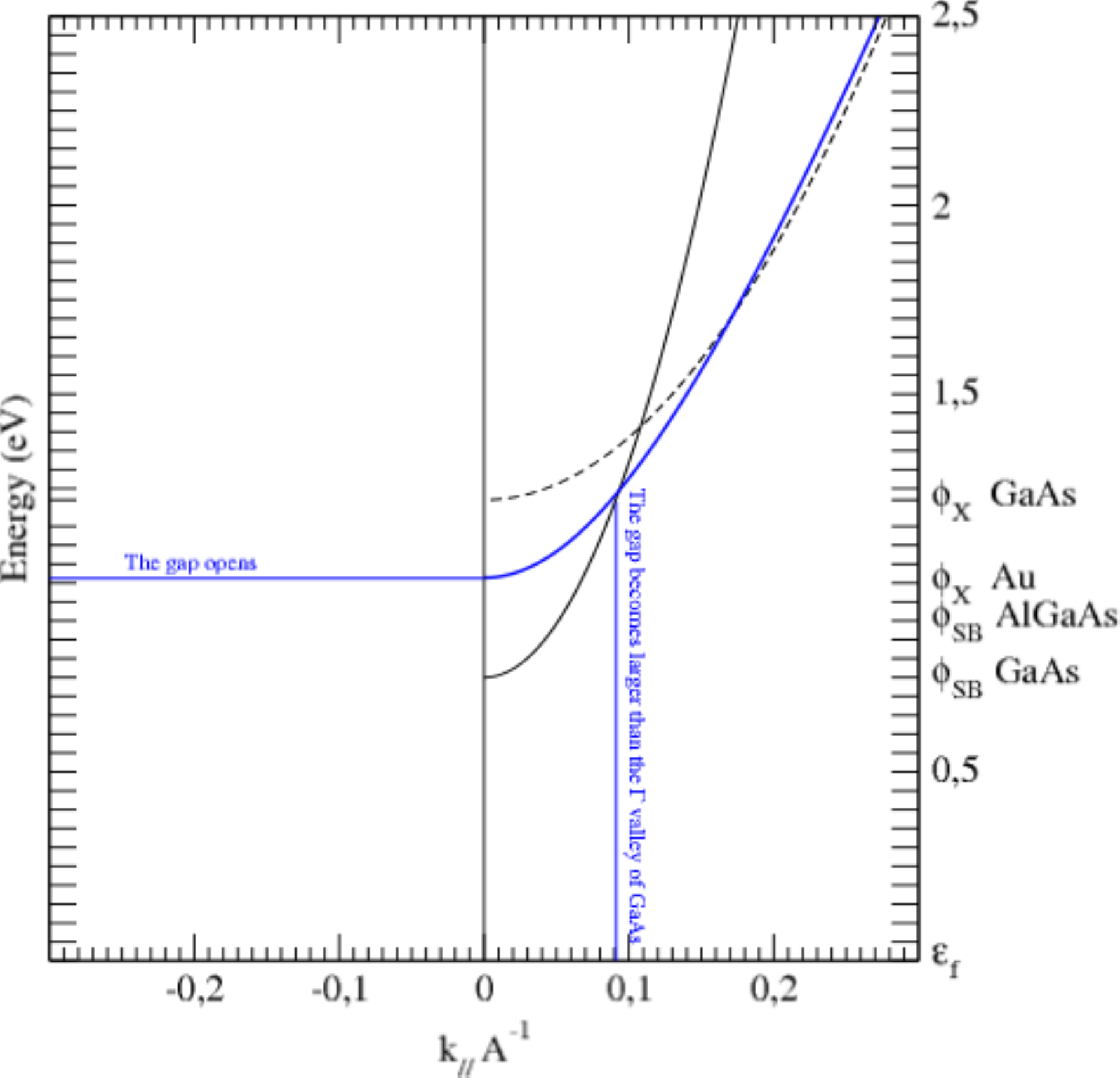}}
      \subbottom[\label{subfig_AlGaAs_valleys}]{\includegraphics[width=0.5\linewidth]{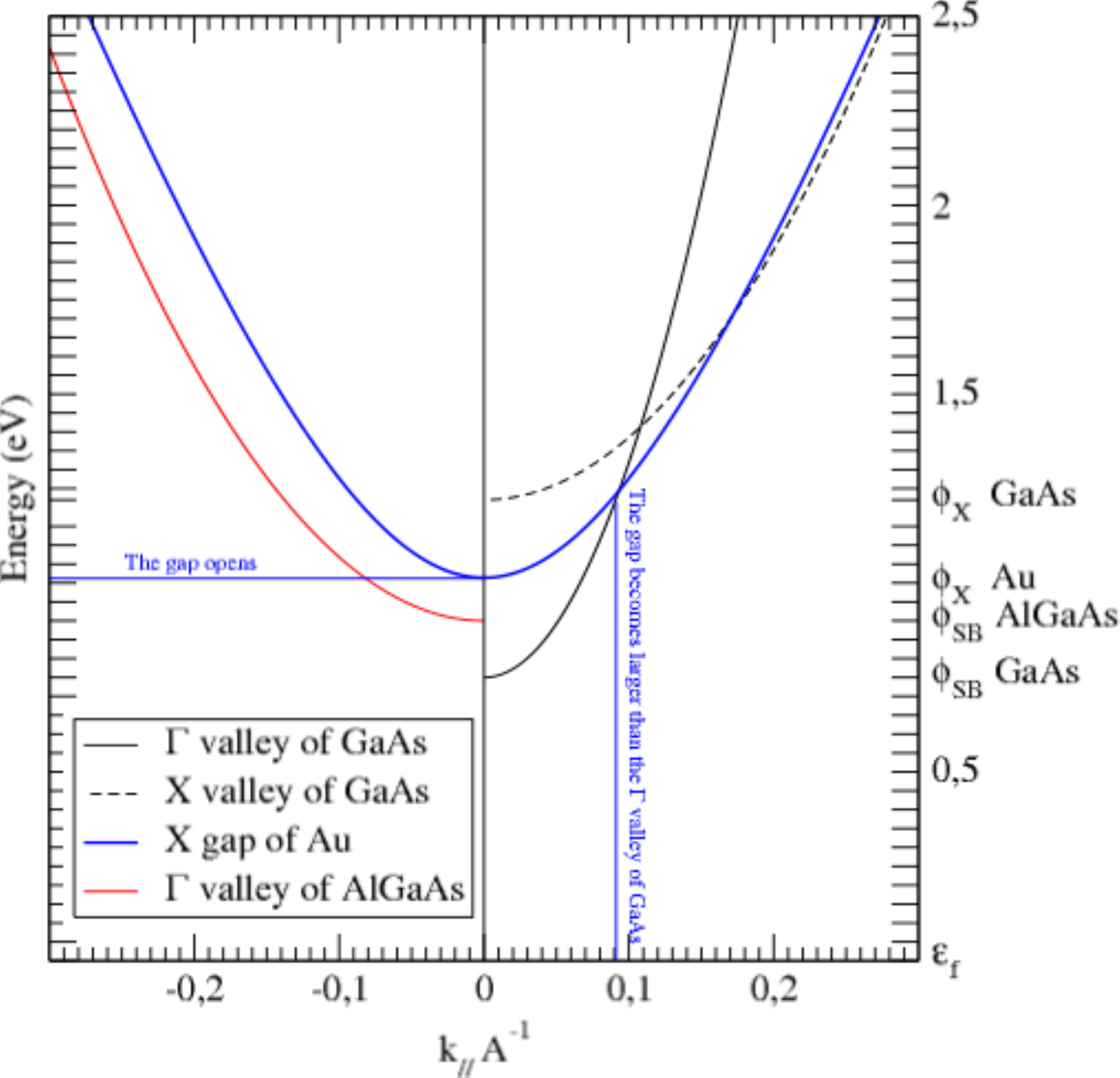}}
      \caption{{The neck of gold is compared to $\Gamma$ and $X$ valleys opening of the conduction band of GaAs and Al$_{0.4}$Ga$_{0.6}$As. Bands are modeled with free-electron bands and the effective mass as we are looking at small $\kpara$ variations.
      \subcaptionref{subfig_XG_valleys} $\Gamma$ and $X$ of GaAs valleys are available for $\kpara=0$.
      \subcaptionref{subfig_gold_neck} The gold neck, in blue, is superposed with the GaAs bands: above the blue curve, no electrons can propagate due to the gap in the Au band structure near $\kpara=0$. Below 1.1 eV, the neck is narrower than the valley opening of $\Gamma$ point: electrons can enter the semiconductor. Between 1.1 and 1.7 eV, the neck is wider then the valley opening: the propagation is forbidden for this wave-vector. Above 1.7 eV, the size of the neck follows the size of the $X$-valley opening.
      \subcaptionref{subfig_AlGaAs_valleys} The $X$ valley of Al$_{0.4}$Ga$_{0.6}$As is represented in red (the minimum of the conduction band at $X$ point is a the same energy than the minimum of the conduction band at $\Gamma$ point). The gap in gold is always narrower than the $X$-valley opening of Al$_{0.4}$Ga$_{0.6}$As that is already available 0.2eV above $\phi_{SB}$: electrons can enter the semiconductor.}}
      \label{superposition}
      \end{center}
      \end{figure}

      Figure \ref{superposition} shows that above 1.1 eV the gap is larger than the opening of $\Gamma$ valley of GaAs (in the free electron approximation around minimum). At higher energies, injection in X valley of GaAs becomes possible, but the size of the gap still matches the opening of the X valley. So for Fe/Au/Fe/GaAs, spin polarized transport might be possible at $\bar{\Gamma}$ point, after switching-off the external magnetic field.

      If we replace now GaAs by Al$_{0.4}$Ga$_{0.6}$As, we see that the size of X valley opening for Al$_{0.4}$Ga$_{0.6}$As is always larger than the gold neck. in other words, The BEEM current should be higher with  Al$_{0.4}$Ga$_{0.6}$As than with GaAs. However, as we shall see in the next subsection, symmetry filtering has also to be taken into account.

    \subsubsection{Wave-function symmetry filtering}\label{sec_wf_sym_filtering}
      Differently of gold characterized by only one band just above the Fermi level, iron has several bands few eV above Fermi level. Because of that, besides $k_\sslash$ conservation, another selection rule at the interface comes from the point-symmetry character of each band. In fact, as we shall detail below, iron bands are characterized by several symmetries ($\Delta_1$,$\Delta_2$,$\Delta_{2'}$ and $\Delta_5$), whereas the gold band around Fermi energy is characterized by just $\Delta_1$ symmetry. 
      For the notation, we remind that $\Delta$ label the $\Gamma X$ direction, whereas indexes 1, 2, $2'$ and 5 refer to the group representations: for example, $\Delta_1$ is totally invariant under all symmetry operations. 
      This is actually the same $\Delta_1$ symmetry that characterizes the conduction band of GaAs, and for this reason the point-symmetry rule was in that case (Au/GaAs) automatically satisfied. 
      This is not the case, however, for iron.
      This rule is a consequence of the fact that if the Hamiltonian describing the metal slab and the semiconductor, as a whole, has a point symmetry, a wave function of the whole system, belonging to a given representation of symmetry group, cannot change representation in passing from the metal slab to the semiconductor slab.

      In order to see which band of Fe(001) is allowed to couple to the $\Delta_1$ conduction band of GaAs(001) in the case of Fe(001)/GaAs(001), we have to find the compatibility between the $C_{2v}$ symmetry group of GaAs(001) and the $C_{4v}$ symmetry group of Fe(001). From the character tables represented in tables \ref{character_table_GaAs} and \ref{character_table_Fe}, it appears that $\Delta_1$ and $\Delta_{2'}$ representations of iron are the only two representations characterized by the same behavior as $\Delta_1$ states of GaAs(001) with respect to the symmetry operations (E, C$_2$, $\sigma_v^x$ and $\sigma_v^y$) common to the intersection of the $C_{4v}$ group of iron and the the $C_{2v}$ symmetry group of GaAs(001). The intersection is clearly the lower-order group, i.e. $C_{2v}$. For this reason, electron transmission from $\Delta_1$ and $\Delta_{2'}$ iron states towards $\Delta_1$ GaAs states is symmetry-allowed. For the same reason, $\Delta_2$ and $\Delta_5$ states of iron are orthogonal to $\Delta_1$ states of GaAs. In other words, only Bloch electrons with s, p$_z$,d$_{3z^2-r^2}$ and d$_{xy}$ orbital character can be transmitted.

\begin{figure}[!btp]
      \centering
      \includegraphics[width=0.8\linewidth]{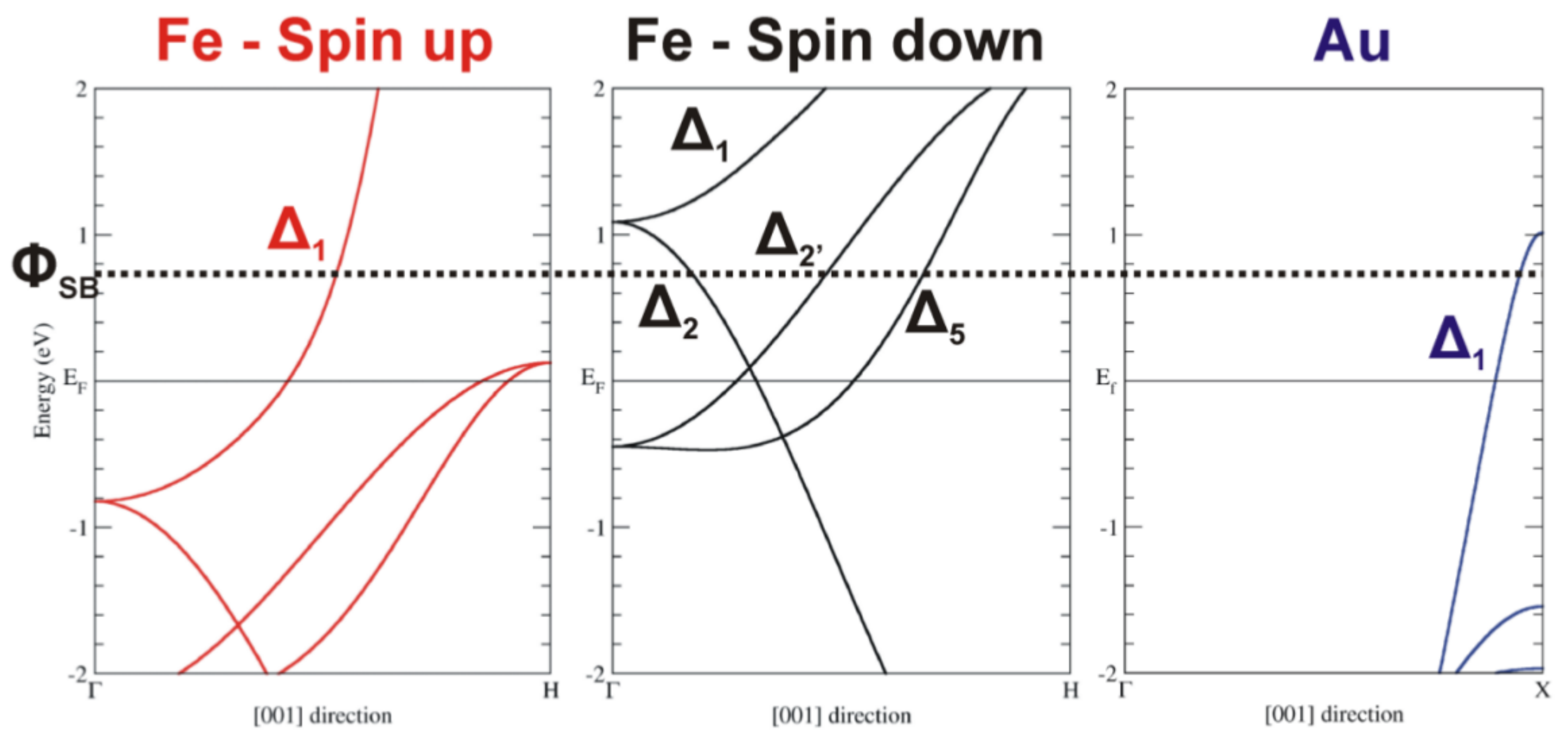}
      \caption{Band structures of Iron and Gold. Only $\Delta_1$-electrons can cross interfaces. Between Schottky barrier and 1.1 eV, only spin-up electrons are injected. Above 1.1 eV, a neck appears in gold band-structure. This gap prevents injection in the spin-valve Fe/Au/Fe/GaAs.}
      \label{fig_Fe_Au_BS}
      \end{figure}

     In the case of the full spin-valve structure Fe/Au/Fe/GaAs, the point-symmetry filtering works already at the level of the first Fe(001)/Au(001) interface. By reminding that gold is rotated by 45\degree\ in the $xy$-plane (see Fig.~\ref{fig_exp_gmr}) $\Delta_2$ and $\Delta_{2'}$ symmetries refer to different orbitals for iron and gold (cf. Table~\ref{character_table}).
     In principle, all irreducible representations should be preserved in passing from Fe(001) to Au(001). However, in the energy range of interest, only the $\Delta_1$ irreducible representation is available for gold, as shown in Fig.~\ref{fig_Fe_Au_BS}. Therefore, only $\Delta_1$ electrons of iron can be transmitted.

      \newcolumntype{g}{>{\columncolor{blue!10}}c} 
      \begin{table}[!hbt]
	\centering
	\subfloat[\label{character_table_GaAs}Character table of $C_{2v}$ group (GaAs(001))]{{\begin{tabular}{|l|cccc|c|}
	      \hline
	      $C_{2v}$             & E & C$_2$ & $\sigma_v^{xy}$ & $\sigma_v^{-xy}$ & orbitals \\ \hline
      \rowcolor{blue!10} $\Delta_1$ (A$_1$)   & \cercle{1} &  \cercle{1}    &  \cercle{1}           &  \cercle{1}           & s,p$_z$,d$_{z^2}$,d$_{xy}$ \\ 
			$\Delta_2$ (A$_2$)   & 1 &  1    & -1           & -1           & d$_{x^2-y^2}$\\ 
			$\Delta_{2'}$ (B$_1$)& 1 & -1    &  1           & -1           & d$_{xz}$, p$_x$ \\ 
			$\Delta_{5} $ (B$_2$)& 1 & -1    & -1           &  1           & d$_{yz}$, p$_y$ \\ \hline
	\end{tabular}}}
	\qquad
	\subfloat[\label{character_table_Fe}Character table of $C_{4v}$ group (Fe(001))]{{
	\begin{tabular}{|l|g|cc|g|cc|g|g|c|}
	      \hline
	      $C_{4v}$             & E        & C$_{4v}^+$&C$_{4v}^-$&C$_2$     &$\sigma_v^x$&$\sigma_v^y$& $\sigma_v^{xy}$& $\sigma_v^{-xy}$& orbitals          \\ \hline
	      $\Delta_1$ (A$_1$)   &\cercle{1}&  1        &  1       &\cercle{1}&  1            &  1             &\cercle{1}   &\cercle{1}   & s,p$_z$,d$_{z^2}$ \\ 
	      $\Delta_1'$ (A$_2$)  & 1        &  1        &  1       &  1       & -1            & -1             & -1          & -1          & R$_z$                                  \\
	      $\Delta_2$ (B$_1$)   &{1}& -1        & -1       &{1}&  1            &  1             & -1          & -1          & d$_{x^2-y^2}$          \\ 
	      $\Delta_{2'}$ (B$_2$)&\cercle{1}& -1        & -1       &\cercle{1}& -1            & -1             &\cercle{1}   &\cercle{1}   &d$_{xy}$           \\ 
	      $\Delta_{5}$ (E)     & 2        &  0        &  0       & -2       &  0            &  0             &  0          &  0          &d$_{xz}$,d$_{yz}$, p$_x$, p$_y$ \\ \hline
	\end{tabular}}}
	\qquad
	\subfloat[\label{character_table_Au}Character table of $C_{4v}$ group (Au(001))]{{
	\begin{tabular}{|l|g|cc|g|g|g|cc|c|}
	      \hline
	      $C_{4v}$             & E        & C$_{4v}^+$&C$_{4v}^-$&C$_2$     &$\sigma_v^{xy}$&$\sigma_v^{-xy}$& $\sigma_v^{x}$& $\sigma_v^{y}$& orbitals          \\ \hline
	      $\Delta_1$ (A$_1$)   &{1}&  1        &  1       &{1}& {1}            &  {1}             &1   &1   & s,p$_z$,d$_{z^2}$ \\ 
	      $\Delta_1'$ (A$_2$)  & 1        &  1        &  1       &  1       & -1            & -1             & -1          & -1          & R$_z$                                  \\
	      $\Delta_2$ (B$_1$)   &1& -1     & -1       &1&  1            & 1             & -1          & -1          & d$_{xy}$      \\ 
	      $\Delta_{2'}$ (B$_2$)&1& -1     & -1       &1& -1            & -1             &1   &1   & d$_{x^2-y^2}$          \\ 
	      $\Delta_{5}$ (E)     & 2        &  0        &  0       & -2       &  0            &  0             &  0          &  0          &d$_{xz}$,d$_{yz}$, p$_x$, p$_y$ \\ \hline
	\end{tabular}}}
	\caption{Electron impinging from one slab  to the other can propagate only if the character of the irreducible representation of the first slab matches that of the second slab. For Fe(001)/GaAs(001) interface, this happens for $\Delta_1$ and $\Delta_{2'}$ band. Therefore, the Fe-bands of $\Delta_2$ and $\Delta_5$ characters are not allowed to pass.
	Notice that the axis-coordinates are expressed in the GaAs basis. As gold is 45\degree\ rotated with respect to iron-BCC and GaAs, its diagonal mirror $\sigma_{xy}^{Au}$ is referred by $\sigma_v^{x}$ and its mirror plane $\sigma_v^{x}$ by $\sigma_v^{xy}$ of GaAs. The symmetries of GaAs are highlight by blue-shaded columns in the case of Fe and Au. The full case of Fe/Au/Fe/GaAs is explained in the text.}\label{character_table}
      \end{table}
      
      The band structure of iron (Fig.\ref{fig_Fe_Au_BS}) shows that, in the energy range between the Schottky barrier and 1.1 eV, only spin-up electrons have a $\Delta_1$ band.
      Therefore, up to 1.1 eV, the Fe(001)/Au(001) interface acts as a spin filter to electron transport.
      Interestingly, such a filter still works with the two final layers (Fe(001) and GaAs(001)) that allow the transmission of $\Delta_1$ electrons. Two counter-checks, experimental and theoretical, of this prediction can be imagined. Experimentally, it should be possible to replace the Au(001) slab by Ag(001), whose $\Delta_1$ band goes higher than 1.1 eV (see Fig.~\ref{fig_silver_harrison}).
      In this way, measuring a current above 1.1 eV would lead to a decrease of the magneto-current, because also Fe minority-spin electrons of $\Delta_1$ character would contribute to the BEEM current, as from Fig.~\ref{fig_Fe_Au_BS}.
      Theoretically, the counter-check would be a fully non-equilibrium calculation of the whole Fe/Au/Fe metal structure with the Keldysh formalism of chapter \ref{chapt_NEPT}. A successful calculation would have proven (or disproved) the symmetry filtering properties of Fe(001)/Au(001) interface. An attempt of this is described in the next section.
      Above 1.1 eV, also spin-down electrons can cross the interface and therefore the current polarization will decrease.

\FloatBarrier
%
%
%
%
%
  \section{Non-equilibrium approach}\label{sec_quantum_approach}
  In this section we present the results obtained with BEEM v3 (Chap.~\ref{chapt_BEEM_prog}) using the formalism introduced in  chapter \ref{chapt_NEPT} based on Keldysh Green functions. In first place, we describe the BEEM current in Au(111) in order to compare the new, finite-slab, approach with the decimation method that had been implemented by P. de Andres group in BEEM v2.1 \cite{reuter-beem_v2.1} (Sec.~\ref{sec_decimation}).  We analyze in this case the effect of the damping parameter $\eta$ and the importance of considering the evolution of the current pattern in $\kpara$ space with the number of layers constituting the finite slab. 
  In particular, we demonstrate that, e.g., after 10 to 20 layers (for $\eta=10$ meV), the surface density of states is the same as the surface density of states of the semi-infinite slab obtained with BEEM v2.1. Then, we move to the Au(111)/Si(001) and Au(111)/Si(111) cases in order to confirm that our model can explain experimental observations. We shall see also that, around a 10-layer slab,  adding or removing one layer can lead to a completely different BEEM current because of subtle interference effects in $\kpara$-space.
  
  Finally, in section \ref{sec_towards_spintronics_results} we turn our interest to the Fe/Au/Fe/GaAs spin-valve already analyzed in section \ref{sec_spinvalve_results1}. Unfortunately, we shall see that the propagation through  Fe(001) presents some numerical hindrances that do not allow completing the task.

  \subsection{Au(111)}
    In several papers \cite{Garcia-Vidal-PhysRevLett.76.807,Reuter-PhysRevB.58.14036,Pedro-ProgSurfS_66_2001}, Fernando Flores and Pedro De Andres' group  used the decimation approach within Keldysh formalism to describe BEEM experiments. In some of these papers, they studied the Au(111)/Si(111) and Au(111)/Si(001) systems and proved that electrons follow the band structure of the metal, differently of the prediction of Kaiser-Bell free-electron theory (see Sec.~\ref{sec_non_eq_calc}). In this section we consider again the Au(111)/Si heterostructures, this time for very thin slabs, and describe the evolution of the BEEM current as a function of the number of layers by using the Keldysh formalism developed in Chap.~\ref{chapt_NEPT} and implemented in BEEM v3. 

    \subsubsection{Surface density of states}
      As a first check, we have calculated the surface density of states for a slab made of 10, 20 and 40 layers of Au(111), without reconstruction.
      The density of states of a system can be obtained by taking the imaginary part of the Green function of this system. In order to get the surface density of states we have to evaluate
      \begin{equation}
	\rho^{(n)}_{1,1}=-\frac{1}{\pi}\Im \Tr \ G^{R(n)}_{1,1}
      \end{equation}

      where $G^{R(n)}_{11}$ is the surface Green function of a n-layer slab that is calculated during the iterative procedure (Sec.~\ref{sec_pert_method_2nd_neighbor}).

      The evolution of the Au(111) surface density of states with respect to the number of layers, for two values of $\eta$, is represented in the Fig.~\ref{surface_dos}. It is interesting to note that here the convergence is reached very quickly: for a 20 layers-thick slab, the surface density of states is the same as the semi-infinite case ($\eta=0.01$ eV) obtained through BEEM v2.1. For a smaller value of $\eta$ more layers are required in order to get the convergence (more than 40 layers for $\eta=1$ meV).
    
    \insertfigure{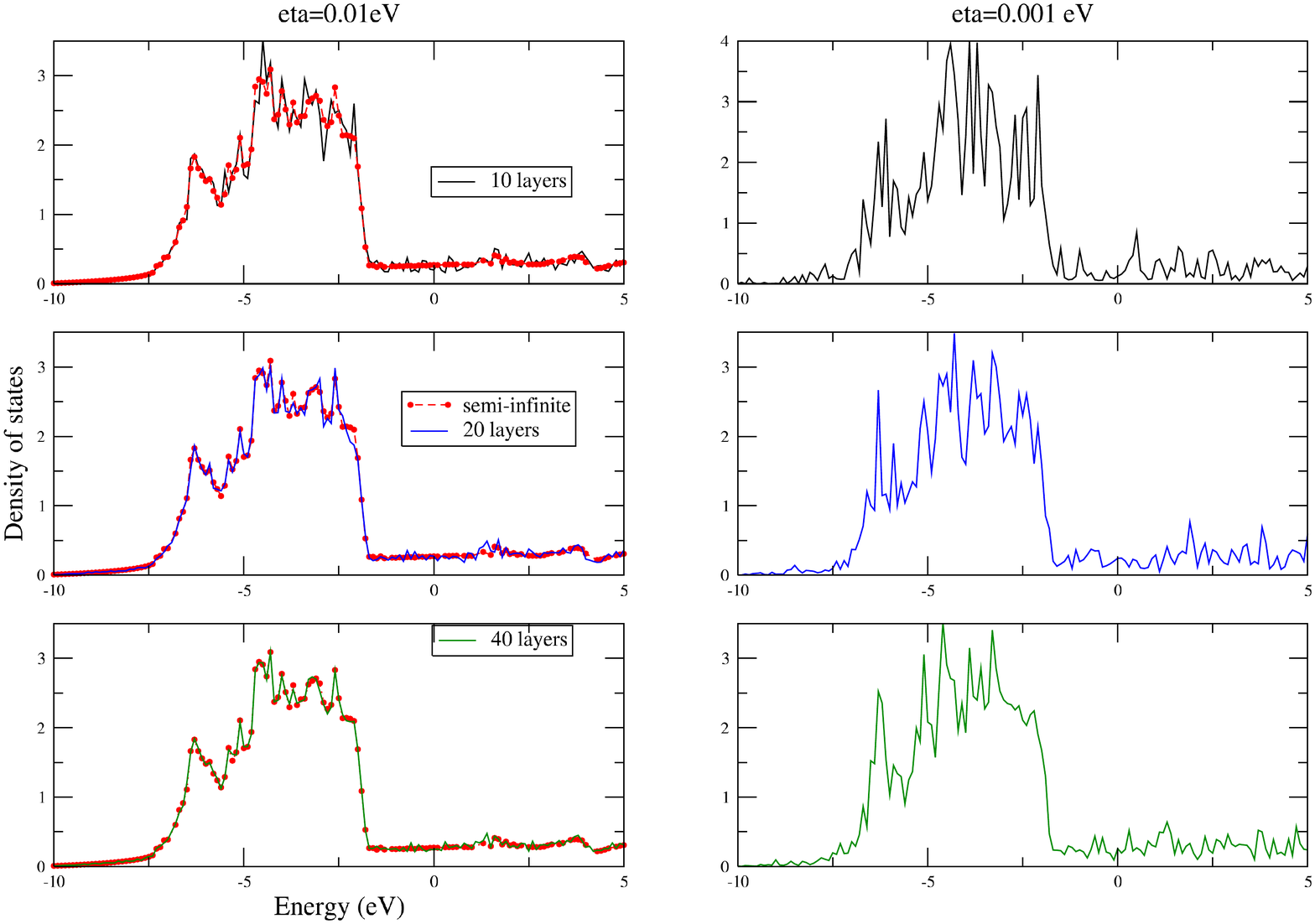}{Evolution of the Au(111) surface density of states with respect to the number of layers for $\eta=0.01$ eV and $\eta=0.001$ eV. The dotted line represents the surface DOS for a semi-infinite slab using decimation technique. Above 20 layers, for $\eta=0.01eV$, the surface density of states obtained with BEEM v3 (finite structure) and with BEEM v2.1 (semi-infinite structure) converge. Already for 10 layers, the surface density of states has almost converged. 
    For a smaller value of $\eta=1$ eV, the convergence is still not reach at 40 layers ($\eta=1$ meV). As the required time of calculation to obtain the surface DOS with BEEM v2.1 is quite long (more than one day for $\eta=0.01$ eV) we did not calculated the DOS of the semi-infinite slab for $\eta=0.001$ eV. Notice that for this value of $\eta$ the peaks are sharper and narrower. This is the analogous effect of the smearing parameter needed for usual DOS calculations.}{\label{surface_dos}}

    \subsubsection{Effect of the damping parameter $\eta$}
          \begin{figure}[!t]
        \centering
        \includegraphics[width=0.7\linewidth]{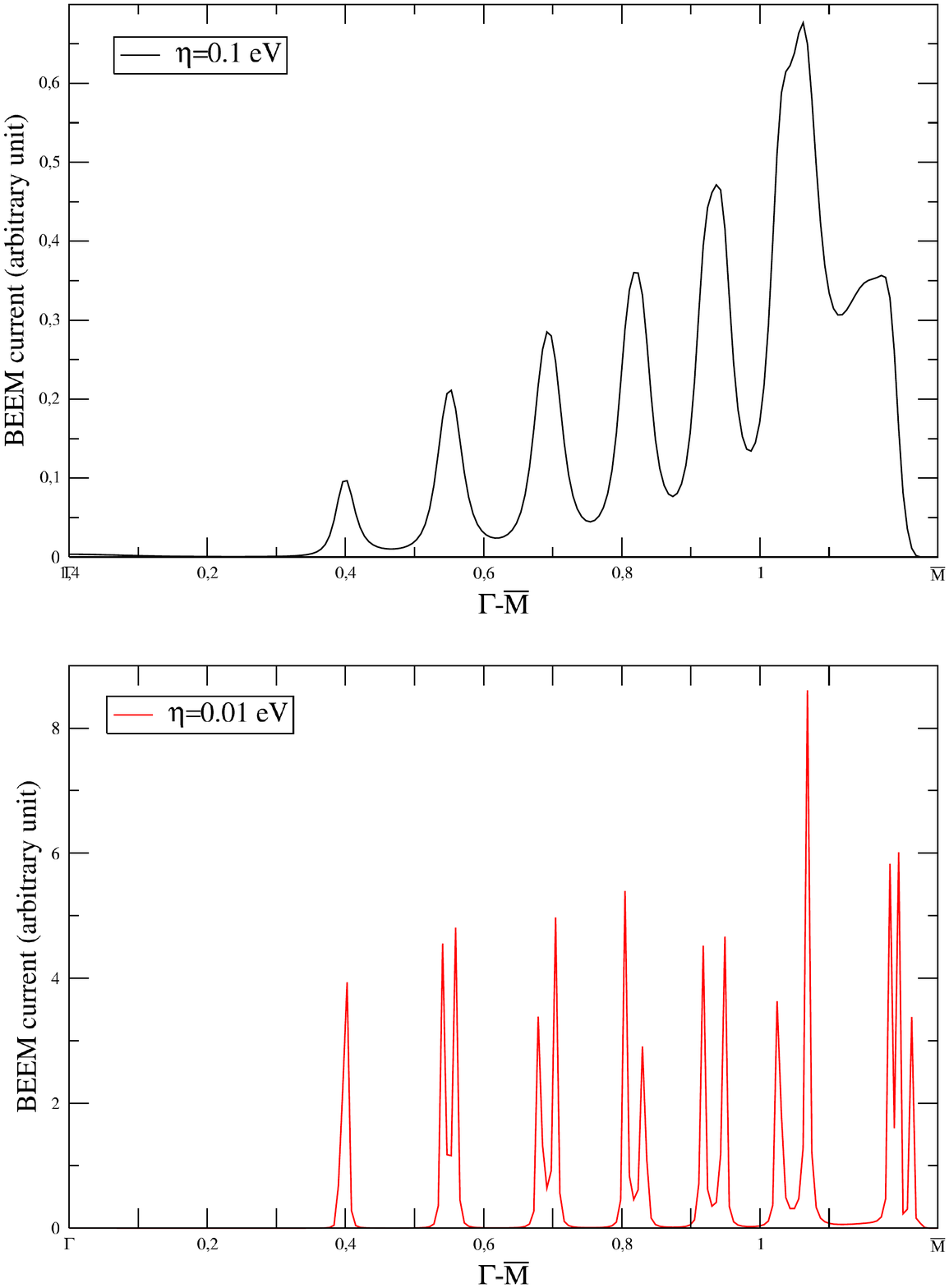}
        \caption{BEEM current along the $\bar{\Gamma}\bar{M}$ path in the Brillouin zone for 20 layers and two values of $\eta$: $\eta=0.1$ eV and $\eta=0.01$ eV. Increasing $\eta$ spreads the peaks.}{\label{effect_of_eta}}
      \end{figure}
      The parameter $\eta$ in $G^{R,A}=[\hbar\omega-\varepsilon\pm\i\eta]^{-1}$ is a damping parameter, as in the classical theory of harmonic oscillator. This means that it takes into account phenomenologically of all the kinds of inelastic effects that lead to the finite mean-free-path of the electrons (but conserving $\kpara$). Figure~\ref{effect_of_eta} shows the intensity profiles of the current, for two different values of $\eta$, along the $\bar{\Gamma}\bar{M}$ direction of the FCC Brillouin zone (depicted in Fig.~\ref{subfig_Au111}) for 20 layers. Small values of $\eta$ mean a large mean free path, i.e., inelastic effects start to be important only after a large number of layers. For example, for 20 layers and $\eta=0.001$eV, the poles of the Green function do not superpose, which leads to a peaked profile.
      For 20 layers and $\eta=0.01$eV instead, the width of the Lorentzian determined by $\eta$ makes the peaks to superpose in $\kpara$-space, which leads to this smooth profile.

      These results imply that the damping parameter might be critical for experiment interpretations. Of course, the integrated current is the same for every $\eta$, but 
      for heterostructures the change of profile in $\kpara$-space can lead to strong differences in the transmitted current, due to the $\kpara$-filtering effects explained above (see Sec.~\ref{sec_spinvalve_results1}).
      Therefore, changing $\eta$ could lead to a completely different BEEM current. 
      
      Consider for instance the Fe/Au/Fe spinvalve: if high current peaks in iron do not overlap with peaks in gold, then the BEEM current is zero. As increasing $\eta$ spreads the peaks, it leads to a higher chance of overlap between the two current map. This reasoning is similar to the one used in the equilibrium approach and does not take into account destructive or constructive interferences that could occur in a full calculation for the Fe/Au/Fe spinvalve.

    \subsubsection{Effect of the number of layers}
      Increasing the number of layers leads to a current profile in $\kpara$-space more ``continuous'', analogously to the increase of the damping parameter $\eta$ for fixed number of layers. In spite of the similar appearence, the two effects are not in close analogy: increasing $\eta$ leads to the increase of the width of each pole of the Green functions (at fixed number of poles), whereas increasing the number of layers corresponds to add more and more poles, closer and closer. In both cases the profile is smoothed.
      
      \begin{figure}[!hbtp]
       \centering
       \includegraphics[width=0.97\linewidth]{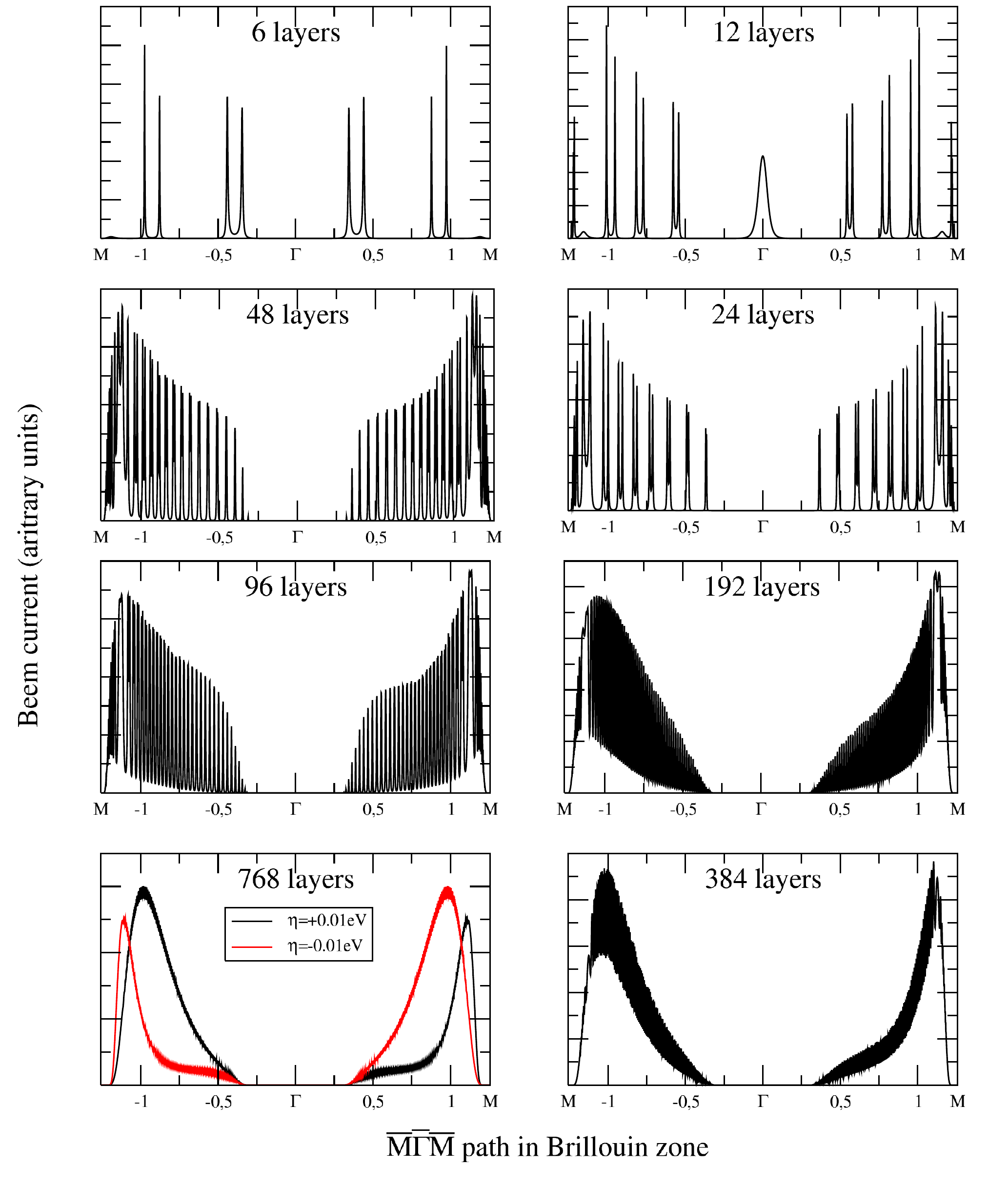}
       \caption{Effect of the number of layers on the current profile, for $\eta=0.01$eV. Above 384 layers, the 6-fold symmetry is lost, and the results converged to the equilibrium approach as found in Ref.~\cite{Reuter-PhysRevB.58.14036} by P. de Andres and F. Flores group. The red curve obtained by taking a negative $\eta$ is a check that by exchanging the retarded by the advanced GF, one finds the symmetric result. }{\label{effect_of_layers}} 
      \end{figure}
      Figure~\ref{effect_of_layers} shows the evolution of the current profiles along $\bar{M}\bar{\Gamma}\bar{M}$ when the number of layers is doubled at each step. The more layers, the more poles in the Green functions and the more the profile is smoothed.
      After a given threshold, between 96 and 192 layers, the sixfold symmetry is lost: such a loss appears in the $\bar{M}\bar{\Gamma}\bar{M}$ section as an asymmetry around the $\bar{\Gamma}$ point that becomes increasingly visible up to around 768 layers, where the equilibrium behavior of Ref.~\cite{Reuter-PhysRevB.58.14036} is found. Notice that unlike the DOS (Fig.~\ref{surface_dos}), the calculation does not converge after 20 layers. It is due to the fact that the surface DOS is extracted from the surface Green function $\green{1,1}{n}$ that is quickly uncoupled from the other extremity of the slab. On the contrary, the calculation of the current is based on Green function of the type $\green{1,n}{n}$. 
      
      We also see from figure \ref{effect_of_layers} that a peak appears at $\kpara=0$ for 12 layers but disappears for 6 and 24 layers. For this reason, we have performed layer-by-layer calculations from 7 to 18 layers, whose results are presented in Fig.~\ref{fig_layer_by_layer}. 
      We have chosen for these calculation the value $\eta=0.005$ eV, half the value of Fig.~\ref{effect_of_layers}, in order to avoid a too big smoothing of the peaks due to the damping parameter.
      The $\kpara=0$-peak appears for 10 layers, reaches its maximum for 11 layers and disappears after 12 layers. 
      This is a very important result because it means that experimentally, if the available density of states in the semi-conductor is located only in $\kpara=0$, then the gold slab must be of 11-layers thickness in order to allow the transport of a BEEM current.
      Morever, it could be a decisive test that gives information of other physical processes (other than elastic scattering) involved in BEEM: if the current is not zero for, e.g., 13 layers, then it could be explained only by inelastic mechanisms leading to non-conservation of $\kpara$, that are not included in the present calculations.

      It is also interesting to note that changing $\eta$ will not change the overall qualitative evolution of the profile by increasing the number of layers. The loss of the six-fold symmetry just appears sooner for a bigger $\eta$ (as also in Ref.~\cite{Reuter-PhysRevB.58.14036}).

      \insertfigure{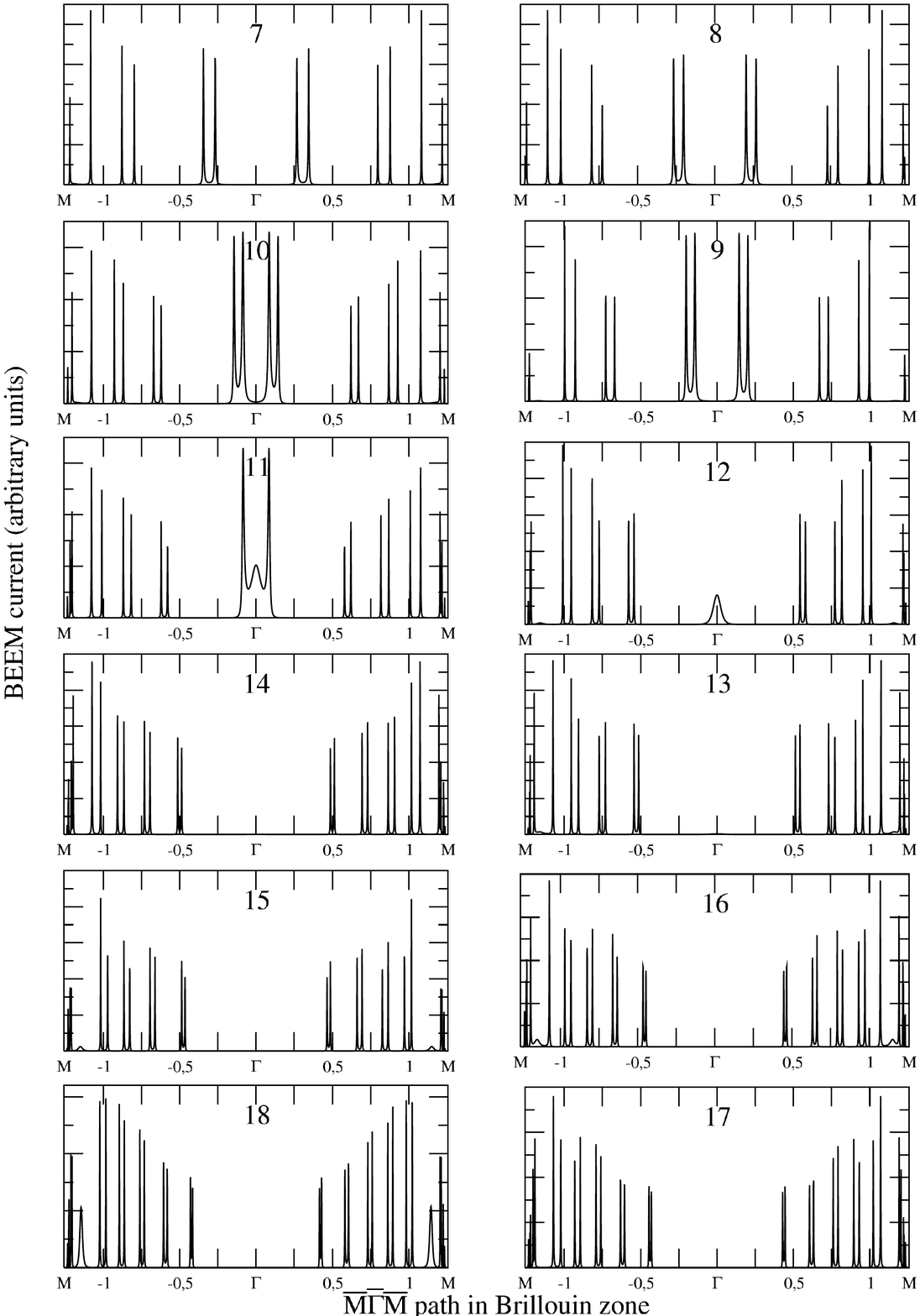}{Evolution of the BEEM current along  $\bar{M}\bar{\Gamma}\bar{M}$ with respect to the number of layers (from 7 to 18 layers) with $\eta=5$meV. A peak at $\kpara=0$ appears for 10 layers, reaches is maximum for 11 layers and disappears after 12 layers.}{\label{fig_layer_by_layer}}

    \subsubsection{Effect of the parametrization}
\begin{figure}[!hbt]
        \begin{center}
          \subbottom[5 layers of Au(111)]{\includegraphics[width=0.48\linewidth]{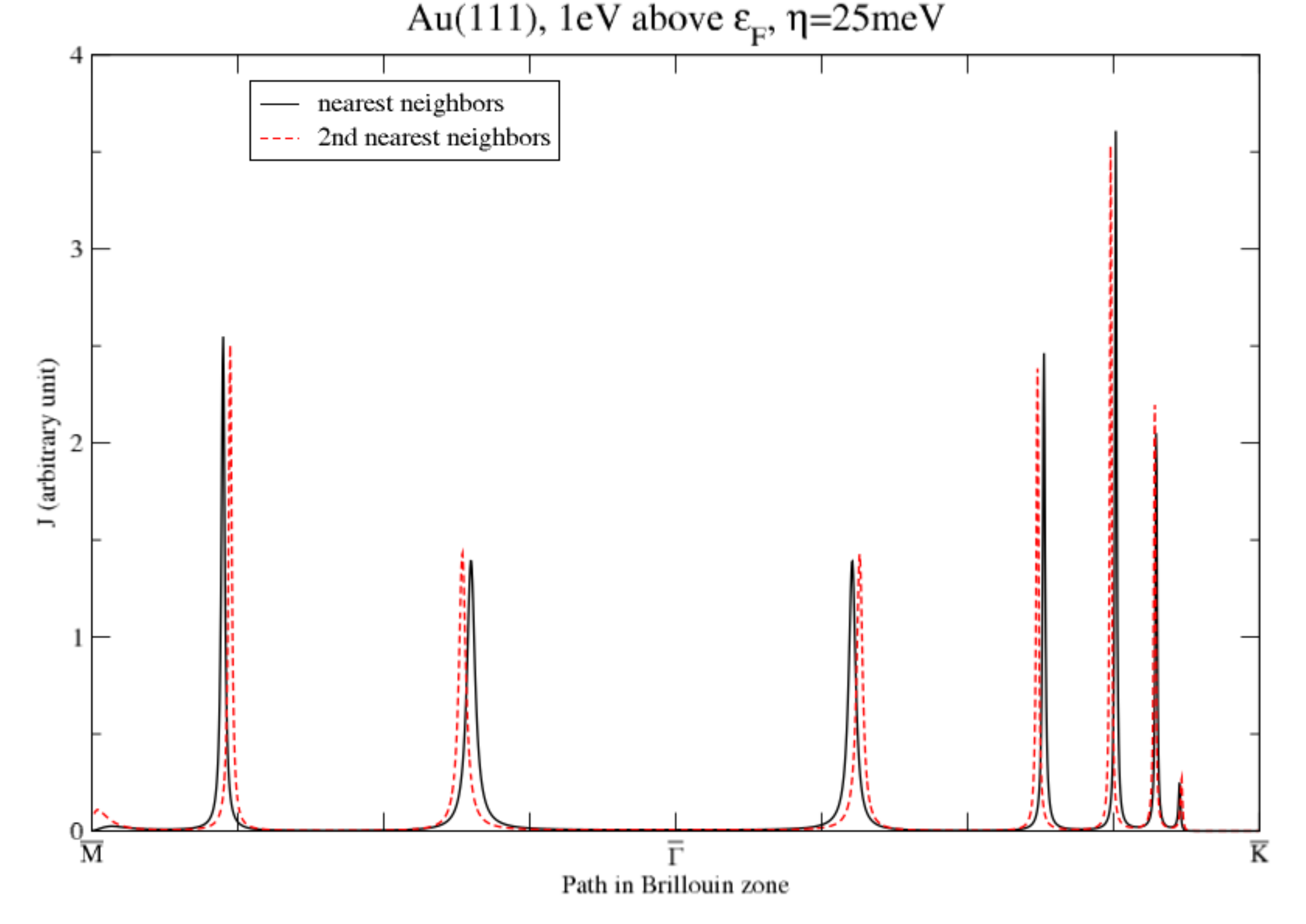}}
          \subbottom[11 layers of Au(111)]{\includegraphics[width=0.48\linewidth]{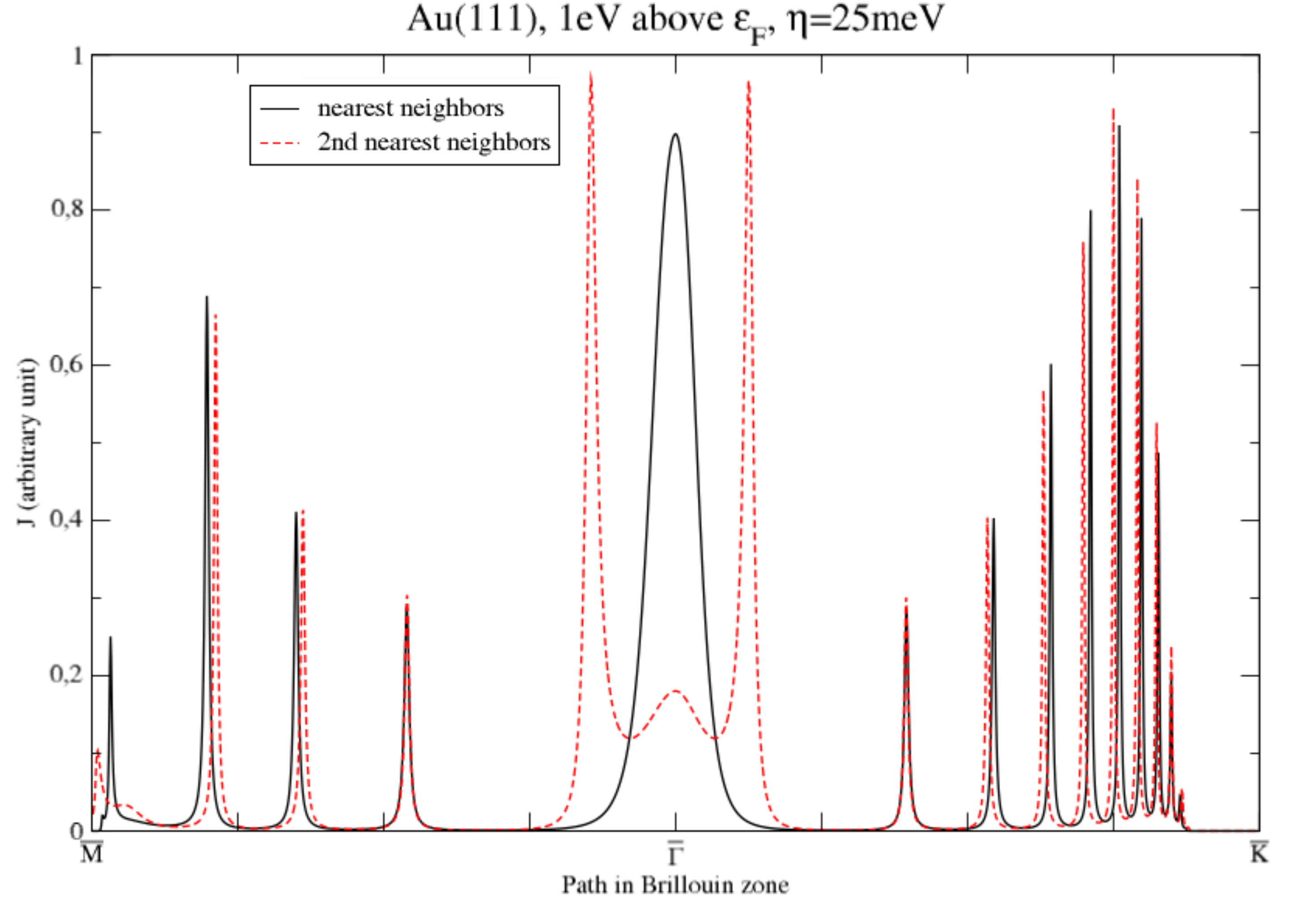}}\\
          \subbottom[50 layers of Au(111)]{\includegraphics[width=0.48\linewidth]{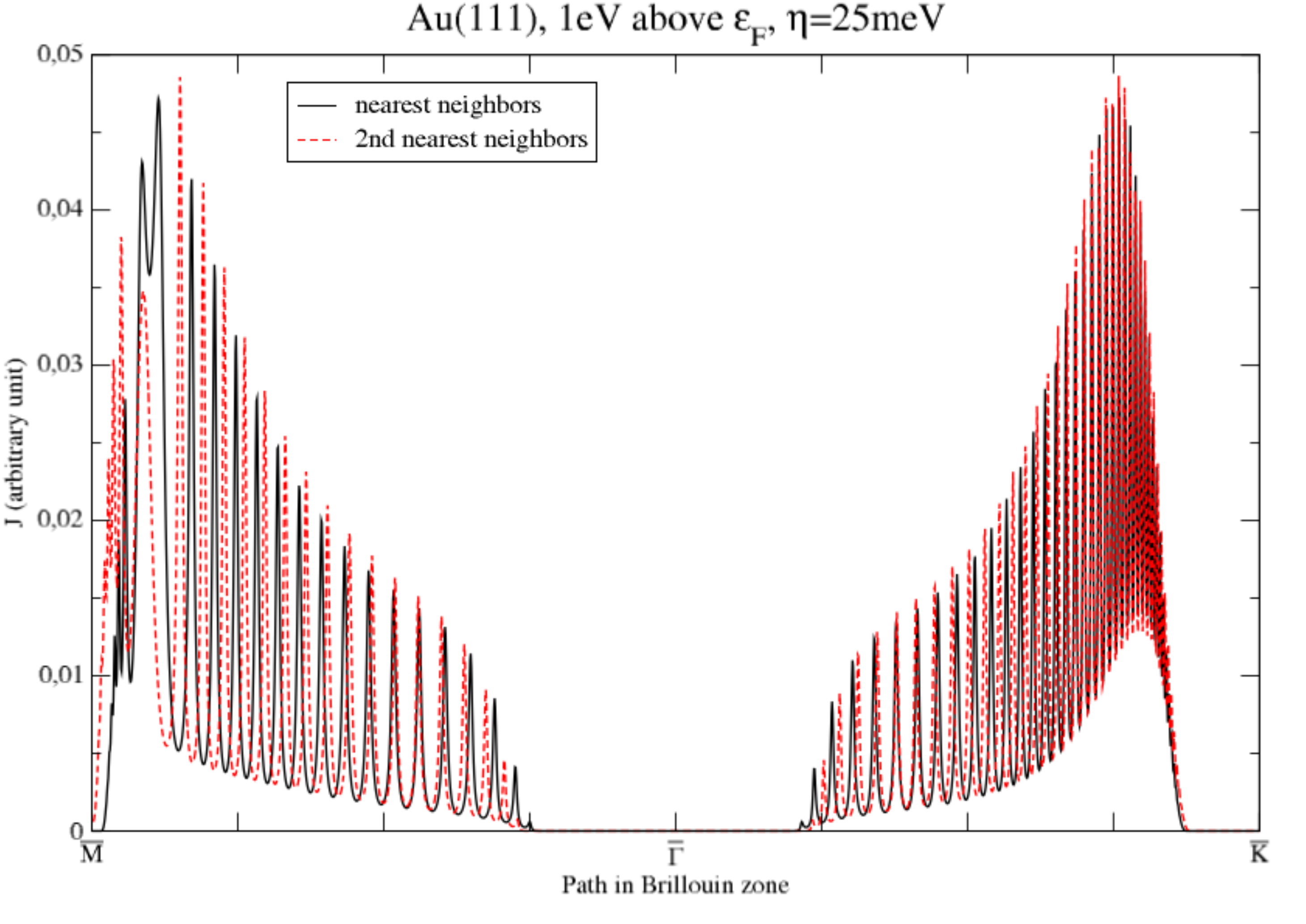}}
          \caption{Nearest-neighbor parametrization from \cite{reuter-beem_v2.1} (in black) vs second-nearest-neighbor parametrization \cite{Papa-handbook} (red dashed-line), 1 eV above Fermi level and $\eta=25$ meV. The peaks are slightly shifted to the $\Gamma$ point, that is critical around 11 layers.}
        \label{fig_effect_parameter}
        \end{center}
      \end{figure}

      We have seen in section \ref{sec_tb_parametrization} that the band structure of FCC crystals like gold are well reproduced considering only nearest neighbor hopping. But because the thickness could be very critical, we have compared the effect of the parametrization on the current with respect to the number of layers, as shown in Fig.~\ref{fig_effect_parameter}. We see that the distributions are very similar but considering second nearest-neighbors shifts the peaks to the $\Gamma$ point, that is very critical around 11 layers, again. However, for a larger number of layer (for instance 50 layers here), the differences are irrelevant given experimental resolutions.
      
      These results suggest that if a critical behavior due to the thickness exist, then the parametrization could be critical too.

    \subsubsection{Au(111)/Si(111) and Au(111)/Si(001)}
\begin{figure}[!hbt]
        \begin{center}
          \subbottom[\label{fig_Au111_Si001_800layer}]{\includegraphics[width=0.4\linewidth]{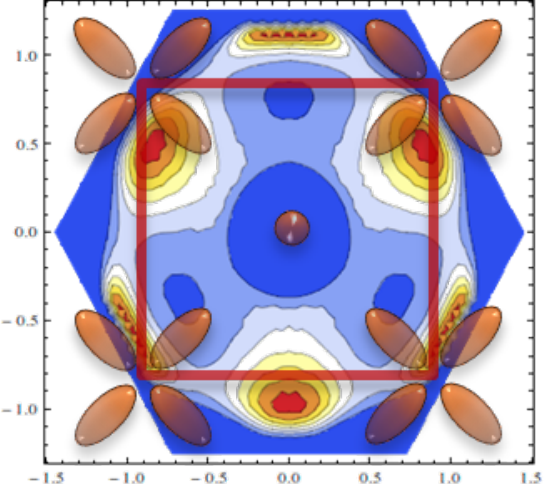}}
          \subbottom[\label{fig_Au111_Si111_800layer}]{\includegraphics[width=0.4\linewidth]{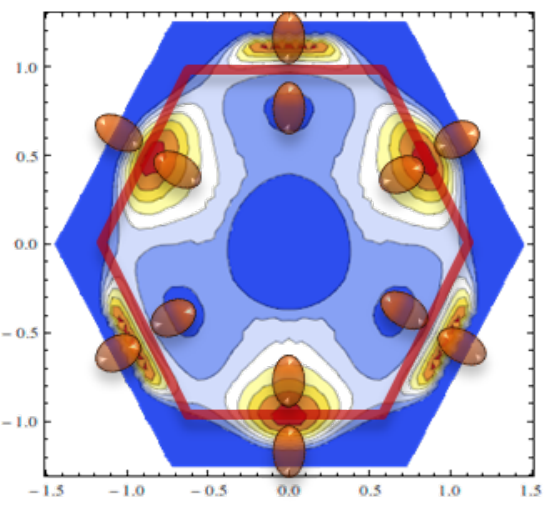}}\\
          \subbottom[\label{fig_Au111_Si001_10layer}] {\includegraphics[width=0.4\linewidth]{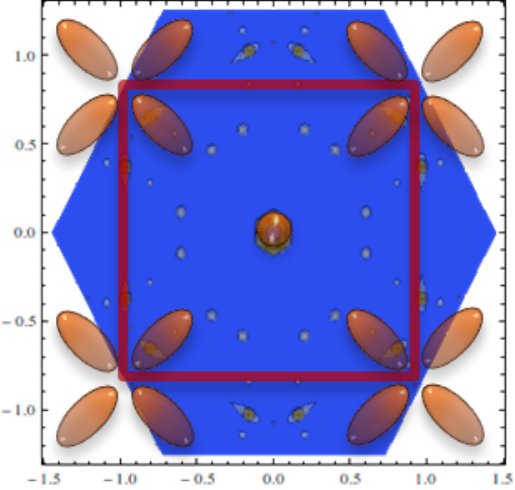}}
          \subbottom[\label{fig_Au111_Si111_10layer}] {\includegraphics[width=0.4\linewidth]{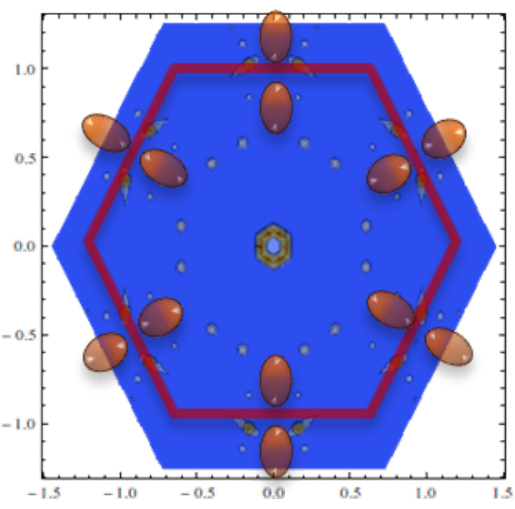}}
          \caption{\subcaptionref{fig_Au111_Si001_800layer} and \subcaptionref{fig_Au111_Si111_800layer} BEEM current projected in the 2D Brillouin-zone for 800 layers of Au(111) on Si(001) and Si(111) respectively, 1 eV above Fermi level. 
          The sixfold symmetry has been lost in favor of a three-fold symmetry, as in the semi-infinite slab \cite{Reuter-PhysRevB.58.14036}. 
          These results show that high-current areas match available density of states for both orientation of silicon. \subcaptionref{fig_Au111_Si001_10layer} and \subcaptionref{fig_Au111_Si111_10layer} BEEM current projected in the 2D Brillouin-zone for 10 layers of Au(111) on respectively Si(001) and Si(111), 1 eV above Fermi level. 
          In that case, the current distribution is discrete. 
          Unlike the 800-layer case, there is a big difference between the two orientations of silicon. In the (001) direction, some available density of states matches high-current areas near $\kpara=0$. This density of states does not exist for the (111) orientation of silicon for which there is no superposition between high-current areas and available density of states.}
        \label{fig_Au_Si_BEEMv3}
        \end{center}
      \end{figure}

      The Au(111)/Si(111) vs  Au(111)/Si(001)  has been the seminal experiment that had led to question the free-electron model. For this reason  we have performed two extreme calculations of the BEEM current corresponding to two extreme cases: 10 layers and 800 layers, 1 eV above Fermi level. Then, we have qualitatively projected the available density of states for Si(001) and Si(111) on the resulting current map. The results are presented in figure \ref{fig_Au_Si_BEEMv3}.
      
      First, we note that the six-fold symmetry, which is present for 10 layers, has been indeed lost for 800 layers in favor of a three-fold symmetry. 
      In second place, we remark that the 800 layers case is very similar to what had been obtained in Refs.~\cite{Reuter-PhysRevB.58.14036,Pedro-ProgSurfS_66_2001} through the decimation technique. 
      The BEEM currents for both orientation of silicon are qualitatively close to each other as both match high current area. For a more quantitative approach, we should include the semiconductor in our non-equilibrium calculation.
      
      However, for 10 layers, the result is completely different (Figs.\ref{fig_Au111_Si001_10layer} and \ref{fig_Au111_Si111_10layer}). The current distribution is discrete and there is almost no match with the available density of states, except near $\kpara=0$ for the (001) orientation of silicon. In other words, for this direction, electrons can enter the semiconductor if their wave-vector is near $\kpara=0$, but they cannot enter the semiconductor in (111) orientation as there is no available density of states at this wave-vector.
      
      This result is again an interesting example showing that thin films may behave differently from thick films for what $\kpara$-filtering properties are concerned. 
      Therefore, we expect a higher BEEM current for Au(111)/Si(001) than for Au(111)/Si(111) if there are only 10 layers of gold. In the seminal experiment \cite{Milliken-AuSi_PhysRevB.46.12826} a thick slab of gold was studied: it should be interesting to reproduce this experiment for a ten layer film of gold.\footnote{Although, this result has to be nuanced because experimentally: the system is not Au/Si but Au/SiO$_2$/Si.}

  \subsection{Towards spintronics: preliminary results on Fe/Au/Fe spinvalve using the non-equilibrium approach}\label{sec_towards_spintronics_results}
    After the band-structure study of section \ref{sec_spinvalve_results1} we have studied the same spinvalve Fe/Au/Fe/GaAs(001) with the non-equilibrium approach of chapter \ref{chapt_NEPT}. However, because of lack of time due to numerical issues, the following are only preliminary results. We have run some calculations for different magnetic-configurations of the spinvalve, without including the semi-conductor in the calculation (as for the previous Au/Si case).

\begin{figure}[!hbtp]
      \begin{center}
      \subbottom[\label{subfig_dndn} spin-down/spin-down configuration]{
	\includegraphics[width=0.45\linewidth]{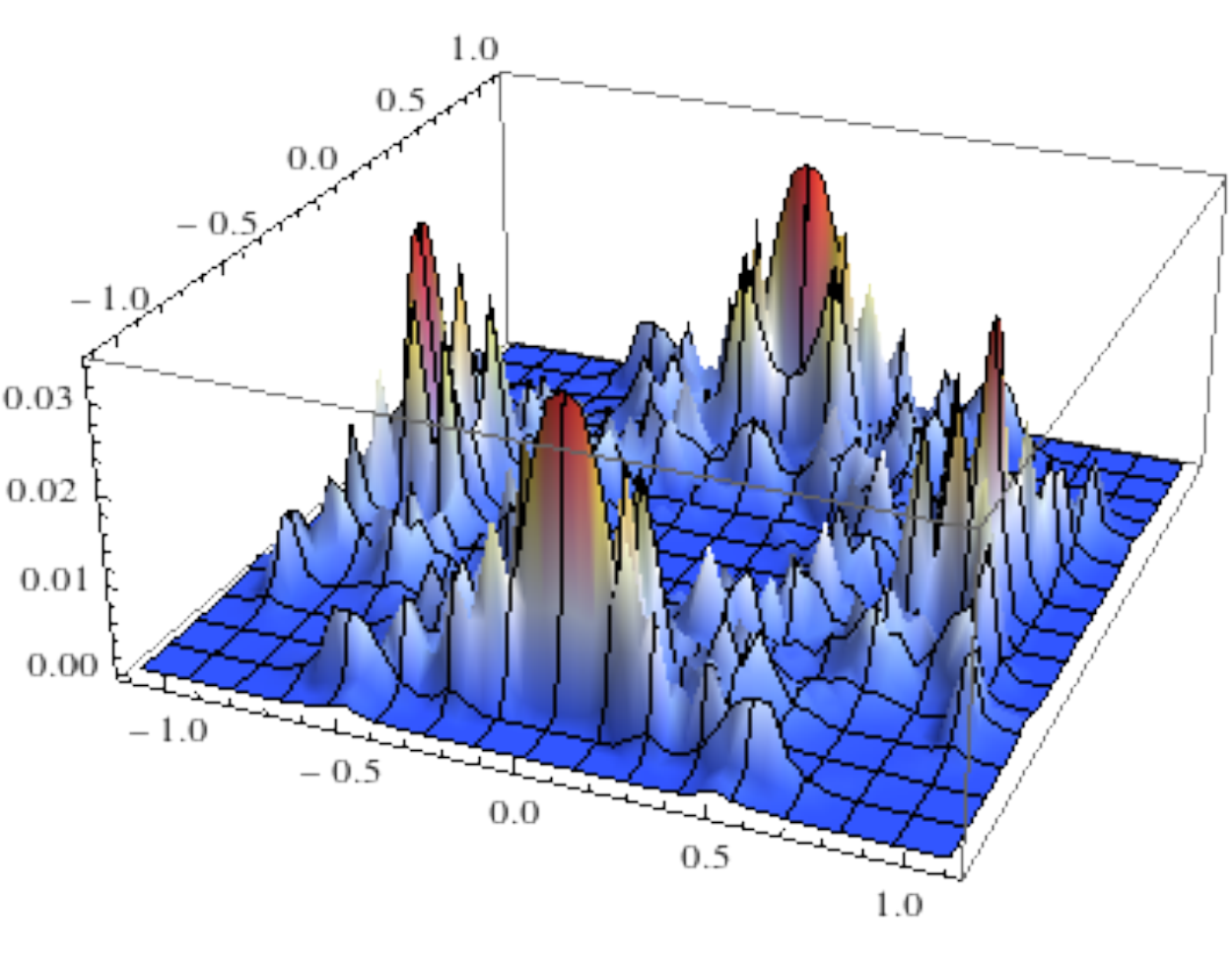}}
      \subbottom[\label{subfig_upup} spin-up/spin-up configuration]{
	\includegraphics[width=0.45\linewidth]{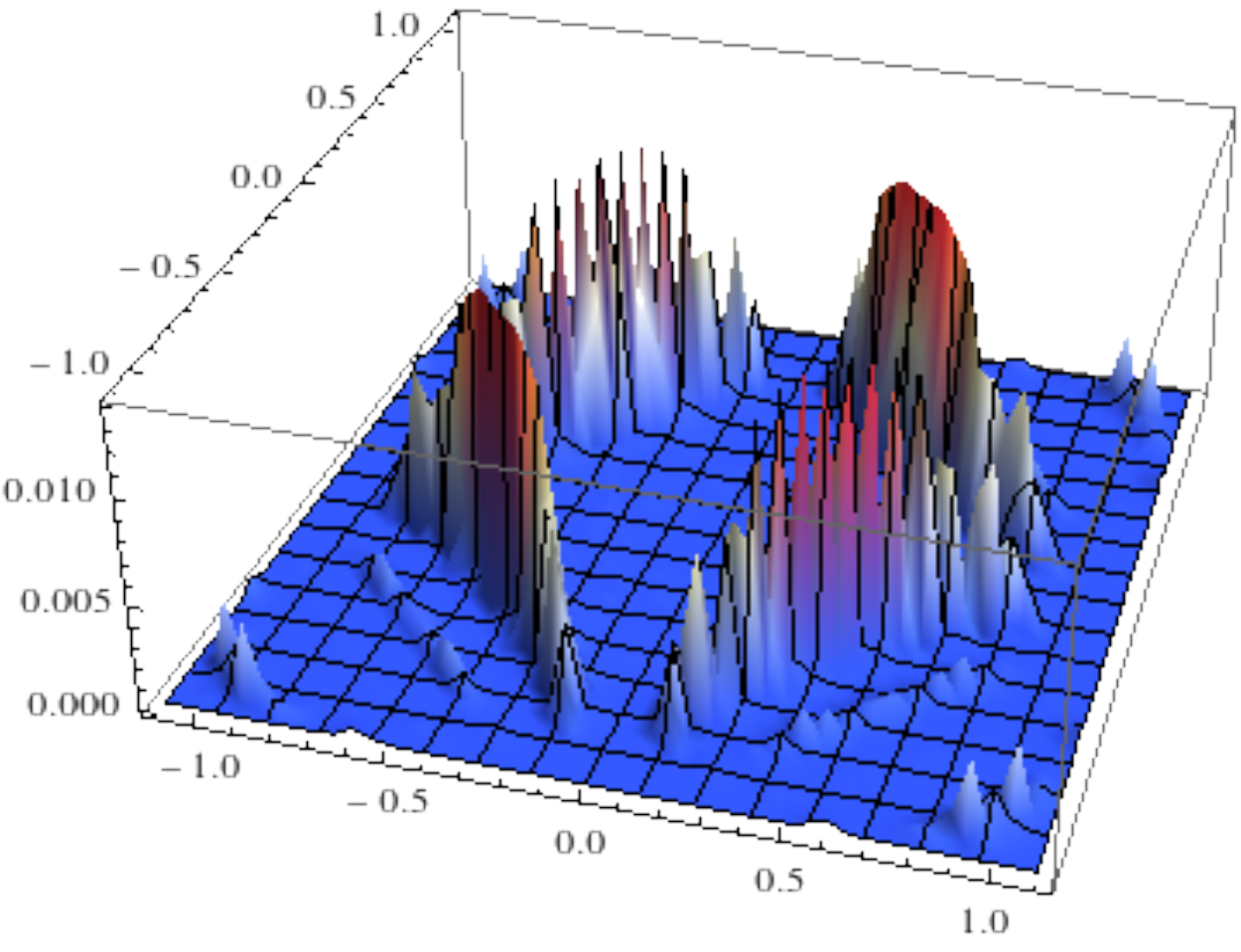}} \\
      \subbottom[\label{subfig_dnup} spin-down/spin-up configuration]{
	\includegraphics[width=0.45\linewidth]{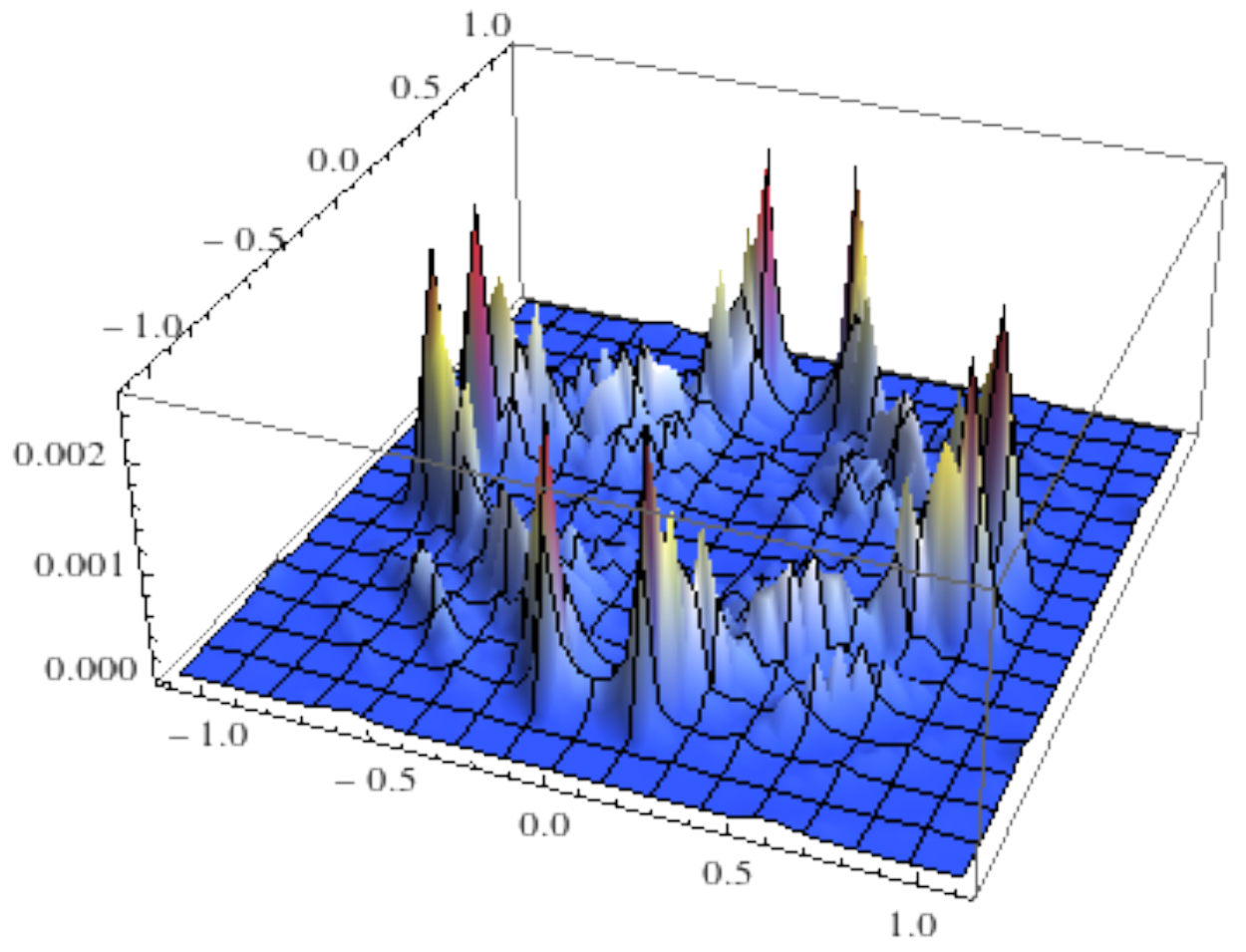}}
      \subbottom[\label{subfig_updn} spin-up/spin-down configuration]{
	\includegraphics[width=0.45\linewidth]{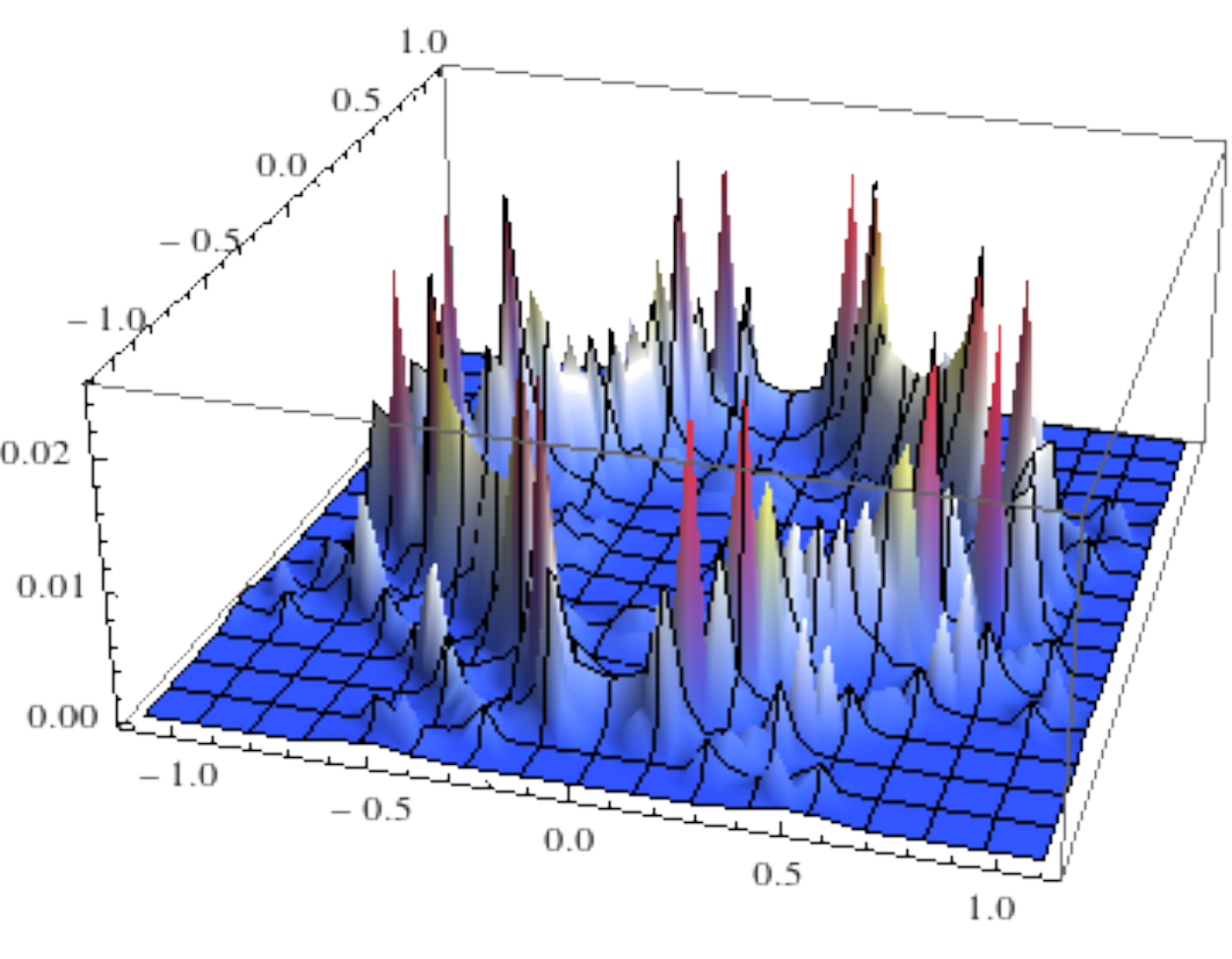}}
      \caption{Four possible spin-configurations for the  Fe(8ml)/Au(30ml)/Fe(9ml) spinvalve \cite{Marie-apl_spin_filtering,Marie-jap113_quantitative_magnetic_imaging}. \subcaptionref{subfig_dndn} and \subcaptionref{subfig_upup} correspond to the parallel magnetic configuration of the spinvalve and \subcaptionref{subfig_dnup} and \subcaptionref{subfig_updn} to the anti-parallel one. $E=\varepsilon_F+1$ eV and $\eta=0.05$ eV.}
      \label{fig_spinvalve_beemv3}
      \end{center}
    \end{figure}

    Figure \ref{fig_spinvalve_beemv3} represents the elastic current 1 eV above the Fermi level for the Fe(8ml)/Au(30ml)/Fe(9ml) spinvalve, that has been studied at IPR \cite{Marie-apl_spin_filtering,Marie-jap113_quantitative_magnetic_imaging}, in the 4 possible spin configurations: spin-up/spin-up, spin-down/spin-down, spin-up/spin-down and spin-down/spin-up. The spin-up/spin-up configuration correspond to the propagation of the majority spin in a parallel magnetic configuration of the two iron electrodes, while spin-down/spin-down describe the propagation of minority electrons in the same parallel configuration. Spin-up/spin-down configuration correspond to the propagation of majority electrons that become minority in an anti-parallel magnetic configuration, while spin-down/spin-up represent the propagation of minority electrons that become majority in the same anti-parallel magnetic configuration. In other word, in order to obtain a magneto-current we have to evaluate (Eq.\ref{eq_GMR}):
    \begin{align}
       MC  &= \frac{J_{\mathrm{P}} - J_{\mathrm{AP}}}{J_{\mathrm{AP}}} \nonumber \\
           &= \frac{\left(J_{\uparrow\uparrow}+ J_{\downarrow\downarrow}\right) - \left(J_{\downarrow\uparrow}+J_{\uparrow\downarrow}\right) }{J_{\downarrow\uparrow}+J_{\uparrow\downarrow}} \\
           &= \frac{(7.63+1.16)-(0.321+3.54)}{3.21+3.54} = 1.27
    \end{align}
    That is to say a magneto-current of 127\%. Notice that this magneto-current is not really a GMR. Indeed, the GMR is due to the difference of mean free path for majority and minority spins. In order to model this behavior, we should use different $\eta$ for minority and majority spins. However, here, as shown by Marie Hervé \emph{et al.}, the magneto-current does not depend on the thickness of iron electrodes \cite{Marie-apl_spin_filtering,Marie-jap113_quantitative_magnetic_imaging,these_marie}. The magneto-current is purely interface filtering due to the band-structure. Notice also that the calculation presented here is not the real magneto-current because we did not consider the semi-conductor.

    These are just preliminary calculations. Before drawing real conclusion we need to:
    \begin{enumerate}
      \item include the semi-conductor,
      \item vary the energy for a given spinvalve,
      \item vary the thickness of gold and/or iron.
    \end{enumerate}

  \subsection{Non-equilibrium calculation conclusion}
    As expected, the non-equilibrium approach allowed us to make more precise predictions. For instance, we have seen that the presence of a high-current peak around $\kpara=0$ makes the BEEM current of a thin slab of gold (11 layers) on Si(001) much bigger than for 14 layers of gold due to a high-current peak around $\kpara=0$. For the same reason, the BEEM current in Au(10 layers)/Si(001) should be bigger than  Au(10 layers)/Si(111) unlike the 800-layer case. These results are interesting as they can, in principle, be confirmed by future experiments, although it requires low temperature experiments and ideal interfaces. Currently, experiments are performed at room temperature and the roughness at Au/Si interfaces forbid layer resolved effects.
    
\begin{figure}[!hbt]
      \begin{center}
      \subbottom[\label{subfig_v2_01} BEEM v2.1: $\eta=0.1$ eV]{
	\includegraphics[width=0.37\linewidth]{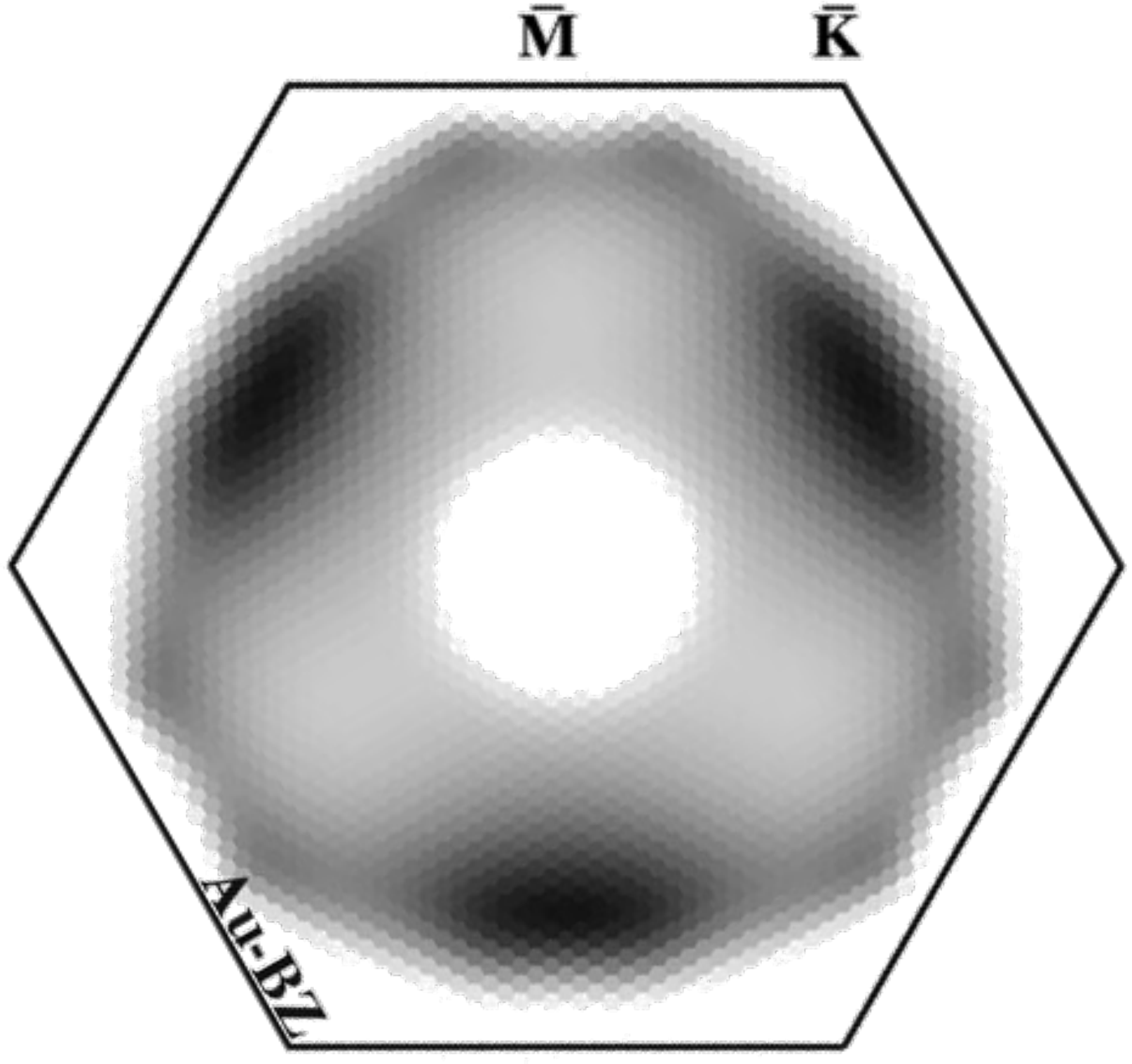}}\qquad\qquad
      \subbottom[\label{subfig_v2_001} BEEM v2.1: $\eta=0.01$ eV]{
	\includegraphics[width=0.37\linewidth]{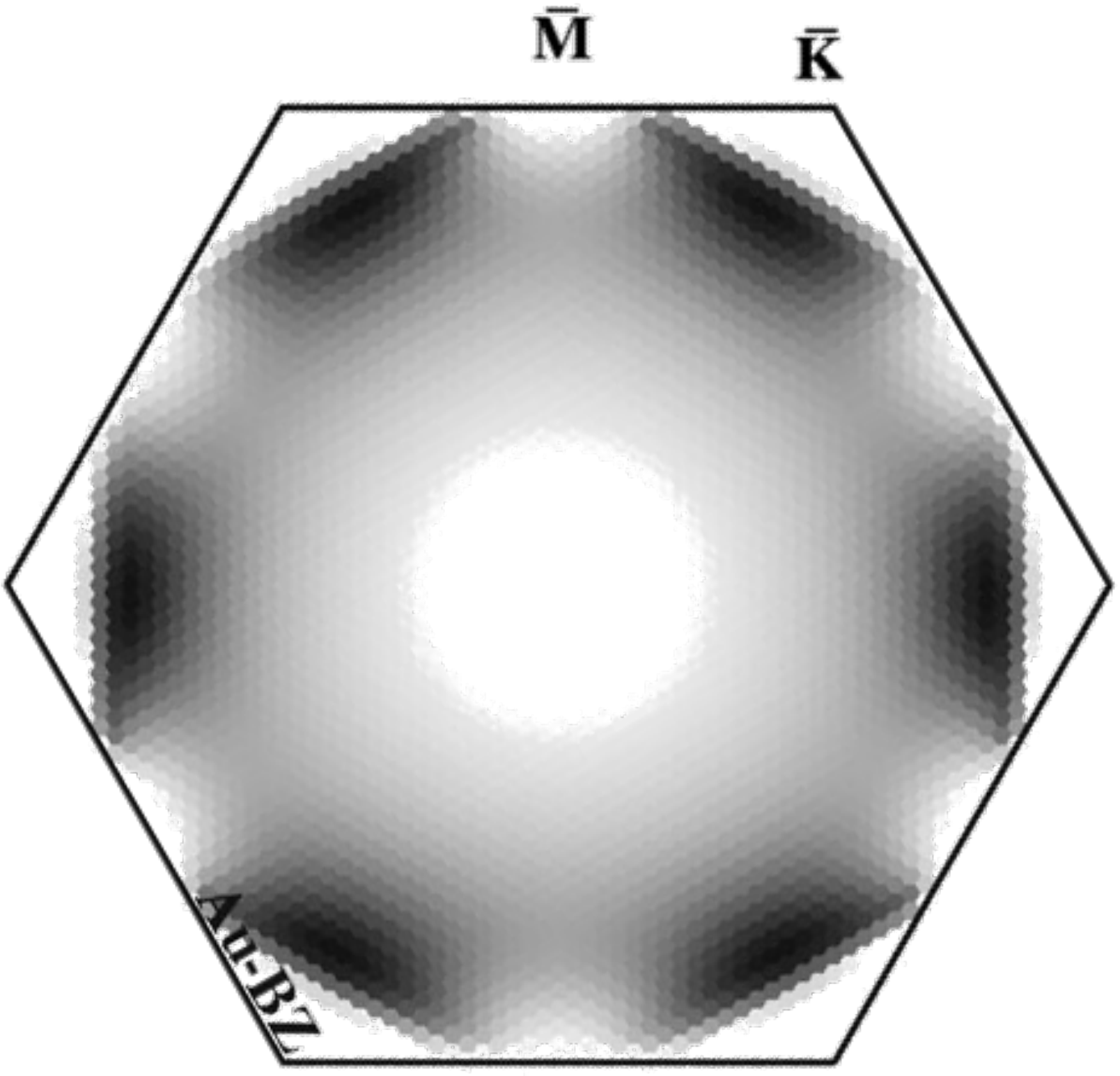}} \\
      \subbottom[\label{subfig_v3_01} BEEM v3: 75 layers, $\eta=0.1$ eV]{
	\includegraphics[width=0.37\linewidth]{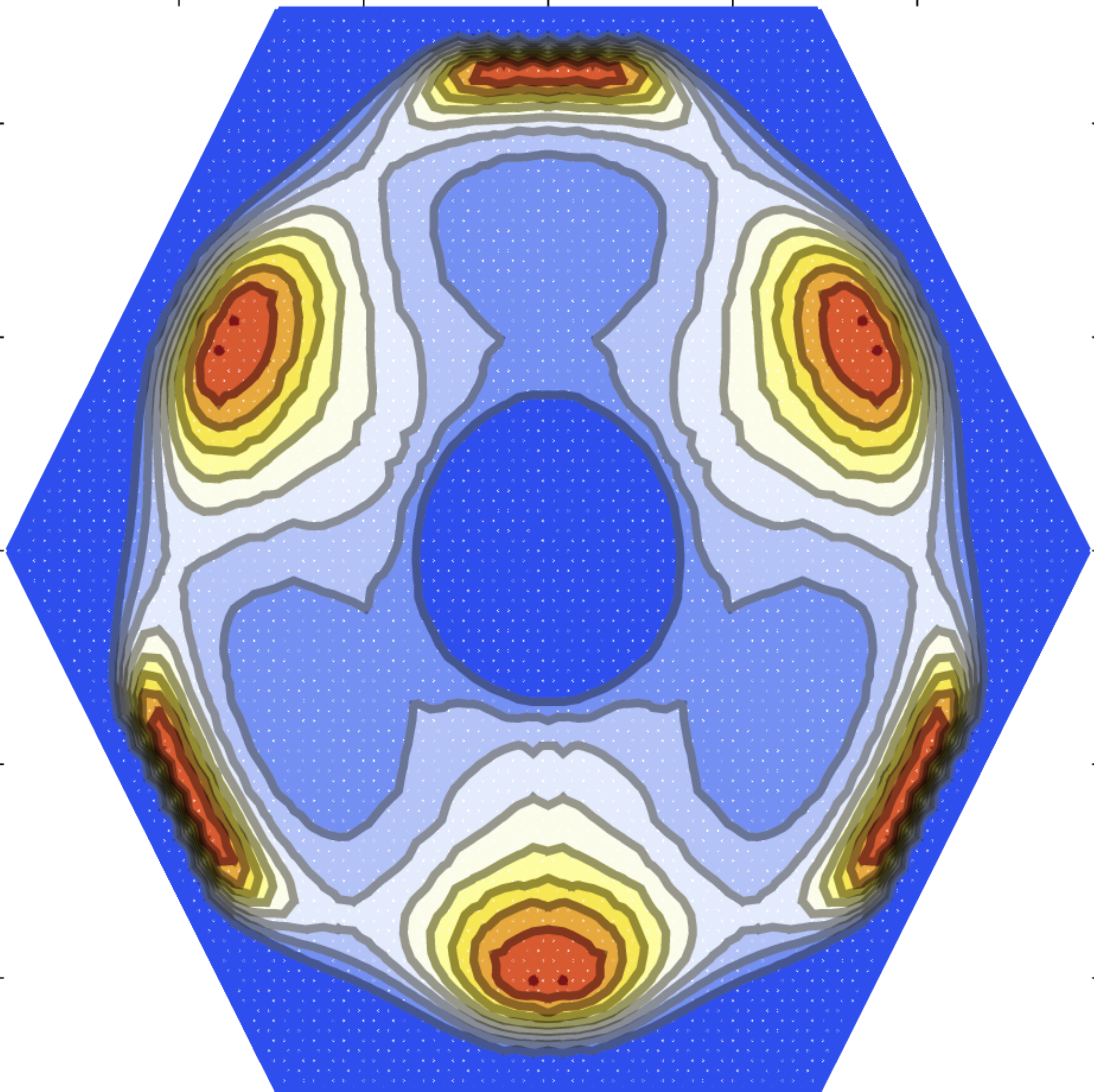}}\qquad\qquad
      \subbottom[\label{subfig_v3_001} BEEM v3: 75 layers, $\eta=0.01$ eV]{
	\includegraphics[width=0.37\linewidth]{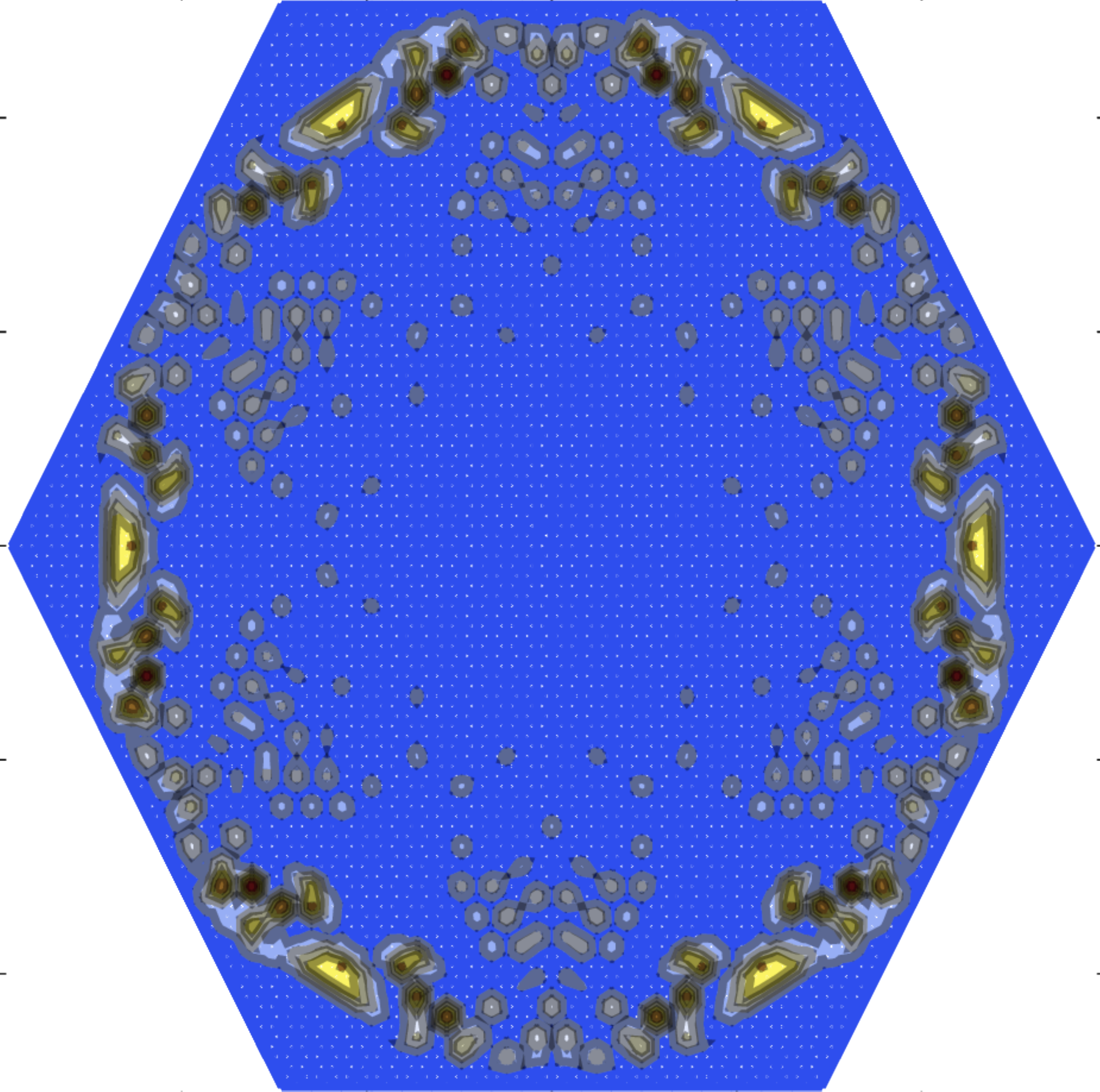}} \\
      \subbottom[\label{subfig_v3_400} BEEM v3: 400 layers, $\eta=0.025$ eV]{
	\includegraphics[width=0.37\linewidth]{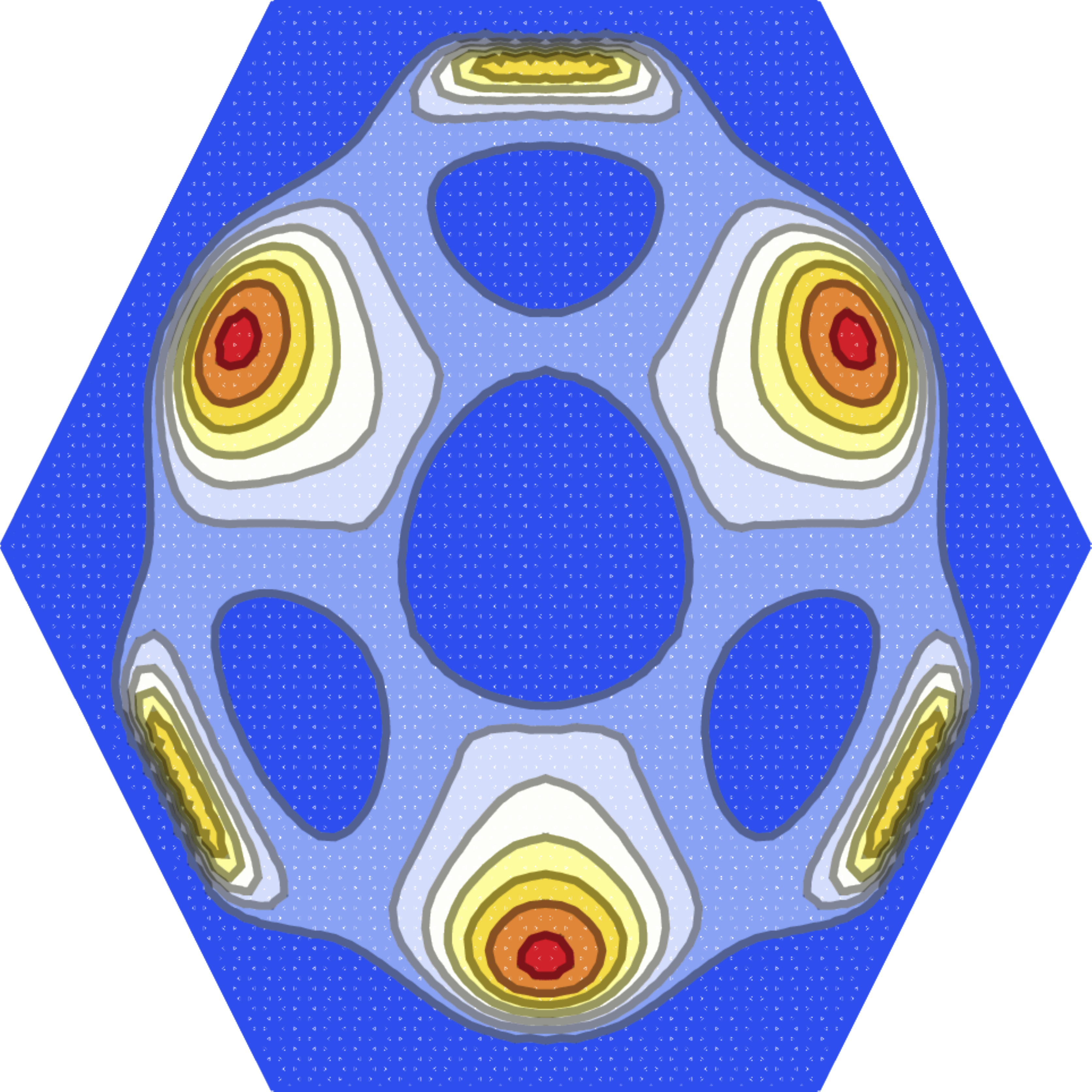}}\qquad\qquad
      \subbottom[\label{subfig_v3_50} BEEM v3: 50 layers, $\eta=0.025$ eV]{
	\includegraphics[width=0.37\linewidth]{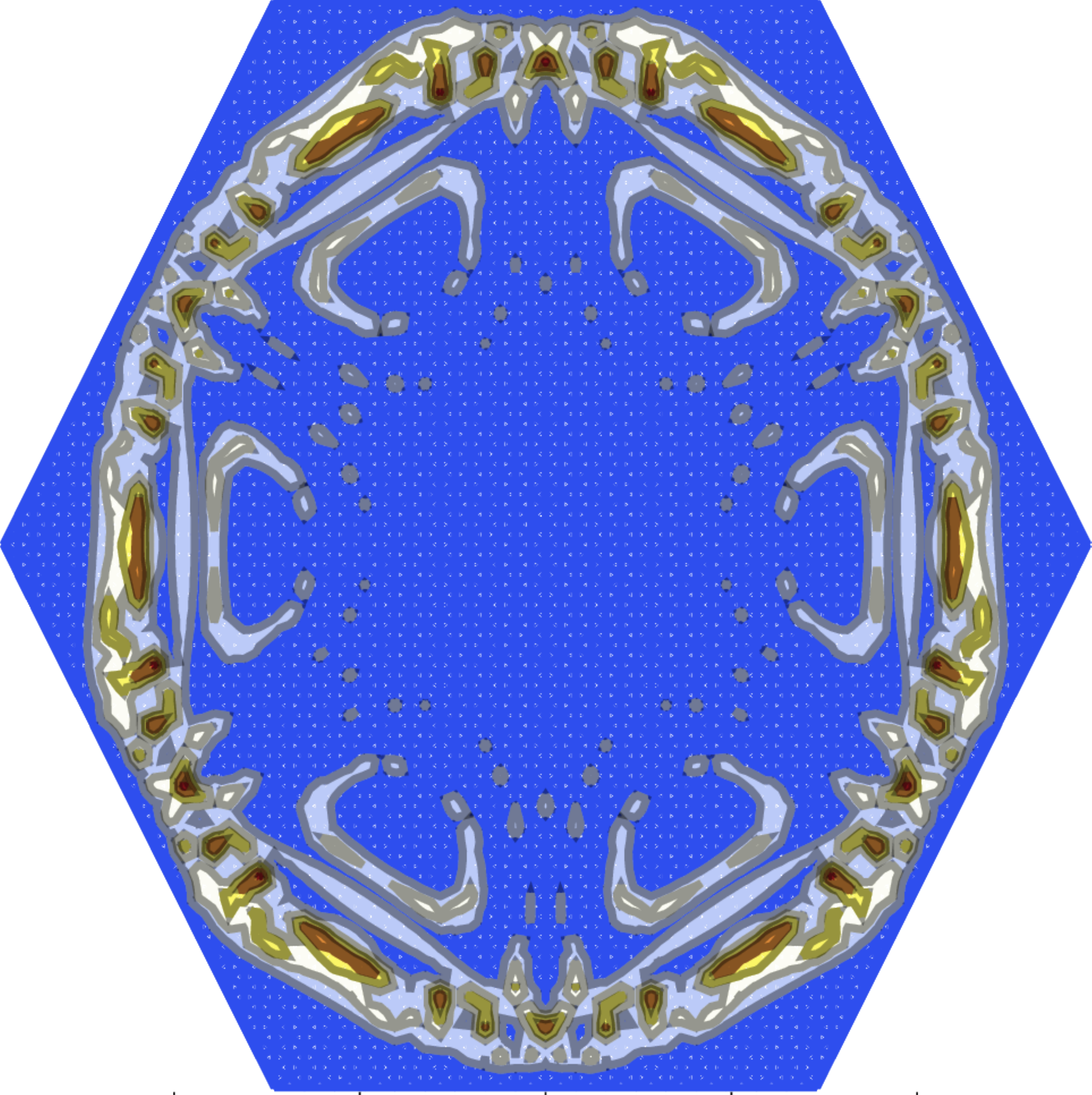}} \\	
      \caption{Distribution of the elastic current, 1 eV above the Fermi level, in Au(111) obtained through BEEM v2.1 and BEEM v3. Increasing $\eta$ or the number of layers leads to the loss of the 6-fold symmetry, that is to say the loss of the time-reversal symmetry, that is similar to equilibrium calculations.}
      \label{fig_summary_v2_vs_v3}
      \end{center}
    \end{figure}

    Regarding the previous semi-infinite approach, we have been able to reproduce similar behavior (Fig.~\ref{fig_summary_v2_vs_v3}): increasing $\eta$ leads to the lost of the time-reversal symmetry and then to the sixfold symmetry, which is similar to equilibrium calculation. And with our new approach, increasing the number of layers also leads to the loss of the time-reversal symmetry.
    
\section{Remark about the DOS-projection method}
  Finally I would like to make a small remark about the DOS projection on the distribution of the elastic current. Both equilibrium and non-equilibrium approach suppose that the BEEM current (not the elastic current) is the overlap between high current area and available density of states within the semi-conductor:
  \begin{equation}\label{eq_integral_dos}
    J_{B}(E) \propto \int \D\kpara \int_{\phi_{SB}}^{U_{gap}} \D E \ J_{n+1}(E,\kpara)\ T_{n+1,n+2}(E,\kpara)\ \rho_{n+2,n+2}(E,\kpara)
  \end{equation} 
  Where $T_{n+1,n+2}(E,\kpara)$ is a transmission coefficient that describes the propagation from the last layer $n+1$ of the metal to the first layer $n+2$ of the semi-conductor and where $\rho_{n+2,n+2}(E,\kpara)$ is the surface DOS of the semi-conductor. 
  
\begin{figure}[!hbt]
      \centering
      \subbottom[10 layers of Au(111), $\varepsilon=\varepsilon_F+ 0.8$eV]{
	\includegraphics[width=0.37\linewidth]{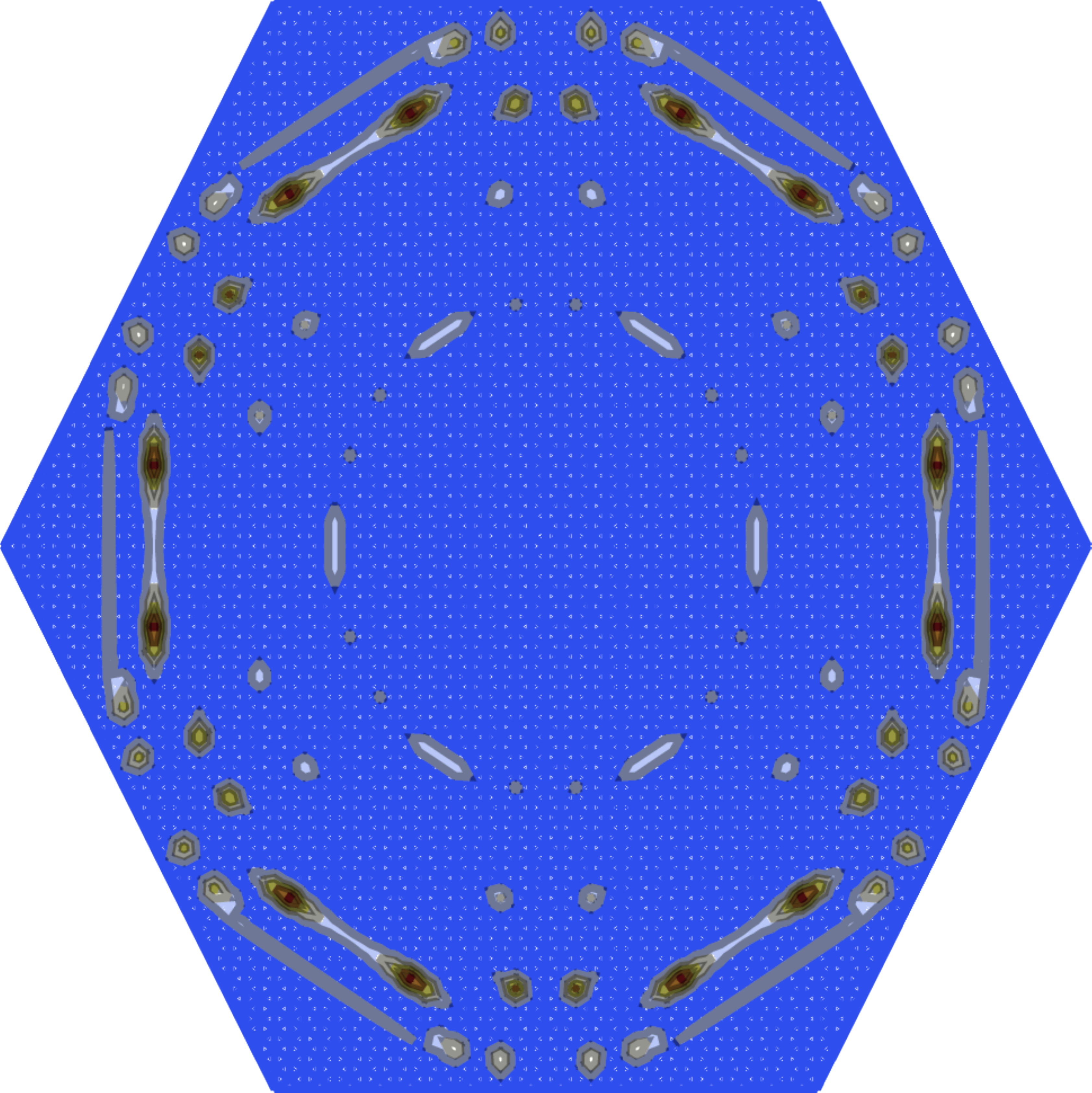}}\qquad\qquad
      \subbottom[200 layers of Au(111),  $\varepsilon=\varepsilon_F+ 0.8$eV]{
	\includegraphics[width=0.37\linewidth]{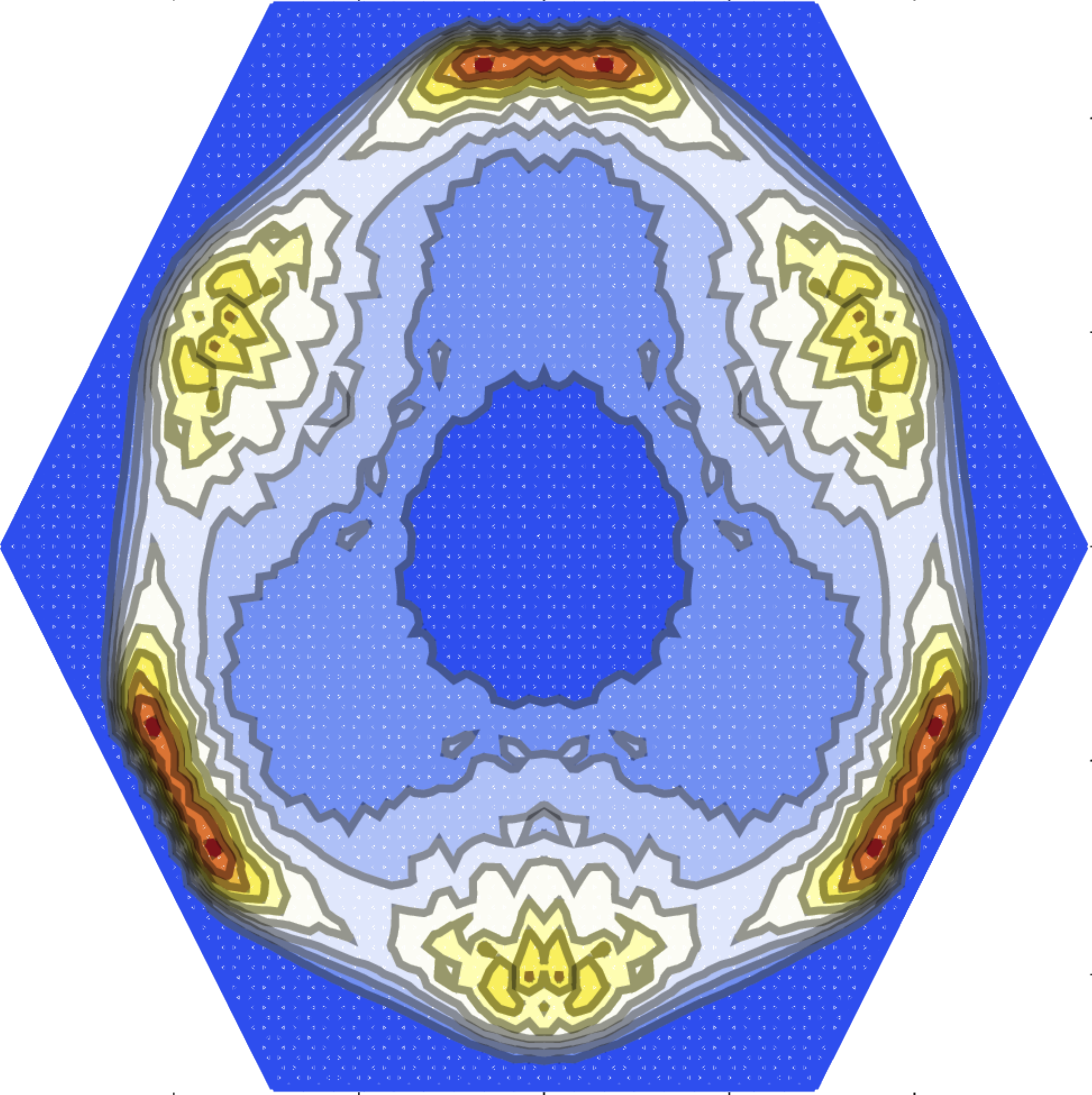}} \\
      \subbottom[10 layers of Au(111),  $\varepsilon=\varepsilon_F+ 1.0$eV]{
	\includegraphics[width=0.37\linewidth]{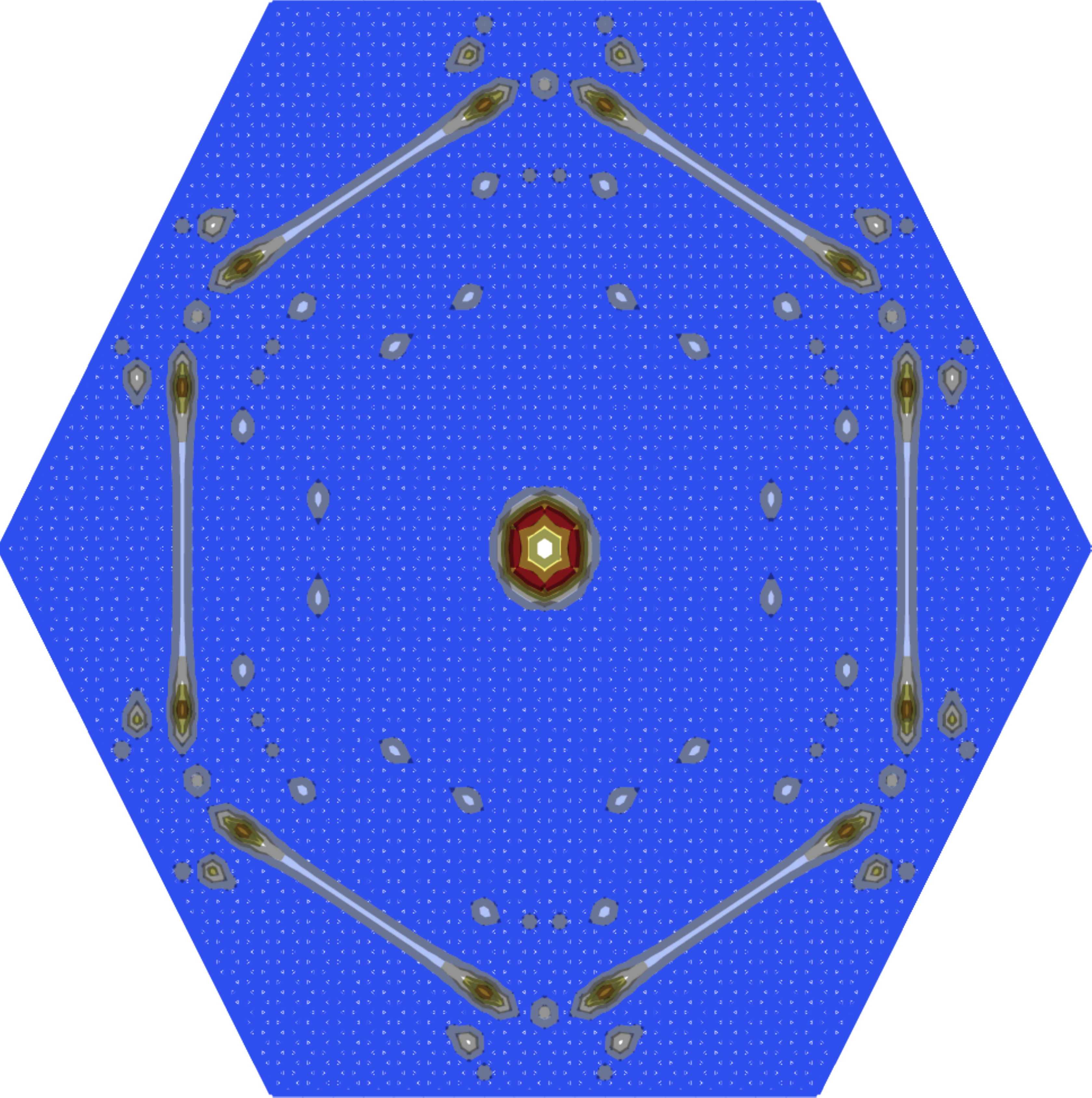}}\qquad\qquad
      \subbottom[200 layers of Au(111,) $\varepsilon=\varepsilon_F+ 1.0$eV]{
	\includegraphics[width=0.37\linewidth]{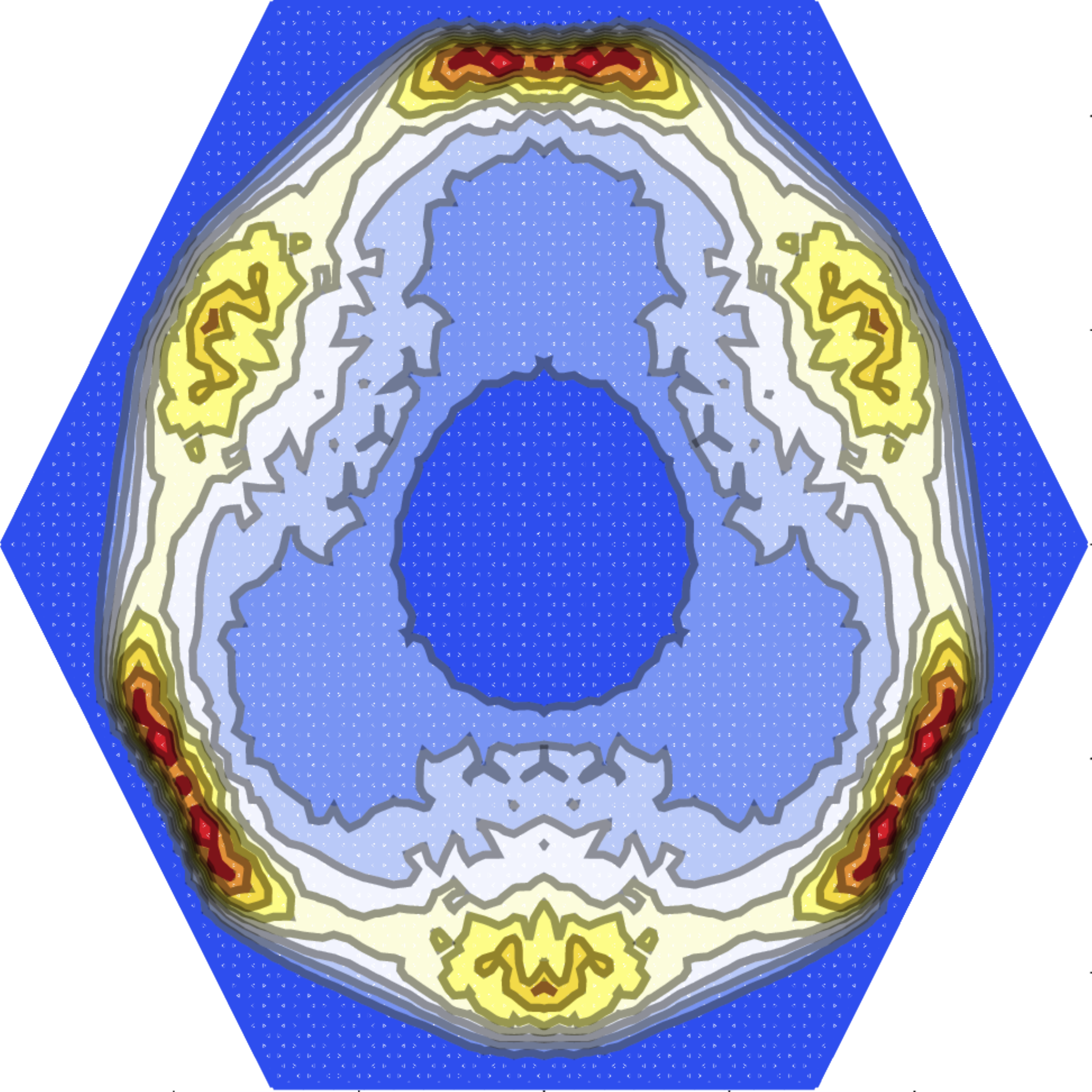}} \\
      \subbottom[10 layers of Au(111),  $\varepsilon=\varepsilon_F+ 1.2$eV]{
	\includegraphics[width=0.37\linewidth]{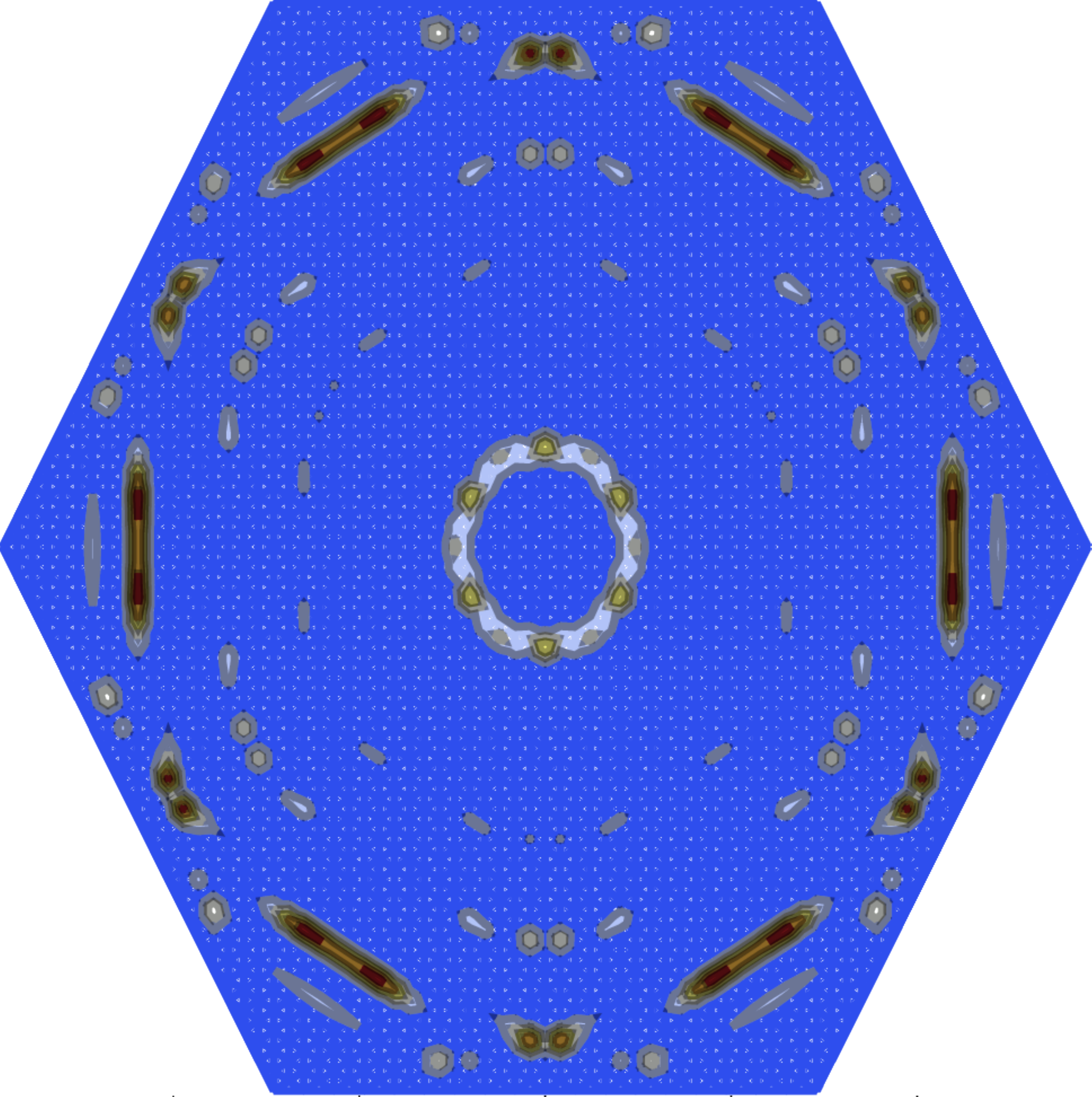}}\qquad\qquad
      \subbottom[200 layers of Au(111), $\varepsilon=\varepsilon_F+ 1.2$eV]{
	\includegraphics[width=0.37\linewidth]{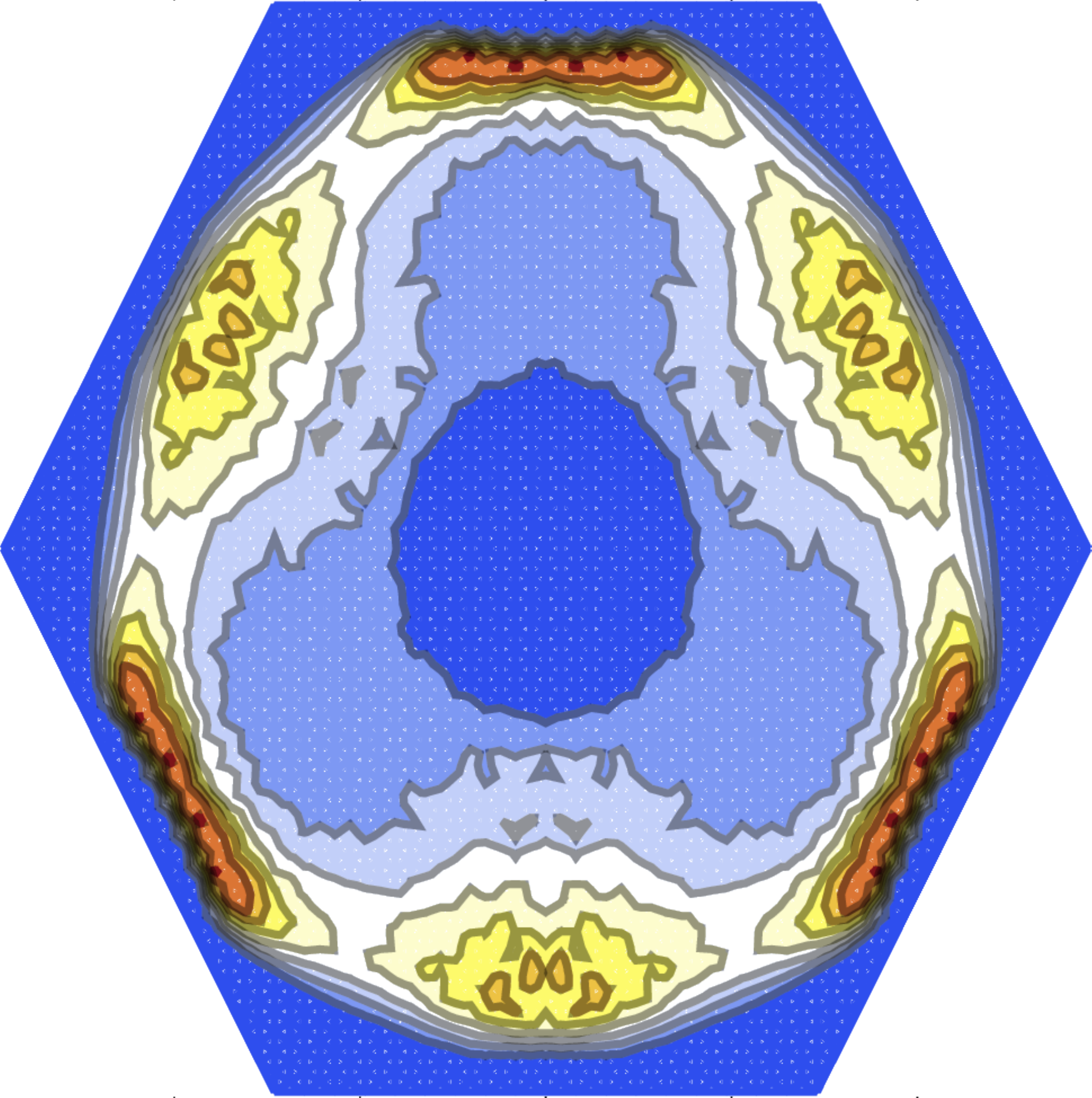}} \\	
      \caption{Distribution of the elastic current with respect to the energy for 10 and 200 layers of Au(111). The current does not vary too much for thick slabs, so it is possible to project the available DOS from the Schottky barrier to the bias, that is not the case for thin structures.}
      \label{fig_energy_variation}
    \end{figure}

  However, when we calculate the BEEM current, as in reference \cite{Reuter-PhysRevB.58.14036}, by summing the elastic current enclosed by the surface DOS, we are actually supposing that the transmission coefficient and the elastic current does not vary with the energy:
  \begin{equation}
    J_{B}(E) \propto \int \D\kpara J_{n+1}(U_{gap},\kpara) T_{n+1,n+2}(U_{gap},\kpara)  \int_{\phi_{SB}}^{U_{gap}} \D E \rho_{n+2,n+2}(E,\kpara)
  \end{equation}
  \emph{A priori} it is not true and we have to check this energy independence before projecting the DOS as we did. Figure \ref{fig_energy_variation} shows that indeed, for thick layers, the distribution of the elastic electrons does not vary too much with the energy. However, for thin films of gold it does. That is why, for this structure we have to evaluate properly the integral \eqref{eq_integral_dos}.
  
\chapter{Conclusions and perspectives}\label{chapt_perspectives}
    \lettrine[lines=3, lhang=0.25, loversize=0.77, findent=2.1em,nindent=-1.6em,slope=-1.8em,ante={\Huge\textquotedblleft}]{P}{lurality must never be posited without necessity''}.\\* Ockham's razor was one of the philosophy of this thesis. We wanted to check up to what limit non-equilibrium perturbation-theory was mandatory to describe Ballistic Electron Emission Microscopy. To this aim, first, we have seen indeed that using the equilibrium approach electrons follow the preferred directions of propagation of -24\degree\ and +30\degree\ with respect to the (111) direction of Au(111). Weighed against the -20\degree/+34\degree\ propagation of electrons inside a semi-infinite slab of Au(111), obtained through the transfer matrix procedure of BEEM v2.1, we can say that equilibrium approach is good enough for experimentalists' needs, in the case of thick films at the present level of experimental sensitivity. In second place, we have succeeded in describing qualitatively the magneto-current in Fe/Au/Fe/GaAs and Fe/Au/Fe/AlGaAs spin-valves using the band structures. Finally, using wave-function symmetry-considerations we have made a prediction on the magne-to-current for the spin-valves Fe/Au/Fe/GaAs versus Fe/Ag/Fe/GaAs. However, this result has to be confirmed experimentally as the spinvalve Fe/Ag/Fe/GaAs has not been studied yet.
  
  Equilibrium approach is a very convenient way to explain experiments or to make predictions just by looking at the band structures and for this it proves to be a very intuitive tool. 
  However, for thin films, one has to calculate the electronic structure of the thin slab. 
  As we cannot represent the band structure for a 2 dimensional crystal, it becomes far less intuitive.
  It is in such cases that the BEEM v3 code becomes extremely useful, as it describes the propagation of electrons also for extremely thin film depths and it allows to deal with the interfaces of different materials through the modified Harrison's approach (presented in chapter~\ref{chapt_results}).
  In particular, we have seen that, for a few-layer slab, the addition of a further layer can change drastically the BEEM current in the $\kpara$-plane due to constructive or destructive interference. For instance, in the case of Au(111), there is a high current at $\kpara=0$ for 12 layers, but there is no propagation at this wave-vector for 14 layers as shown in section \ref{sec_quantum_approach}. 
  In order to make predictions for thin films, it appears that this non-equilibrium approach gives new results that are not obtainable from the band structure. This layer-resolved effect could be very interesting to confirm the existence of inelastic effects or the conservation of $\kpara$ at interfaces: as for thin films of Au(111)/Si(111), no BEEM current is expected, if experimentally a current is measured, it could be only due to inelastic effects or to the non conservation of $\kpara$ at metal/semi-conductor interface. However this has to be confirm with low temperature experiments. As it is not that easy to perform low temperature experiments, we have also planned to consider electron/phonon interactions as presented in reference \cite{Andres-phonon_PhysRevB.66.075411} in order to see if these layer-resolved effects are smoothed by the temperature. We should also nuance the fact that the strong variations of the elastic current with respect to the number of layers could be observed only for ideal interfaces. The rugosity of Fe/Au interface is actually too large. For this reason, we think that it could be interesting to study Pb/Si as lead grows atomically flat on silicon. Currently it is probably the best candidate to see layer-resolved effects.

  Unfortunately, it has not been possible to end the study of the Fe/Au/Fe/GaAs spinvalve. Although the first preliminary results seem to show that, indeed, the magneto-current that has been observed experimentally is strongly influenced by interface effects, we have to vary the thickness of iron and/or gold slabs to see how their thickness influences the elastic current. In principle, the lower mean free path of minority spins is not critical here, as we are dealing with very thin films, but besides interface effects, we should also describe GMR by using different $\eta$ for minority and majority spins, while in our work we have used a constant $\eta$.

  Another issue that we could not deal with in this thesis, and that has to be considered at a later stage, concerns the tip. In this work, we have considered that the density of states of the tip was constant as well as the probability for electrons to jump from the tip to the sample. Moreover, we considered the tip localized at the origin. As in our calculation we worked in reciprocal space, the Fourier transform of this delta function is the identity. Therefore, in our work the tip only provided the bias and was a constant in $\kpara$-space. A more realistic tip (already foreseen at the level of the 2.1 version of the code) taking into account extended tunneling effects might be conceived.

  Finally, along the same guideline of a more realistic treatment, the BEEM v3 program should be extended in order to describe the metal/semiconductor interface through a tight-binding parametrization and by modeling the (thick) semiconductor with the decimation procedure of BEEM v2.1. In this way also the propagation through the semiconductor, neglected in this work where we focused on interface filtering effects, would be described.
  
  I would like to conclude coming back again on the adopted tight-binding approach through Green functions method. One of the elements of flexibility of this approach is that it allows turning on the electronic correlations by moving to the Hubbard (or Hubbard-like) Hamiltonian of section \ref{sec_hubbard_model}. Such a model is fundamental if we want to extend the BEEM current calculations to transition-metal oxides or quantum dots, or to some organic molecules grafted on the semi-conducting surface.
  The present approach allows for a straightforward generalization, once the retarded and advanced Green functions for the Hubbard model have been evaluated. This was a part of the initial plan that we could not complete: in fact, to this aim, during the work on the Keldysh formalism, I had worked in parallel to study the Hubbard model (two articles have been written on the model, one published, the other submitted, both reported in the Appendix).

\FloatBarrier
\appendix
  \chapter{The formalism of the second quantization for fermions}\label{appendix_second_quantization}
  \section{Definition}
  The creation operator acts on the Fock space by changing a state with $n$ particles in a state with $n+1$ particles. The action of the creation operator on a Slater determinant is defined by:
  \begin{equation}\label{creation}
  c_{\varphi}^{\dag}\Ket{\varphi_1 \ldots \varphi_n}_s = \sqrt{n+1} \Ket{\varphi\varphi_1\ldots\varphi_n}_s,
  \end{equation} 
  where $\Ket{\varphi_1 \ldots \varphi_n}_s$ is a Slater determinant of $n$ particles built from $\varphi_i$ orbitals from single-particle space.
  A fermion is added in the state $\Ket{\varphi_1}$ to a system of $n$ fermions without modifying the respective states.

  The annihilation operator is the adjoint of the creation operator:
  \begin{equation}
  c_{\varphi}=\left[c_{\varphi}^{\dag}\right]^{\dag}
  \end{equation}
  His action on a state with $n$ particles $\Ket{\varphi_1\ldots\varphi_n}_s$ is obtained by calculating the scalar product:
  \begin{equation}\label{eq_A3}
  \Braket{\psi_1\ldots\psi_m| c_{\varphi} | \varphi_1\ldots\varphi_n}_s=\sqrt{m+1}\Braket{\varphi\psi_1\ldots\psi_m | \varphi_1\ldots\varphi_n}_s
  \end{equation}

  For states with 
  $m\neq n-1$, this scalar product is zero. And in particular, for all $m$ if $n=0$. In other word, the vacuum state is an eigenvector of $c_{\varphi}$ with eigenvalue 0.
  \begin{equation}
  c_{\varphi}\Ket{0}=0
  \end{equation}

  For $m=$ ($n\neq 0$), by switching $\varphi$ with $\psi_1\ldots \psi_{n-1}$ and by noting $\varphi=\psi_n$ Eq.~\eqref{eq_A3} becomes:
  \begin{equation}\label{avant_permut}
  \Braket{\psi_1\ldots\psi_{n-1}| c_{\varphi} | \varphi_1\ldots\varphi_n}_s = \sqrt{n} (-1)^{n-1} \Braket{\psi_1\ldots\psi_n| \varphi_1\ldots\varphi_n}_s
  \end{equation}
  The Slater determinant guarantees the anti-symmetrization of the wave function: by switching two row or column, the sign of the determinant is changed. In addition, the scalar product of states resulting from a projection written
  \begin{align}\label{etat_projecte}
  \Braket{\psi_1\ldots\psi_n| \varphi_1\ldots\varphi_n}_s &= \Braket{\psi_1 \otimes \ldots \otimes \psi_n | S^*_- S_- | \varphi_1 \otimes \ldots \otimes \varphi_n} \nonumber\\
  &= \Braket{\psi_1 \otimes \ldots \otimes \psi_n | S^*_- S_- |rewritten \varphi_1 \otimes \ldots \otimes \varphi_n} \nonumber\\
  &= \frac{1}{n!}\sum_{\pi\in \mathcal{P}_n} (-1)^{\pi} \Braket{\psi_1 \otimes \ldots \otimes \psi_n | \varphi_{\pi(1)} \otimes \ldots \otimes \varphi_{\pi(n)}} \nonumber\\
    &= \frac{1}{n!}\sum_{\pi\in \mathcal{P}_n} (-1)^{\pi} \Braket{\psi_1|\varphi_{\pi(1)} }\ldots \Braket{ \psi_n | \varphi_{\pi(n)}}.
  \end{align}
  Where $\pi$ is an element of the permutation group $\mathcal{P}_n$ of $n$ elements and where
  \begin{equation}\label{antisymmetrisation}
  S_-=\frac{1}{n!}\sum_{\pi\in \mathcal{P}_n} (-1)^{\pi} P_{\pi}
  \end{equation} 
  is the anti-symmetrization operator. Eq.~(\ref{avant_permut}) is then :
  \begin{multline}\label{apres_permut}
  \Braket{\psi_1\ldots\psi_{n-1}| c_{\varphi} | \varphi_1\ldots\varphi_n}_s = \frac{1}{\sqrt{n}} \frac{(-1)^{n-1}}{(n-1)!} \sum_{\pi\in \mathcal{P}_n} (-1)^{\pi} \Braket{\psi_1|\varphi_{\pi(1)} }\ldots \Braket{ \psi_n | \varphi_{\pi(n)}}.\\
  = \frac{1}{\sqrt{n}}\sum_{i=1}^n (-1)^{i-1} \Braket{\psi_n| \varphi_i} \frac{1}{(n-1)!} \sum_{\pi\in \mathcal{P}_n} (-1)^{\pi} \Braket{\psi_1|\varphi_{\pi(1)} }\ldots \Braket{ \psi_{n-1} | \varphi_{\pi(n-1)}}
  \end{multline}
  considering that the sum over the permutation  $\mathcal{P}_n$ is equal to the sum over the $(n-1)!$ permutations such as $\pi(n)=i$ and then to sum from $i=1$ to $n$. Hence, after the last term of  (\ref{apres_permut}), $\{\pi(1),\ldots,\pi(n-1)\}$ is a permutation of $\{1,\ldots,i-1,i+1,\ldots, n\}$. Using (\ref{etat_projecte}), one obtains:
  \begin{multline}
    \Braket{\psi_1\ldots\psi_{n-1}| c_{\varphi} | \varphi_1\ldots\varphi_n}_s = \\ \frac{1}{\sqrt{n}}\sum_{i=1}^n (-1)^{i-1}\Braket{\psi_1\ldots\psi_{n-1} | \varphi_1\ldots \varphi_{i-1}\varphi_{i+1}\ldots, \varphi_n}_s \Braket{\varphi| \varphi_i}
  \end{multline} 
  
  Finally, this relation being valid for all states $\Ket{\psi_1\ldots\psi_{n-1}}$, the annihilation operator reduces the number of particle by one by keeping the symmetry of the state:
  \begin{equation}\label{annihilation}
  c_{\varphi}\Ket{\varphi_1 \ldots \varphi_n}_s = \frac{1}{\sqrt{n}} \sum_i (-1)^{i-1} \Braket{\varphi|\varphi_i} \Ket{\varphi_1 \ldots \varphi_{i-1}\varphi_{i+1} \ldots \varphi_n}_s 
  \end{equation}

  \section{Anti-commutation rules}\label{section_anticommutation}
    In this language
    \begin{align}
      \creation{\varphi}\creation{\psi} \Ket{0} &= \frac{1}{\sqrt{2}}(\Ket{\varphi}\Ket{\psi}-\Ket{\psi}\Ket{\varphi})
      \\                                               &= \Ket{\varphi,\psi} = -\Ket{\psi,\varphi}
    \end{align}
    The state on the right hand side is normalized and antisymmetric. There are two copies of the one particle Hilbert space. In one component of the wave-function, the particle in the first copy is at $\Ket{\varphi}$, in the other component it is at $\Ket{\psi}$. The two-body wave-function $\Braket{r,r'|\phi}$ is antisymmetric and in the case where there are only two one-particle states that are occupied, it is a Slater determinant.      Clearly, that can become a mess. In term of creation and annihilation operators however, all one needs to know is that by definition of these operators,
    \begin{align}
      \creation{\varphi}\creation{\psi}+\creation{\psi}\creation{}(\varphi) & =0
      \\ \boxedalign{ \left\{ \creation{\varphi},\creation{\psi} \right\} &= 0}\label{eq_A14}
    \end{align}
    and by taking the adjoint
    \begin{equation}\fbox{$
      \left\{\annihilation{\varphi},\annihilation{\psi} \right\} = 0$}\label{eq_A15}
    \end{equation}
    The missing relation is
    \begin{equation}\fbox{$
      \left\{\annihilation{\varphi},\creation{\psi} \right\} = \Braket{\phi|\psi}\identite$}\label{eq_A16}
    \end{equation} 
    These three anti-commutation rules are demonstrated below.
    
    Demonstration of the the anti-commutation rule of the creation operator:
    \begin{align}
    c_{\varphi}^{\dag}c^{\dag}_{\psi} \Ket{\varphi_1 \ldots \varphi_n}_s = \sqrt{(n+1)(n+2)}\Ket{\varphi \psi \varphi_1\ldots\varphi_n}_s
    \end{align}
    Because the Slater determinant guarantees the anti-symmetry of the wave function,  the sign changes by switching $\varphi$ and $\psi$:
    \begin{align}\label{commutation_creation}
    c^{\dag}_{\psi}c_{\varphi}^{\dag} \Ket{\varphi_1 \ldots \varphi_n}_s &= \sqrt{(n+1)(n+2)}\Ket{\psi \varphi \varphi_1\ldots\varphi_n}_s \nonumber\\
      &= - \sqrt{(n+1)(n+2)}\Ket{\varphi \psi \varphi_1\ldots\varphi_n}_s \nonumber\\
      &= -c_{\varphi}^{\dag} c^{\dag}_{\psi} \Ket{\varphi_1\ldots\varphi_n}_s
    \end{align}
    From this we recover Eq.~\eqref{eq_A14}:
    \begin{equation}\fbox{$
      c_{\varphi}^{\dag}c^{\dag}_{\psi} + c^{\dag}_{\psi}c_{\varphi}^{\dag} = 0$}
    \end{equation}
    and by taking the adjoint Eq.~\eqref{eq_A15}:
    \begin{equation}\label{commutation_annihilation}\fbox{$
      c_{\varphi}c_{\psi} + c_{\psi}c_{\varphi} = 0$}
    \end{equation}
   These rules imply the Pauli principle: it is not possible to create two fermions in the same states.
   
    For the last commutation rule:
    \begin{align}
      c_{\varphi}^{\dag}c_{\psi} \Ket{\varphi_1 \ldots \varphi_n}_s &= \frac{1}{\sqrt{n}} c_{\varphi}^{\dag} \sum_i (-1)^{i-1} \Braket{\psi|\varphi_i} \Ket{\varphi_1 \ldots \varphi_{i-1} \varphi_{i+1}\ldots \varphi_n}_s \nonumber\\
	&= \sum_i (-1)^{i-1} \Braket{\psi|\varphi_i} \Ket{\varphi \varphi_1 \ldots \varphi_{i-1} \varphi_{i+1}\ldots \varphi_n}_s
    \end{align}
    Making $(i-1)$ permutations:
    \begin{align}\label{ccroixc}
    c_{\varphi}^{\dag}c_{\psi} \Ket{\varphi_1 \ldots \varphi_n}_s &= \sum_i \Braket{\psi|\varphi_i} \Ket{\varphi_1\ldots \varphi_{i-1} \varphi \varphi_{i+1}\ldots \varphi_n}_s
    \end{align}
    then applying the opposite combination:
    \begin{align}
      c_{\psi}c_{\varphi}^{\dag}\Ket{\varphi_1 \ldots \varphi_n}_s &= \sqrt{n+1} c_{\psi} \Ket{\varphi \varphi_1\ldots \varphi_n}_s \nonumber\\
      &= \Braket{\psi|\varphi }\Ket{\varphi_1 \ldots \varphi_n}_s + \sum_i (-1)^i \Braket{\psi|\varphi_i} \Ket{ \varphi \varphi_1\ldots \varphi_{i-1} \varphi_{i+1}\ldots \varphi_n}_s \nonumber\\
    \end{align}
    Making again $i-1$ permutations:
    \begin{align}
      c_{\psi}c_{\varphi}^{\dag}\Ket{\varphi_1 \ldots \varphi_n}_s  &= \Braket{\psi|\varphi }\Ket{\varphi_1 \ldots \varphi_n}_s - \sum_i \Braket{\psi|\varphi_i} \Ket{ \varphi_1\ldots \varphi_{i-1} \varphi \varphi_{i+1}\ldots \varphi_n}_s
    \end{align} 
    The final relation is Eq.~\eqref{eq_A15}:
    \begin{align}\label{commutation}\fbox{$
      c_{\psi}c_{\varphi}^{\dag} + c_{\varphi}^{\dag} c_{\psi} = \Braket{\psi|\varphi} \identite$}
    \end{align}

    \section{Change of basis}
      A useful formula for these field operators is the formula for the change of basis. Considering a new complete basis of single-particle states $\alpha$, then the change of basis is:
      \begin{equation}
       \Ket{r} = \sum_{\alpha} \Ket{\alpha}\Braket{\alpha | r}
      \end{equation} 
      Given the definition of the creation operator, the creation operator $\psi^{\dagger}(r)$ for a particle in state $r$ is related to the creation operator $\creation{\alpha}$ for a particle in state $\alpha$ by the analogous formula
      \begin{equation}
       \psi^{\dagger}(r) = \sum_{\alpha} \creation{\alpha}\Braket{\alpha | r}
      \end{equation}

    \section{Second quantization Hamiltonian}
      \subsection{One body operator}
	Define a single-particle observable $A$ and the one-body operator $A(n)=\sum_i A(\vec{r}_i)$ associated to the $n$ particle system. As  $A(n)$ commute with the permutation operators, the action of the observable $A$ on the Fock space's states is 
	\begin{align}\label{Aket}
	  A\Ket{\varphi_1 \ldots \varphi_n}_s = \sum_j \Ket{\varphi_1, \ldots, \varphi_{j-1}, A\varphi_i, \varphi_{j+1}, \ldots, \varphi_n}_s \nonumber\\
	\end{align}
	And as
	\begin{equation}\label{phi_i}
	  A\Ket{\varphi_j}= \sum_i \underbrace{\Braket{\varphi_i| A | \varphi_j}}_{A_{ij}} \Ket{\varphi_i}
	\end{equation} 
	where the basis  $\{\Ket{\varphi_j}\}$ is complete, one can write (\ref{Aket}) as:
	\begin{equation}\label{aijket}
	  A\Ket{\varphi_1 \ldots \varphi_n}_s = \sum_{i,j} A_{ij} \Ket{\varphi_1, \ldots, \varphi_{j-1}, \varphi_j, \varphi_{j+1}, \ldots, \varphi_n}_s
	\end{equation} 
	Using the properties of linearity of Slater determinant. Identifying eq.~\ref{ccroixc}, (\ref{aijket}) becomes
	\begin{equation}
	  A\Ket{\varphi_1 \ldots \varphi_n}_s = \sum_{i,j} A_{ij} c^{\dag}_i c_j \Ket{\varphi_1 \ldots \varphi_n}_s 
	\end{equation} 
	The one-body operators, then, can be written with ladder operators:
	\begin{equation}\label{operateur_un_corps}
	  A=\sum_{i,j} A_{ij} c^{\dag}_i c_j
	\end{equation} 
	If $\varphi_i$ in Eq.~(\ref{phi_i}) are choosen to be eigenvectors of $A$, then (\ref{operateur_un_corps}) is simply:
	\begin{equation}
	  A=\sum_i \alpha_i c^{\dag}_i c_i
	\end{equation}

      \subsection{Two-body operator}
	Consider an operator $V$ which acts on two-particle space. The two-body observable of a $n$ particle state of Fock space is given by:
	\begin{equation}\label{observable_2corps}
	  V\Ket{\varphi_1\ldots \varphi_n}_s = \sum_{i<j} V_{ij} \Ket{\varphi_1\ldots \varphi_n}s 
	\end{equation} 
	We want to show that $V$ can be written in term of ladder operator:
	\begin{equation}\label{operateur_2corps}
	  V=\frac{1}{2} \sum_{d,e,f,g} \Braket{de| V | fg} c^{\dag}_d c^{\dag}_e c_g c_f
	\end{equation} 
	where 
	\begin{align}\label{representation_reelle}
	  \Braket{de| V | fg} &=\Braket{\psi_d \otimes \psi_e | V | \varphi_f \otimes \varphi_g}\nonumber\\
	                      &=\int \D \vec{r} \int \D \vec{r}~' \, \psi^*_d(\vec{r})\psi^*_e(\vec{r}~') V(\vec{r}-\vec{r}~') \varphi_f(\vec{r})\varphi_g(\vec{r}~')
	\end{align}
	
	In this way, applying twice (\ref{creation}) and (\ref{annihilation}):
	\begin{multline}
	  c^{\dag}_d c^{\dag}_e c_f c_g \Ket{\varphi_1 \ldots \varphi_n}_s = c^{\dag}_d c^{\dag}_e \\
	   \cdot \left\{ \sum_{i=1}^n \sum_{j=1}^{i-1} (-1)^{i-1} (-1)^{j-1} \Ket{\varphi_1\ldots\varphi_{j-1}\varphi_{j+1} \ldots \varphi_{i-1} \varphi_{i+1} \ldots \varphi_n}_s \Braket{\psi_f| \varphi_i}\Braket{\varphi_g| \varphi_j}    + \right.\\
	   \left. \sum_{i=1}^n \sum_{j=i+1}^n (-1)^{i-1} (-1)^{j-2} \Ket{\varphi_1\ldots\varphi_{i-1}\varphi_{i+1} \ldots \varphi_{j-1} \varphi_{j+1} \ldots \varphi_n}_s \Braket{\psi_f| \varphi_i}\Braket{\varphi_g| \varphi_j} \right\} \\
	  =\sum_{i=1}^n \sum_{j=1}^{i-1} \Ket{\varphi_1\ldots\varphi_{j-1}\psi_e\varphi_{j+1} \ldots \varphi_{i-1}\psi_d \varphi_{i+1} \ldots \varphi_n}_s \Braket{\psi_f\otimes \psi_g| \varphi_i\otimes\varphi_j} + \\
	   \sum_{i=1}^n \sum_{j=i+1}^n \Ket{\varphi_1\ldots\varphi_{i-1}\psi_d\varphi_{i+1} \ldots \varphi_{j-1}\psi_e \varphi_{j+1} \ldots \varphi_n}_s\Braket{\psi_f\otimes \psi_g| \varphi_i\otimes\varphi_j}
	\end{multline}
	and multiplying the term for which $j<i$ by matrix elements $ \Braket{de| V | fg}$ and summing over $d,e,f,g$
	\begin{multline}\label{terme_ij}
	  \sum_{d,e,f,g} \Ket{\varphi_1\ldots\varphi_{j-1}\psi_e\varphi_{j+1} \ldots \varphi_{i-1}\psi_d \varphi_{i+1} \ldots \varphi_n}_s \\ \Braket{\psi_d\otimes\psi_e | V | \psi_f \otimes \psi_g}  \Braket{\psi_f\otimes \psi_g| \varphi_i\otimes\varphi_j}
	\end{multline}
	The basis $\{\Ket{\psi_f \otimes \psi_g}\}$ being complete ($ \sum_k \Ket{\psi_k} \Bra{\psi_k} = \identite$), it gives
	\begin{equation}\label{permutation_particules}
	  (\ref{terme_ij}) = \sum_{d,e} \Ket{\varphi_1\ldots\varphi_{j-1}\psi_e\varphi_{j+1} \ldots \varphi_{i-1}\psi_d \varphi_{i+1} \ldots \varphi_n}_s \Braket{\psi_d\otimes\psi_e | V | \varphi_i\otimes\varphi_j}
	\end{equation} 
	On the other hand, $V$ is invariant under permutation of two particles $$\Braket{\psi_d\otimes\psi_e | V | \psi_f \otimes\psi_g}=\Braket{\psi_e\otimes\psi_d | V | \psi_g\otimes \psi_f}$$. 
	\begin{equation}\label{somme_de}
	  (\ref{permutation_particules}) = \sum_{d,e} \Ket{\varphi_1\ldots\varphi_{j-1}\psi_d\varphi_{j+1} \ldots \varphi_{i-1}\psi_e \varphi_{i+1} \ldots \varphi_n}_s \Braket{\psi_d\otimes\psi_e | V | \varphi_j\otimes\varphi_i}
	\end{equation} 
	And as after (\ref{representation_reelle})
	\begin{multline}
	  \Ket{\varphi_1\ldots\varphi_{j-1}\psi_d\varphi_{j+1} \ldots \varphi_{i-1}\psi_e \varphi_{i+1} \ldots \varphi_n}_s =\\
	  S_-\left[\varphi_1(\vec{r}_1)\ldots \varphi_{j-1}(\vec{r}_{j-1})\psi_d(\vec{r}_j) \varphi_{j+1}(\vec{r}_{j+1}) \ldots \varphi_{i-1}(\vec{r}_{i-1})\psi_e(\vec{r}_i)\varphi_{i+1}(\vec{r}_{i+1})\ldots \varphi_n(\vec{r}_n)\right]
	\end{multline}
	(\ref{somme_de}) reads
	\begin{align}\label{svket}
	  &\int \D \vec{r} \int \D \vec{r}~' \, V(\vec{r}-\vec{r}')\, S_-\bigg[ \varphi_1(\vec{r}_1)\ldots \varphi_{j-1}(\vec{r}_{j-1})\overbrace{\bigg( \sum_d\psi^*_d(\vec{r})\psi_d(\vec{r}_j)\bigg)}^{\delta(\vec{r}-\vec{r}_j)}\varphi_{j+1}(\vec{r}_{j+1}) \ldots \nonumber\\  &\qquad\quad \ldots \varphi_{i-1}(\vec{r}_{i-1})\underbrace{\bigg(\sum_e\psi^*_e(\vec{r}~')\psi_e(\vec{r}_j)\bigg)}_{\delta(\vec{r}'-\vec{r}_i)}\varphi_{i+1}(\vec{r}_{i+1})\ldots \varphi_n(\vec{r}_n) \bigg]\nonumber\\
	  &=\,V(\vec{r}_j -\vec{r}_i)\varphi_1(\vec{r}_1)\ldots \varphi_{j-1}(\vec{r}_{j-1})\varphi_j(\vec{r}_j) \varphi_{j+1}(\vec{r}_{j+1})\ldots \nonumber\\& \quad \ldots\varphi_{i-1}(\vec{r}_{i-1})\varphi_i(\vec{r}_i)\varphi_{i+1}(\vec{r}_{i+1})\ldots \varphi_n(\vec{r}_n) \nonumber\\
	  &=S_- V_{ji} \Ket{\varphi_1 \otimes \ldots \otimes \varphi_n}_s 
	\end{align}
	Here once again, the condition of the complete basis has been used. A similar result holds for the other part of the sum $j>i$. Combining  (\ref{operateur_2corps}), (\ref{somme_de}) and (\ref{svket}), and by remembering that $V_{ij}=V_{ji}$, one obtains:
	\begin{align}
	  V\Ket{\varphi_1\ldots\varphi_n}_s &= \left[ \frac{1}{2} \sum_{i=1}^n \sum_{j=1}^{i-1} V_{ji} \Ket{\varphi_1 \otimes \ldots \otimes \varphi_n}_s + \frac{1}{2}\sum_{i=1}^n \sum_{j=i+1}^{n} V_{ij} \Ket{\varphi_1 \otimes \ldots \otimes \varphi_n}_s \right]\nonumber\\
	  &= \frac{1}{2} \sum_{i\neq j}^n V_{ji} \Ket{\varphi_1 \otimes \ldots \otimes \varphi_n}_s 
	\end{align}
	which is the same as (\ref{observable_2corps}) hence $\sum_{i\neq j}^n V_{ji}$ commute with all the permutation operators.
	
	As a consequence, a two-body operator writes
	\begin{equation}\label{terme_2corps}
	  V=\frac{1}{2} \sum_{d,e,f,g} \Braket{de| V | fg} c^{\dag}_d c^{\dag}_e c_g c_f
	\end{equation}
	within the second quantization formalism.

\chapter{Mathematical tricks}

  \section{Fourier transform of a Green's function}\label{appendix_green_fourier_transform}

  The Fourier transform is defined by
  \begin{align}
      G(\omega) &=  \int_{-\infty}^{+\infty} G(t) \e^{\i\omega t} \D t
    \\ G(t)      &= \frac{1}{{2\pi}} \int_{-\infty}^{+\infty} G(\omega) \e^{-\i\omega t} \D \omega
  \end{align}
  If we can exchange time-derivative and $\omega$-integral, we get: 
  \begin{align}
    \i\hbar\partial_t G(t) &= \frac{\i\hbar}{{2\pi}} \int_{-\infty}^{+\infty} -i\omega G(\omega)  \e^{-\i\omega t} \D \omega
    \\                     &= \frac{1}{{2\pi}} \int_{-\infty}^{+\infty}\hbar\omega G(\omega)  \e^{-\i\omega t} \D \omega
  \end{align}

  The Fourier transform of Dirac delta function is
  \begin{align}\label{eq_A54}
    \delta(\omega) &= \int_{-\infty}^{+\infty}  \e^{\i\omega t} \D t
  \end{align}
  From the definition of $\delta(t)$.  Eq.~\eqref{eq_A54} is the origin of the finite jump in the Green functions at $t-t'=0$.

\section{Alternative derivation of Dyson equation for retarded and advanced Green functions}\label{appendix_alternate_gf}

    It exists a simple way to find the expansion for the retarded (and advanced) Green function. Though this simplicity hides all the subtleties of perturbation theory, for completeness, we present it here. 
    
    If the Hamiltonian is diagonalizable then, the retarded Green function can be obtained through
    \begin{equation}\label{eq_Gr_omega}
      \hat{G}^{R}(\omega) = \frac{1}{\omega-\hat{H}+\i\eta}
    \end{equation}
    We want to develop a perturbation method that allows to evaluate the retarded Green function which describes the propagation of an electron within a potential, in the case where one part of the Hamiltonian, $\hat{H}_0$, can be diagonalized while the other part, $\hat{V}$, cannot. The present approach is limited to the case where both $\hat{H}_0$ and the perturbation $V$ are time independent. We start from Eq.~\eqref{eq_Gr_omega} written as follows:
    \begin{equation}
      (\omega + \i\eta - \hat{H}_0 -\hat{V}) \hat{G}^R(\omega) = 1
    \end{equation}
    Putting the perturbation on the right-hand side and using the definition of the unperturbed Green function
    \begin{equation}
      \hat{g}_0^R(\omega) = \frac{1}{\omega-\hat{H}_0+\i\eta}
    \end{equation}
    we have
    \begin{equation}
      \left[\hat{g}_0^R(\omega)\right]^{-1} \hat{G}^R(\omega) = 1 + \hat{V}\hat{G}^R(\omega)
    \end{equation}
    by multiplying by $\hat{g}_0^R(\omega)$ it gives the equation
    \begin{equation}
      \hat{G}^R(\omega) = \hat{g}_0^R(\omega) + \hat{g}_0^R(\omega)  \hat{V}\hat{G}^R(\omega)
    \end{equation}
    whose solution is:
    \begin{equation}
      \hat{G}^R(\omega) = \left[{1-\hat{g}_0^R(\omega) \hat{V}}\right]^{-1}\hat{g}_0^R(\omega)
    \end{equation} 
    Its perturbation expansion writes:
    \begin{equation}
      \hat{G}^R(\omega) = \hat{g}_0^R(\omega) + \hat{g}_0^R(\omega)  \hat{V} \hat{g}_0^R(\omega) +\hat{g}_0^R(\omega)  \hat{V} \hat{g}_0^R(\omega)  \hat{V} \hat{g}_0^R(\omega) +\dots
    \end{equation}
    which coressponds to Dyson equation of section \ref{sec_pert_expansion}.

\section{Heisenberg's equation of motion of the particle number operator $\hat{n}$}\label{appendix_commutator}

Solving the equation of motion of the occupation number by calculating the commutator between the occupation number operator and a tight binding hamiltonian:
\begin{align}
 J_{lm} &= -e \braket{\partial_t \hat{n}_l}
 \\    &= \frac{\i e}{\hbar} \braket{[\hat{n}_l,H]}
 \\  \boxedalign{ &= \frac{\i e}{\hbar} \braket{[c^{\dagger}_{l}c_{l},\sum_{m,n}T_{m,n}c^{\dagger}_{m}c_{n}]}} \label{eq_of_motion_n_appendix}
\end{align}
In order to solve this equation of motion, we have to express the last formula with anti-commutators of 2 operators. It can be done by using the following identities:
\begin{align}
 [AB,CD] &= AB\,CD-CD\,AB
 \\      &= A\,BC\,D-BC\,A\,D+B\,CA\,D-CA\,B\,D+C\,AB\,D-C\,D\,AB
 \\      &= [A,BC]D+[B,CA]D+C[AB,D] \label{commutator_4op}
\end{align}
\begin{itemize}
  \item the two first commutators of 3 operator can be expressed as
        \begin{align}
        [A,BC] &= ABC-BCA
        \\     &= ABC-BAC+BAC-BCA
        \\     &= [A,B]C + B[A,C] \label{commutator_3op}
        \end{align}
        Which can be rewritten with anti-commutators:
        \begin{equation}\label{anti-commutator_3op}
        ABC-BAC+BAC-BCA = \{A,B\}C-B\{C,A\}
        \end{equation}
  \item The third commutator of 3 operators in eq. (\ref{commutator_4op}) can be rewrite as
        \begin{align}
        [AB,D] &= ABD-DAB \nonumber
        \\     &= ABD -BDA+BDA -DAB \nonumber
        \\     &= [A,BD]+[B,DA]
        \end{align}
        using eq. (\ref{commutator_3op}) and (\ref{anti-commutator_3op}) it gives
        \begin{equation}
        [AB,D] =  \{A,B\}D-B\{D,A\} + \{B,D\}A-D\{A,B\}
        \end{equation}

        Then, eq. (\ref{commutator_4op}) becomes
        \begin{align}
        [AB,CD] =& \{A,B\}CD-B\{C,A\}D + \{B,C\}AD-C\{A,B\}D+C\{A,B\}D \notag
        \\       & -CB\{D,A\} +C\{B,D\}A-CD\{A,B\} \nonumber
        \\      =& \{A,B\}CD - CD\{A,B\} -B\{C,A\}D +C\{B,D\}A + \{B,C\}AD -CB\{D,A\}
        \end{align}
\end{itemize}

Applying those identities to eq. (\ref{eq_of_motion_n_appendix})
\begin{align}
 J_{lm} = \frac{\i e}{\hbar}           &\braket{[c^{\dagger}_{l}c_{l},\sum_{m,n}T_{m,n}c^{\dagger}_{m}c_{n}]}
  \nonumber\\     = \frac{\i e}{\hbar} &\Bigg(\underbrace{\underbrace{\{c^{\dagger}_{l},c_{l}\}}_{1} \sum_{m,n}T_{m,n}c^{\dagger}_{m}c_{n}
                                             -\sum_{m,n}T_{m,n}c^{\dagger}_{m}c_{n}\underbrace{\{c^{\dagger}_{l},c_{l}\}}_{1}}_{0}
 \nonumber\\                           &     -\sum_{m,n}T_{m,n}c_{l} \underbrace{\{c^{\dagger}_{m},c^{\dagger}_{l}\}}_{0} c_{n}
                                             +\sum_{m,n}T_{m,n}c^{\dagger}_{m} \underbrace{\{c_{l},c_{n}\}}_{0} c^{\dagger}_{l} \notag
  \nonumber\\                          &     +\sum_{m,n}T_{m,n} \underbrace{\{c_{l},c^{\dagger}_{m}\}}_{\delta_{l,m}} c^{\dagger}_{l}c_{n}
                                             -\sum_{m,n}T_{m,n} c^{\dagger}_{m}c_{l} \underbrace{\{c_{n},c^{\dagger}_{l}\}}_{\delta_{l,n}}
                             \Bigg)
  \nonumber\\    = \frac{\i e}{\hbar}& \left(\sum_{n}T_{l,n} c^{\dagger}_{l}c_{n} - \sum_{m}T_{m,l} c^{\dagger}_{m}c_{l}
                             \right)
\end{align}
Finally, the solution of the equation of motion is
\begin{equation}\label{equation_of_motion_n}\color{red}\fbox{$
 -e \braket{\partial_t \hat{n}_l} = \frac{\i e}{\hbar} \sum_m \left( T_{l,m} c^{\dagger}_{l}c_{m} - T_{m,l} c^{\dagger}_{m}c_{l} \right)$}
\end{equation}

\chapter{Scientific production and resume}
This appendix references all my activities and my scientific production that I have done during My PhD.

\section{Collaboration}
 All this work has been done in collaboration with Fernando Flores (Departamento de Física Teórica de la Materia Condensada, Universidad Autonoma de Madrid)  and Pedro de Andres (Instituto de Ciencia de Materiales de Madrid, Consejo Superior de Investigaciones Científicas) groups. In total, I have spent 6 weeks in Madrid.

\section{Conferences}
During this 3 years, I have attended several conferences:
\begin{description}
  \item[January 2012:] Journées Surfaces et Interfaces (Paris, France)
  \item[January 2013:] Journées Surfaces et Interfaces (Orléans, France)
  \item[June 2013:] GdR co-DFT (guidel, France)
  \item[July 2014:] International Conference on Advanced Materials Modelling (Nantes, France)
\end{description}

\section{Formations}
I have also followed several formations:
\begin{description}
  \item [March 2013:] Brittany Synchrotron Radiation School (Rennes, France)
  \item [July 2013:] Interpersonal communication (Rennes, France, by Sud Performance)
  \item [May 2014:] Sherbrooke International summer school on Computational Methods (Jouvence, Canada)
  \item [September 2014:] Initiation to Python (Rennes, France)
\end{description}

\section{Teaching and popularization}
Besides my research activities, I have also been involved in teaching and popularization:
\begin{description}
  \item [2012 Exercises:] crystallography (2nd year of Licence, physics)
  \item [2012/2013 Lecture/Exercises]: mathematics (1st year of Licence, biology)
  \item [2012/2013 Practicals:] LibreOffice  (1st year of Licence, physics)
  \item [2013 Exercises:] electromagnetism in matter (3rd year of Licence, physics)
  \item [2013 Lectures in High school:] The photography at the light of physics (Rennes)
  \item [2013 stand animation:] ``Trip in the nano-world'' (fête de la science, Betton, France), in particluar: wave/particle duality explained using bouncing oil-dropplets
  \item [2013 public conference:] The photography at the light of physics (Fête de la Science, Dinan \& Montgermont, France)
  \item [2014 Practicals:] waves, particles and relativity (2nd year of Licence, physics)
  \item [2014 Lectures in high-school]: Aurorae (Saint-Brieuc, France)
\end{description}

\section{Articles}
I have two papers that have been published and one submitted (see below). One has been written by experimentalists of our department and is about the BEEM. The two others concern a parallel work about Hubbard model. We planned to submit four other papers during the next year (2015): 
\begin{enumerate}
  \item one where we compare our new approach for Au/Si as presented in this thesis, but we want to include the semi-conductor at the same level as the metal in the calculation (not only by projecting the DOS),
  \item one for Fe/Au/Fe/GaAs,
  \item one for the code itself (in Computer Physics Communication). We also plan to propose our code on our team's website\\* (http://ipr.univ-rennes1.fr/d3/them?lang=fr\&mtop=dpt3),
  \item one for the equilibrium approach.
\end{enumerate}



# Mean-field solution of the Hubbard model: the magnetic phase diagram

## Y Claveau, B Arnaud and S Di Matteo

Groupe Théorie, Département Matériaux Nanosciences, Institut de Physique de Rennes
UMR UR1-CNRS 6251, Université de Rennes 1, F-35042 Rennes Cedex, France

E-mail: yann.claveau@univ-rennes1.fr, brice.arnaud@univ-rennes1.fr and
sergio.dimatteo@univ-rennes1.fr

**Abstract**
The present paper is based on our graduate lectures in condensed-matter physics. We found that the mean-field solution of the Hubbard model is an excellent tool to stimulate students' reflections towards the treatment of realistic magnetic interactions. We show by detailed analytical and numerical calculations how to find the mean-field solution of the model on a square lattice. We then interpret the physical implications of the ground-state magnetic phase diagram in terms of the electron density and the ratio between the Coulomb repulsion and the electron-structure bandwidth.

Keywords: Hubbard model, physics condensed matter, magnetism

(Some figures may appear in colour only in the online journal)

## 1. Introduction

In our graduate lectures in condensed-matter physics (second semester of master 1 or first semester of master 2) we found the mean-field solution of the Hubbard model to be a very useful tool for approaching realistic descriptions of materials. What is required is a general knowledge of the second-quantization formalism, with creation and annihilation operators that graduate students often find easier to visualize than the corresponding first-quantization wave functions. The mean-field solution of the Hubbard model is then obtained in a straightforward way through a Fourier transform to $\vec{k}$-space and a matrix diagonalization. In spite of the relatively small amount of work involved, the lesson that a student can learn is very rich: he can construct, by himself, a magnetic phase diagram and understand in this way why ferromagnetism or antiferromagnetism can be determined by the interplay of Coulomb repulsion, band energy and average electron density—an excellent way to start going beyond the independent-electron approximation, towards the complexity of real materials.

Though the literature on the Hubbard model is vast, the model is usually dealt with only in the so-called two-pole approximation like in the original Hubbard papers [1–3], where

the use of rather complex mathematical tools like Green-function equations-of-motion is mandatory. To the contrary, our mean-field solution allows dealing with continuity rather than discontinuity aspects compared to the usual single-particle approach: this might allow filling the gap between the latter and the more advanced research treatments of condensed-matter physics.

The present paper is organized as follows: in section 2 we introduce the Hubbard Hamiltonian and our notation. Section 3 is devoted to the solution of the model in the mean-field approximation on a square lattice. We have chosen the square lattice in order to fix a realistic case (e.g., copper sites in $CuO_2$-planes of superconducting cuprates) by keeping a simple geometry. In section 4 we describe the computational details needed to obtain the ground-state phase diagram and discuss it with respect to the physical parameters of interest. Finally, in section 5 we linger on possible generalizations as a long-term exercise for students and draw our conclusions.

## 2. The Hubbard model

### 2.1. Definitions

The Hubbard Hamiltonian in the simplest case of a non-degenerate band [1], i.e., with one orbital per site, can be expressed in the second-quantization formalism [4] as:

$$\hat{H}_H = - \sum_{ij\sigma} t_{ij}\hat{c}_{i\sigma}^{\dagger}\hat{c}_{j\sigma} + U \sum_i \hat{n}_{i\uparrow}\hat{n}_{i\downarrow} \equiv \hat{H}_t + \hat{H}_U. \tag{1}$$

Here, as usual, $\hat{c}_{i\sigma}^{\dagger}$ is an operator representing the creation of an electron of spin $\sigma$ $(=\uparrow, \downarrow)$ at site $i$, $\hat{c}_{j\sigma}$ is the annihilation of an electron of spin $\sigma$ at site $j$ and $t_{ij}$ is the amplitude of the process, the so-called hopping amplitude from site $j$, where the electron is destroyed, to site $i$, where the electron is created. We finally notice that the sum over $i, j$ is unrestricted and $t_{ij} = (t_{ji})^*$, see equation (2): therefore the Hamiltonian is Hermitian, as it should be. The term $t_{ij}$ is the translation in the second-quantization language of both the kinetic energy and the crystal-potential energy associated with an electron at site $i$:

$$t_{ij} \equiv t_{ij,\sigma} = \int \mathrm{d}^3\vec{r}\, \varphi_{\vec{R}_i,\sigma}^*(\vec{r}) \left( -\frac{\hbar^2\nabla^2}{2m} + V(\vec{r}) \right) \varphi_{\vec{R}_j,\sigma}(\vec{r}). \tag{2}$$

Wannier wave functions [5] $\varphi_{\vec{R}_i,\sigma}(\vec{r})$ are centred at site $\vec{R}_i$. We suppose that the hopping amplitude does not depend on the spin variable and therefore we drop the $\sigma$ label. The term $V(\vec{r})$ represents the periodic crystal-potential energy. As stated above, one approximation that is usually employed for the hopping term $t_{ij}$ is to consider it as being different from zero only when $i$ and $j$ are nearest-neighbour sites ($t_{ij} = t$), as their overlap is usually the largest. In this case the conventional minus sign associated to the hopping term in equation (1) allows having a minimum at the $\Gamma$-point in the reciprocal space ($\vec{k} = 0$) for positive $t$.

Operators $\hat{n}_{i\sigma} \equiv \hat{c}_{i\sigma}^{\dagger}\hat{c}_{i\sigma}$ count the number of particles at site $i$ with spin $\sigma$. They are projection operators, i.e., $\hat{n}_{i\sigma}^2 = \hat{n}_{i\sigma}$ (either there is one electron with spin $\sigma$ at site $i$ or there are no electrons). The expectation value $\langle \hat{n}_{i\sigma} \rangle \equiv \langle \Psi_0 | \hat{n}_{i\sigma} | \Psi_0 \rangle$ in the many-body ground-state $|\Psi_0\rangle$ represents the electron density $n_{i\sigma}$ at site $i$ with spin $\sigma$ (mean occupation number). The physical origin of $\hat{H}_U$ is the Coulomb repulsion of the electrons: when at site $i$ both spin-up and spin-down electrons are present, from equation (1) they contribute to the total energy with a term $+U$, as both $\hat{n}_{i\uparrow}$ and $\hat{n}_{i\downarrow}$ in $\hat{H}_U$ give one. If, on the other side, the two electrons belong to two separate atoms, they do not feel any Coulomb repulsion (this is of course a strong constraint).

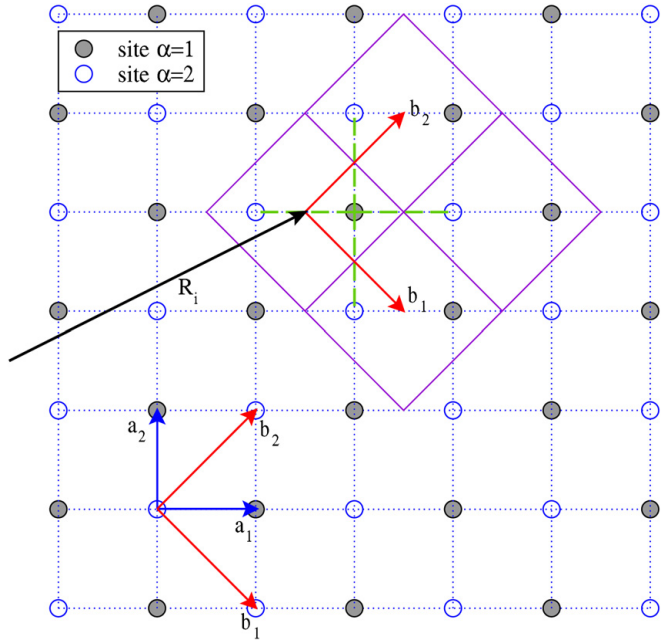

**Figure 1.** Single $(\vec{a}_1, \vec{a}_2)$ and double $(\vec{b}_1, \vec{b}_2)$ cells for the square lattice. Nearest neighbours of $\alpha = 1$, $\vec{R}_i$ are highlighted by the green dashed lines. The four violet cells represent nearest-neighbour cells in the $(\vec{b}_1, \vec{b}_2)$-basis.

Formally, the on-site Coulomb repulsion can be written as:

$$U \equiv \frac{e^2}{4\pi\varepsilon_0} \int \mathrm{d}^3\vec{r}\,\mathrm{d}^3\vec{r}'\; |\varphi_{\vec{R}_i,\sigma}(\vec{r})|^2 \frac{1}{|\vec{r}-\vec{r}'|} |\varphi_{\vec{R}_i,\bar{\sigma}}(\vec{r}')|^2. \tag{3}$$

The Coulomb term does not depend on the site label, $i$, if we suppose the system homogeneous. Notice that in the extreme limit $U/t = 0$, we recover a purely band-like (tight-binding) picture, with just kinetic energy and crystal periodic potential, and in the opposite limit $t/U = 0$, we find a purely atomic picture.

### 2.2. The square lattice

The bi-dimensional square lattice is drawn in figure 1. The unit cell is spanned by the two vectors $\vec{a}_1$ and $\vec{a}_2$, of common length $a$. However, for our calculations, we consider the cell of double area spanned by the two vectors $\vec{b}_1$ and $\vec{b}_2$, with the idea of looking for possible antiferromagnetic ground-states. Such a cell encloses two atomic sites that can in principle be inequivalent (e.g., spin ↑ and spin ↓). In what follows, we adopt the site label $\vec{R}_i$ for the double unit cell and, within each cell, the two atoms are labelled by an extra index $\alpha = 1, 2$. For example, as shown in figure 1, nearest neighbours of $\alpha = 1$ sites are necessarily $\alpha = 2$ and vice-versa (bipartite lattice). The general creation (annihilation) operator for an electron

at site $\vec{R}_i$, position $\alpha$, spin $\sigma$ can be written as: $\hat{c}^\dagger_{i\alpha\sigma}$ ($\hat{c}_{i\alpha\sigma}$). Equation (1) should be changed accordingly:

$$\hat{H}_H = -\sum_{ij\alpha\alpha'\sigma} t^{\alpha\alpha'}_{ij} \hat{c}^\dagger_{i\alpha\sigma} \hat{c}_{j\alpha'\sigma} + U\sum_{i\alpha} \hat{n}_{i\alpha\uparrow}\hat{n}_{i\alpha\downarrow} \equiv \hat{H}_{\bar{t}} + \hat{H}_{\bar{U}}. \tag{4}$$

The structure constants $t^{\alpha\alpha'}_{ij}$ for our problem (nearest-neighbour hopping) can be written as follows:

$$
\begin{aligned}
t^{\alpha\alpha'}_{ij} = {}&-t\delta_{\alpha,1}\delta_{\alpha',2}[\delta_{\vec{R}_i,\vec{R}_j} + \delta_{\vec{R}_i+\vec{b}_1,\vec{R}_j} + \delta_{\vec{R}_i+\vec{b}_2,\vec{R}_j} + \delta_{\vec{R}_i+\vec{b}_1+\vec{b}_2,\vec{R}_j}] \\
&-t\delta_{\alpha',1}\delta_{\alpha,2}[\delta_{\vec{R}_i,\vec{R}_j} + \delta_{\vec{R}_i-\vec{b}_1,\vec{R}_j} + \delta_{\vec{R}_i-\vec{b}_2,\vec{R}_j} + \delta_{\vec{R}_i-\vec{b}_1-\vec{b}_2,\vec{R}_j}].
\end{aligned}
\tag{5}
$$

In spite of the cumbersome form, the meaning of terms appearing in (5) is quite straightforward: the first line represents the four nearest-neighbour hopping energies (represented by dashed green lines in figure 1) when site $\vec{R}_i$ is of $\alpha = 1$-type and site $\vec{R}_j$ is of $\alpha = 2$-type, and the second line the opposite case. As we employ the reciprocal lattice of the $(\vec{b}_1, \vec{b}_2)$ direct lattice, we must express the nearest neighbours in terms of linear combinations of $\vec{b}_1, \vec{b}_2$ vectors.

## 3. Mean-field solution

In order to derive the full mean-field solution, we first discuss the tight-binding solution, for $U = 0$, and then analyse the action of the $\hat{H}_{\bar{U}}$ term in the mean field. In both cases we can perform a Fourier transform to the $\vec{k}$-space, in order to gain full advantage of the two-dimensional periodicity. As stated above, we use the reciprocal space of the $(\vec{b}_1, \vec{b}_2)$ cell. The Fourier transform of the creation operator is: $\hat{c}^\dagger_{i\alpha\sigma} = \frac{1}{\sqrt{N}}\sum_{\vec{k}} e^{-i\vec{k}\cdot\vec{R}_i}\hat{c}^\dagger_{\vec{k}\alpha\sigma}$. That of the annihilation operator is the Hermitian conjugate: $\hat{c}_{i\alpha\sigma} = \frac{1}{\sqrt{N}}\sum_{\vec{k}} e^{i\vec{k}\cdot\vec{R}_i}\hat{c}_{\vec{k}\alpha\sigma}$. Here, $N$ is the number of unit cells spanned by the vectors $\vec{b}_1$ and $\vec{b}_2$.

### 3.1. Tight-binding solution: $U = 0$

Consider first the hopping part of the Hamiltonian, $\hat{H}_{\bar{t}}$. By inserting the $\vec{k}$-transformed operators, we get the usual tight-binding band structure [5]:

$$
\begin{aligned}
\hat{H}_{\bar{t}} &= -\frac{1}{N}\sum_{ij\alpha\alpha'\sigma} t^{\alpha\alpha'}_{ij} \sum_{\vec{k}\vec{k}'} e^{-i\vec{k}\cdot\vec{R}_i}\hat{c}^\dagger_{\vec{k}\alpha\sigma} e^{i\vec{k}'\cdot\vec{R}_j}\hat{c}_{\vec{k}'\alpha'\sigma} \\
&= -\frac{1}{N}\sum_{\vec{k}\vec{k}'\alpha\alpha'\sigma} \hat{c}^\dagger_{\vec{k}\alpha\sigma}\hat{c}_{\vec{k}'\alpha'\sigma} \sum_{ij} t^{\alpha\alpha'}_{ij} e^{-i\vec{k}\cdot\vec{R}_i} e^{i\vec{k}'\cdot\vec{R}_j} \\
&= -\sum_{\vec{k}\vec{k}'\alpha\alpha'\sigma} \hat{c}^\dagger_{\vec{k}\alpha\sigma}\hat{c}_{\vec{k}'\alpha'\sigma} \sum_{j} t^{\alpha\alpha'}_{\vec{\eta}_j} e^{i\vec{k}'\cdot\vec{\eta}_j} \frac{1}{N}\sum_i e^{-i(\vec{k}-\vec{k}')\cdot\vec{R}_i} \\
&= \sum_{\vec{k}\alpha\alpha'\sigma} \hat{c}^\dagger_{\vec{k}\alpha\sigma}\hat{c}_{\vec{k}\alpha'\sigma}\, \varepsilon^{\alpha\alpha'}_{\vec{k}}
\end{aligned}
\tag{6}
$$

where we defined the matrix Hamiltonian in the $\alpha$-$\alpha'$ basis as: $\varepsilon^{\alpha\alpha'}_{\vec{k}} = -\sum_j t^{\alpha\alpha'}_{\vec{\eta}_j} e^{i\vec{k}\cdot\vec{\eta}_j}$. In passing from the second to the third line of (6) we used the translational invariance of the structure factor $t^{\alpha\alpha'}_{ij}$. This means that if we write $\vec{R}_j = \vec{R}_i + \vec{\eta}_{ij}$, the vectors $\vec{\eta}_{ij}$, and therefore the way of counting nearest neighbours, are independent of the starting point $\vec{R}_i$. So, we can write: $\vec{\eta}_{ij} \to \vec{\eta}_j$. Moreover we used the fact that $\frac{1}{N}\sum_i e^{-i(\vec{k}-\vec{k}')\cdot\vec{R}_i} = \delta_{\vec{k}\vec{k}'}$.

As we have two atoms per unit cell, the matrix Hamiltonian is a $2 \times 2$ matrix ($\alpha, \alpha' = 1, 2$) that should be diagonalized in order to have the band structure $\varepsilon_{0\vec{k}}^{\pm}$. From the explicit expression of the structure constants (5), we get the matrix:

$$H_{\vec{k}} = \begin{bmatrix} 0 & t\gamma_{\vec{k}} \\ t\gamma_{\vec{k}}^* & 0 \end{bmatrix} \tag{7}$$

where $\gamma_{\vec{k}} = -\{1 + e^{-i\vec{k}\cdot(\vec{b}_1+\vec{b}_2)} + e^{-i\vec{k}\cdot\vec{b}_1} + e^{-i\vec{k}\cdot\vec{b}_2}\}$. Its diagonalization leads to two bands given by:

$$\varepsilon_{0\vec{k}}^{\pm} = \pm t|\gamma_{\vec{k}}| = \pm t \left| 2\cos\left(\vec{k}\cdot\frac{\vec{b}_1+\vec{b}_2}{2}\right) + 2\cos\left(\vec{k}\cdot\frac{\vec{b}_2-\vec{b}_1}{2}\right) \right|. \tag{8}$$

The band structure and density of states (DOS) per unit cell are depicted in figure 2, together with those for the single cell ($\vec{a}_1, \vec{a}_2$), for comparison. Bandwidth is $W = 8t$. At half filling, we have a nested Fermi surface, leading to a van Hove logarithmic singularity in the DOS [5]. We remind readers that nesting is the property by which any point of the Fermi surface is related to another point of the Fermi surface by a fixed vector, in this case the vector ($\pi/a, \pi/a$), because the Fermi surface is a square at half filling, as shown in figure 2(b). Clearly, at this level, changing the choice of the unit cell (a pure convention) does not change our results. In fact, the second band is just the folding of the first band of the single cell, as can be also seen by considering that $\frac{\vec{b}_1+\vec{b}_2}{2} = \vec{a}_1$ and $\frac{\vec{b}_2-\vec{b}_1}{2} = \vec{a}_2$. This comes from the fact that the point $M_a \equiv (\pi/a, \pi/a)$ in the reciprocal cell of the direct cell ($\vec{a}_1, \vec{a}_2$) becomes equivalent to $\Gamma_b \equiv (0, 0)$ in the reciprocal cell of the direct cell ($\vec{b}_1, \vec{b}_2$). The two DOS of figure 2 are the double of each other just because they are normalized per unit cell and there are two atoms per unit cell in the case of the double cell.

### 3.2. Introduction of $\hat{H}_U$ in the tight-binding solution

The presence of $\hat{H}_U$, equation (4), changes these simple results even at a mean-field level. We are reminded that the mean-field approximation corresponds to neglecting the fluctuations around the mean density. Such fluctuations are defined as $\Delta\hat{n}_{i\alpha\sigma} \equiv \hat{n}_{i\alpha\sigma} - \langle\hat{n}_{i\alpha\sigma}\rangle$, i.e., the difference between the exact number operator $\hat{n}_{i\alpha\sigma}$ and the mean occupation number $\langle\hat{n}_{i\alpha\sigma}\rangle$. From this relation, we get: $\hat{n}_{i\alpha\sigma} \equiv \langle\hat{n}_{i\alpha\sigma}\rangle + \Delta\hat{n}_{i\alpha\sigma}$. If we suppose homogeneity of the system, then $\langle\hat{n}_{i\alpha\sigma}\rangle$ is independent of the cell position $\vec{R}_i$, and we can drop the label $i$ and write:

$$\begin{aligned} \hat{n}_{i\alpha\uparrow}\hat{n}_{i\alpha\downarrow} &= [\Delta\hat{n}_{i\alpha\uparrow} + \langle\hat{n}_{\alpha\uparrow}\rangle] \cdot [\Delta\hat{n}_{i\alpha\downarrow} + \langle\hat{n}_{\alpha\downarrow}\rangle] \\ &= \Delta\hat{n}_{i\alpha\uparrow}\Delta\hat{n}_{i\alpha\downarrow} + \sum_\sigma \hat{n}_{i\alpha\sigma}\langle\hat{n}_{\alpha\bar{\sigma}}\rangle - \langle\hat{n}_{\alpha\uparrow}\rangle\langle\hat{n}_{\alpha\downarrow}\rangle. \end{aligned} \tag{9}$$

In a mean-field approximation we can neglect the first term of the previous equation, quadratic in the fluctuations. This implies that $\hat{H}_U$ becomes:

$$\hat{H}_U^{\mathrm{MF}} = \underbrace{U \sum_{i\alpha\sigma} \hat{n}_{i\alpha\sigma}\langle n_{\alpha\bar{\sigma}}\rangle}_{\hat{H}_U} - \underbrace{UN \sum_\alpha \langle n_{\alpha\uparrow}\rangle\langle n_{\alpha\downarrow}\rangle}_{E_U}. \tag{10}$$

The second term $E_U$ is a constant for a given magnetic configuration and number of particles, as it does not depend on creation or annihilation operators but only on their average values. However, it must be integrated in the calculation of the magnetic phase diagram, as such a term is advantageous for paramagnetic (PM) configurations with respect to ferromagnetism and antiferromagnetism ($E_U$ is negative and the product $\langle n_{\alpha\uparrow}\rangle\langle n_{\alpha\downarrow}\rangle$, for fixed number of particles per site, is maximum when $\langle n_{\alpha\uparrow}\rangle = \langle n_{\alpha\downarrow}\rangle$).

Y Claveau *et al*

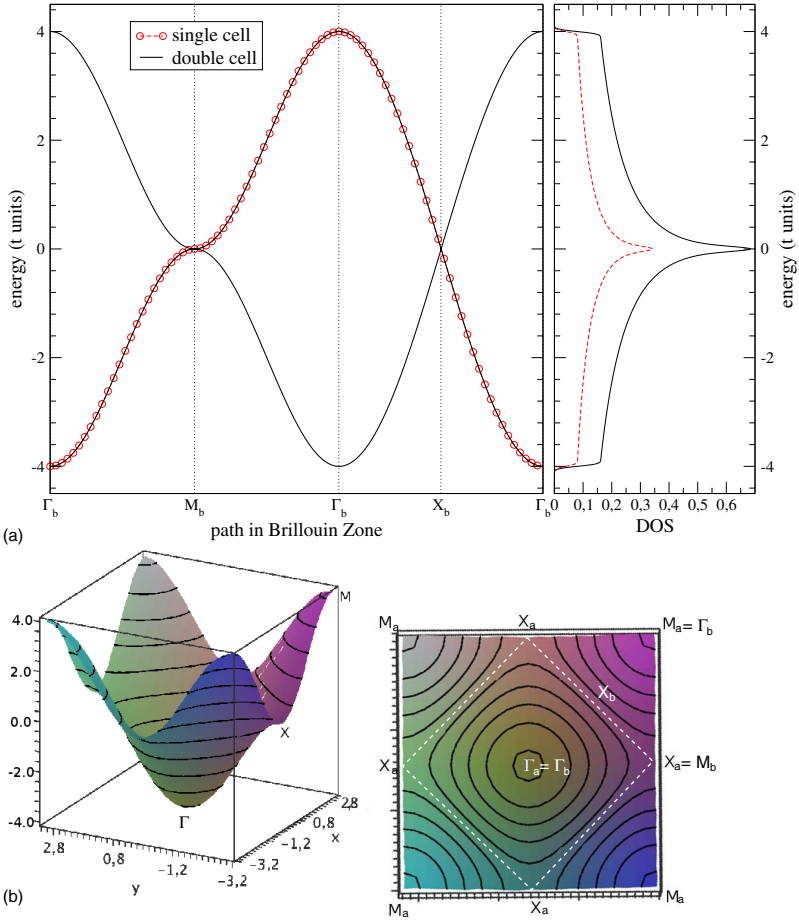

(a)

(b)

**Figure 2.** (a) Band structure in tight-binding approach ($U = 0$) for single and double cells of figure 1. For the double cell a second band appears due to the folding of the first band, as both points $\Gamma_a$ and $M_a$ of the reciprocal space of the single cell correspond to point $\Gamma_b$ of the reciprocal space of the double cell. (b) Full band structure for the single cell. A Van-Hove singularity (the white dashed line which connects the $X_a$ points) appears at half filling.

By using the Fourier transform as for equation (6), the first term of (10) becomes:
$\tilde{H}_U = U \sum_{i\alpha\sigma} \langle n_{\alpha\bar{\sigma}} \rangle \hat{c}^\dagger_{i\alpha\sigma} \hat{c}_{i\alpha\sigma} = U \sum_{\vec{k}\alpha\sigma} \langle n_{\alpha\bar{\sigma}} \rangle \hat{c}^\dagger_{\vec{k}\alpha\sigma} \hat{c}_{\vec{k}\alpha\sigma}$, diagonal in $\alpha$. Therefore the energy per $\vec{k}$-point for a spin-$\sigma$ electron is obtained by diagonalizing the $2 \times 2$ matrix:

$$\begin{bmatrix} U\langle n_{1\bar{\sigma}} \rangle & t\gamma_{\vec{k}} \\ t\gamma^*_{\vec{k}} & U\langle n_{2\bar{\sigma}} \rangle \end{bmatrix}, \tag{11}$$

where $\gamma_{\vec{k}}$ has been defined after (7). This Hamiltonian can be easily diagonalized analytically, as the associated eigenvalue problem leads to the second-order algebraic equation for the eigenenergies $\varepsilon_{\vec{k}}^{\pm}$:

$$\varepsilon_{\vec{k}}^2 - \varepsilon_{\vec{k}} \left( U \langle n_{1\bar{\sigma}} \rangle + U \langle n_{2\bar{\sigma}} \rangle \right) + U^2 \langle n_{1\bar{\sigma}} \rangle \langle n_{2\bar{\sigma}} \rangle - t^2 |\gamma_{\vec{k}}|^2 = 0. \tag{12}$$

The solutions of this second-order equation provide the mean-field band energies per $\vec{k}$-point for a spin-$\sigma$ electron, once we add again the term $E_U$ (divided by $N$, because $\varepsilon_{\vec{k}}$ is the energy per spin-$\sigma$ electron):

$$\begin{aligned}
\varepsilon_{\vec{k}\sigma}^{\pm} &= U \left( \frac{\langle n_{1\bar{\sigma}} \rangle + \langle n_{2\bar{\sigma}} \rangle}{2} \right) - U (\langle n_{1\uparrow} \rangle \langle n_{1\downarrow} \rangle + \langle n_{2\uparrow} \rangle \langle n_{2\downarrow} \rangle) \\
&\quad \pm \frac{1}{2} \sqrt{ U^2 (\langle n_{1\bar{\sigma}} \rangle - \langle n_{2\bar{\sigma}} \rangle)^2 + 16 t^2 \left( \cos \left[ \vec{k} \cdot \frac{\vec{b}_1 + \vec{b}_2}{2} \right] + \cos \left[ \vec{k} \cdot \frac{\vec{b}_1 - \vec{b}_2}{2} \right] \right)^2 } \\
&= U \left( \frac{\langle n_{1\bar{\sigma}} \rangle + \langle n_{2\bar{\sigma}} \rangle}{2} \right) - U (\langle n_{1\uparrow} \rangle \langle n_{1\downarrow} \rangle + \langle n_{2\uparrow} \rangle \langle n_{2\downarrow} \rangle) \\
&\quad \pm \frac{1}{2} \sqrt{ U^2 (\langle n_{1\bar{\sigma}} \rangle - \langle n_{2\bar{\sigma}} \rangle)^2 + 4 \left( \varepsilon_{0\vec{k}}^{\pm} \right)^2 }.
\end{aligned} \tag{13}$$

From the last line we see that $\hat{H}_U^{\mathrm{MF}}$ contributes to the total energy as a density-dependent and spin-dependent (because of $\bar{\sigma}$) renormalization of the tight-binding energy $\varepsilon_{0\vec{k}}^{\pm}$. This double dependence on the density and on the spin is at the basis of the richness in the phase diagram shown in figure 6, allowing to get PM, ferromagnetic (FM) and antiferromagnetic (AFM) phases.

## 4. Results and discussion

### 4.1. Computational details

In order to find the ground-state energy for a given magnetic configuration, we need to employ a self-consistent scheme as, for a given spin $\sigma$, the energy depends on the mean occupation number of opposite spin, $\langle \hat{n}_{\alpha\bar{\sigma}} \rangle$, which in turns depends on the eigenvectors of the energy matrix itself. As shown in figure 3, we therefore start from a given configuration of input parameters (frames 1 and 2), we follow steps 3–6, up to the self-consistency condition expressed in frames 7 and 8, and finally move back to frame 2 or end to frame 9, depending on whether the condition is not satisfied or satisfied, respectively. The condition is satisfied when the output occupation number is the same as the input occupation number within a threshold that we fixed to $10^{-5}$.

In more detail, we proceeded as follows: for practical reasons we found it simpler to numerically diagonalize the Hamiltonian of (11) through the Lapack libraries [6] (frame 3) in order to determine eigenvalues $\varepsilon_{j\sigma}(\vec{k})$ and eigenvectors $|\Psi_{\vec{k}\sigma}^j\rangle = \sum_\alpha A_{\alpha,\sigma}^j(\vec{k})|\alpha\vec{k}\rangle$ (with $j = 1, 2$). We actually work with the chemical potential in order to fix the particle density *a posteriori*: $\hat{H}_H^\mu = \hat{H}_H - \mu \sum_{i\alpha\sigma} \hat{n}_{i\alpha\sigma}$. The calculations performed in frames 4–6 are based on equations (14)–(16) given below. The first of these equations expresses the DOS per unit cell, $\rho(\varepsilon)$:

$$\rho(\varepsilon) = \frac{1}{N} \sum_{\vec{k}, j, \sigma} \delta[\varepsilon - \varepsilon_{j\sigma}(\vec{k})]. \tag{14}$$

The Dirac delta function in (14) is numerically calculated through a Gaussian function whose broadening is optimized after a convergence study. If the width of the Gaussian is too

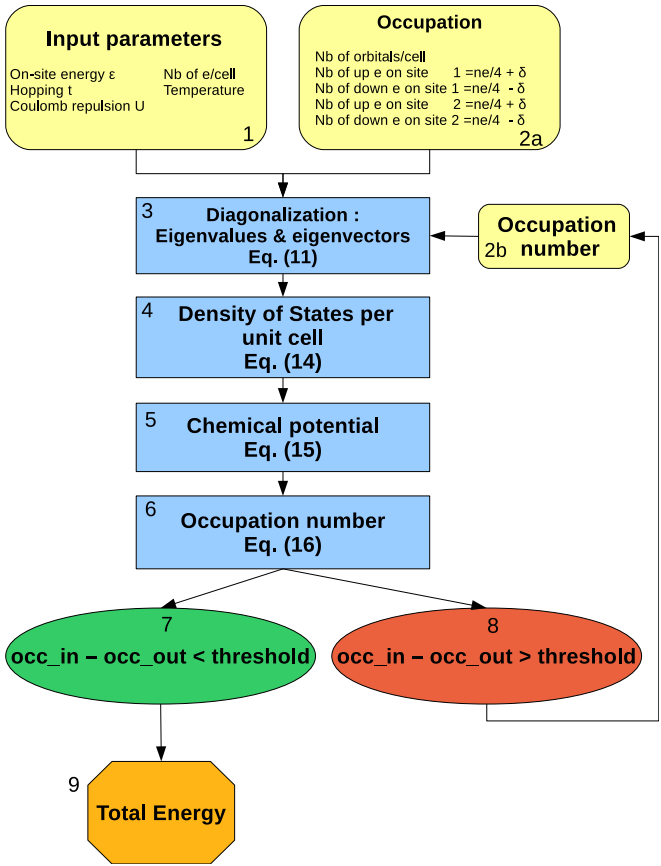

**Figure 3.** Scheme of the self-consistent algorithm leading to the ground-state phase diagram of figure 6. Equations are described in the text

wide compared to the step between two energy points in $\varepsilon$, the DOS is too smooth, whereas if the width is too narrow, artificial oscillations appear. A more elegant approach might be to use the Methfessel–Paxton method [7], usually implemented in more advanced packages for electronic structure calculations.

As a second step we determine the chemical potential $\mu$ by the implicit equation:

$$\int_{-\infty}^{+\infty} d\varepsilon \, \rho(\varepsilon) \frac{1}{\exp[\beta(\varepsilon - \mu)] + 1} = n_e. \tag{15}$$

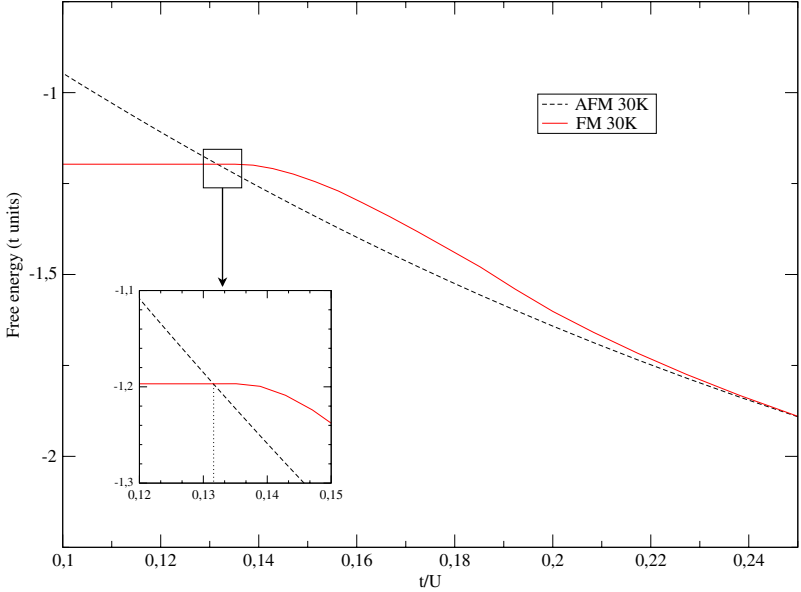

**Figure 4.** A phase transition occurs where magnetic-energy curves cross each other. For $n = 0.8$ the system becomes ferromagnetic below $t/U = 0.13$. We have chosen an inverse temperature $\beta \sim 0.003$ eV.

Here $\beta = 1/k_B T$ is the Boltzmann factor and we have defined the total number of electrons per unit cell (as used in figures 3, 5, 6): $n_e \equiv \sum_{\alpha,\sigma} \langle n_{\alpha\sigma} \rangle$. Finally, the average number of particles per site $\alpha$ and per spin $\sigma$ at a given temperature $T$ is given by:

$$\langle n_{\alpha\sigma} \rangle = \frac{1}{N} \sum_{\vec{k}, j=1,2} |A^j_{\alpha,\sigma}(\vec{k})|^2 \frac{1}{\exp[\beta(\varepsilon_{j\sigma}(\vec{k}) - \mu)] + 1}. \tag{16}$$

At finite temperature, the thermodynamic variable to minimize is of course not the total energy, $E$, but the free energy, $F = E - TS$, what implies a calculation of the entropy of the system, through the equation:

$$S(T) = -k_B \int_{-\infty}^{\infty} \mathrm{d}\varepsilon\, \rho(\varepsilon)\{f(\varepsilon)\ln[f(\varepsilon)] + [1 - f(\varepsilon)]\ln[1 - f(\varepsilon)]\}. \tag{17}$$

Here $f(\varepsilon)$ is the Fermi–Dirac distribution. However, in what follows, we are interested in the ground-state phase diagram of the model, i.e., at $T = 0$. Therefore, we minimized the total energy and used the parameter $T$ of equations (15) and (16) for convergence purposes only. It is in fact common practice to use a Fermi–Dirac distribution, characterized by a parameter $T$ different from zero, even in the $T \to 0$ limit, in order to smooth the step function that appears in equations (15) and (16) for $T = 0$.

In general, we computed the total energy for each magnetic phase as a function of $t/U$. A phase transition is then characterized by the crossing of two (or more) free-energy curves: for example, we have represented the magnetic free energy as a function of $t/U$ in figure 4 for a

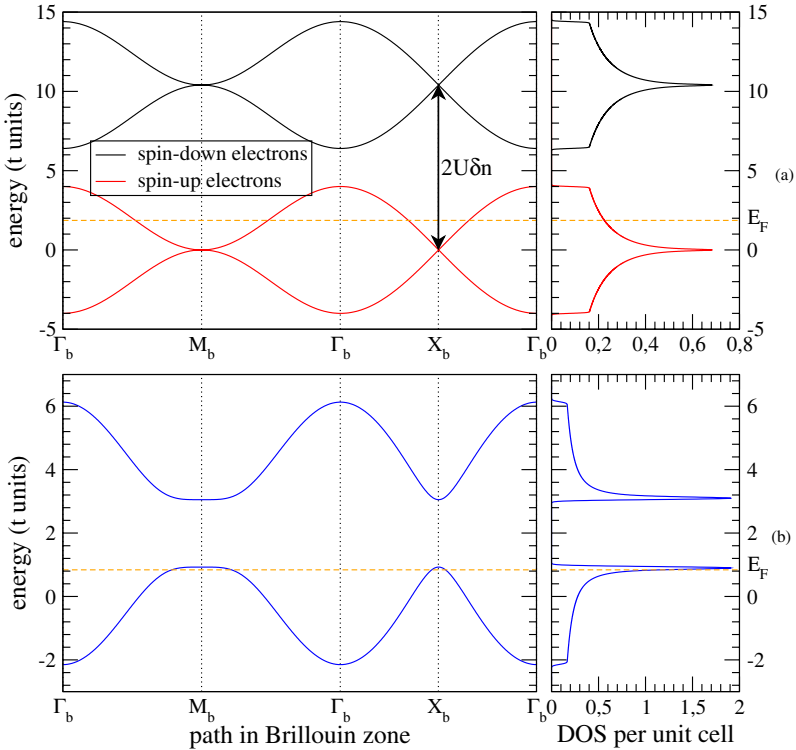

**Figure 5.** Band structure and DOS for: (a) FM configuration: $t/U = 0.077$ ; $n_e = 1.6$ ($n_{1\uparrow} = n_{2\uparrow} = 0.8$ and $n_{1\downarrow} = n_{2\downarrow} = 0$) and (b) AFM configuration: $t/U = 0.2$ ; $n_e = 1.6$ ($n_{1\uparrow} = n_{2\downarrow} = 0.62$ and $n_{1\downarrow} = n_{2\uparrow} = 0.18$). In the FM case the exchange splitting is $2U\delta n$. In the AFM case a Slater gap appears, leading to an insulator for $n_e = 2.0$. The DOS has been obtained by modelling equation (14) with a normalized Gaussian of width $0.05\,t$ and with a $500 \times 500$ $k$-point grid.

specific occupation number ($n = 0.8$), in order to highlight a magnetic (AFM/FM) transition. In this case the phase transition from AFM to FM phase is obtained at $t/U = 0.13$, the value at which the total FM energy becomes smaller than the AFM energy.

### 4.2. Discussion of results: phase diagram

In figure 5, we have drawn the band structures for both FM and AFM phases to highlight the differences with the PM case. FM bands are shifted rigidly with respect to PM bands by about $\pm U\delta n$, where $\delta n$ is half the spin unbalance (see below). AFM bands are instead characterized by the opening of a gap that can be related to the difference $U(|\langle n_{1\bar{\sigma}}\rangle - \langle n_{2\bar{\sigma}}\rangle|)$ in (13). This gap leads to an insulating phase for $n_e = 2.0$. It is important to notice that such a gap is not related to a metal–insulator transition (MIT) of the Mott–Hubbard kind, as it is not related

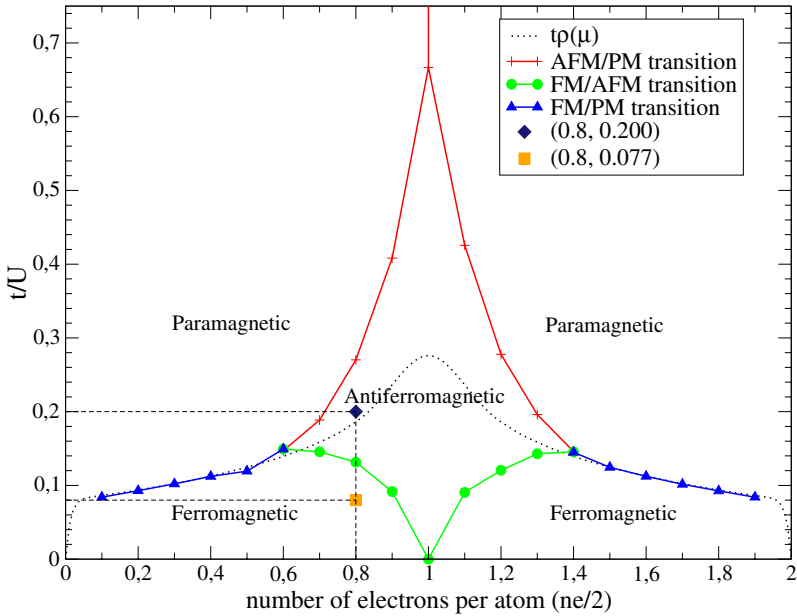

**Figure 6.** Ground-state phase diagram of the Hubbard model on a square lattice as a function of the ratio $t/U$ and of the electron filling. The physical relevance of the curve $t\rho(\mu)$ is detailed in the text. We used an inverse-temperature value $\beta = 0.03t$ in the equations. The two points (0.8, 0.077) and (0.8, 0.2) correspond to the band structure depicted in figure 5.

to the electronic correlations (that are absent by definition in a mean-field calculation), but to magnetism. This MIT is called a Slater MIT, in honour of Slater, who foresaw it in 1951 [8]. In fact, unlike Mott, who did not originally ascribe his MIT to magnetic interactions, Slater thought that the origin of the metal-to-insulator transition was determined by the onset of AFM long-range order, exactly as in the scheme described in the present paper. Therefore a Slater insulator is characterized by a band gap determined by a superlattice modulation of the magnetic periodicity. This is not the case of a Mott insulator.

Our main result is the ground-state phase diagram, drawn in figure 6 as a function of the number of electrons per site and of $t/U$. Such a phase diagram has been already obtained in the literature in 1985 by Hirsch [9], though in a different context and without providing all the details of the derivation that can be found here. As a general feature, the phase diagram shows a clear symmetry around half filling, i.e., one electron per site, where antiferromagnetism is the lowest-energy configuration. This symmetry had to be expected, since it is a symmetry of the Hubbard Hamiltonian for nearest-neighbour hopping. Far from half filling, paramagnetism has the advantage of a high value of $t/U$, whereas a low value of $t/U$ leads instead to ferromagnetism. This tendency for the PM/FM phases can be easily understood: the ground-state of $n$ non-interacting electrons ($U = 0$) is PM because the minimum-energy constraint in combination with the Pauli principle forces to fill all the energy levels from the lowest ($\varepsilon_{min}$)

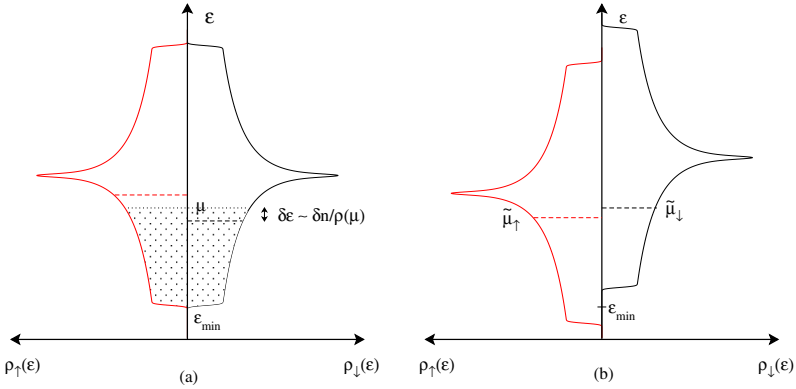

**Figure 7.** (a) If $\delta n \ll n$ spin-down electrons are moved to spin-up DOS, for $U = 0$, this leads to a shift of the chemical potential $\pm \delta \varepsilon \sim \pm \delta n / \rho(\mu)$: PM state is always favoured. (b) If $U \neq 0$, then the spin-up DOS is lowered in energy by $-U\delta n + U(\delta n)^2/n$ and the spin-down DOS is raised in energy by $U\delta n + U(\delta n)^2/n$. Therefore, depending on $\rho(\mu)$, FM stability can be obtained for sufficiently high $U$ (see text).

to the highest ($\mu$) with an equal number of $n/2$ up and $n/2$ down electrons. In the opposite extreme case, for $U/t \to \infty$, the system can gain energy by a total magnetization (say, all electrons with spin up), in order to minimize the energy term $U\langle n_\downarrow \rangle$. In the intermediate $t/U$ cases, only the numerical study of equations (11), (14)–(16) can provide us with the magnetic phase of the system.

There exists however a criterion that allows us to foresee the stability of the PM phase versus the FM phase just on the base of the two parameters $U$ and $\rho(\mu)$, the DOS at the Fermi energy ($\varepsilon_F = \mu$ at $T = 0$). This is the Stoner criterion [10]. Start from the PM phase, represented in grey in figure 7(a) ($n_\uparrow = n_\downarrow = n/2$) and move $\delta n$ electrons from spin-down to spin-up states, so that $n_\uparrow = n/2 + \delta n$ and $n_\downarrow = n/2 - \delta n$. This leads to a shift of the chemical potential by $\pm \delta \varepsilon \simeq \pm \delta n / \rho(\mu)$ (if $\delta n \ll n$). The system is not at equilibrium, as $\mu_\uparrow = \mu + \delta n / \rho(\mu) \neq \mu_\downarrow = \mu - \delta n / \rho(\mu)$. Yet, because of the change in $\delta n$ and of the mean-field Coulomb energy (10), the minority-spin band shifts as a whole upwards by $U\delta n + U(\delta n)^2/n$ and majority-spin band shifts as a whole downwards by $U\delta n - U(\delta n)^2/n$ as shown in figure 7(b).

In this situation, if $\tilde{\mu}_\uparrow = \mu + \delta n / \rho(\mu) - U\delta n + U(\delta n)^2/n$ is less than $\tilde{\mu}_\downarrow = \mu - \delta n / \rho(\mu) + U\delta n + U(\delta n)^2/n$, then it is favourable to still increase the number of spin-up electrons until $\tilde{\mu}_\uparrow = \tilde{\mu}_\downarrow$. Therefore, ferromagnetism appears when $\tilde{\mu}_\uparrow \leqslant \tilde{\mu}_\downarrow$. This leads to the Stoner criterion for ferromagnetic stability:

$$U\rho(\mu) \geqslant 1. \tag{18}$$

Equation (18) can also be written as $\frac{t}{U} \leqslant t\rho(\mu)$. The corresponding equality, that marks the phase transition, has been reproduced in figure 6. All our calculated points for the PM/FM transition lie on the theoretical curve $\frac{t}{U} = t\rho(\mu)$ represented by a dotted line in figure 6. We infer from the Stoner criterion that an FM instability is expected in materials showing a high DOS at Fermi level. This is indeed the case for Fe, Co and Ni [11]. It is also possible to find a similar criterion for the AFM/PM and AFM/FM transitions, but its derivation is

technically more involved because it is based on a Bogoliubov transformation. This leads to a gap equation formally equivalent to the one of the BCS theory of superconductivity. A rather detailed description of this generalization can be found in [12].

## 5. Generalizations and conclusion

In the previous section we have analysed the formalism and phase diagram of the one-band Hubbard model on a square lattice in the mean-field approximation. For completeness, in this section we give a brief overview of four possible generalizations of the model to provide the link to the more advanced literature, with applications to real materials.

The first modification that can be dealt with concerns the application of (1) to a different lattice. This primarily leads to a different DOS than the one shown in figure 2, thereby modifying quantitatively, but not necessarily qualitatively, the phase diagram. The fact that quantitative changes in the DOS do not necessarily imply qualitative (topological) modifications of the phase diagram can be understood by comparing our case with the example given in [12], where a free-electron-gas parabolic DOS is employed. The resulting phase diagram is topologically similar to ours. Qualitatively similar behaviour is also obtained in all cases where bipartite lattices are considered, as the honeycomb lattice. We are reminded that a lattice is called bipartite if two atoms of kinds A and B can be accommodated in it in such a way that any atom of kind A is only surrounded by atoms of kind B and vice versa. On the contrary, qualitatively and quantitatively different results are obtained in the case of frustrated lattices, like the triangular lattice. This is the case because AFM interactions can be depleted by geometrical frustrations and paramagnetism is generally advantageous [13] in lattices that are not bipartite.

As a second modification, it is possible to extend the hopping term beyond nearest neighbours. For example next-nearest-neighbour hopping is not necessarily zero as supposed in the present paper. In the square lattice such a term corresponds to a hopping integral between the two atoms along the directions of $\vec{b}_1$ and $\vec{b}_2$ of figure 1. The effect of this term in the DOS of a square lattice is to remove the nesting property of a Fermi surface at half filling. For a two-dimensional square lattice such a calculation in the mean-field approximation has been performed, e.g., by Hirsch [14], who indeed found a deformed phase diagram with respect to that of figure 6, without the symmetry around half filling.

Thirdly, instead of two-dimensional systems, we could move to three-dimensional lattices. In the case of a cubic lattice, for example, this leads to the removal of the van Hove singularity, determined by the two-dimensional square-lattice topology with nearest neighbours, and therefore to the removal of the logarithmic singularity at half filling in the DOS of figure 2. All these modifications can of course be combined together to get a final phase diagram that can be substantially different from the one presented in this paper even in the mean-field approximation. A further possible generalization concerns the search for ground states that are not commensurate with the crystal lattice [15].

One final modification that applies to realistic systems, is to introduce multi-orbital Hubbard models and/or multi-band Hubbard models. Models where d or f orbitals are introduced belong to the first kind. In this case a further index $m$ up to five for d orbitals and up to seven for f orbitals must be introduced to deal with electron creation and annihilation operators for different wave functions (e.g., $d_{xy}$, $d_{yz}$, $d_{x^2-y^2}$, etc). Hopping terms are then modified in a similar way as to when we moved from (1) to (4). However, the extra labels, $\alpha$, $\alpha'$ in (4) and $m$, $m'$ in (19) below have different physical interpretations, $m$, $m'$ representing two different d or f (or sometimes p) orbitals on the same atom. The multi-orbital Hubbard

Hamiltonian is written (see, e.g., [2] or section [2] of [16]):

$$H = \sum_{ijmm'\sigma} t_{ij}^{mm'} \hat{c}_{im\sigma}^{\dagger} \hat{c}_{jm'\sigma} + \sum_{imm'\sigma\sigma'} U^{mm'} \hat{n}_{im\sigma} \hat{n}_{im'\sigma'}$$
$$+ J \sum_{m \neq m'} \left( \hat{c}_{im\uparrow}^{\dagger} \hat{c}_{im\downarrow}^{\dagger} \hat{c}_{im'\downarrow} \hat{c}_{im'\uparrow} - \hat{c}_{im\uparrow}^{\dagger} \hat{c}_{im\downarrow} \hat{c}_{im'\downarrow}^{\dagger} \hat{c}_{im'\uparrow} \right). \tag{19}$$

It is important to underline that in this case, several intra-atomic Coulomb terms appear, depending on whether intra-orbital ($U^{mm'}$, with $m = m'$) or inter-orbital ($U^{mm'}$, with $m \neq m'$) Coulomb repulsion is concerned. Moreover, because of the multi-dimensional orbital degree of freedom, also Hund's exchange $J$ appears, for $m \neq m'$. Interestingly, this implies the appearance of an exchange term in the mean-field approximation: the Hartree approximation in this paper would become an Hartree–Fock approximation.

Finally, multi-band Hubbard models are those where several atomic species are present; not all necessarily characterized by the same Hubbard $U$ (that can also be zero in some cases). Probably the most famous of this kind are the Anderson periodic model [17], used to describe the interaction of a localized electron (e.g. an f electron) with a 'Fermi sea', or the pd-model [18, 19] used to describe $CuO_2$-planes in superconducting cuprates, where two kinds of p bands and 1 d band are introduced.

This rapid overview shows the potential applications that the generalizations of a simple mean-field solution of the Hubbard model can have. We would like to stress, again, that many of these generalizations [9, 13–19] are not just academic exercises and can bring the interested student very close to real research in condensed-matter physics. At the same time, based on our experience, we found out that the calculations and the physical concepts presented in this paper are on average understood by graduate students. Last but not least, the self-consistent numerical procedure used in section [4] to diagonalize the Hamiltonian and find the phase diagram can represent a useful tool for students to make the link between formal implicit formulas and the way in which useful figures have to be derived.

As a final remark, we should also remind readers that the recent literature about the Hubbard model is extremely vast and, in the last 30 years, many developments have been made in strong synergy with the discovery of new strongly-correlated electron materials, mainly, transition-metal oxides [20]. An excellent review on this aspect is [21]. A book where the analytical properties of the Hubbard model are developed, especially in relation with the so-called two-pole approximation (or expansion around the atomic limit), is [22]: the approach is rather mathematical and many of the exact results of the model, together with the original references, are discussed in detail. Concerning the numerical approaches related to band-structure calculations, nowadays no PhD (and post-doc) student should neglect the study of the two methods (in order of difficulty) known as DFT+U (density functional theory + Hubbard U) [23–25] and DFT+DMFT (DFT + dynamical mean-field theory) [26, 27]. DFT+U is based on the mean-field approach developed in the present paper, with the only difference being that it is orbitally and site-dependent. Dynamical mean-field approaches, instead, allow more complex dynamics of the system, with shifts of the spectral weight and non-Fermi-liquid behaviour. We refer readers to the comprehensive review work [26] for a deeper introduction.

# An exact sum-rule for the Hubbard model.

Y. Claveau, B. Arnaud, and S. Di Matteo[*]
*Groupe théorie, Département Matériaux Nanosciences,*
*Institut de Physique de Rennes UMR UR1-CNRS 6251,*
*Université de Rennes 1, F-35042 Rennes Cedex, France*
(Dated: October 7, 2014)

We derive an exact integral equation for the Green function of the Hubbard model, valid in any number of dimensions, through a limit procedure in the equations of motion. The key point in the procedure to obtain this exact equation is to use the translational invariance of the Green functions with respect to the time difference t-t' and to derive the second equation of motion of the infinite Hubbard chain with respect to t', instead of t as usually done. Though our exact integral equation does not allow to solve the Hubbard model, it represents a strong constraint on its approximate solutions. We show some examples of its possible uses and discuss a real self-energy solution that can be used in the place of the Hubbard I solution.

PACS numbers: To insert

## I. INTRODUCTION

As widely known, the Hubbard model [1–3], in spite of its apparent formal simplicity, can be solved exactly only in one [4] and infinite [5] dimensions. The behaviour of its ground-state phase diagram in the most important cases of two and three dimensions is rather deduced by means of approximate solutions, usually determined by the competing actions of the hopping electron-energy ($t$) and of the Coulomb electron repulsion ($U$). A review of several approximate solving schemes in use today can be found in Ref. [6]. Historically, the model had been attacked around the atomic limit, by means of a Green function formalism based on the equation-of-motion approach [1–3, 7–11]. Within this approach a succession of coupled equations of motion for the Green functions is written down and then some decoupling procedure is introduced to close it. The usual decoupling procedure consists in expressing a Green function in terms of another (or more) of the set: of course the specific approximation that is performed on the Green function determines the subsequent dynamics of the system, as was the case for the original Hubbard I [1] and Hubbard III [3] approximations as well as for several others following during the next years [8–11]. However, in spite of the historical importance of the equation-of-motion decoupling procedure, such an approach was finally abandoned as it could not guarantee a proper control on the kind of approximation performed.

The aim of the present paper is to show that, instead, an exact integral equation for the Green function of the Hubbard model, Eq. (9) below, can be derived from the set of equations of motion. The key feature allowing us to derive the exact sum rule is a limiting procedure in the time domain, together with the time translation invariance of all Green functions $G(t - t')$, so that their

derivative with respect to $t$ is equal to the opposite of the one with respect to $t'$. Unfortunately, our sum rule does not allow to recover the full Green function. However, it can be used as a constraint on the available solutions, constraint that expresses, for example, the correct physical limit of the solution in the high-energy region. This in practice means that we can check, among all approximated solutions proposed in the past, like the mean-field, Hubbard I or Hubbard III solutions, which one respects the exact constraint imposed by Eq. (9) and which one should be discarded.

The plan of the paper is the following: section II is devoted to the mathematical derivation of our main result, Eq. (9). In section III we discuss the implications of our result on some of the usually adopted solutions of the Hubbard model and underline the reason why it is important to fulfill it. Finally, in section IV, we discuss a real-self-energy solution compatible with our sum rule, that can be considered as an improvement to the Hubbard I solution, that is much more sounding from a physical point of view. Its behaviour is finally compared with the LDA+U approach.

## II. DERIVATION OF THE SUM RULE

The Hubbard Hamiltonian on a crystal lattice, in its simplest version with one orbital per site [1], can be expressed in standard notation as:

$$\hat{H}_H = \sum_{ij\sigma} t_{ij} \hat{c}_{i\sigma}^\dagger \hat{c}_{j\sigma} + U \sum_i \hat{n}_{i\uparrow} \hat{n}_{i\downarrow} \qquad (1)$$

We define the causal Green function in the usual way: $i\hbar G_{ij\sigma}^{(c)}(t - t') \equiv \langle \hat{T} \left( \hat{c}_{i\sigma}(t) \hat{c}_{j\sigma}^\dagger(t') \right) \rangle$, by introducing the time-ordering operator $\hat{T}$. Retarded Green functions ($G^{(R)}$) are defined following, e.g., Ref. [12]. As it is well known [12], the time-frequency Fourier transform of these Green functions is not properly defined unless a

---
[*]Electronic address: sergio.dimatteo@univ-rennes1.fr

convergence factor $e^{\pm(t-t')\eta^+}$ is introduced, where $\eta$ is a small imaginary part of the frequency and $\eta \to 0$ at the end of the calculation. This allows to handle the Fourier transforms of all distributions involved (Dirac-$\delta$, Heaviside-$\theta$ and Green functions). All expectation values are calculated, for finite temperature, through the grand-canonical statistical weight, $e^{-\beta(H_H - \mu\hat{N})}$, where $\hat{N}$ is the number operator and $\mu$ the chemical potential, as in [12].

Both $G^{(c)}$ and $G^{(R)}$ satisfy the same equations of motions, as their difference only comes from the boundary conditions related to the imaginary part, and equations of motion determine the real part. If we write the equation of motion of the Green function in the time-domain, instead of the frequency ($\omega$)-domain as in Hubbard papers, we get:

$$i\hbar\partial_t G_{ij\sigma}(t-t') = i\hbar\delta_{ij}\delta(t-t') - \mu G_{ij\sigma}(t-t') \\ + \sum_l t_{il}G_{lj\sigma}(t-t') + U\Gamma_{ij\sigma}(t-t') \quad (2)$$

where $\Gamma_{ij\sigma}(t-t') \equiv \left\langle \left( \hat{T}\hat{n}_{i\bar{\sigma}}(t)\hat{c}_{i\sigma}(t)\hat{c}_{j\sigma}^\dagger(t') \right) \right\rangle$.

We use now the two key points that allow to derive our result are: (a) contrary to what done in the Hubbard papers, and usually in the literature, we write down the equation of motion of $\Gamma(t-t')$ with respect to $t'$ instead of $t$. The reason for choice (a) comes from the translational invariance of any Green function due to the dependence on $(t - t')$. This implies that for any Green function $\partial_t\Gamma_{ij\sigma}(t-t') = -\partial_{t'}\Gamma_{ij\sigma}(t-t')$; (b) we perform a limit procedure in the time-domain for the on-site Green functions $G_{ii\sigma}(t-t')$ and $\Gamma_{ii\sigma}(t-t')$.

By using remark (a), we write, after Eq. (2):

$$i\hbar\partial_t\Gamma_{ij\sigma}(t-t') = -i\hbar\partial_{t'}\Gamma_{ij\sigma}(t-t') = \delta_{ij}n_{i\bar{\sigma}}\delta(t-t') \\ -\mu\Gamma_{ij\sigma}(t-t') + \sum_l t_{il}\Gamma_{lj\sigma}(t-t') + UM_{ij\sigma}(t-t') (3)$$

The main difference with the usual equation-of-motion approach is that the hopping contribution does not lead to a new set of functions as in the usual chain [1, 3], but to the same function $\sum_l t_{il}\Gamma_{lj\sigma}(t - t')$, that can be easily diagonalised by a Fourier transform. This time the unknown is the new Green function in $M_{ij\sigma}^{(c)}(t-t') \equiv \left\langle \left( \hat{T}\hat{n}_{i\bar{\sigma}}(t)\hat{c}_{i\sigma}(t)\hat{n}_{j\bar{\sigma}}(t')\hat{c}_{j\sigma}^\dagger(t') \right) \right\rangle$ (for the causal case). Of course writing down the equation of motion for $M_{ij\sigma}(t-t')$ would continue the infinite set in a different way from the usual one, but, though some exact relations amongst higher-order Green functions can be deduced, no further insight is provided.

We can now apply the second part (b) of our strategy to equation (3): it is possible to find a point in space-time where for all kinds of Green functions (causal, retarded, etc.) $M_{ij\sigma}(t-t') = \Gamma_{ij\sigma}(t-t')$: this is the case for $i = j$ and $t-t' \to 0^\pm$. Mathematically, we have, for the causal Green function:

$$\lim_{t-t'\to 0^+} M_{ii\sigma}(t-t') = \langle \hat{n}_{i\bar{\sigma}}(t)\hat{c}_{i\sigma}(t)\hat{n}_{i\bar{\sigma}}(t)\hat{c}_{i\sigma}^\dagger(t)\rangle$$
$$= \langle \hat{n}_{i\bar{\sigma}}(t)\hat{n}_{i\bar{\sigma}}(t)\hat{c}_{i\sigma}(t)\hat{c}_{i\sigma}^\dagger(t)\rangle = \langle \hat{n}_{i\bar{\sigma}}(t)\hat{c}_{i\sigma}(t)\hat{c}_{i\sigma}^\dagger(t)\rangle$$
$$= \lim_{t-t'\to 0^+} \Gamma_{ii\sigma}(t-t') \quad (4)$$

In the third step of the equation, we have used the commutativity at equal times of $\hat{n}_{i\bar{\sigma}}(t)$ and $\hat{c}_{i\sigma}(t)$ and in the fourth step the projection property of the number operator $\hat{n}_{i\bar{\sigma}} = \hat{n}_{i\bar{\sigma}}^2$. A similar derivation can be easily set in the opposite limit $t - t' \to 0^-$ as well as for the retarded case.

Physically, this property is easily understandable: $M_{ij\sigma}(t-t')$ represents the probability that a particle, created in $j$ at time $t'$ on a doubly occupied site, is destroyed at the site $i$ that is also doubly occupied at time $t$. Instead, $\Gamma_{ij\sigma}(t-t')$ represents the probability that a particle, created in $j$ at time $t'$ (in this case $j$ can be either singly or doubly occupied) is destroyed at the site $i$ that is doubly occupied at time $t$. Of course, if $i = j$ and $t - t' \to 0^\pm$, we are dealing with the same site at the same time, which implies that if the site is doubly occupied for the destruction, it is also doubly occupied for the creation, thereby leading to the equality of the two Green functions, $\Gamma$ and $M$.

Our sum-rule, Eq. (9), can be derived from equation (3), by performing the $\lim_{t-t'\to 0^+}$ of both the left-hand-side and the right-hand-side. The safest way to handle the limit procedure on Green functions and delta function, that are distributions with a discontinuity at $t = t'$, is to move to the Fourier transform first, and then perform the $\lim_{t-t'\to 0^+}$ by keeping track of the convergence factor $e^{-(t-t')\eta}$, $\eta$ being, e.g., for $G^R$, a small positive quantity forcing the displacement of the poles out of the real $\omega$ (frequency) axis, thereby allowing the integrability of the Fourier transform [12]. With this in mind, we can continue from equation (3) with $i = j$:

$$\lim_{t-t'\to 0^+} \left( i\hbar\partial_t \int \frac{d^s\vec{k}}{(2\pi)^s} \int \frac{d\omega}{2\pi} \Gamma_{\vec{k}\sigma}(\omega)e^{-i[\omega(t-t')]} \right) =$$
$$\lim_{t-t'\to 0^+} n_{i\bar{\sigma}} \int \frac{d^s\vec{k}}{(2\pi)^s} \int \frac{d\omega}{2\pi} e^{-i[\omega(t-t')]}$$
$$+ \lim_{t-t'\to 0^+} \left( (U-\mu) \int \frac{d^s\vec{k}}{(2\pi)^s} \int \frac{d\omega}{2\pi} \Gamma_{\vec{k}\sigma}(\omega)e^{-i[\omega(t-t')]} \right)$$
$$+ \sum_l t_{il} \int \frac{d^s\vec{k}}{(2\pi)^s} \int \frac{d\omega}{2\pi} \Gamma_{\vec{k}\sigma}(\omega)e^{-i[\omega(t-t')-\vec{k}\cdot(\vec{R}_l-\vec{R}_i)]} \right) (5)$$

where $\vec{k}$-integrals are extended over the Brillouin zone, $\omega$-integrals from $-\infty$ to $+\infty$ and $s$ represents the dimensionality of the system (the present derivation is valid for any dimensions). If we first perform the time-derivative of the left-hand-side and then the limit of both sides, we obtain an integral expression for $\Gamma_{\vec{k}\sigma}(\omega)$:

3

$$\int \frac{d^s\vec{k}}{(2\pi)^s} \int \frac{d\omega}{2\pi} \left[(\hbar\omega + \mu - t_{\vec{k}} - U)\Gamma_{\vec{k}\sigma}(\omega) - n_{\bar{\sigma}}\right] = 0 \quad (6)$$

where we have supposed an homogeneous system, so that $n_{i\bar{\sigma}} = n_{\bar{\sigma}}$. We can now Fourier transform the equation (2) for the Green function:

$$(\hbar\omega + \mu - t_{\vec{k}})G_{\vec{k}\sigma}(\omega) = 1 + U\Gamma_{\vec{k}\sigma}(\omega) \qquad (7)$$

solve for $\Gamma_{\vec{k}\sigma}(\omega)$:

$$\Gamma_{\vec{k}\sigma}(\omega) = \frac{1}{U}(\hbar\omega + \mu - t_{\vec{k}})G_{\vec{k}\sigma}(\omega) - \frac{1}{U} \qquad (8)$$

and replace in (6):

$$\int \frac{d^s\vec{k}}{(2\pi)^s} \int \frac{d\omega}{2\pi} \left[(\hbar\omega + \mu - t_{\vec{k}} - U)(\hbar\omega + \mu - t_{\vec{k}})G_{\vec{k}\sigma}(\omega) - (\hbar\omega + \mu - t_{\vec{k}} - U(1 - n_{\bar{\sigma}}))\right] = 0 \qquad (9)$$

This is the main result of our work, an exact integral equation for $G_{\vec{k}\sigma}(\omega)$, to be used with the appropriate boundary conditions. Though it does not allow a full determination of the Green function, our integral equation acts as a constraint that must be fulfilled by the exact solution. From now on, to simplify notations, we shall measure the energy from the chemical potential, thereby replacing $t_{\vec{k}} - \mu$ with $t_{\vec{k}}$.

### III. COMPARISON WITH PREVIOUS RESULTS

As a first check, it is possible to verify that Eq. (9) is satisfied both in the atomic limit, i.e., when $t_{\vec{k}} \to t_0$ and in the band limit, i.e., when $U = 0$. In the first case, considering for example the retarded Green function, we get $G^{(R)}_{\text{at},\sigma}(\omega) = (1 - n_{\bar{\sigma}})/(\hbar\omega - t_0 + i\eta) + n_{\bar{\sigma}}/(\hbar\omega - t_0 - U + i\eta)$ and in the second case $G^{(R)}_{0\vec{k}\sigma}(\omega) = 1/(\hbar\omega - t_{\vec{k}} + i\eta)$. Both satisfy Eq. (9), as can be seen by direct inspection. Even if we consider the non-local correlations that should be present in a proper atomic limit for $G^{(R)}_{\text{at},\sigma}(\omega)$ (see section IV of Ref. [8]), Eq. (9) is still satisfied. The equation is however not satisfied by the mean-field Green function and the Hubbard I and Hubbard III Green functions. In fact, if we replace in Eq. (9) the mean-field retarded Green function $G^{(R)}_{\vec{k}\sigma}(\omega) = 1/(\hbar\omega - t_{\vec{k}} - Un_{\bar{\sigma}} + i\eta)$, we get the result $i\pi U^2 n_{\bar{\sigma}}(1 - n_{\bar{\sigma}})$ instead of zero. The calculation is performed by reminding that $(x + i\eta)^{-1} = Px^{-1} - i\delta(x)$,

where $P$ is the integral principal part and $\delta$ is the Dirac distribution: though the integral of the real, principal part is zero, there remains a contribution from the imaginary term.

A slightly more complex calculation, because of the $\vec{k}$-dependence, shows that neither the Hubbard I nor the Hubbard III Green functions satisfy Eq. (9) (or Eq. (10)). In the case of Hubbard I solution, the Green function is: $G^{(R)}_{\vec{k}\sigma}(\omega) = \frac{A^-_{\vec{k}}}{\hbar\omega - E^-_{\vec{k}} + i\eta} + \frac{A^+_{\vec{k}}}{\hbar\omega - E^+_{\vec{k}} + i\eta}$, where $E^{\pm}_{\vec{k}} = (U + t_{\vec{k}} \pm \sqrt{(U - t_{\vec{k}})^2 + 4Ut_{\vec{k}}n_{\bar{\sigma}}})/2$ is the energy spectrum and $A^{\pm}_{\vec{k}} = (E^{\pm}_{\vec{k}} - U(1 - n_{\bar{\sigma}}))/(E^{\pm}_{\vec{k}} - E^{\mp}_{\vec{k}})$ are the spectral weights. Replacement of this Green function in Eq. (9) gives again a non-zero imaginary part, that, in the simplifying case $n_{\bar{\sigma}} = 1$, is given by $i\pi(\frac{U^2}{2} + \int \frac{d^s\vec{k}}{(2\pi)^s} \frac{t^2_{\vec{k}}}{2}) \neq 0$.

Finally, in the case of the Hubbard III solution, the analysis is more complex because an explicit expression for the Green function is not available, as $G_{\vec{k}\sigma}(\omega)$ in this case is determined by means of a self-consistent calculation on five equations (Eqs. (57) to (61) in Ref. [3]). However, it is possible in this case to compare the limiting behaviour for $\omega \to \infty$ of the Hubbard III self-energy with a modified form of Eq. (9). In fact, if we write the Green function in terms of its self-energy $\Sigma_{\vec{k}\sigma}(\omega)$ as $G_{\vec{k}\sigma}(\omega) = 1/(\hbar\omega - t_{\vec{k}} - \Sigma_{\vec{k}\sigma}(\omega))$, then we can rewrite Eq. (9) as a constraint on the self-energy as:

$$\int \frac{d^s\vec{k}}{(2\pi)^s} \int \frac{d\omega}{2\pi} \frac{\left[(\Sigma_{\vec{k}\sigma}(\omega) - Un_{\bar{\sigma}})\hbar\omega - \Sigma_{\vec{k}\sigma}(\omega)(t_{\vec{k}} + U(1 - n_{\bar{\sigma}})) + t_{\vec{k}}Un_{\bar{\sigma}}\right]}{\hbar\omega - t_{\vec{k}} - \Sigma_{\vec{k}\sigma}(\omega)} = 0 \qquad (10)$$

It should be remembered that, as from the general theory of Green functions [12] $\lim_{\omega \to \infty} G_{\vec{k}\sigma}(\omega) = 1/(\hbar\omega)$,

then it follows that $\lim_{\omega\to\infty}\Sigma_{\vec{k}\sigma}(\omega)$ is a constant. Such a constant can be determined from Eq. (10), as in order to have a finite integral, it implies that in the limit $\omega\to\infty$ the coefficient of $\hbar\omega$ at the numerator must be zero. This gives the following constraint on the self-energy: $\lim_{\omega\to\infty}\Sigma_{\vec{k}\sigma}(\omega)=a_{\vec{k}}Un_{\bar\sigma}+b_{\vec{k}}$, where $\int\frac{d^s\vec{k}}{(2\pi)^s}a_{\vec{k}}=1$ and $\int\frac{d^s\vec{k}}{(2\pi)^s}b_{\vec{k}}=0$. However, this constraint is not fulfilled in the Hubbard III solution: in the notation of Ref. [3] the self-energy is $\Sigma_{\vec{k}\sigma}(E)=E-F^\sigma(E)$, with $E=\hbar\omega$ the energy and $F^\sigma(E)$ is given by Eq. (59) of Ref. [3]. The previous constraint is fulfilled if $\lim_{E\to\infty}F^\sigma(E)=E-Un_{\bar\sigma}$ is fulfilled, and a direct calculation from Eq. (59) of Ref. [3] (we remind that Hubbard $U$ was called $I$ by Hubbard) shows that this is not the case.

Interestingly, if we suppose the self-energy $\vec{k}$-independent, as in the case of DMFT [13], then $\lim_{\omega\to\infty}\Sigma_\sigma(\omega)=Un_{\bar\sigma}$, ie, dynamical mean-field exactly reduces to static mean-field in the infinite-frequency limit, as it should [14]. It is important to remind that having the correct limit for the high-energy region allows reproducing the right behaviour for the formation of the upper and lower Hubbard bands, as shown in [14], so that the former constraint should always be verified in DMFT calculations.

We finish this section by reminding that an equation equivalent to imaginary part of Eq. (9) had been derived in the literature [15], with a method based on the calculation of the spectral function. However, our full Eq. (9) cannot be obtained by this method, because the knowledge of a sum-rule involving the imaginary part does not allow to derive the real part by means of Kramers-Kronig transformations, as in this case the imaginary part over the whole frequency range is needed and not just its first and second momenta.

## IV. A REAL SELF-ENERGY SOLUTION.

Though Eq. (9) does not provide us with sufficient information to find the Green function, among all its possible solutions there is one with a two-pole structure with real self-energy (infinite time-life of the two quasiparticles), therefore compatible with our exact sum-rule. For the retarded Green function it is:

$$G^R_{\vec{k}\sigma}(\omega)=\frac{1-n_{\bar\sigma}+f_{\vec{k}}(t/U,n_{\bar\sigma})}{\hbar\omega-t_{\vec{k}}+i\eta}+\frac{n_{\bar\sigma}-f_{\vec{k}}(t/U,n_{\bar\sigma})}{\hbar\omega-t_{\vec{k}}-U+i\eta}\quad(11)$$

where $f_{\vec{k}}(t/U,n_{\bar\sigma})$ is not determined by Eq. (9) except for having a null integral in the Brillouin zone:

$$\int\frac{d^s\vec{k}}{(2\pi)^s}f_{\vec{k}}(t/U,n_{\bar\sigma})=0\qquad(12)$$

The self-energy associated to this solution is: $\Sigma_{\vec{k}}(\omega)=U\frac{(\hbar\omega-t_{\vec{k}})}{)}(n_{\bar\sigma}-f_{\vec{k}})\hbar\omega-t_{\vec{k}}-U(1-n_{\bar\sigma}+f_{\vec{k}})$. It is interesting

to notice that, provided Eq. (12) is fulfilled, the function $f_{\vec{k}}$ could be even $\omega$-dependent: in this case we would deal with quasiparticle inelastic scattering.

If we limit ourselves to the real self-energy case, Eq. (11), we propose it to be used as a first-order solution in the place of the Hubbard I approximation, not suffering of its drawbacks. In the present section, we analyze the advantages of this solution and compare it, and its orbitally-degenerate version, with the well-known LDA+U approach, highlighting the analogies and the differences. Physically, for $f_{\vec{k}}=0$, such a solution represents non-mixing Hubbard quasiparticles, i.e., electrons moving in a singly occupied band and electrons moving in a doubly occupied band, respectively, without intercrossing. In this case, Eq. (11) is the solution that we would have had from Eq. (3) by putting $M_{ij}=\Gamma_{ij}$ identically (it corresponds to the hypothesis that the creation of a doubly-occupied site at $j$, at time $t'$, destroyed at time $t$ at a site $i$ that is still doubly occupied, is the same as the creation of a doubly-occupied site at $j$, at time $t'$, destroyed at time $t$ at the site $i$, independently of its occupancy). It is therefore at an analogous level of approximation of the Hubbard I solution, characterized by infinite lifetime for the two sub-bands quasiparticles, but with the correct high-energy limit, by respecting Eq. (9).

In order to analyze the behaviour of this solution, we can evaluate the density of states from the imaginary part of the retarded Green function. Taking the specific case of a square-lattice band, $t_{\vec{k}}=2t(\cos(k_xa)+\cos(k_ya))$, $a$ the lattice unit, the behaviour of the spectrum, represented in Fig. 1, is calculated through the following formula:

$$\rho_\sigma(\varepsilon)=\frac{1}{N}\sum_{\vec{k}}\left\{(1-n_{\bar\sigma})\,\delta\left[\varepsilon-\varepsilon_{\vec{k}}^{\mathrm{LHB}}\right]\right.$$
$$\left.+n_{\bar\sigma}\delta\left[\varepsilon-\varepsilon_{\vec{k}}^{\mathrm{UHB}}\right]\right\}\qquad(13)$$

where $\varepsilon_{\vec{k}}^{\mathrm{LHB}}=t_{\vec{k}}$ and $\varepsilon_{\vec{k}}^{\mathrm{UHB}}=t_{\vec{k}}+U$ are the lower Hubbard band and the upper Hubbard band, respectively.

When $t/U$ is below a critical value $(t/U)_c$, for appropriate values of $n_{\bar\sigma}$, two bands are formed from the two-pole atomic solution, as shown in Fig. 1: at half-filling the Fermi energy is within the gap and the system is a Mott insulator [16]. If instead $t/U$ is above the critical value, the two bands merge, though at different $\vec{k}$ values and the system behaves like a metal. For a square lattice $(t/U)_c=0.125$, as the bandwidth is $W=8t$. It is however important to remark that even in the metallic state of figure (1b), such a metal still keeps one of the main features of the atomic behaviour as is clear from the analysis of the $\vec{k}$-dependence at a given energy: at the energy $\varepsilon_{\vec{k}}^{\mathrm{LHB}}$, each $\vec{k}$ point has a spectral weight $(1-n_{\bar\sigma})$, less than one, as the remaining $n_{\bar\sigma}$ weight is associated to the other branch of the spectrum. This is a marked difference with the mean-field approximation, with weight 1 for each $\vec{k}$, and it corresponds to

5

the non-Fermi-liquid behaviour of our solution that does not fulfill the Luttinger's theorem [17], which necessarily characterizes Fermi liquids when the interaction is adiabatically switched on from the Fermi gas. We should also notice that the present solution is different of the Hubbard-I solution: though they are both characterized by a two-pole solution leading to a MIT of Mott-Hubbard kind, the Hubbard I MIT is found even for an infinitesimal value of $U$, what is quite unphysical. Our solution more realistically gives $U/W \sim 1$.

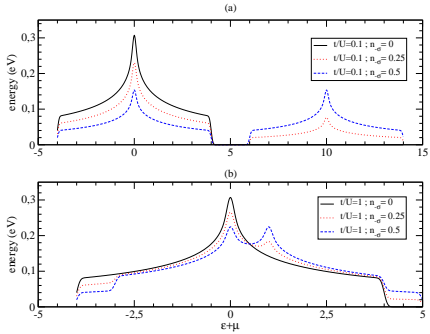

FIG. 1: Density of states corresponding to Eq. (13), showing the Mott-Hubbard metal-insulator transition.

A different behaviour appears when we consider the $f_{\vec{k}}$-term, that allows inter-band crossing by shifting some spectral weight, at any given $\vec{k}$, depending on the filling and the ration $t/U$. The specific form of the function $f_{\vec{k}}$ cannot be fixed unambiguously. However, two conditions should be respected: mathematically, the condition represented by Eq. (12) on the integral in the Brillouin zone in order to fulfill Eq. (9). Physically, the condition first evidenced in Ref. [7] on the dependence of interband hopping on the filling: if we consider the probability that a spin-$\sigma$ electron put randomly in the lattice falls on a $\bar{\sigma}$-occupied site (so as to give double occupancy), this probability is, in a frozen configuration (no interband crossing), clearly proportional to $n_{\bar{\sigma}}$. However, by re-

laxing the unphysical condition of no interband-crossing, the hopping from a doubly occupied configuration to a singly occupied configuration is allowed, and, even, advantaged in the case of less than half-filled band. This implies a reduction of the spectral width of the upper Hubbard band from the 'frozen' $n_{\bar{\sigma}}$ value that should be taken into account by the function $f_{\vec{k}}$.

We would like to end this section by highlighting that our real self-energy solution, Eq. (11) shows an interesting complementarity with the well-known LDA+U method [18, 19]. The latter is a multiorbital unrestricted mean-field approach that advantages orbital (and charge) separation, compared to a bare LDA calculation, because of the $U$-term: once the average Coulomb energy is subtracted by a LDA calculation, an energy shift $U$ is attributed, compared to the LDA band $t_{\vec{k}}$, to doubly occupied orbitals so as to reduce the fractional occupancy of these orbitals in favour of an integer one. This can force the system towards an insulating state but, in order to do that, a multiband material is needed. The latter is the main difference with our approach: with Eq. (11) a metal-insulator transition is possible with just one band, as shown in Fig. 1: this is a consequence of the loss of validity of Luttinger's theorem, leading to two different quasiparticles for a given $\vec{k}$. In order to perform a more complete comparison of the two approaches, our solution should be extended to the multiorbital case. However, this will be the scope of a future work.

## V. CONCLUSIONS

Exact results for the Hubbard model are rare [4, 5, 15]. The exact sum rule that we have found in Eq. (9), valid for any number of dimensions, has been given a simple physical interpretation and some of its possible uses as a constraint to be put on the general solution have been provided. Past solutions not respecting this constraint have been highlighted. The behaviour of several approximate solutions respecting Eq. (9) will be the subject of a future analysis, as well as its generalization to the orbitally-degenerate case, that will be compared to the well-known LDA+U results.

## C.6   Resume

### Yann Claveau

Born the 26/03/1987
2 children

**Contact**

mail : yann.claveau@gmail.com

## Curriculum vitae

### Formation at Université de Rennes 1, France

- 2011-2014  **PhD**
  Modeling Ballistic Electron Emission Microscopy on metal thin films
  Advisor: Sergio Di Matteo

- 2005-2011  **Master 2 internship (2011), Physics**
  Experimental study and modeling hot-electrons transport in Fe/Au/Fe/GaAs spin-valve

  **Master 1 internship (2009), Physics**
  Mean-field solution of the Hubbard model on a square lattice

  **B.Sc. internship (2008), Physics**
  DFT-LDA study of a Al crystal (ABINIT & nearly-free electron model)

### Professional experiences

- 2014-2015  **Postdoctoral position,** CEMES, MC2 team.
  ANR project NAIADE: Modeling III-V semi-conductor/semi-conductor interfaces through DFT
  Leaders: Anne Ponchet and Hao Tang

- 2012-2014  **Teaching**

  Exercises: electromagnetism in matter (B.Sc.) ; crystallography (L2)

  Lecture/exercices: mathematics (L1 biology)

  Practicals: libreOffice (L1) ; wave, particles and relativity (L2)

  Lecture in high schools: Aurorae ;
    The photography at the light of physics

- 2010  **Volontaire Civil à l'Aide Technique (2010)**
  (military service for young civilian scientists)
  Winterizing a year in the Kerguelen Islands
  (Indian Ocean, 0 inhabitant)
  Responsible of Seismological Observatory and
  responsible of Geomagnetic Observatory

- 2000-2007  **Seasonal jobs**

## Other experiences and formations

- 2014    **Initiation to Python** (Rennes)

  **International summer school on Computational Methods for Quantum Materials** (Jouvence, Québec)

  Density Functional Theory,
  Dynamical Mean-Field Theory,
  Continuous-Time Quantum Monte Carlo,
  Density Matrix Renormalization Group,
  Quantum Cluster Approaches

- 2013    **Interpersonal communication** (Rennes)

  **Brittany Synchrotron Radiation School** (Rennes)

## Publications and communications

- **publications**    **M. Hervé et al., Appl. Phys. Lett. 103, 202408 (2013)**
  « k-space spin filtering effect in the epitaxial Fe/Au/Fe/GaAs(001) spin-valve »

  **Y. Claveau et al., Eur. J. Phys. 35 035023 (2014)**
  « Mean-field solution of the Hubbard model: the magnetic phase diagram. »

- **Posters**    ICAMM 2014

  Internationnal Summer school on Computational Methods for Quantum Materials (2014)

  GDR coDFT (2013)

  Journées Surfaces et Interfaces Orléans (2013)

  Journées Surfaces et Interfaces Saclay (2012)

- **Conferences**    The photography at the light of physics (Fête de la science 2013)

- **Animations**    stand animator: « Trip in the nano-world »
  (fête de la science 2013)

## Collaborations

| **Fernando Flores** | **Pedro De Andres** |
|---|---|
| (Departamento de Física Teórica de la Materia Condensada,Universidad Autonoma de Madrid) | (Instituto de Ciencia de Materiales de Madrid, Consejo Superior de Investigaciones Científicas) |



 \newpage \thispagestyle{empty}
\strut
  
\FloatBarrier

\bibliographystyle{plainnat}
\bibliography{these_A5}

\newpage \thispagestyle{empty}
\strut
\newpage\thispagestyle{empty}
\strut
\newpage\thispagestyle{empty}

\thispagestyle{empty}
\calccentering{\unitlength}                         
\changetext{14cm}{}{}{}{}
\changepage{}{-1.9cm}{}{}{}{}{}{}{}
 \begin{adjustwidth*}{\unitlength}{-\unitlength}     
     \begin{adjustwidth}{-0.5cm}{-0.5cm}                 

\begin{center}
\Huge\textbf{Summary}
 
 \phantom{}
\end{center}

After the discovery of Giant Magneto-Resistance (GMR) by Albert Fert and Peter Grünberg, electronics had a breakthrough with the birth of a new branch called spintronics. 
This discipline, while still young, exploits the spin of electrons, for instance to store digital information. Most quantum devices exploiting this property of electrons consists of alternating magnetic and nonmagnetic thin layers on a semiconductor substrate.

One of the best tools used for characterizing these structures, invented in 1988 by Kaiser and Bell, is the so-called Ballistic Electron Emission Microscope (BEEM). Originally, this microscope, derived from the scanning tunneling microscope, was dedicated to the imaging of buried (nanometer-scale) objects and to the study of the potential barrier (Schottky barrier) formed at the interface of a metal and a semiconductor when placed in contact.
   With the development of spintronics, the BEEM became an essential spectroscopy technique but still fundamentally misunderstood.
   It was in 1996 that the first realistic model, based on the non-equilibrium Keldysh formalism, was proposed to describe the transport of electrons during BEEM experiments. In particular, this model allowed to explain some experimental results previously misunderstood. However, despite its success, its use was limited to the study of semi-infinite structures through a calculation method called decimation of Green functions. 
   
   In this context, we have extended this model to the case of thin films and hetero-structures like spin valves: starting from the same postulate that electrons follow the band structure of materials in which they propagate, we have established an iterative formula allowing calculation of the Green functions of the finite system by tight-binding method.
   This calculation of Green's functions has been encoded in a FORTRAN 90 program, BEEM v3, in order to calculate the BEEM current and the surface density of states. 
   
    In parallel, we have developed a simpler method which allows to avoid passing through the non-equilibrium Keldysh formalism. 
    Despite its simplicity, we have shown that this intuitive approach gives some physical interpretation qualitatively similar to the non-equilibrium approach. However, for a more detailed study, the use of ``non-equilibrium approach'' is inevitable, especially for the detection of thickness effects linked to layer interfaces.
  
   Both tools should be useful to experimentalists, especially for the Surfaces and Interfaces team of our department.
     \end{adjustwidth}
 \end{adjustwidth*}

\end{document}